\documentclass[12pt,a4paper,twoside,openright,dvips]{report}
\usepackage{a4}
\usepackage{axodraw}
\usepackage{amsmath}
\usepackage{amsfonts}
\usepackage{amssymb}
\usepackage{color}
\usepackage{graphicx}
\usepackage{subfigure}
\usepackage{longtable}
\usepackage[ps2pdf,bookmarks,pdfpagelabels]{hyperref}
\usepackage{mcite}
\usepackage{bbm}
\usepackage{slashed}
\usepackage[latin1]{inputenc}

\definecolor{darkred}{rgb}{0.2,0,0}
\definecolor{darkgreen}{rgb}{0,0.3,0}
\hypersetup{colorlinks,linkcolor=black,citecolor=black,urlcolor=black,
    pdftitle=Quantum Effects in Higgs-Boson Production Processes at Hadron Colliders,
    pdfauthor=Michael Rauch}

\let\originalcleardoublepage=\cleardoublepage
\renewcommand{\cleardoublepage}
             {\newpage{\pagestyle{empty}\originalcleardoublepage}}

\newcommand{\di}{\mathrm{d}}
\newcommand{\fslash}[1]{\slashed{#1}}  % Feynman slash
\newcommand{\La}{\mathcal{L}}
\newcommand{\M}{\mathcal{M}}
\newcommand{\Mfi}{\mathcal{M}_{fi}}
\newcommand{\MSbar}{\ensuremath{\overline{\textrm{MS}}}}
\newcommand{\DRbar}{\ensuremath{\overline{\textrm{DR}}}}
\newcommand{\Order}[1]{\mathcal{O}\left(#1\right)}
\newcommand{\Cmplx}{\mathbb{C}}
\newcommand{\R}{\mathrm{Re}}
\newcommand{\Id}{\mathbbm{1}}
\newcommand{\GeV}{\textrm{ GeV}}
\newcommand{\TeV}{\textrm{ TeV}}

\newcommand{\ket}[1]{\left|#1\right>}
\newcommand{\abs}[1]{\left|#1\right|}
\newcommand{\spa}{\ensuremath\text{SPS1a}^\prime}
\newcommand{\mssmstrut}{\rule[-1.75ex]{0pt}{5ex}}
\newcommand{\mssmbigstrut}{\rule[-3.5ex]{0pt}{8ex}}
\newcommand{\pvechat}{{\vec{\hat p}\,}}

\DeclareMathOperator{\Tr}{Tr}
\DeclareMathOperator{\diag}{diag}
\DeclareMathOperator{\sign}{sign}
\DeclareMathOperator{\artanh}{artanh}

\newcommand{\eq}[1]{eq.~(\ref{#1})}
\newcommand{\fig}[1]{Fig.~\ref{#1}}
\newcommand{\chap}[1]{chapter~\ref{#1}}
\newcommand{\app}[1]{appendix~\ref{#1}}
\newcommand{\program}[1]{\mbox{#1}}
\newcommand{\HadCalc}{\program{HadCalc}}
\newcommand{\FeynArts}{\program{FeynArts}}
\newcommand{\FormCalc}{\program{FormCalc}}
\newcommand{\LoopTools}{\program{LoopTools}}
\newcommand{\LHAPDF}{\program{LHAPDF}}

\newcommand{\var}[1]{\textit{#1}}
\newcommand{\param}[1]{\texttt{#1}}
\newcommand{\command}[1]{\texttt{#1}}

\begin{document}
\pagestyle{empty}
\begin{titlepage}
\begin{center}
{
\Large
\noindent Technische Universit\"at M\"unchen

\vspace{0.4cm}
\noindent Max-Planck-Institut f\"ur Physik\\
(Werner-Heisenberg-Institut)

\vspace{1.0cm}
\Huge
{\bfseries Quantum Effects in \\
                \mbox{Higgs-Boson
                Production Processes}\\
                at Hadron Colliders
}

\vspace{1.5cm} \Large
\noindent Michael~Rauch

\vspace{1.5cm}
\normalsize
\noindent Vollst\"andiger Abdruck der von der Fakult\"at f\"ur Physik\\
\noindent der Technischen Universit\"at M\"unchen\\
\noindent zur Erlangung des akademischen Grades eines\\
\noindent {\bfseries Doktors der Naturwissenschaften (Dr.\ rer.\ nat.)}\\
\noindent genehmigten Dissertation.

\vspace{1.5cm}
\large
  \begin{tabular}{lll}
    Vorsitzender :              &    & Univ.-Prof.\ Dr.\ L.\ Oberauer \\[16pt]
    Pr\"ufer der Dissertation : & 1. & Hon.-Prof.\ Dr.\ W.\ F.\ L.\ Hollik \\
                                & 2. & Univ.-Prof.\ Dr.\ A.\ J.\ Buras \\
  \end{tabular} \newline
\vspace{1.5cm}

\noindent Die Dissertation wurde am 31.\,01.\,2006 \\
\noindent bei der Technischen Universit\"at M\"unchen eingereicht \\
\noindent und durch die Fakult\"at f\"ur Physik am 14.\,03.\,2006 angenommen.

}
\end{center}
\end{titlepage}
\cleardoublepage

% settings for TOC and TOC itself
\setcounter{page}{0}
\pagenumbering{roman}
\makeatletter
\immediate\write\@outlinefile{\string\BOOKMARK  [0][-]{section*.1}{Contents}{}}
\makeatother
\phantomsection
\tableofcontents

% now the settings for the main part
\cleardoublepage
\pagestyle{headings}
\let\MakeUppercase\relax
\setcounter{page}{0}
\pagenumbering{arabic}

\chapter{Introduction}
\label{intro}

The quest for the fundamental building blocks and laws of the world surrounding us 
has been a driving force to mankind since its early days. The idea that nature 
consists of small, invisible constituents was first expressed by the ancient Greek
Democritus in the fifth century BC. It was not until the nineteenth century AD that this 
idea was picked up again and embedded in a scientific context. Over time experiments
discovered ever smaller substructures, from atoms to electrons and hadrons, 
and thereon to quarks. From a theoretical point of view, the aim is to embed these
experimental results in a model which is based on as few assumptions as possible
and can explain all other physical effects. 

%%% Standard Model
The currently established model which performs
this task is the Standard Model of elementary particle
physics~\cite{Glashow:1961tr,*Weinberg:1967tq,*Salam:1968rm,Glashow:1970gm}. 
It is one of the best-tested theories of contemporary physics. All known elementary 
particles are accommodated in this model. Solely the scalar Higgs
boson~\cite{Higgs:1964ia,*Higgs:1964pj,*Higgs:1966ev,*Englert:1964et,*Kibble:1967sv} 
is included in the theory, but could not be found in experiments so
far~\cite{Heister:2001kr}. It is this particle which is assumed to be responsible for 
the masses of the fermions and weak gauge bosons.

%%% supersymmetry
In spite of its success, the Standard Model also has its insufficiencies, and 
new theories are searched for, which might provide an even better description
of nature. One of the most popular ones is supersymmetry~\cite{Wess:1974jb}. 
It extends the two,
fundamental symmetries of the Standard Model, the Poincar\'e group 
and the non-Abelian gauge group 
$SU(3)_C \otimes SU(2)_L \otimes U(1)_Y$ of strong, weak and 
electromagnetic interactions, by an anticommuting operator which induces
an equal number of bosonic and fermionic states.

%%% LHC & Perturbation theory
The search for supersymmetry and the Higgs boson are main tasks of the 
Large Hadron Collider (LHC) at CERN. It will start operation in mid-2007 and
provide a wealth of data. To verify or falsify theories and to relate this data 
to parameters of a model, it is necessary to calculate precise theoretical
predictions, which match the accuracy which LHC will be able to obtain.
As both the Standard Model and its supersymmetric extension are defined
as perturbative theories with a series expansion in Planck's constant $\hbar$,
the inclusion of effects beyond leading order is often necessary.

%%% scope of the thesis
In this thesis production processes for Higgs bosons in the Standard Model
and its supersymmetric extension, the Minimal Supersymmetric Standard Model
(MSSM)~\cite{Haber:1984rc}, at hadron colliders are considered. The 
calculations are performed at the one-loop level and include the 
SUSY-QCD corrections, i.e.\ corrections with squarks and gluinos running in the 
loop, for the MSSM Higgs bosons.

%%% outline
The outline of this thesis is as following. First, a short introduction to the 
Standard Model (SM) is given in \chap{sm}. Special emphasis is put on the 
Higgs sector of the SM. Here also a possible extension including higher-order
operators is discussed. Despite being a well-tested theory, the Standard Model
also has its shortcomings, which are mentioned in the last section of this chapter.

Out of the possible extensions of the Standard Model which aim to solve these 
deficencies, supersymmetry is the most popular one, as it is appealing from both 
an experimental and a theoretical point of view. Its discussion in \chap{susy}
of this dissertation starts with the basic principles of the theory. After 
the necessary ingredients to build a phenomenologically viable model are 
investigated, the focus is put on the 
simplest supersymmetric extension of the Standard Model, the Minimal
Supersymmetric Standard Model (MSSM)~\cite{Haber:1984rc}. 
The Lagrangian of the MSSM after supersymmetry breaking is written down and 
the particle content of the model is explained.

Chapter~\ref{renorm} is concerned with the methods of regularization 
and renormalization. The first one is necessary to cancel the divergences which
appear in the calculation of one-loop cross sections, and renders the 
amplitudes finite. Renormalization then restores the physical meaning of the 
calculated cross sections. After a general introduction to the concepts,
the renormalization of the strong coupling constant $\alpha_s$ in the way it is used 
in this thesis, is presented. The chapter concludes with a discussion of 
the bottom-quark Yukawa coupling. Here the mass counter term introduces
large one-loop corrections to the cross
section~\cite{Carena:1999py,Guasch:2001wv}. They are universal, so they 
can be included in an effective tree-level coupling. Additionally, they are a 
one-loop exact quantity, so a resummation to all orders in perturbation theory
is possible.

The next chapter deals with the calculation of hadronic cross 
sections. The underlying theory, QCD and the parton model, is briefly 
introduced. Then explicit formulae for the calculation of integrated and
differential hadronic cross sections are given. An important technique to
improve the cross-section ratio of signal over background processes
and to enable the reconstruction of particular event-types in the detector
is the application of cuts to final-state particles. 
The implementation of these formulae in computer code is done in a program,
called \HadCalc{}, which is developed by the author
of this thesis and which is lastly presented. It is based on the tools
\FeynArts{}~\cite{Kublbeck:1990xc,*Hahn:2000kx,*Hahn:2001rv,FeynArtsmanual},
\FormCalc{}~\cite{Hahn:1998yk,Hahn:2005vh,FormCalcmanual} and
\LoopTools{}~\cite{Hahn:1998yk,Hahn:2006MR,LoopToolsmanual}. The latter is
extended to include now the five-point loop integrals, 
such that a complete one-loop 
calculation of $2\rightarrow3$ processes is possible.
\HadCalc{} completes the tool set to provide a largely automated way of calculating 
hadronic cross sections. 
%The scope of the program is the convolution
%with the parton distribution functions to obtain hadronic cross sections out of
%partonic ones.

In the subsequent chapters, this program is applied to the calculation of 
processes which
contain supersymmetric Higgs bosons in the final state. The full one-loop SUSY-QCD
corrections, i.e.\ corrections with squarks and gluinos running in the loop, are 
included in the numerical results.

The associated production of a charged Higgs boson $H^\pm$ and a $W$ boson
via bottom quark--anti-quark annihilation is studied in \chap{bbWH}. The 
discovery of a charged Higgs boson would be a clear sign of physics beyond 
the Standard Model. The above-mentioned universal corrections to the 
bottom-quark Yukawa coupling are expected to yield a numerically large and dominant 
contribution for certain regions of the MSSM parameter space, but the size of the 
SUSY-QCD corrections in the other regions is not known and requires a full 
one-loop calculation, which is presented in this thesis.

In \chap{vbf} the production of the lighter CP-even neutral Higgs boson $h^0$ 
via vector-boson fusion is investigated. This process has a clear final state of two 
jets in the forward region of the detector and forms an important 
$h^0$-production mode with small theoretical uncertainties.
For the corresponding Standard Model process with a Standard Model 
Higgs boson $H$ in the final state, the Standard-QCD corrections
are already known. They are the same as for $h^0$-production in the MSSM 
up to the replacement of the Higgs coupling. In the MSSM case additional
SUSY-QCD corrections appear. In this thesis the complete one-loop 
SUSY-QCD corrections are calculated and their effect on 
the total cross section is discussed.
In the last section of this chapter a background to the vector-boson-fusion process, 
$h^0$-production with two outgoing jets and one or two gluons in the initial state, 
is considered and its numerical impact studied. 

The SUSY-QCD corrections to $h^0$-production in association with 
heavy, i.e.\ bottom or top, quarks are presented in \chap{hq}. Besides
being additional discovery channels for the Higgs boson, these processes
can also be used to extract the respective quark Yukawa couplings from
the data. This task can only be performed if the theoretical uncertainty of the 
cross section is small. The Standard-QCD corrections to these processes
are available in the literature and greatly reduce the dependence on the 
renormalization and factorization scale. Additionally, 
there are SUSY-QCD corrections which can also yield large corrections and must be
taken into account. Therefore, a full calculation of the one-loop 
SUSY-QCD corrections is necessary, which is presented in this dissertation.

Lastly, the possibility to measure the quartic Higgs coupling at hadron colliders 
is analyzed in \chap{quartic}. For this purpose triple-Higgs production via gluon fusion 
is studied at the leading one-loop order. In this chapter not the MSSM is used as the 
underlying model, but an effective theory based on the Standard Model where the 
trilinear and quartic Higgs self-couplings are left as free parameters.

% The results of this dissertation are summarized in \chap{concl}.

In \app{param} the numerical values of the Standard Model parameters, which were
kept fixed for all calculations in this thesis, and of the MSSM parameters for the
reference point $\spa$~\cite{spa}
are noted. Appendix~\ref{appsusy} contains the definitions 
of mathematical quantities which are used throughout the dissertation, and 
\app{ps} the parametrization of the phase space for two- and three-particle
final states.

In loop calculations integrals over the loop momentum appear which can be solved
analytically. The definition of these integrals is given in \app{loopint}. Special attention
is paid to the five-point integrals which have not been implemented in the package
\LoopTools{}~\cite{Hahn:1998yk,Hahn:2006MR,LoopToolsmanual} before. 
The numerical method of Gaussian elimination, which is used
to further improve the stability of the loop-integral calculation, is presented in
\app{numerics}. 

Finally, the complete user manual of \HadCalc{} is attached in
\app{hadcalc}. The program itself can be obtained from the 
author\footnote{email: \texttt{mrauch@mppmu.mpg.de}}.

\chapter{Standard Model}
\label{sm}

\section{Structure of the Standard Model}

The Standard Model (SM) of elementary particle physics~%
\cite{Glashow:1961tr,*Weinberg:1967tq,*Salam:1968rm,Glashow:1970gm}
is one of the best
tested theories in physics. It consists of an outer symmetry of the Poincar\'e
group of space-time transformations and a non-Abelian gauge group of the
inner direct product $SU(3)_C \otimes SU(2)_L \otimes U(1)_Y$. 
$SU(3)_C$ is the color gauge group and describes the strong interactions 
by the theory of QCD.
The product $SU(2)_L \otimes U(1)_Y$ specifies the electroweak interactions which 
unify the electromagnetic and weak interactions. The Higgs mechanism, which will
be described in \chap{sm:higgs}, breaks this symmetry spontaneously, thereby
leaving a $U(1)_Q$ symmetry of electromagnetic interactions which is described 
by QED.
The one remaining interaction, gravitational interaction, is beyond the scope of the SM. 
In fact, a consistent theory which formulates general relativity in terms
of a quantum field theory is not known until today. At the center-of-mass energies 
used at present or at planned future colliders, which are maximally of the order of
a few hundred TeV, the effects due to gravitational interactions are negligibly small.
The Standard Model therefore provides an excellent approximation to describe collider
physics.

The fermionic sector of the SM consists of spin-$\frac12$ 
leptons ($\nu_e$,$\nu_\mu$,$\nu_\tau$,$e$,$\mu$,$\tau$) 
and quarks ($u$,$c$,$t$,$d$,$s$,$b$) which appear in 
three different generations. The particles of each generation have the
same quantum numbers but a different coupling to the Higgs field which will 
be introduced below. Left-handed fermions transform as a 
doublet under $SU(2)_L$ where the upper component forms the 
neutrinos ($\nu_e$,$\nu_\mu$,$\nu_\tau$) and up-type quarks ($u$,$c$,$t$), 
respectively, and the lower component the electron-type leptons ($e$,$\mu$,$\tau$)
and the down-type quarks ($d$,$s$,$b$). Right-handed fermions transform
as a singlet under $SU(2)_L$ the only exception being that there are no 
right-handed neutrinos at all.
For each group generator a spin-1 gauge boson exists which transforms under 
the adjoint representation of the respective group. Consequently there are 
eight gauge bosons for $SU(3)_C$, the gluons, three gauge bosons for $SU(2)_L$,
the $W$ bosons, and one for $U(1)_Y$, called $B$.

Experiments show that not all gauge bosons are massless~\cite{Eidelman:2004wy}. 
Adding an explicit mass term for these gauge bosons is not possible for 
renormalizable quantum field theories. Such terms are forbidden due to the postulate 
that the Lagrangian should be invariant under gauge transformations. 
Otherwise the resulting theory would be non-renormalizable. For this reason another 
way of giving masses to the gauge bosons is needed. This is achieved by the
Higgs mechanism which will be described in the next chapter.

\section{Higgs mechanism}
\label{sm:higgs} 

\subsection{Standard Model Higgs sector}

As mentioned above, it is a difficult task to construct a gauge theory which is 
renormalizable and has massive gauge bosons. In the Standard Model this problem
is solved by the Higgs mechanism~%
\cite{Higgs:1964ia,*Higgs:1964pj,*Higgs:1966ev,*Englert:1964et,*Kibble:1967sv}.
The idea is to add additional terms to the Lagrangian, such that the Lagrangian
is invariant under the $SU(2)_L \otimes U(1)_Y$ gauge transformations with a
ground state which does not share this invariance. 
To realize this idea one introduces a new complex scalar field, 
the Higgs field $\Phi$, which 
behaves like a doublet under $SU(2)_L$ gauge transformations 
and has hypercharge $Y=+1$. Its 
ground state acquires a vacuum expectation value $v$, such that a 
$U(1)_Q$ symmetry of electromagnetic interactions is preserved. 
The electromagnetic charge is defined as $Q = I_3 + \frac{Y}2$, where $I_3$ 
is the quantum number of the third component of the weak isospin operator. 
Therefore only the lower component of the doublet can have a 
vacuum expectation value, as assigning 
a vacuum expectation value to the upper component would also 
break the $U(1)_Q$. 
The Higgs field can be parametrized as
\begin{equation}
\Phi(x) = 
\begin{pmatrix} \phi^+(x)\\ \phi^0(x) \end{pmatrix} = 
\begin{pmatrix}
G^+(x) \\ v + \frac1{\sqrt{2}}\left( H(x) + i G^0(x) \right)
\end{pmatrix} \quad ,
\label{sm:higgsvev}
\end{equation}
where $G^+$ is a complex and $H$ and $G^0$ are two real scalar fields.
The Higgs potential, i.e.\ the non-kinematic part of the SM Lagrangian which 
contains only Higgs fields, can be written as
\begin{align}
V(\Phi) =& - \frac{m_H^2}2 \left( \Phi^\dagger \Phi \right) 
  + \frac{m_H^2}{2 v^2} \left( \Phi^\dagger \Phi \right)^2 .
\label{sm:higgspot}
\end{align}

The breaking of a continuous global symmetry leads to massless scalar particles,
the Goldstone bosons~%
\cite{Nambu:1960xd,*Goldstone:1961eq,*Goldstone:1962es}. One Goldstone boson
occurs for each broken generator of the symmetry group. In case of a broken
continuous local symmetry, like a gauge symmetry, these Goldstone bosons
are unphysical. They can be eliminated by an appropriate choice of gauge, the 
unitary gauge. Their degrees of freedom are ``eaten up'' by the gauge bosons 
which become massive. Once ``eaten up'', the Goldstone bosons form the 
longitudinal modes of the gauge bosons.
For electroweak symmetry breaking there are three broken generators leading
to three ``would-be'' Goldstone bosons $G^\pm$ and $G^0$. 
Only the field $H$ in \eq{sm:higgsvev}
is physical. It is the field of the Higgs boson which has not been discovered yet. 
Its mass $m_H$ is a free parameter of the theory. It is bounded from below by 
experimental searches $m_H \ge 114.4 \GeV$~\cite{Barate:2003sz} 
and from above by electroweak precision data where a best fit yields
$m_H = 114^{+69}_{-45} \GeV$~\cite{Group:2005di}.

After electroweak symmetry breaking the gauge boson triplet 
$W_\mu^i, i=1\dots3,$ of $SU(2)_L$ and
the gauge boson $B_\mu$ ($U(1)_Y$) no longer form the mass 
eigenstates of the theory. The mass eigenstates are obtained by rotations
\begin{align}
W_\mu^\pm =& \frac1{\sqrt{2}} \left( W_\mu^1 \mp i W_\mu^2 \right) , &
Z_\mu =& c_W W_\mu^3  - s_W B_\mu , &
A_\mu =& s_W W_\mu^3 + c_W B_\mu .
\end{align}
$s_W$ and $c_W$ denote the sine and cosine of the electroweak
mixing angle, the Weinberg angle.
The photon field $A_\mu$ stays massless and can be interpreted as 
the gauge boson of the remaining $U(1)_Q$ symmetry of
electromagnetic interactions. The electromagnetically neutral $Z$
and the charged $W$ bosons receive a mass, which is proportional
to the vacuum expectation value of the Higgs field:
\begin{align}
m_Z =& \frac{e}{2 s_W c_W} v , & 
m_W =& \frac{e}{2 s_W} v 
\end{align}
where $e$ is the electromagnetic unit charge.
As $W$ and $Z$ have already been found in experimental searches
these equations determine the Weinberg angle and 
the scale of electroweak symmetry breaking $v=247$ GeV.

The Goldstone bosons $G^\pm$ and $G^0$ of \eq{sm:higgsvev} are absorbed 
by the $W$ and $Z$ bosons, respectively. In this thesis the 't Hooft-Feynman gauge
is used which has technical advantages for loop calculations since the gauge 
boson propagators in this gauge take a simpler form.
In the 't Hooft-Feynman gauge the Goldstone bosons appear explicitly as internal 
propagators with a mass equal to that of the associated gauge boson.
For external propagators their contribution is accounted for in
the longitudinal component of the polarization vector of the respective gauge boson.

In analogy to the inclusion of massive gauge bosons into a renormalizable 
quantum field theory there is no possibility to introduce fermion mass terms directly.
To generate fermion masses one introduces Yukawa interactions which couple
the fermions to the Higgs field
\begin{align}
\La_{\text{Yukawa}} =&
  -\lambda_e^{IJ} \overline{L}_I \Phi e_{R,J}
  -\lambda_u^{IJ} \overline{Q}_I \Phi^c u_{R,J}
  -\lambda_d^{IJ} \overline{Q}_I \Phi d_{R,J} 
  + h.c.
\end{align}
with
\begin{equation}
\Phi^c = i \sigma_2 \Phi^* = 
\begin{pmatrix} {\phi^0}^*\\ -{\phi^+}^* \end{pmatrix} 
\end{equation}
which is also an $SU(2)_L$ doublet but has hypercharge $Y=-1$.
The vacuum expectation value $v$ in the decomposition of $\Phi$ 
(\eq{sm:higgsvev}) leads to terms
which are bilinear in the fermion fields, i.e.\ to mass terms for the fermions.
The $\lambda_f^{IJ}$ are $3 \times 3$ Yukawa coupling matrices. They 
parameterize the masses of the quarks and further mixing effects in 
the quark sector.

\subsection{Higher-dimensional operators}
\label{sm:higherdimhiggs}
The realization of the Higgs sector in the SM is minimal in the sense that
it contains just enough additional parameters and fields to give a consistent
theory of the particles known nowadays. In extensions of the SM additional 
terms are possible, which lead to the following general parameterization
of the Higgs potential with one doublet 
$\Phi$~\cite{Barger:2003rs,*Grojean:2004xa,*Kanemura:2004mg,Plehn:2005nk}:
\begin{equation}
\label{sm:generalhiggspot}
V\left( \Phi\right)  = 
\sum_{n\ge0}\frac{\tilde{\lambda}_n}{\Lambda^{2n}} 
  \left( \Phi^\dagger \Phi - \frac{v^2}2 \right) ^{2+n} = 
\tilde{\lambda}_0 \left(  \Phi^\dagger \Phi - \frac{v^2}2 \right) ^2
+ \Order{\frac1{\Lambda^2}} .
\end{equation}
The expansion for $n=0$ on the right-hand side is identical to 
the SM Higgs potential \eq{sm:higgspot} with 
$\tilde{\lambda}_0 = \frac{m_H^2}{2v^2}$ up to the
constant term which is not a physical observable and leaves the equations 
of motion unchanged.
The additional terms for $n>0$ contain operators of
mass dimension 6 and higher. Such terms are non-renormalizable 
but can be considered as effective terms of an extended theory.
They are suppressed by the scale $\Lambda$ which is the scale where
new physics sets in. The only requirement \eq{sm:generalhiggspot}
has to fulfill is that its highest non-vanishing coefficient $\tilde{\lambda}_i$
is positive so that the potential is bounded from below.

\section{Problems of the Standard Model}

Despite its large success there are both experimental and theoretical hints that the 
SM is only the low-energy limit of a more general theory. 

%%% g-2(muon)
An experimental clue is the measured value of the anomalous magnetic moment of the 
muon~\cite{Bennett:2002jb}. This observable is known to an extremely
high precision from both experiment and theory, where the uncertainty
stems from unknown higher-loop contributions and experimental errors on the 
input parameters. The deviation from the SM prediction is about 
0.7-3.26~standard deviations~\cite{Passera:2005mx}.

%%% CDM
Another evidence comes from the dark matter problem in the
universe~\cite{Bertone:2004pz}.
Looking at the rotation of galaxies as a function of the distance from the center
shows that for large distances the circular velocity is constant, whereas the 
observed radiating matter would result in a decrease of the velocity
with the distance. This implies that there is some fraction of matter which 
is contributing to the overall mass density of the galaxy, but not emitting
electromagnetic radiation, hence the name \emph{dark matter}.
Precision measurements of the cosmic microwave 
background~\cite{Spergel:2003cb,*MacTavish:2005yk}
yield an average density of the universe that is very close to the so-called 
critical density, where the curvature of the universe vanishes.
Combining these data with our current understanding how the universe 
emerged and evolves requires that the total matter content of the universe
which contributes to this density is about 27\%. The rest is some 
form of energy, so-called \emph{dark energy}.
Of these 27\% of matter content, only about 4\% of the total matter content
consist of the usual 
baryonic matter, i.e.\ of matter built up of protons and neutrons. 
The remaining 23\% must be made of non-baryonic, only weakly-interacting 
matter.
The only particles in the SM which fulfill this requirement are the neutrinos.
Current upper limits on their masses~\cite{Eidelman:2004wy} imply however
that they cannot account for the whole required dark matter density.

%%% unification of coupling constants
One of the theoretical clues is the unification of coupling constants in 
Grand Unified Theories (GUT), where all three SM gauge groups merge in a 
single gauge group. 
Possible GUT gauge groups are
$SU(5)$~\cite{Georgi:1974sy,*Dimopoulos:1981zb,*Sakai:1981gr}, which is 
experimentally not viable due to a too large proton decay rate~\cite{Amaldi:1991cn} 
or $SO(10)$~\cite{Fritzsch:1974nn}.
Via the renormalization group equations the coupling constants of the three SM
gauge groups can be written as running coupling constants which depend on 
the energy. GUT theories predict that at a high energy scale, typically of the order 
$M_{GUT} \approx 10^{15} \GeV$, all three gauge couplings unify.
Such a unification does not occur in the SM, even if one takes into account that
new particles at the GUT scale might slightly modify the running.

%%% hierarchy problem
Another hint is the so-called hierarchy problem. 
If one considers one-loop corrections to 
the mass of the Higgs boson quadratic divergences appear~\cite{Drees:1996ca}.
These divergences can be erased by renormalization. One finds that the corrections
are of the order of the largest mass in the loop. If the SM is indeed the ultimate 
theory up to arbitrary high energies, this heaviest particle is the top quark and the
corrections are well under control. But if the SM is replaced by a new theory at 
higher energies, like a Grand Unified Theory which unifies the electroweak with the 
strong interactions or a quantum theory which includes gravity, new particles
with masses of the order of this new theory will appear, typically with masses of the 
Planck scale $M_{Planck} \approx 10^{19} \GeV$. In such new models
extreme fine-tuning
is necessary to get a Higgs mass of the order of the electroweak scale, as
is predicted by electroweak precision data~\cite{Group:2005di}. 
In particular there is no symmetry, neither conserved nor broken, which would
explain such a fine-tuning in a natural way.

%%% neutrinos
The last problem concerns the neutrino sector.
Neutrinos are assumed to be massless in the SM. It is known from
the observation of neutrino oscillations~\cite{Smy:2002hr} that neutrinos
possess a tiny mass. There is no conceptual problem to introduce
such a mass in the SM. As neutrinos are not of importance for the work presented
in this dissertation the exact formulation of the neutrino sector can be ignored.

%%% => SUSY
To solve the problems mentioned above various models
have been proposed. The model widely believed to be the most
promising candidate is supersymmetry. This extension of the Standard Model
was studied in this thesis and will be introduced in the following.

\chapter{Supersymmetry}
\label{susy}

\section{Basic principles}

It was shown by Coleman and Mandula~\cite{Coleman:1967ad} that combining
space-time and internal symmetries is only possible in a trivial way. In the proof of this 
theorem only general assumptions on the analyticity of scattering
amplitudes and the assumption that the S-matrix is invariant under 
Lorentz transformations are made.

Later it was realized~\cite{Haag:1975qh} that besides of Lie-algebras, 
which are defined via commutation relations, one can also use so-called superalgebras,
which also contain anticommutators. Then a new type of operators $Q$ is allowed
which has the following properties~\cite{Haber:1984rc,wb,Martin:1997ns}:
\begin{gather}
\left\{ {Q_\alpha}^A,\bar{Q}_{\dot{\beta}B} \right\} = 2
 \sigma^\mu_{\alpha\dot\beta} P_\mu {\delta^A}_B \nonumber\\
\left\{ {Q_\alpha}^A, {Q_\beta}^B \right\} = \left\{
\bar{Q}_{\dot{\alpha}A}, \bar{Q}_{\dot{\beta}B}\right\} = 0 \nonumber\\
\left[ P_\mu, {Q_\alpha}^A \right] = \left[ P_\mu,
\bar{Q}_{\dot{\alpha}A}\right] = 0
\label{susy:comm_qp}
\end{gather}
The supersymmetry generators $Q$ and $\bar Q$ carry Weyl spinor indices
$\alpha$, $\dot{\alpha}$, $\beta$ and $\dot{\beta}$ which run from $1$ to $2$,
where the undotted indices transform under the $(0,\frac12)$ representation of 
the Poincar\'e group and the dotted ones under the $(\frac12,0)$ conjugated
representation. The indices $A$ and $B$ refer to an internal space and run 
from $1$ to a number $N\ge1$. For $N>1$ chiral fermions are not
allowed~\cite{Sohnius:1985qm}.
These are necessary to construct the observed parity violation via 
$SU(2)_L$, where left- and right-handed fermions carry different quantum numbers.
Therefore only ($N=1$)-supersymmetries are relevant for phenomenologically
interesting energy ranges and in the following only such supersymmetries will
be considered.
$P_\mu$ denotes the generator of Lorentz translations, the 
energy-momentum operator, and 
$\sigma^\mu_{\alpha\dot\beta} = (1,\sigma^i_{\alpha\dot\beta})$ is 
the four-dimensional generalisation of the Pauli matrices.
The first line of 
\eq{susy:comm_qp} shows the entanglement of space-time symmetry
and the internal symmetry. The last line indicates the invariance of 
supersymmetry under Lorentz transformations.

As the operators anticommute with themselves, they have half-integer
spin according to the spin-statistics theorem. A detailed calculation shows that
their spin is always $\frac12$. Therefore we have
\begin{align}
Q\left|\text{boson}\right> &= \left|\text{fermion}\right> &
Q\left|\text{fermion}\right> &= \left|\text{boson}\right>.
\end{align}

The one-particle states belong to irreducible representations of the
supersymmetry algebra, the so-called supermultiplets. Each 
supermultiplet includes both bosonic and fermionic states which
are called superpartners to each other. They can be transformed
into each other by applying $Q$ and $\bar Q$.

Each supermultiplet contains the same number of bosonic and fermionic
degrees of freedom. For example the simplest
supermultiplet incorporates a Weyl fermion with two helicity states,
hence two degrees of freedom. Its bosonic partners are two
real scalars each with one degree of freedom, which can also
be combined into one complex scalar field. This is called the 
scalar or chiral supermultiplet.

The next possibility is a spin-1 vector boson. To guarantee the
renormalisability of the theory this has to be a gauge boson which 
is massless and contains two degrees of freedom. It follows that 
the partner is a massless Weyl fermion. A spin-$\frac32$ fermion
would render the theory non-renormalisable, so it must be a 
spin-$\frac12$ fermion. This is called a gauge or vector supermultiplet.

From \eq{susy:comm_qp} follows
\begin{equation}
\left[ P_\mu P^\mu, {Q_\alpha} \right] = 
\left[ P_\mu P^\mu, \bar{Q}_{\dot{\alpha}}\right] = 0 \quad . 
\label{susy:samemass}
\end{equation}
$P_\mu P^\mu$ is just $M^2$, the squared mass of a state in the supermultiplet.
Applying the supersymmetry operator therefore does not change the
mass of the state  and all states in a supermultiplet have the same mass if 
supersymmetry is unbroken. This will be important later on when
the Lagrangian is constructed.

\section{Superfields}
\label{mssm:superfields}

Starting from the supermultiplets one can construct superfields. To simplify
the notation Grassmann variables are introduced. These are anticommuting
numbers whose properties are defined in \chap{appsusy:grassmann}.
The superalgebra can now be written in terms of commutators
\begin{gather}
\left[ {\theta^\alpha Q_\alpha},
          \bar{Q}_{\dot{\beta}} \bar{\theta}^{\dot{\beta}} \right] 
= 2  \theta^\alpha \sigma^\mu_{\alpha\dot\beta} \bar{\theta}^{\dot{\beta}} P^\mu\\
\left[ \theta^\alpha {Q_\alpha}, \theta^\beta {Q_\beta} \right] = \left[
\bar{Q}_{\dot{\alpha}} {\bar\theta}^{\dot{\alpha}},
  \bar{Q}_{\dot{\beta}}{\bar\theta}^{\dot{\beta}}\right] = 0 \quad .
\end{gather}

In general a finite supersymmetric transformation is given by the
group element
\begin{equation}
G\left(x_\mu,\theta^\alpha,{\bar\theta}_{\dot\alpha}\right) = 
e^{i\left\{ x_\mu P^\mu + \theta^\alpha Q_\alpha + 
{\bar\theta}_{\dot\alpha}\bar{Q}^{\dot{\alpha}} \right\}} \quad ,
\end{equation}
in complete analogy to a general non-Abelian gauge transformation
$e^{i \phi_a T^a}$ with the group generators $T^a$.
$P^\mu$, $Q_\alpha$ and $\bar{Q}^{\dot{\alpha}}$ are the
generators of the supersymmetry group. The coordinates can be combined into
a tuple which represents a superspace coordinate
$z = \left(x_\mu, \theta^\alpha, {\bar\theta}_{\dot\alpha}\right)$. 
The set of all possible coordinates spans the superspace.

The fields on which these generators operate must then also be a function 
of $\theta$ and $\bar\theta$ besides $x_\mu$. These are the so-called superfields
$\Phi(x_\mu, \theta, {\bar\theta})$. 
In superspace one can obtain an explicit representation of 
$Q_\alpha$ and $\bar{Q}^{\dot{\alpha}}$ as differential operators.
For that purpose one considers a supersymmetry transformation of $\Phi$
\begin{equation}
G(y_\mu,\xi,\bar\xi) \Phi(x,\theta,\bar\theta) .
\end{equation}
Taking the parameters as infinitesimal and performing a Taylor expansion 
one obtains the following explicit representation of the supersymmetric generators
\begin{align}
Q_\alpha &= \frac{\partial}{\partial\theta^\alpha} - i \sigma^\mu_{\alpha\dot\beta}
    \bar\theta^{\dot\beta} \frac{\partial}{\partial x^\mu}\label{susy:q}\\
\bar{Q}_{\dot\alpha} &= -\frac{\partial}{\partial\bar\theta^{\dot\alpha}} 
   + i \theta^\beta \sigma^\mu_{\beta\dot\alpha} \frac{\partial}{\partial x^\mu}\label{susy:qbar}\\
P_\mu &= i \frac{\partial}{\partial x^\mu}.
\end{align}

For the further treatment it is sufficient
to consider only infinitesimal supersymmetric transformations which have
the following form
\begin{equation}
\delta_G(\xi, {\bar\xi}) \Phi(x_\mu, \theta, {\bar\theta})= 
\left[
  \xi \frac{\partial}{\partial\theta}+
  \bar\xi \frac{\partial}{\partial\bar\theta}-
  i \left( \xi \sigma_\mu \bar\theta - \theta\sigma_\mu \bar\xi
    \right) \frac{\partial}{\partial x_\mu}
\right] \Phi(x_\mu, \theta, {\bar\theta}),
\end{equation}
where $\xi$ and $\bar\xi$ are also Grassmann variables.
Contracted indices which are summed over have been suppressed
in this equation.

Analogously to the covariant derivative in gauge theories one also
introduces covariant derivatives $D_\alpha$ and $\bar{D}_{\dot\alpha}$
with respect to the supersymmetry generators. These derivatives
have to be invariant under $Q$ and $\bar{Q}$, which is equivalent
to the postulate
\begin{equation}
\{D_\alpha,Q_\alpha\} = \{\bar{D}_{\dot\alpha},Q_\alpha\} = 
\{D_\alpha,\bar{Q}_{\dot\alpha}\} = 
\{\bar{D}_{\dot\alpha},\bar{Q}_{\dot\alpha}\} = 0.
\end{equation}
Thus the covariant derivatives are
\begin{align}
D_\alpha &= \frac{\partial}{\partial\theta^\alpha} + i \sigma^\mu_{\alpha\dot\beta}
    \bar\theta^{\dot\beta} \frac{\partial}{\partial x^\mu}\label{susy:d}\\
\bar{D}_{\dot\alpha} &= -\frac{\partial}{\partial\bar\theta^{\dot\alpha}} 
   - i \theta^\beta \sigma^\mu_{\beta\dot\alpha} \frac{\partial}{\partial x^\mu}\label{susy:dbar}.
\end{align}
From eqs.\ (\ref{susy:q}, \ref{susy:qbar}, \ref{susy:d}, \ref{susy:dbar}) one can 
also deduce 
that the Grassmann variables $\theta$ and $\xi$ have spin $-\frac12$, 
while $D$ and $Q$ have spin $+\frac12$.

Superfields can be expanded into component fields. The general expansion
of superfields in terms of Grassmann variables is
\begin{equation}
\begin{split}
\Phi(x,\theta,\bar\theta) =& f(x) + \theta \phi(x) + \bar\theta\bar\chi(x)\\
& + \theta\theta m(x) + \bar\theta\bar\theta n(x) + \theta \sigma^\mu
\bar\theta v_\mu(x) \\ 
& + \theta\theta\bar\theta\bar\lambda(x) +
\bar\theta\bar\theta\theta\psi(x) + \theta\theta\bar\theta\bar\theta d(x).
\end{split}\label{susy:phigen}
\end{equation}
Due to the anticommuting properties of Grassmann variables this 
expansion is complete, i.e.\ it truncates with the last shown term.

Up to now all expressions have been written out for general superfields. 
To construct a supersymmetric Lagrangian only two special types
of superfields are needed. They are irreducible representations of the
supersymmetry algebra. One obtains them by imposing
covariant restrictions on a general superfield. In this way they still
span a representation space of the algebra but have less components.

\subsection{Chiral Superfields}

One possibility are chiral superfields. They are defined by applying
the covariant derivative $\bar{D}_{\dot\alpha}$ on the scalar 
superfield~$\Phi$ as defined in \eq{susy:phigen}
\begin{equation}
\bar{D}_{\dot\alpha}\Phi(z) = 0.
\label{susy:chiralsuperfield}
\end{equation}
The solution of this differential equation leads to a chiral superfield
which can be expressed in general component fields as
\begin{equation}
\begin{split}
 \Phi =& A(x) + i \theta \sigma^\mu \bar\theta \partial_\mu A(x) + \frac14
 \theta\theta\bar\theta\bar\theta \partial_\mu \partial^\mu A(x) \\
 & + \sqrt2 \theta \psi(x) - \frac{i}{\sqrt2} \theta\theta \partial_\mu
 \psi(x) \sigma^\mu \bar\theta + \theta\theta F(x) .
\end{split}\label{susy:chiralcomp}
\end{equation}
$A$ is a complex scalar field, $\phi$ a complex Weyl spinor and
$F$ an auxiliary complex scalar field which has mass dimension two. It 
transforms under supersymmetry transformations into a total space-time 
derivative and therefore does not represent a physical, propagating
degree of freedom.
The product of chiral superfields is again a chiral superfield. For two
chiral superfields $\Phi_1$ and $\Phi_2$ this follows directly from the
product rule for derivatives
\begin{equation}
\bar{D}_{\dot\alpha} \left( \Phi_1 \Phi_2 \right)
= \left( \bar{D}_{\dot\alpha} \Phi_1\right) \Phi_2
   + \Phi_1 \left( \bar{D}_{\dot\alpha} \Phi_2\right) 
= 0.
\end{equation}

Analogously one can define an antichiral superfield $\Psi$ by the equation
\begin{equation}
D_\alpha \Psi(z) = 0.
\end{equation}
In particular the hermitian conjugate $\Phi^\dagger$ of a chiral 
superfield $\Phi$ is an antichiral superfield.

\subsection{Vector Superfields}

The second special type of superfields are vector superfields. They 
are derived from a general scalar superfield $V$ by demanding it to be real:
\begin{equation}
V^\dagger(z) = V(z) . \label{susy:vectorsuperfield}
\end{equation}
The name vector superfield stems from the fact that in the expansion a real vector field
appears as a component field and that these fields are used as generalized 
gauge fields when supersymmetric gauge theories are constructed.

The complete expansion in terms of component fields is
\begin{equation}
\begin{split}
V(x,\theta,\bar\theta) =& C(x) + i \theta \chi(x) - i
\bar\theta\bar\chi(x) + \frac i2 \theta\theta \left[ M(x)+iN(x)\right] -
\frac i2 \bar\theta\bar\theta\left[M(x)-iN(x)\right] \\
&-\theta\sigma^\mu\bar\theta v_\mu(x) + i\theta\theta\bar\theta\left[
\bar\lambda(x) + \frac i2 \bar\sigma^\mu \partial_\mu\chi(x)\right]
-i \bar\theta\bar\theta\theta \left[ \lambda(x) + \frac i2 \sigma^\mu
\partial_\mu \bar\chi(x) \right] \\ & + \frac12
\theta\theta\bar\theta\bar\theta \left[ D(x) + \frac12 \partial_\mu
\partial^\mu C(x) \right] .
\end{split}
\label{susy:vectorcomp}
\end{equation}
$C$, $D$, $M$ and $N$ are scalar fields and $v_\mu$ is the vector field
which gives the name to this type of superfields.
They all have to be real to fulfill \eq{susy:vectorsuperfield}. 
$\lambda$ and $\chi$ are Weyl spinors.

For the vector superfield we can now define a supersymmetric gauge
transformation which is in the general non-Abelian case described by
\begin{equation}
e^{gV} \rightarrow e^{-ig\Phi^\dagger} e^{gV} e^{ig\Phi} \quad ,
\end{equation}
where $\Phi$ denotes again a chiral superfield.
This simplifies in the Abelian case to
\begin{equation}
V \rightarrow V + i \left( \Phi - \Phi^\dagger \right) .
\end{equation}
Using this gauge transformation we can simplify \eq{susy:vectorcomp}
and choose
\begin{equation}
\chi(x) = C(x) = M(x) = N(x) \equiv 0
\end{equation}
thereby eliminating unphysical degrees of freedom.
This choice of gauge is called Wess-Zumino gauge~\cite{Wess:1974jb}.
As we have used only three of the four bosonic degrees of freedom
in $\Phi$ the ``ordinary'' gauge freedom of an Abelian gauge group 
is still present and the Wess-Zumino gauge is compatible with the
usual gauges. 

The vector superfield is now simplified to
\begin{equation}
 V = -\theta \sigma^\mu\bar\theta v_\mu(x) + i
  \theta\theta\bar\theta\bar\lambda(x) -
i\bar\theta\bar\theta\theta\lambda(x) +
\frac12\theta\theta\bar\theta\bar\theta D(x)
\end{equation}
with the scalar auxiliary field $D$ with mass dimension two.
As in the case of chiral superfields this auxiliary
field turns into a total derivative under supersymmetry transformations
and does not contribute to the propagating degrees of freedom.

Now we have all building blocks to construct a
supersymmetric extension of the Standard Model.

\section{A Supersymmetric Lagrangian}

A supersymmetric Lagrangian requires the action to remain unchanged
under supersymmetry transformations
\begin{equation}
\delta_G \int \di^4 x \La(x) = 0 \quad . 
\end{equation}
This requirement is fulfilled if the Lagrangian $\La$ turns into a total
space-time derivative under supersymmetry transformations. 
A comparison with the transformation properties of chiral and
vector superfields shows that the $F$ and $D$ terms of 
\eq{susy:chiralcomp} and (\ref{susy:vectorcomp})
show exactly this behavior. Schematically the Lagrangian can be
written simply as
\begin{equation}
\La = \int \di^2\theta \La_F + \int \di^2 \theta \di^2 \bar\theta \La_D \quad .
\end{equation}

As was noted already in the previous chapter the product  of two chiral
superfields is again a chiral superfield. Explicit multiplication of the
component fields yields a term proportional to $\psi_i \psi_j$ which
has the form of a fermion mass term. The product of three chiral 
superfields which is by induction also a chiral superfield contains
terms of the type $\psi_i \psi_j A_k$ which describe Yukawa-like
couplings between two fermions and a scalar. Products of four or 
more chiral superfields would lead to terms with a mass dimension
greater than four and yield a Lagrangian which is no longer renormalizable.
Thus the terms which can contribute to a supersymmetric
Lagrangian can be written in a compact way with the superpotential
\begin{equation}
 W(\Phi_i) = \lambda_i \Phi_i + \frac12 m_{ij} \Phi_i \Phi_j + \frac1{3!}
  g_{ijk} \Phi_i \Phi_j \Phi_k \quad .
\end{equation}

The product $\Phi \Phi^\dagger$ of a chiral superfield with its 
hermitian conjugate is self-conjugate. Therefore it is a vector superfield
according to the definition \eq{susy:vectorsuperfield} and a 
possible candidate for a contribution to $\La_D$:
\begin{equation}
 \int \di^2\theta\di^2\bar\theta \Phi \Phi^\dagger = F F^* - A
  \partial_\mu \partial^\mu A^\dagger - i \bar\psi\sigma_\mu\partial^\mu \psi \quad .
\end{equation}
The expression contains terms for the kinetic energy of both the scalar
and the fermionic component. The auxiliary fields $F$ do not have any
kinematic terms so they can be integrated out.

Gauge interactions are introduced by a supersymmetric generalization
of the ``minimal coupling'' 
$\Phi^\dagger \Phi \rightarrow \Phi^\dagger e^{2gV}\Phi $
with a vector superfield $V$ with $V = T^a V^a$, where $T^a$ are the 
generators of the gauge group. Written in component fields one can
replace the ordinary derivatives by covariant derivatives
$D_\mu = \partial_\mu + i g v_\mu^a T_a$.

The terms for the kinetic energy of the gauge fields can also be 
expressed in terms of a superpotential
\begin{equation}
 W_\alpha = -\frac14 \left({\bar D} {\bar D} \right) e^{-2gV} D_\alpha e^{2gV} \quad .
\end{equation}
The product $W_a W^a$ is gauge invariant and also a chiral superfield, so
its $\theta\theta$-term can appear in the supersymmetric Lagrangian.
Again only the gauge bosons and their superpartners, the gauginos, obtain
kinetic terms, but not the auxiliary fields, so we can eliminate them.

Therefore the most general form of a supersymmetric Lagrangian
has the following form:
\begin{equation}
 \La_{\text{SUSY}} = \int \di^2\theta \left[ \left( 
  \frac1{16 g^2} W^a_\alpha W^{a\alpha} + W\left(\Phi\right) \right) 
   + \text{h.c.} \right]
   + \int \di^2\theta \di^2\bar\theta \left(\Phi^\dagger e^{2gV}\Phi\right) \quad .
  \label{susy:lagr_allg}
\end{equation}

As the two auxiliary fields $F$ and $D$ do not have any terms for the
kinetic energy, their equations of motion have a simple form
\begin{align}
 \frac{\partial {\cal L}}{\partial F_i} &= 0 & 
 \frac{\partial {\cal L}}{\partial D_a} &= 0 .
\end{align}
Solving these equations for the $F$ and $D$ fields
\begin{align}
 F_i &= - \left[\frac{\partial W(A_i)}{\partial A_j} \right]^* &
 D_a &= - g A_i^* T^{ij}_a A_j
\end{align}
and inserting these expressions into \eq{susy:lagr_allg} the 
Lagrangian can be completely expressed in terms of
physical fields.

\section{Supersymmetry breaking}
\label{susy:softbreaking}

As shown in \eq{susy:samemass} all members of a supermultiplet
must have the same mass. This means if the Standard Model
was supersymmetrized by just replacing the fields with 
their respective superfields there would exist for example
a supersymmetric partner to the electron with a mass of
511 keV/$c^2$. This partner particle is a boson with spin 0, 
but otherwise with the same quantum numbers as the electron, 
i.e.\ particularly with a charge of one negative elementary charge.
Such a particle would have been discovered experimentally
a long time ago.

This problem can be circumvented by requiring that supersymmetry
is broken. In this way one can give a mass to the supersymmetric partners which
is beyond the current experimental limits. In analogy to 
spontaneous symmetry breaking in the electroweak sector
the Lagrangian itself should be invariant under supersymmetry
transformations, but have a vacuum expectation value which
is not invariant under such transformations. For supersymmetry
this problem is somewhat complicated because additional 
constraints appear which have to be fulfilled simultaneously.
Such a constraint follows immediately from the definition of 
the supersymmetry algebra \eq{susy:comm_qp} which implies
\begin{equation}
 H \equiv P^0 = \frac14 \left( {\bar Q}_1 Q_1 + Q_1 {\bar Q}_1 + 
{\bar Q}_2 Q_2 + Q_2 {\bar Q}_2 \right) \ge 0 .
\end{equation}
Applying the Hamiltonian $H$ onto a state $\ket{\Psi}$ leads
to the result that supersymmetry is broken if neither the
$D$ nor the $F$ term can be made zero simultaneously.

The Fayet-Iliopoulos mechanism~\cite{Fayet:1974jb} achieves 
supersymmetry breaking by adding a $D$ term to the Lagrangian
which is linear in the auxiliary field,
while O'Raifeartaigh models~\cite{O'Raifeartaigh:1975pr} do 
this via chiral supermultiplets and a superpotential such that
not all auxiliary fields $F$ can be made zero at the same time.
Both mechanisms are phenomenologically not viable because
they can lead to color breaking or the breaking of electromagnetism,
or need an unacceptable fine-tuning~\cite{Haber:1993wf}. 

%%% AMSB, GMSB, SUGRA
Hence one expects that supersymmetry is not broken directly 
by renormalizable tree-level couplings, but indirectly or radiatively.
For these purposes one introduces a hidden sector of particles
in which supersymmetry is broken and which has only small or no
direct couplings at all to the normal visible sector.
The two sectors however share some common interaction
which mediates the breaking from the hidden to the visible sector
and leads to additional supersymmetry breaking terms.
Two possible scenarios for this mediation are widely discussed
in the literature~\cite{Nilles:1983ge}. 
The first one is gravity-mediated supersymmetry breaking. At the Planck
scale gravity is anticipated to become comparable in size to the gauge 
interactions. The mediating interaction is associated with the new
gravitational interactions which enter at this scale. Because of the 
flavor blindness of gravity these gravitational interactions are 
expected to be flavor-blind as well.
A second possibility is that the mediating interactions are the ordinary 
QCD and electroweak gauge interactions. They connect the visible and 
the hidden sector via loop diagrams involving messenger particles.
This scenario is called gauge-mediated supersymmetry breaking.

For a phenomenological analysis it is often not relevant
what the exact way of supersymmetry breaking is but only 
which additional terms in the Lagrangian are generated.
Thereby the cancellation of quadratic divergences should 
remain valid, such that the solution of the naturalness problem
of the Standard Model is not lost.
Terms which do not spoil the cancellation 
are called \emph{soft} supersymmetry breaking terms.
It was shown~\cite{Girardello:1982wz} that only the
following terms are soft supersymmetry breaking up to all
orders in perturbation theory:
\begin{align*}
\bullet\quad&\text{scalar mass terms} && m^2_{ij} A_i^* A_j \\
\bullet\quad&\text{trilinear scalar interactions} && 
 t_{ijk} A_i A_j A_k + h.c. \\
\bullet\quad&\text{mass terms for gauge particles} && 
 \frac12 m_l {\bar\lambda}_l \lambda_l \\
\bullet\quad&\text{bilinear terms} && 
 b_{ij} A_i A_j + h.c. && \\
\bullet\quad&\text{linear terms} && l_i A_i \quad .
\end{align*}

Now all building blocks are in place and we can turn to building a 
supersymmetric version of the Standard Model.

\section{Minimal Supersymmetric Standard Model}

The simplest possibility of a supersymmetric extension of
the Standard Model is called Minimal Supersymmetric 
Standard Model (MSSM). The underlying algebra
is an (N=1)-supersymmetry with soft supersymmetry
breaking.
As in the Standard Model the MSSM shall have a local
gauge symmetry with respect to the gauge group
$SU(3)_C \otimes SU(2)_L \otimes U(1)_Y$, which describe
the strong, weak and electromagnetic interactions.
Its particle content is obtained by replacing all fields
with their corresponding superfields. 

Each matter field is assigned a chiral superfield.
Its fermionic part describes
the usual fermions of the Standard Model and its bosonic 
part contains the ``scalar fermions'', the \emph{sfermions}, 
as superpartners.
For each gauge group a vector superfield is introduced 
whose vector bosons form the usual gauge bosons of the
Standard Model, and the fermionic superpartners are
two-component Weyl spinors, in general called 
\emph{gauginos}.
The nomenclature of the new particles usually follows the
convention that the bosonic superpartners carry the name 
of the fermion with a prefix ``s'', which is short for ``scalar'',
and the fermionic superpartners carry the name of the 
gauge boson with the suffix ``-ino''. 

In the Higgs sector of the MSSM it is not sufficient to replace 
the scalar field by a vector superfield. 
One would need both the field $H$ and its hermitian conjugate
$H^*$ to give mass to both up- and down-type quarks. 
This is forbidden by the requirement that the
superpotential must be analytic and so one needs a second
Higgs doublet with negative hypercharge.
Additionally the fermion which emerges from the single Higgs 
superfield would carry a non-vanishing
hypercharge $Y$. This hypercharge contributes to the chiral
anomaly~\cite{Adler:1969er,*Bell:1969ts} which is not compensated 
by other particles. The quantized version of such a theory would
be inconsistent. In the MSSM the two fermions, one from each Higgs
doublet, have opposite hypercharge and their contribution 
to the anomaly exactly cancels.

Table~(\ref{susy:mssmparticles}) gives an overview of the
particle content of the MSSM in the interaction basis.
\begin{table}
\begin{tabular}{|l|c||c|c||c|c|c|}
\hline
 & \multicolumn{3}{|c||}{fields} & \multicolumn{3}{|c|}{group representation} \\
\hline
 & superfield & fermion field & boson field & $SU(3)_C$ & $SU(2)_L$ & $U(1)_Y$ \\
\hline\hline
\multicolumn{7}{|l|}{\textbf{matter sector}}\\
\hline
Quarks & $\hat{Q}_I$ & $\begin{pmatrix}u_{L,I}\\d_{L,I}\end{pmatrix}$ & 
    $\begin{pmatrix}\tilde{u}_{L,I}\\\tilde{d}_{L,I}\end{pmatrix}$  & 
    $\mathbf{3}$ & $\mathbf{2}$ & $\phantom{-}\frac13$ \mssmbigstrut \\
  & $\hat{U}_I$ & $ u_{R,I}^c$ & $ \tilde{u}_{R,I}^*$ & 
    $\mathbf{\bar{3}}$ & $\mathbf{1}$ & $-\frac43$ \mssmstrut \\
  & $\hat{D}_I$ & $ d_{R,I}^c$ & $ \tilde{d}_{R,I}^*$ & 
    $\mathbf{\bar{3}}$ & $\mathbf{1}$ & $\phantom{-}\frac23$ \mssmstrut \\
\hline
Leptons & $\hat{L}_I$ & $\begin{pmatrix}\nu_{L,I}\\e_{L,I}\end{pmatrix}$ & 
    $\begin{pmatrix}\tilde{\nu}_{L,I}\\\tilde{e}_{L,I}\end{pmatrix}$  & 
    $\mathbf{1}$ & $\mathbf{2}$ & $-1$ \mssmbigstrut \\
  & $\hat{E}_I$ & $ e_{R,I}^c$ & $ \tilde{e}_{R,I}^*$ & 
    $\mathbf{1}$ & $\mathbf{1}$ & $\phantom{-}2$ \mssmstrut \\
\hline\hline
\multicolumn{7}{|l|}{\textbf{gauge sector}}\\
\hline
$SU(3)_C$ & $\hat{G}^a$ & $\tilde{\lambda}_{G}^a$ & $ G^a_\mu $ & 
    $\mathbf{8}$(adj.) & $\mathbf{1}$ & $\phantom{-}0$ \mssmstrut \\
$SU(2)_L$ & $\hat{W}^i$ & $\tilde{\lambda}_{W}^i$ & $ W^i_\mu $ & 
    $\mathbf{1}$ & $\mathbf{3}$(adj.) & $\phantom{-}0$ \mssmstrut \\
$\phantom{S}U(1)_Y$ & $\hat{B}$ & $\tilde{\lambda}_{B}$ & $ B_\mu $ & 
    $\mathbf{1}$ & $\mathbf{1}$ & $\phantom{-}0$ \mssmstrut \\
\hline\hline
\multicolumn{7}{|l|}{\textbf{Higgs sector}}\\
\hline
  & $\hat{H_1}$ & $\begin{pmatrix}\tilde{H}_1^1\\\tilde{H}_1^2\end{pmatrix}$ &
    $\begin{pmatrix}H_1^1\\H_1^2\end{pmatrix}$  &
    $\mathbf{1}$ & $\mathbf{2}$ & $-1$ \mssmbigstrut \\
  & $\hat{H_2}$ & $\begin{pmatrix}\tilde{H}_2^1\\\tilde{H}_2^2\end{pmatrix}$ &
    $\begin{pmatrix}H_2^1\\H_2^2\end{pmatrix}$  &
    $\mathbf{1}$ & $\mathbf{2}$ & $\phantom{-}1$ \mssmbigstrut \\
\hline
\end{tabular}
\caption{Superfields and particle content of the MSSM in the interaction basis. Superfields
are denoted with a hat and the superpartners all carry a tilde. The generation
index $I$ of the quarks and leptons runs from 1 to 3. For the gauge fields the color 
index~$a$ runs from 1 to 8 and the weak isospin index~$i$ from 1 to 3. The bold numbers in the group 
representation of the non-Abelian groups $SU(3)_C$ and $SU(2)_L$ denote the dimension
of the representation, where $\mathbf{1}$ is the trivial representation and the gauge bosons
are in the adjoint representation of the group.
The number for the Abelian group $U(1)_Y$ denotes the hypercharge of the 
particle.}
\label{susy:mssmparticles}
\end{table}
For the gauge superfields we have the following field strengths in the MSSM
\begin{align}
{W_C}^a_\alpha &= -\frac14 \left({\bar D} {\bar D} \right) e^{-2 g_s \hat{G}} D_\alpha e^{2 g_s \hat{G}} 
  \label{susy:qcd fieldstrength}\\
{W_L}^i_\alpha &= -\frac14 \left({\bar D} {\bar D} \right) e^{-2 g_w \hat{W}} D_\alpha e^{2 g_w \hat{W}} \\
{W_Y}_\alpha &= -\frac14 \left({\bar D} {\bar D} \right) e^{-2 g_y \hat{B}} D_\alpha e^{2 g_y \hat{B}} 
  = - \frac{g_y}4 {\bar D} {\bar D} D_\alpha \hat{B} \quad . 
\end{align}
Additionally the superpotential must be fixed. In the MSSM it is defined as
\begin{equation}
W_{\text{MSSM}} = \epsilon^{ij} \left( \lambda_e^{IJ} \hat{H}_1^i \hat{L}^{jI} \hat{E}^J
  - \lambda_u^{IJ} \hat{H}_2^i \hat{Q}^{jI} \hat{U}^J
  + \lambda_d^{IJ} \hat{H}_1^i \hat{Q}^{jI} \hat{D}^J
  - \mu \hat{H}_1^i \hat{H}_2^j \right) \quad ,
\label{susy:superpotential} 
\end{equation}
where $\lambda_e$, $\lambda_u$ and $\lambda_d$ are $3\text{x}3$ Yukawa coupling matrices and 
$I$ and $J$ denote the generation index.

Inserting eqs.~(\ref{susy:qcd fieldstrength})-(\ref{susy:superpotential}) into \eq{susy:lagr_allg}
and adding the $F$ terms to the Lagrangian yields the supersymmetric part
of the MSSM Lagrangian
\begin{equation}
\begin{split}
\La_{\text{SUSY}} =& \int \di^2\theta \left[ \left(
  \frac1{16 g_s^2} {W_C}^a_\alpha {W_C}^{a\alpha} + 
  \frac1{16 g_w^2} {W_L}^i_\alpha {W_L}^{i\alpha} \right. \right. \\
  &\left. \left. + 
  \frac1{16 g_y^2} {W_Y}_\alpha {W_Y}^{\alpha}
   + W_{\text{MSSM}} \right) 
   + \text{h.c.} \right] \\
   &+ \int \di^2\theta \di^2\bar\theta \left[
  \hat{L}^\dagger e^{2g_w \hat{W} + 2 g_y \hat{B}}\hat{L}
   + \hat{E}^\dagger e^{2 g_y \hat{B}}\hat{E} \right. \\
  &+ \hat{Q}^\dagger e^{2g_s \hat{G} + 2g_w \hat{W} + 2 g_y \hat{B}}\hat{Q}
   + \hat{U}^\dagger e^{-2g_s \hat{G}^T + 2 g_y \hat{B}}\hat{U}
   + \hat{D}^\dagger e^{-2g_s \hat{G}^T + 2 g_y \hat{B}}\hat{D}\\
  &\left.+ \hat{H}_1^\dagger e^{2g_w \hat{W} + 2 g_y \hat{B}}\hat{H}_1
   + \hat{H}_2^\dagger e^{2g_w \hat{W} + 2 g_y \hat{B}}\hat{H}_2
\right] \quad .
\end{split}
\end{equation}

Supersymmetry in the MSSM is broken explicitly by soft 
supersymmetry breaking terms, i.e.\ only the terms 
mentioned at the end of \chap{susy:softbreaking} are allowed.
This leads to the following contributions to the
MSSM Lagrangian:
\begin{align}
\intertext{\begin{itemize}
\item Majorana mass terms for all gauginos
\end{itemize}
}
\La_{\text{soft,majoranamass}} =& 
    \frac12\left( M_3 \overline{\tilde{\lambda}_{G}^a} \tilde{\lambda}_{G}^a 
    + M_2 \overline{\tilde{\lambda}_{W}^i} \tilde{\lambda}_{W}^i  
    + M_1 \overline{\tilde{\lambda}_{B}} \tilde{\lambda}_{B} \right) +h.c. \\
\intertext{\begin{itemize}
\item mass terms for all scalar superpartners of the Standard Model fermions
 and for the scalar Higgs fields
\end{itemize}
}
\La_{\text{soft,scalarmass}} =&
    -M^2_{\tilde{L}} \left( \tilde{\nu}^*_{L,I} \tilde{\nu}_{L,I} + \tilde{e}^*_{L,I} \tilde{e}_{L,I}\right)
    -M^2_{\tilde{E}} \tilde{e}^*_{R,I} \tilde{e}_{R,I} \nonumber\\
  &-M^2_{\tilde{Q}} \left( \tilde{u}^*_{L,I} \tilde{u}_{L,I} + \tilde{d}^*_{L,I} \tilde{d}_{L,I}\right)
    -M^2_{\tilde{U}} \tilde{u}^*_{R,I} \tilde{u}_{R,I}
    -M^2_{\tilde{D}} \tilde{d}^*_{R,I} \tilde{d}_{R,I} \nonumber\\
  &-m_1^2 \left| H_1 \right|^2
    -m_2^2 \left| H_2 \right|^2 \\
\intertext{\begin{itemize}
\item bilinear term which couples the two scalar Higgs fields
\end{itemize}
}
\La_{\text{soft,bilinear}} =& m_{12}^2 \left( \epsilon_{ij}H_1^i H_2^j + h.c. \right) \\
\intertext{\begin{itemize}
\item trilinear interaction terms for the scalar superpartners
of the Standard Model fermions
\end{itemize}
}
\La_{\text{soft,trilinear}} =& - \epsilon_{ij} \left( 
    \lambda_e^{IJ} A_e H_1^i \tilde{L}^{jI} \tilde{E^J}
   -\lambda_u^{IJ} A_u H_2^i \tilde{Q}^{jI} \tilde{U^J}
   +\lambda_d^{IJ} A_d H_1^i \tilde{Q}^{jI} \tilde{D^J}
\right) \nonumber\\&+ h.c. \quad .
\end{align}
In the general case the Yukawa couplings $\lambda_e$, $\lambda_u$ and $\lambda_d$
as well as the trilinear couplings $A_e$, $A_u$ and $A_d$ are 
complex $3 \times 3$ matrices.
The scalar mass parameters $M_{\tilde{L}}$, $M_{\tilde{E}}$, $M_{\tilde{Q}}$,
$M_{\tilde{U}}$, $M_{\tilde{D}}$, are hermitian $3 \times 3$ matrices. The scalar 
Higgs mass parameters $m_1$ and $m_2$ are real numbers, and the 
gaugino mass parameters $M_1$, $M_2$ and $M_3$ 
as well as the bilinear Higgs coupling $m_{12}$ are complex numbers.

There is an additional possibility for soft-breaking trilinear 
couplings~\cite{Rosiek:1990rs,*Rosiek:1995kg} which has the form
\begin{equation}
\La_{\text{soft,tri2}} = 
\left(
    {A_e^\prime}^{IJ} H_2^{i*} \tilde{L}^{iI} \tilde{E^J}
   -{A_u^\prime}^{IJ} H_1^{i*} \tilde{Q}^{iI} \tilde{U^J}
   +{A_d^\prime}^{IJ} H_2^{i*} \tilde{Q}^{iI} \tilde{D^J}
\right) + h.c. \quad .
\end{equation}
This expression involves charge-conjugated Higgs fields which, in contrast to
the superpotential, are possible for soft supersymmetry-breaking terms.
However, it turns out that in most scenarios of supersymmetry breaking
such terms are not generated. Therefore they are normally not 
considered and will also be neglected in this thesis.

The complete soft supersymmetry breaking Lagrangian
is given by
\begin{equation}
\La_{\text{soft}} = \La_{\text{soft,majoranamass}}
  + \La_{\text{soft,scalarmass}} + \La_{\text{soft,bilinear}} + \La_{\text{soft,trilinear}}
\end{equation}

As next step gauge fixing terms must be added to the Lagrangian. This is 
required so that all Green functions are still calculable. In this dissertation the 
$R_\xi$- or 't Hooft gauge is used
\begin{equation}
\begin{split}
\La_{\text{gauge-fixing}} =& 
  -\frac1{2\xi}\left( \partial^\mu G_\mu^a \right)^2
  -\frac1{2\xi} \left( \partial^\mu W^1_\mu + \frac{i}{\sqrt{2}} m_W \xi \left(G^+ - G^-\right) \right)^2  \\
 &-\frac1{2\xi} \left( \partial^\mu W^2_\mu - \frac{1}{\sqrt{2}} m_W \xi \left(G^+ + G^-\right) \right)^2 \\
 &-\frac1{2\xi} \left( \partial^\mu W^3_\mu + c_W m_Z\xi G^0\right)^2
  -\frac1{2\xi}\left( \partial^\mu B_\mu - s_W m_Z \xi G^0 \right)^2 .
\end{split}
\end{equation}
$G^\pm$ and $G^0$ are the Goldstone bosons which were already described 
for the Standard Model case in \chap{sm:higgs} and appear in the MSSM in 
the same way.

Setting $\xi=1$ yields the 't Hooft-Feynman gauge which is advantageous for 
one-loop calculations, because the propagators take a very simple shape, while the
Goldstone bosons appear explicitly in the calculation. This kind of gauge is used 
throughout this thesis.

Finally unphysical modes which were introduced by the gauge-fixing terms are
compensated by Faddeev-Popov ghost terms $\La_{\text{ghost}}$~\cite{Faddeev:1967fc}. 

Adding up all contributions gives the complete Lagrangian of the MSSM
\begin{equation}
\La_{\text{MSSM}} = \La_{\text{SUSY}} + \La_{\text{soft}}
  + \La_{\text{gauge-fixing}} + \La_{\text{ghost}} .
\end{equation}

%%% R parity %%%
Additional terms could be added to the superpotential in \eq{susy:superpotential}
which are also gauge-invariant and analytic in the superfields, but violate
lepton or baryon number conservation which has not been observed experimentally 
so far. Such terms include the coupling of three lepton or quark superfields or the
coupling of lepton to quark superfields. The strictest limits on lepton and baryon 
number violation are obtained by searching for a possible decay of the proton 
which violates each baryon and lepton number by one unit. 
Experiments have established a lower limit on the proton lifetime of 
$10^{29}$ years~\cite{Eidelman:2004wy} while general violating terms 
predict a decay time in the order of minutes or hours.
Thus a mechanism must exist which forbids or at least heavily suppresses these
terms. The simplest possibility is to postulate a conservation of baryon and lepton
number. Such a postulate would be a regression with respect to the Standard Model.
There the conservation is fulfilled automatically and a consequence of the fact that there
are no renormalizable lepton and baryon number violating terms. 
Furthermore, postulating lepton and
baryon number conservation as a fundamental principle of nature is generally 
not viable. It is known that there are non-perturbative effects in the electroweak 
sector which do violate lepton and baryon number conservation, although their 
effect is negligible for the energy ranges of current experiments.

Instead a symmetry should be introduced which has the conservation of these 
quantum numbers 
as a natural consequence. So in the MSSM as a perturbative theory baryon and lepton 
number conservation is again guarantueed while the existence of non-perturbative 
effects is not contradicted by demanding a fundamental symmetry.
Such a symmetry is given by 
\emph{$R$-parity}~\cite{Farrar:1978xj}.
A new quantum number $R$ is introduced and from that a so-called $R$-parity 
$P_R = (-1)^R$ is derived. It is induced by the generators of supersymmetry,
stays intact after spontaneous supersymmetry breaking and is multiplicatively
conserved.
$R=0$ for all Standard Model particles and the additional Higgs scalars and 
$R=1$ for all supersymmetric partners. The link to lepton and baryon 
number conservation is obvious if one writes the $R$-parity quantum number 
in terms of baryon number $B$, lepton number $L$ and spin $s$
\begin{equation}
P_R = (-1)^{2s+3(B-L)} \quad .
\end{equation}
$B$ is $+\frac13$ for the left-handed chiral quark superfield $Q_I$, 
$-\frac13$ for the right-handed quark superfields $\hat{U}_I$ and $\hat{D}_I$, 
and $0$ for all other particles.
Analogously $L$ is $+1$ for the left-handed lepton superfield $\hat{L}_I$, 
$-1$ for the right-handed lepton superfield $\hat{E}_I$, and $0$ for all other particles.
Then all Standard Model particles and the Higgs scalars have $P_R=+1$ 
and the supersymmetric partners have an odd $R$-parity of $P_R=-1$.

An interesting consequence of this is that each interaction vertex 
connects an even number of supersymmetric particles. Therefore 
they can only be produced in pairs and the lightest supersymmetric
particle (LSP) must be stable.

\section{Particle content of the MSSM}

\subsection{Higgs and Gauge bosons}

As in the Standard Model the $SU(2)_L \otimes U(1)_Y$ symmetry 
is broken by the vacuum expectation values of the Higgs fields
in such a way that a $U(1)_Q$ symmetry of electromagnetic 
interactions remains. Its associated conserved quantum number 
is the usual electromagnetic charge.
As shown before the Higgs sector
of the MSSM must consist of two scalar isospin doublets
\begin{align}
H_1 =& \begin{pmatrix}v_1 + \frac1{\sqrt{2}}\left( \phi_1^0-i \chi_1^0\right)\\
 -\phi_1^-\end{pmatrix}&
H_2 =& \begin{pmatrix}\phi_2^+ \\ 
  v_2 + \frac1{\sqrt{2}}\left( \phi_2^0+i \chi_2^0\right)
\end{pmatrix}
\label{susy:higgsparam}
\end{align}
with opposite hypercharge. $\phi_1^0$, $\phi_2^0$, $\chi_1^0$ and $\chi_2^0$ are 
real scalar fields, and $\phi_1^-$ and $\phi_2^+$ are complex scalar fields.
In \eq{susy:higgsparam} 
an expansion around the vacuum expectation values 
has been performed, which satisfy the 
equation
\begin{align}
\left< H_1 \right> =& \begin{pmatrix}v_1\\0\end{pmatrix}&
\left< H_2 \right> =& \begin{pmatrix}0\\v_2\end{pmatrix} \quad .
\end{align}

Collecting all terms in the Lagrangian which contain only the Higgs fields
we have contributions to the Higgs potential from the $F$ terms in the 
superpotential, from the $D$ terms and finally from the soft 
supersymmetry breaking terms
\begin{equation}
\begin{split}
V_{\text{Higgs}} =& \abs{\mu}^2 \left( \abs{H_1}^2 + \abs{H_2}^2 \right) \\
 & + \frac18 \left( g_w^2 + g_y^2 \right) \left( \abs{H_1}^2 - \abs{H_2}^2 \right)^2
  + \frac12 g_w^2 \abs{H_1^\dagger H_2}^2 \\
 & + m_1^2 \abs{H_1}^2 + m_2^2 \abs{H_2}^2 
  - m_3^2 \left(\epsilon_{ij} H_1^i H_2^j + h.c. \right) \quad .
\end{split}
\end{equation}
This equation shows the close entanglement between supersymmetry breaking
and electroweak symmetry breaking. Only including the soft breaking terms it 
is possible that the minimum of the Higgs potential is not at the origin and the
fields acquire a vacuum expectation value.

The mass matrices of the Higgs fields are obtained by differentiating twice
with respect to the fields $\phi$ and $\chi$. This leads to four uncoupled 
real $2\times 2$ matrices.
To obtain the mass eigenstates these matrices have to be diagonalized
by unitary matrices. In the case of a real $2\times 2$ matrix this is simply 
a rotation matrix.
We obtain
\begin{align}
\begin{pmatrix}G^\pm\\H^\pm\end{pmatrix} &=
  \begin{pmatrix}c_\beta&s_\beta\\-s_\beta&c_\beta\end{pmatrix}
  \begin{pmatrix}\phi_1^\pm\\\phi_2^\pm\end{pmatrix} \\
\begin{pmatrix}G^0\\A^0\end{pmatrix} &=
  \begin{pmatrix}c_\beta&s_\beta\\-s_\beta&c_\beta\end{pmatrix}
  \begin{pmatrix}\chi_1^0\\\chi_2^0\end{pmatrix} \\
\begin{pmatrix}H^0\\h^0\end{pmatrix} &=
  \begin{pmatrix}c_\alpha&s_\alpha\\-s_\alpha&c_\alpha\end{pmatrix}
  \begin{pmatrix}\phi_1^0\\\phi_2^0\end{pmatrix} \quad .
\end{align}
$c_\beta$, $s_\beta$, $c_\alpha$ and $s_\alpha$ is a short-hand notation
for $\cos\beta$, $\sin\beta$, $\cos\alpha$ and $\sin\alpha$, respectively.
Similar abbreviations will also be used for the other angles in this thesis,
as well as $t_\beta$ denoting $\tan\beta$.
The mixing angle $\beta$ is defined as the ratio of the two 
vacuum expectation values
\begin{align}
t_\beta &= \frac{v_2}{v_1} & \text{with} \qquad 0 < \beta < \frac{\pi}2 \quad .
\end{align}
$t_\beta$ is a free parameter of the MSSM.
The mixing angle $\alpha$ is determined by the relation
\begin{align}
t_{2\alpha} &= t_{2\beta} \frac{m_A^2 + m_Z^2}{m_A^2-m_Z^2} 
  & \text{with} \qquad  -\frac{\pi}2 < \alpha < 0 \quad .
\end{align}
The restriction on the given interval determines $\alpha$ uniquely
and is chosen such that always $m_{h^0} < m_{H^0}$.
By electroweak symmetry breaking three group generators are
broken and therefore as in the Standard Model three unphysical 
would-be Goldstone bosons $G^\pm$ and $G^0$ emerge. 
The five remaining Higgs bosons are physical ones. 
There are two electrically neutral CP-even Higgs bosons
$h^0$ and $H^0$, one CP-odd $A^0$ and two electrically
charged ones $H^\pm$. The mass of the CP-odd Higgs boson $m_A$
is usually chosen to be the second free parameter of the MSSM
Higgs sector. The masses of the other Higgs bosons at 
tree-level are then
\begin{align}
m_{h^0,H^0} =& \frac12 \left( m_A^2 + m_Z^2 
  \mp \sqrt{\left( m_A^2 + m_Z^2 \right)^2 - 4 m_A^2m_Z^2 c_{2\beta}^2} \right)\\
m_{H^\pm} =& m_A^2 + m_W^2 \quad .
\end{align}
These relations receive large corrections at higher orders which must be taken 
into account when one wants to obtain realistic predictions.
The one-loop corrections are known 
completely~%
\cite{Ellis:1990nz,*Okada:1990vk,*Haber:1990aw,Brignole:1992uf,Chankowski:1992ej,
*Chankowski:1992er,Dabelstein:1995js,*Dabelstein:1994hb}.
On the two-loop level the calculation of the supposedly dominant corrections 
in the Feynman diagrammatic approach \cite{Degrassi:2002fi} of 
$\Order{\alpha_t \alpha_s}$~%
\cite{Hempfling:1993qq,Heinemeyer:1998jw,*Heinemeyer:1998kz,Heinemeyer:1998np,
Zhang:1998bm,*Espinosa:1999zm,Degrassi:2001yf}, $\Order{\alpha_t^2}$~%
\cite{Hempfling:1993qq,Espinosa:2000df,Brignole:2001jy}, $\Order{\alpha_b\alpha_s}$~%
\cite{Brignole:2002bz,Heinemeyer:2004xw} and $\Order{\alpha_t\alpha_b+\alpha_b^2}$~%
\cite{Dedes:2003km}, a calculation in the effective potential approach~%
\cite{Martin:2001vx,*Martin:2002iu,*Martin:2002wn,*Martin:2003qz,*Martin:2003it}
and the evaluation of momentum-dependent effects~\cite{Martin:2004kr} have 
been performed.
As these expressions are rather lengthy they are not written out here.
For the numerical evaluation the expressions given in~%
\cite{Heinemeyer:1999be} have been used.

As in the Standard Model, electroweak symmetry breaking 
turns the $W^i$ and $B$ gauge bosons into the mass eigenstates 
$W^\pm$, $Z$ and the photon $\gamma$. 
$W$ and $Z$ bosons acquire a mass, where the single 
vacuum expectation value 
of \eq{sm:higgsvev} is replaced by $v=\sqrt{v_1^2+v_2^2}$.

The gauge bosons of $SU(3)_C$ are the eight massless 
gluons. Their mass eigenstates are identical to the
interaction eigenstates $g^a_\mu = G^a_\mu$.

\subsection{Higgsinos and Gauginos}
All particles which have the same quantum numbers can mix with each
other. As the $SU(2)_L \otimes U(1)_Y$ symmetry is broken, only the
$SU(3)_C$ and $U(1)_Q$ quantum numbers have to match.

In the sector of non-colored, charged particles there are the
Winos $\tilde{W}^\pm$ 
and the charged Higgsinos $\tilde{H}_1^+$ and $\tilde{H}_2^+$
with
\begin{align}
\tilde{W}^\pm =& 
  \begin{pmatrix} -i \tilde{\lambda}_W^\pm \\ i \overline{\tilde{\lambda}_W^\mp}\end{pmatrix} &
\tilde{H}_1^+ =& 
  \begin{pmatrix} \tilde{H}_2^1 \\ \overline{\tilde{H}_1^2}\end{pmatrix} &
\tilde{H}_2^+ =& 
  \begin{pmatrix} \tilde{H}_1^2 \\ \overline{\tilde{H}_2^1}\end{pmatrix} 
\end{align}
As for the $W$ bosons the relation
\begin{equation}
\tilde{\lambda}_W^\pm = \frac1{\sqrt{2}} \left( \tilde{\lambda}_W^1 \mp i \tilde{\lambda}_W^2 \right) 
\end{equation}
holds.

These four two-component Weyl spinors combine into two
four-component Dirac fermions called charginos. Their mass matrix
is diagonalized by
\begin{equation}
U^* 
\begin{pmatrix}M_2&\sqrt{2}m_W s_\beta \\ \sqrt{2} m_W c_\beta&\mu\end{pmatrix}
V^\dagger = 
\begin{pmatrix}m_{\tilde{\chi}_1^+}&0\\0&m_{\tilde{\chi}_2^+}\end{pmatrix} \quad .
\end{equation}
$U$ and $V$ are two unitary matrices which are chosen such that
$m_{\tilde{\chi}_{1,2}^+}$ are both positive and 
$m_{\tilde{\chi}_1^+} \le m_{\tilde{\chi}_2^+}$.
The chargino mass eigenstates are given by
\begin{equation}
\tilde{\chi}_i^+ = 
\begin{pmatrix}\chi_i^+\\ \overline{\chi_i^-}\end{pmatrix} = 
\begin{pmatrix}
  V \begin{pmatrix}- i \lambda_W^+ \\ H_2^1\end{pmatrix} \\
  U \begin{pmatrix}- i \lambda_W^- \\ H_1^2\end{pmatrix} \\
\end{pmatrix} \quad .
\end{equation}

The uncolored neutral higgsinos and gauginos also mix among each other. 
We have the two neutral Higgsinos $\tilde{H}_1^1$ and $\tilde{H}_2^2$,
the Zino $\tilde{Z}$ and
the Photino $\tilde{A}$
\begin{align}
\tilde{H}_1^0 =& 
  \begin{pmatrix} \tilde{H}_1^1 \\ \overline{\tilde{H}_1^1}\end{pmatrix} &
\tilde{H}_2^+ =& 
  \begin{pmatrix} \tilde{H}_2^2 \\ \overline{\tilde{H}_2^2}\end{pmatrix} &
\tilde{Z} =& 
  \begin{pmatrix} -i \tilde{\lambda}_Z \\ i \overline{\tilde{\lambda}_Z}\end{pmatrix} &
\tilde{A} =& 
  \begin{pmatrix} -i \tilde{\lambda}_A \\ i \overline{\tilde{\lambda}_A}\end{pmatrix}  \quad .
\end{align}
The latter two are obtained, as in the case of $Z$ and $\gamma$,
by rotating $\tilde{\lambda}_W^3$ and $\tilde{\lambda}_B$ by the Weinberg angle
\begin{align}
\tilde{\lambda}_Z =& c_W \tilde{\lambda}_W^3 - s_W \tilde{\lambda}_B & 
\tilde{\lambda}_A =& s_W \tilde{\lambda}_W^3 + c_W \tilde{\lambda}_B \quad .
\end{align}

The four Weyl spinors form four Majorana fermions, called neutralinos, whose 
mass matrix is also diagonalized by a unitary matrix $N$
\begin{multline}
N^* 
\begin{pmatrix}
  M_1 & 0 & -m_Z s_W c_\beta & m_Z s_W s_\beta \\
  0 & M_2 & m_Z c_W c_\beta & -m_Z c_W s_\beta \\
  -m_Z s_W c_\beta & m_Z c_W c_\beta & 0& -\mu \\
  m_Z s_W s_\beta & -m_Z c_W s_\beta & -\mu & 0 \\
\end{pmatrix}
N^\dagger \\ 
=
\begin{pmatrix}
  m_{\tilde{\chi}_1^0}&0&0&0\\
  0&m_{\tilde{\chi}_2^0}&0&0\\
  0&0&m_{\tilde{\chi}_3^0}&0\\
  0&0&0&m_{\tilde{\chi}_4^0}\\
\end{pmatrix} \quad .
\end{multline}
Again the remaining freedom in the choice of $N$ is used to order
the neutralino masses such that
$m_{\tilde{\chi}_1^0} \le m_{\tilde{\chi}_2^0} 
  \le m_{\tilde{\chi}_3^0} \le m_{\tilde{\chi}_4^0}$.
The neutralino mass eigenstates are given by 
\begin{equation}
\begin{pmatrix}
  \tilde{\chi}_1^0 \\ \tilde{\chi}_2^0 \\ \tilde{\chi}_3^0 \\ \tilde{\chi}_4^0 
\end{pmatrix}
= N 
\begin{pmatrix}
  -i \tilde{\lambda}_B \\ -i \tilde{\lambda}_W^3 \\ \tilde{H}_1^1 \\ \tilde{H}_2^2
\end{pmatrix} \quad .
\end{equation}

The gauginos of $SU(3)_C$, the gluinos, do not mix with other particles 
as they are the only fermions which are subject to the
strong interaction exclusively. There are eight gluinos with mass 
$m_{\tilde{g}}=\abs{M_3}$.
Gluinos are Majorana particles and have the following form 
\begin{equation}
\tilde{g}^a = \begin{pmatrix}-i\tilde{\lambda}_{G}^a\\i\overline{\tilde{\lambda}_{G}^a}\end{pmatrix} \quad .
\end{equation}

\subsection{Leptons and Quarks}

Leptons and quarks have similar properties as in the Standard Model.
The Weyl spinors of left- and right-handed fermions can be combined
into one Dirac spinor
\begin{align}
e_I =& \begin{pmatrix} e_{L,I} \\ \overline{e_{R,I}^c} \end{pmatrix} & 
u_I =& \begin{pmatrix} u_{L,I} \\ \overline{u_{R,I}^c} \end{pmatrix} & 
d_I =& \begin{pmatrix} d_{L,I} \\ \overline{d_{R,I}^c} \end{pmatrix} \quad ,
\end{align}
where $I$ is again the generation index.
The down-type quarks $d_I$ are not exact mass eigenstates.
A rotation 
\begin{equation}
d_I^\prime = V_{CKM}^{IJ} d_J
\end{equation}
by a unitary matrix, the 
Cabibbo-Kobayashi-Maskawa(CKM)-matrix $V_{CKM}$~\cite{Cabibbo:1965vd,*Kobayashi:1973fv},
is required to turn the 
flavor eigenstates $d_J$ into mass eigenstates $d_I^\prime$. As the 
CKM-matrix is close to a unity matrix and flavor-mixing effects do not 
play any role in the processes which are calculated in this thesis
effects induced by the CKM-matrix will be neglected and the CKM-matrix 
is set to exactly the unity matrix.

Leptons and quarks receive their masses via the Yukawa terms in
the superpotential which are bilinear in the lepton and quark fields:
\begin{align}
m_e =& \lambda_e v_1 &
m_u =& \lambda_u v_2 &
m_d =& \lambda_d v_1 \quad .
\end{align}
These equations are often rewritten such that the Yukawa couplings are 
expressed in terms of the fermion masses and the mass of the $W$
boson
\begin{align}
\lambda_e =& \frac{m_e e}{\sqrt{2} m_W c_\beta} &
\lambda_u =& \frac{m_u e}{\sqrt{2} m_W s_\beta} &
\lambda_d =& \frac{m_d e}{\sqrt{2} m_W c_\beta} \quad , 
\end{align}
$e$ denoting the elementary charge.

\subsection{Sleptons and Squarks}

In the sfermion sector mixing between different interaction
eigenstates can occur in the same way as for the gauginos.
In general the $3 \times 3$ trilinear coupling matrices and mass matrices
in the soft supersymmetry breaking part of the MSSM Lagrangian
can be fully occupied, thus leading to mixing between the
sfermions of different generations. Such mixing results in 
contributions to flavor changing neutral currents (FCNCs)
besides the contribution of the CKM-matrix which is already present in the
Standard Model.
Experimental limits~\cite{Eidelman:2004wy} show that such additional 
contributions have to be small~\cite{Gabbiani:1996hi}. 
Additionally, most popular models of supersymmetry breaking mediate this 
breaking from the hidden sector by flavor-blind interactions.
Therefore the soft breaking mass matrices and trilinear couplings are chosen
purely diagonal.
Then the mass matrices of the electron-like sleptons and the squarks decompose 
into $2\times 2$ matrices where only the left- and right-handed fields of each generation mix.
These can be written as
\begin{equation}
M_{\tilde{f}} = 
\begin{pmatrix}
M_{\tilde{f}}^{LL} + m_f^2 & m_f \left(M_{\tilde{f}}^{LR}\right)^*\\
m_f M_{\tilde{f}}^{LR} & M_{\tilde{f}}^{RR} + m_f^2
\end{pmatrix}
\end{equation}
with
\begin{align}
M_{\tilde{f}}^{LL} =& m_Z^2 \left( I_3^f - Q_f s_W^2\right) c_{2\beta} + 
  \begin{cases}
    M_{\tilde{L}}^2 & \text{for left-handed sleptons} \\
    M_{\tilde{Q}}^2 & \text{for left-handed squarks}
  \end{cases} \label{susy:sfermionsll}\\
M_{\tilde{f}}^{RR} =& m_Z^2 \left(Q_f s_W^2\right) c_{2\beta} + 
  \begin{cases}
    M_{\tilde{R}}^2 & \text{for right-handed electron-like sleptons} \\
    M_{\tilde{U}}^2 & \text{for right-handed up-like squarks} \\
    M_{\tilde{D}}^2 & \text{for right-handed down-like squarks} 
  \end{cases} \label{susy:sfermionsrr}\\
M_{\tilde{f}}^{LR} =& A_f - \mu^*
  \begin{cases}
    \frac1{t_\beta} & \text{for up-like squarks} \\
    t_\beta & \text{for electron-like sleptons and down-like squarks}
  \end{cases} \quad .
\label{susy:sfermionslr}
\end{align}
$Q_f$ is the electromagnetic charge of the sfermion. $I_3^f$ denotes the 
quantum number of the third component of the weak isospin operator $T_3$
which is $+\frac12$ for up-like squarks and $-\frac12$ for down-like squarks and 
electron-like sleptons.
These mass matrices can again be diagonalized by a unitary matrix
\begin{equation}
U_{\tilde{f}} M_{\tilde{f}} U_{\tilde{f}}^\dagger = 
\begin{pmatrix} m_{\tilde{f}_1}^2&0\\ 0&m_{\tilde{f}_2}^2 \end{pmatrix} \quad .
\end{equation}
The fields then transform as
\begin{equation}
\begin{pmatrix} \tilde{f}_1 \\ \tilde{f}_2\end{pmatrix}
= U_{\tilde{f}}
\begin{pmatrix} \tilde{f}_L \\ \tilde{f}_R\end{pmatrix}  \quad .
\end{equation}
In the sneutrino sector only left-handed fields exist. For this reason the mass 
matrix consists only of the $M_{\tilde{f}}^{LL}$ element given in 
\eq{susy:sfermionsll}. $M_{\tilde{f}}^{LL}$ is therefore a free 
parameter of the theory
which directly gives the squared mass of the sneutrinos according to
\begin{equation}
m_{\tilde{\nu}_I}^2 = \frac12 m_Z^2 c_{2\beta} + M_{\tilde{L}}^2  \quad .
\end{equation}
The interaction eigenstates $\tilde{\nu}_I$ are identical to the mass eigenstates.

\chapter{Regularization and Renormalization}
\label{renorm} 

% Why renormalization
In general the Lagrangian of a model contains free parameters which are not fixed 
by the theory, but must be determined in experiments. On tree-level these
parameters can be identified directly with physical observables like
masses or coupling constants. If one goes to higher-order perturbation theory
these relations are modified by loop contributions. Additionally the integration
over the loop momenta is generally divergent which further complicates the situation.
To achieve a mathematically consistent treatment it is necessary to regularize
the theory before predictions can be made. This introduces a cutoff in the 
relations between the parameters 
and the physical observables. As a consequence, the parameters appearing in 
the basic Lagrangian, the so-called ``bare'' parameters, have no longer a 
physical meaning. This physical meaning is then restored via renormalization.
The renormalized parameters obtained in this way are again finite. Their value 
is fixed by renormalization conditions.

The details of this procedure are described in the following sections.

\section{Regularization}
\label{renorm:regul}
The ultra-violet divergences appearing in the integration over loop momenta 
must be treated via a regularization scheme. Therefore a regularization 
parameter is introduced into the theory which leads to finite expressions, 
but leaves the expressions dependent on the renormalization parameter.

There exist different regularization schemes, three of which are shortly described
in the following:

\paragraph{Pauli-Villars Regularization}{~}\\
This regularization scheme~\cite{Pauli:1949zm} is very simple. 
Originally the integration region over the four-dimensional loop momentum 
ranges from plus to minus infinity. In this scheme it is restricted such that 
the absolute value of the loop momentum is below a certain finite value.
This cutoff parameter must be much larger than any other mass scale 
appearing in the theory. Performing a regularization in this way 
usually destroys gauge symmetry, so it is not used for practical calculations 
and not further taken into account in this dissertation.

\paragraph{Dimensional Regularization}{~}\\
Loop integrals are divergent if the dimension of the integration is exactly four. 
Dimensional 
regularization~\cite{Bollini:1972ui,*Ashmore:1972uj,*'tHooft:1972fi} exploits
this fact. If one shifts the dimension of the loop momentum by 
an infinitesimal value and performs the integration in $D=4-2\epsilon$ dimensions,
the integral becomes finite. The divergences now appear as poles in the 
infinitesimal parameter $\epsilon$.
Additionally, the dimensions of all fields are also set to
$D$ dimensions and the gauge couplings are multiplied by $\mu^{2\epsilon}$.
The parameter $\mu$ has the dimension of a mass and specifies the regularization
scale. It is introduced to keep the coupling constants dimensionless. 
This scheme is normally used in Standard Model calculations as it preserves
gauge symmetry. It does, however, not preserve supersymmetry. As the fields
are $D$-dimensional, additional degrees of freedom are introduced so that
the number of fermionic degrees of freedom no longer equals the number 
of bosonic degrees of freedom and therefore supersymmetry is broken.

\paragraph{Dimensional Reduction}{~}\\
This scheme~\cite{Siegel:1979wq,Capper:1979ns} is similar to dimensional
regularization in the respect that the loop integration is performed in 
$D$ dimensions and the divergences are recovered as poles 
in $\epsilon$. In this scheme the fields are kept four-dimensional in order 
to avoid explicit supersymmetry breaking.
The mathematical consistency of dimensional reduction has long been 
questioned~\cite{Siegel:1980qs}, but recently a consistent 
formulation~\cite{Stockinger:2005gx} could be established.
It could be shown that supersymmetry is conserved for matter fields at least up 
to the two-loop order.

\section{Renormalization}
The dependence on the unphysical scale $\mu$ can be removed via
renormalization. It consists of a set of rules which consistently replaces 
the bare parameters in the Lagrangian by new finite ones.

There exist different degrees of renormalizability. 
%%% WRONG: For unbroken SUSY still wave-function renormalization needed
% There are theories
%in which renormalization is not necessary. A model with an unbroken 
%supersymmetry would for example be such a model. This can be most easily 
%seen in the case of self-energy corrections to scalar particles. These corrections
%consist of a particle loop which is either inserted into the propagator via two
%three-particle vertices or attached to it via a four-particle vertex. As each fermion
%has a bosonic superpartner with the same quantum numbers
%and mass, their contributions to loop diagrams are the same. 
%One of Feynman's rules states that every closed fermion loop 
%receives an additional minus sign while bosonic loops are not 
%further modified, so fermionic and bosonic contributions exactly cancel.
%Similar arguments hold for the other loop corrections.
%Hence the born-level Lagrangian stays the same to all orders in 
%perturbation theory and the bare parameters can be identified as 
%the renormalized ones.
%
%The next possibility are super-renormalizable theories. 
One possibility are super-renormalizable theories. 
They are 
characterized by the fact that the coupling has positive 
mass dimension. In these theories only a finite number of 
basic Feynman diagrams diverge. These divergences can, however, 
appear as subdiagrams at every order in perturbation theory. 
An example for such a theory is scalar $\phi^3$-theory. Here
apart from vacuum polarization graphs only the one- and 
two-loop tadpoles and the one-loop self-energy diagram are 
divergent. 

In renormalizable theories only a finite number of amplitudes diverge,
but these divergences occur at all orders of perturbation theory. 
In such theories there are also dimensionless couplings but none
with a mass dimension smaller than zero.
To cancel the divergences a finite set of rules is necessary.
Non-Abelian gauge theories like the Standard Model and the MSSM 
belong to this category. Their renormalizability was first proven
in ref.~\cite{'tHooft:1971fh,*'tHooft:1971rn}.

Finally a theory can be non-renormalizable. In this case all 
amplitudes are divergent if the order of perturbation theory is 
sufficiently high. 
The set of rules to absorb the divergences is infinite and new 
ones appear at each order of perturbation theory. This means that the 
theory loses its predictive power. It might at first sight look like
such models would be useless, but this is not the case.
Non-renormalizable models are often used for effective theories.
Here operators of a mass dimension greater than four appear in 
the Lagrangian. As the final expression in the Lagrangian must be 
of mass dimension four, this is compensated by an appropriate
power of a cut-off mass appearing in the denominator. This cut-off 
mass defines the energy scale up to which the effective theory is 
valid and above which it must be replaced by the full renormalizable theory.
In the overlap region where both theories give a useful result, a matching
between the two is performed, thereby fixing the 
renormalization conditions and allowing meaningful predictions.

\subsection{Counter terms}

One of the most popular renormalization approaches nowadays is multiplicative 
renormalization with counter terms. 
In this scheme the bare parameters $g_0$, i.e.\ couplings and masses
appearing in the Lagrangian, are replaced by renormalized ones
$g$, which are related to the bare ones via the renormalization constant $Z_g$
\begin{equation}
g_0 = Z_g g = \left( 1 + \delta Z_g^{(1)} + \delta Z_g^{(2)} + \dots \right) g \quad ,
\label{renorm:renconst}
\end{equation}
where on the right-hand side the renormalization constant has been expanded
in orders of perturbation theory and the order is denoted by the superscript. 
The renormalized $g$ have a finite value.
The $\delta Z_g^{(i)}$ absorb the divergences which appear in the loop
integrals and are parametrized in the regularization parameter. Therefore they
remove the dependence on the unphysical regularization parameter.
Additionally, finite parts can be absorbed in the renormalization constants, as 
the decomposition in \eq{renorm:renconst} is not unique. Which finite parts
are absorbed in the renormalization constants depends on the chosen
renormalization scheme, which will be discussed below.
If one also adds the wave function renormalization of external particles, 
the renormalization of the parameters is sufficient to obtain finite 
S-matrix elements. To achieve the finiteness of off-shell Green functions, the
fields must be renormalized as well. Therefore the bare fields $\Phi_0$ are 
replaced by the renormalized ones $\Phi$ and the field 
renormalization constant $Z_\Phi$
\begin{equation}
\Phi_0 = \sqrt{Z_\Phi} \Phi = \left( 1 + \frac12 \delta Z_\Phi^{(1)} 
 - \frac18 \delta {Z_\Phi^{(1)}}^2 +\frac12 \delta Z_\Phi^{(2)} + \dots \right) \Phi
  \quad .
\end{equation}
Again on the right-hand side the field renormalization constant is
written out as an expansion in orders of perturbation theory. Thereby, the
term containing the squared of the one-loop renormalization 
constant ($- \frac18 \delta {Z_\Phi^{(1)}}^2$) is part of the two-loop
contribution. Similarly, for higher orders the orders of all renormalization
constants which appear in a term must be added up to give the loop order
to which the term contributes.

Using both parameter and field renormalization all Green functions are
finite. We can now insert the renormalized parameters and fields into
the bare Lagrangian
\begin{equation}
\La \left(g_0, \Phi_0\right) = \La \left(Z_g g, \sqrt{Z_\Phi}\Phi\right)
 = \La \left(g, \Phi\right) + \La_{CT} \left(g, \Phi, Z_g, Z_\Phi \right)
\end{equation}
and write it as a sum of the renormalized Lagrangian 
$\La \left(g, \Phi\right)$ and the counter-term part which can be 
expanded in terms of the loop order
\begin{equation}
\begin{split}
\La_{CT} \left(g, \Phi, Z_g, Z_\Phi \right) =& 
\La_{CT}^{(1)} \left(g, \Phi, \delta Z_g^{(1)}, \delta Z_\Phi^{(1)} \right) +\\
 & \La_{CT}^{(2)} \left(g, \Phi, \delta Z_g^{(1)}, \delta Z_\Phi^{(1)}, 
    \delta Z_g^{(2)}, \delta Z_\Phi^{(2)}\right) + \dots \quad .
\end{split}
\end{equation}
In this thesis only one-loop corrections to processes are considered.
So only the one-loop counter terms $\delta Z^{(1)}$ enter the
calculations, hence for simplicity the superscript $(1)$ on the $\delta Z$ will be
dropped from now on.

\subsection{Renormalization Schemes}

The finite part of the renormalization constants is not fixed by the 
divergences, but can be chosen in a suitable way. The definition 
of these finite parts together with an independent set of parameters
comprises a renormalization scheme and therefore defines the 
relation between the observables and the parameters of the theory.
If one adds up all orders of perturbation theory the result is 
independent of the chosen renormalization scheme. The value 
of the input parameters, however, still depends on the renormalization
scheme and must be chosen appropriately. For actual calculations 
only a finite number of orders can be taken into account. The resulting
dependence on the renormalization scheme is then a measure for 
the theoretical uncertainty which is induced by the missing 
higher-order terms.

The simplest renormalization scheme is the minimal-subtraction scheme or short 
MS-scheme~\cite{'tHooft:1973mm}.
It is based on dimensional regularization as regularization scheme.
In this scheme the counter terms absorb just the divergent 
$\tfrac1{\epsilon}$-terms but no finite contributions. 
This scheme is actually a whole set of schemes, as the scale $\mu$, 
which was introduced in the regularization step, is still present. This scale
is now taken as the renormalization scale $\mu_R$ and for specifying 
a concrete renormalization scheme $\mu_R$ must be fixed as well.

A commonly used variant of the MS-scheme is the
modified minimal subtraction scheme or short
\MSbar{} scheme~\cite{Bardeen:1978yd,*Buras:1979yt,
Passarino:1989ey,Marciano:1980be,*Sirlin:1989uf}. It is based on the
observation that the $\tfrac1{\epsilon}$-terms are always associated with 
other constant terms that emerge from the continuation of the loop momentum
in $D$ dimensions and are denoted by $\Delta^n$, where $n$ is the loop order. 
At one-loop order it has the following explicit form
\begin{equation}
\Delta = \frac1{\epsilon} - \gamma_E + \ln 4\pi \quad ,
\end{equation}
where $\gamma_E$ denotes the Euler-Mascheroni constant.
The absorption of the numerical constants $\gamma_E$ and $\ln 4\pi$
corresponds to a redefinition of the renormalization scale
\begin{equation}
{\mu_R^2}^{\MSbar} := \mu_R^2  e^{\ln 4\pi - \gamma_E} .
\end{equation}

If dimensional reduction is used as the regularization scheme, the
renormalization scheme is called \DRbar. Apart from that the procedure 
is identical to the \MSbar{} scheme. The $\Delta^n$ terms are 
subtracted by the renormalization constants but no other finite parts. 
As before, this corresponds to a redefinition of the renormalization scale
${\mu_R^2}^{\DRbar}$. On the one-loop level the counter terms are 
identical, while at higher orders they differ because the two regularization 
schemes induce different finite parts.

Another, distinct possibility is the on-shell scheme 
(OS scheme)~\cite{Bohm:1986rj,*Hollik:1993cg,Denner:1991kt}. 
The expression on-shell means that the renormalization conditions are
set for particles which are on their mass shell. The mass of a particle 
which is on-shell is given by the real part of the pole of the propagator 
and can be interpreted as its physical mass. In the OS scheme
the real parts of all loop contributions to the propagator pole and consequently
to this mass are absorbed in the mass counter terms. Hence the counter terms in the 
OS scheme also have a non-vanishing finite part and the dependence on the
regularization scale $\mu$ is completely eliminated in this scheme.
Coupling constants are renormalized in the OS scheme by demanding
that the coupling constants stay unchanged if all particles coupling to the vertex
are on-shell. This means that all corrections to the vertex are compensated by
the counter term of the coupling constant.
For the on-shell renormalization of fields one demands that the propagators 
are correctly normalized, i.e.\ the residue of the renormalized on-shell 
propagator is equal to one. 

The renormalization of $t_\beta$, the ratio of the two 
Higgs vacuum expectation values, is performed via \DRbar{} also 
when otherwise the OS scheme is used~\cite{Frank:2002qf}.
As $t_\beta$ does not receive any SUSY-QCD corrections at one-loop order,
its renormalization is not necessary for the calculations of this thesis.
Also the strong coupling constant $\alpha_s$ is always renormalized
in the \MSbar{} or \DRbar{} scheme. The details of the
renormalization of $\alpha_s$ are presented in the next section.

A complete expression of all Standard Model one-loop counter terms in the 
OS scheme was given in ref.~\cite{Denner:1991kt}. Its extension to the MSSM
was performed in ref.~\cite{Fritzsche:2005da}. In this thesis the same conventions
as in these two references are used.

\subsection{Renormalization of the strong coupling constant}
\label{renorm:alphas}

As every other parameter, the coupling constant $g_s$ of the strong interaction
receives divergent loop corrections. These divergences must be removed by
renormalization. As shown in the previous section, 
$g_s \equiv \sqrt{4\pi \alpha_s}$ is renormalized multiplicatively such that
\begin{equation}
g_s^0 = Z_{g_s} g_s \overset{\textrm{one-loop}}{=} 
  \left( 1 + \delta Z_{g_s} \right) g_s \quad .
\end{equation} 
The explicit form of $\delta Z_{g_s}$ depends on the renormalization scheme. 
Choosing the OS scheme for this task, however, is not possible. 
If the renormalization condition for $g_s$ is formulated
completely analogous to the renormalization of the electromagnetic coupling constant,
one must demand that the corrections to the gluon--quark--anti-quark vertex vanish
in the limit of zero-momentum transfer. To formulate this condition the value of
$g_s$ would be needed in a region which is below the QCD scale $\Lambda_{QCD}$.
Coming from values above, $g_s$ formally reaches infinity at this scale 
and perturbative methods are no longer defined. As the OS scheme is based on 
the validity of perturbation theory this would lead to a self-contradiction.

Instead another renormalization scheme must be used, which avoids the dependence
on $g_s$ at zero-momentum transfer. The \MSbar{} and \DRbar{} schemes share
this property. In these schemes the counter term $\delta Z_{g_s}$ is fixed by the
condition that the gluon--quark--anti-quark vertex is finite. Due to a Slavnov-Taylor 
identity, which guarantees the universality of $g_s$, this automatically results in 
finite three- and four-gluon vertices. The counter term has the following explicit form
\begin{equation}
\delta Z_{g_s} = - \frac{\alpha_s}{4 \pi} 
  \left(11 - \frac23 n_f - 2 - \frac13 n_f \right) \Delta \quad ,
\end{equation}
where the contributions to the sum originate from gluons, quarks, 
gluinos and squarks. $n_f=6$ denotes the number of quark flavors.
The last two terms originate from the supersymmetric particles and are 
not present in the Standard Model.

The behavior of $g_s$ with respect to higher-order corrections can be
improved by the use of renormalization group equations (RGE). The one-loop RGE 
sum up all leading-log contributions which have the 
form $g_s^{2n}\left(\mu_R \right) \left( \ln \mu_R\right)^n$.
Their application leads to the following expression for the strong coupling
constant\footnote{Here the formula for $\alpha_s$ is quoted as this constant is 
normally used in calculations and also the experimental value of the coupling constant 
is given in terms of $\alpha_s$. The corresponding expression for $g_s$ can simply be 
derived from the relation $\alpha_s\equiv \frac{g_s^2}{4 \pi}$.}~\cite{spa}:
\begin{equation}
\alpha_s^{\DRbar}\left( \mu_R \right) = 
  \frac{\alpha_s^{\DRbar}\left( m_Z \right)}%
    {1-\frac{3}{2\pi} \alpha_s^{\DRbar}\left( m_Z \right) \ln\frac{m_Z}{\mu_R}} \quad .
\label{renorm:alphasrun}
\end{equation}
The experimental value~\cite{Eidelman:2004wy} for $\alpha_s$ is given in the
$\MSbar${} scheme at the scale $m_Z$ and using the Standard Model RGE for
extracting $\alpha_s^{\MSbar}$ from the data. This must be converted to 
$\alpha_s^{\DRbar}$ via
\begin{gather}
\alpha_s^{\DRbar}\left( m_Z \right) = 
  \frac{\alpha_s^{\MSbar}\left( m_Z \right)}{1-\Delta \alpha_s}
\intertext{with}
\Delta \alpha_s = \frac{\alpha_s^{\MSbar}\left( m_Z \right)}{2\pi}
\left( \frac12 - \frac23 \ln\frac{m_t}{m_Z} - 2 \ln \frac{m_{\tilde g}}{m_Z} 
  - \frac16 \sum_{\mathrm{squarks}}\left( 
    \ln\frac{m_{\tilde q_1}}{m_Z} + 
    \ln\frac{m_{\tilde q_2}}{m_Z} \right)
  \right) \quad .
\end{gather}
The $\ln \frac{m}{m_Z}$ terms in the last equation decouple the particles heavier 
than $m_Z$ from the running of $\alpha_s$.

Also the finite part of the one-loop contribution to the gluon--quark--anti-quark 
vertex depends on the renormalization scale $\mu_R$. It should best be 
chosen in a way that the error, which is induced by missing higher-order corrections, 
is as small as possible~\cite{Alam:1996mh,*Sullivan:1996ry,Berge:1999}. 
Since $R$-parity is conserved in 
the MSSM, the one-loop diagrams decompose into two distinct sets, where the 
loop either consists solely of SUSY particles or does not contain any 
supersymmetric particles at all.
The latter ones form the corrections which also appear in the Standard Model. Except
for the top quark, which is decoupled, they take part in the running of $\alpha_s$. 
For these contributions the same renormalization scale $\mu_R$ should be used as in 
\eq{renorm:alphasrun} which is typically of the order of the energy scale of the
considered process. 

For the additional SUSY contributions another, special value $\tilde{\mu}_R$ is
chosen~\cite{Alam:1996mh,*Sullivan:1996ry,Berge:1999}. This is 
possible because the two sets of diagrams are distinct and all supersymmetric
particles are decoupled from the running of $\alpha_s$. The scale is chosen such 
that the contribution of these diagrams vanishes at zero-momentum transfer. 
Under this condition $g_s$ is taken at the scale $\mu_R$, so the procedure 
is well-defined. It is fulfilled if
\begin{equation}
2 \ln\frac{m_{\tilde g}^2}{\tilde{\mu}_R^2} + 
  \frac16 \sum_{\mathrm{squarks}} \left( 
    \ln\frac{m_{\tilde q_1}^2}{\tilde{\mu}_R^2} +
    \ln\frac{m_{\tilde q_2}^2}{\tilde{\mu}_R^2}
  \right) = 0 \quad .
\end{equation}
Solving for $\tilde{\mu}_R$ yields
\begin{equation}
\tilde{\mu}_R = \sqrt{\smash[b]{m_{\tilde g}}}
  \prod_{\mathrm{squarks}} \left(m_{\tilde q_1} m_{\tilde q_2} \right)^{\frac1{24}}
 \quad .
\label{renorm:tildemur}
\end{equation}
This procedure reduces the numerical value of the one-loop corrections and therefore 
makes the calculation more stable against the theoretical uncertainty from missing
higher-order terms.

\section{Bottom-quark Yukawa Coupling}
\label{renorm:deltamb}
% large SUSY-QCD corrections due to the m_b counter term: 
% Origin: b -> g~ b~ -> b
% contributions up to O(1)
% pert. theory?
% no higher order contrib.
% => resum
% redefinition of Yukawa coupling
% for comparison with one-loop expr. unresummed
% 
% bbh vertices: add also terms prop. tan(alpha)
% similar for tth, but suppressed with 1/tan(beta)

The mass of the bottom quark and its Yukawa coupling to the Higgs particles are
intimately related. They originate from the same term in the unbroken Lagrangian.
After the Higgs fields have acquired a vacuum expectation value, 
the vacuum-expectation-value component yields the mass
term of the bottom quark in the Lagrangian and the other components describe the
Yukawa coupling of the bottom quark to the various Higgs particles. This relation
can be modified by loop corrections, and it turns out that these are very large in 
the case of bottom quarks~\cite{Carena:1999py,Guasch:2001wv}. 
A resummation of the leading corrections to all orders
in perturbation theory can be performed, which greatly reduces the theoretical
uncertainty originating from unknown higher-order corrections.

\begin{figure}
\begin{center}
 \includegraphics[scale=0.6]{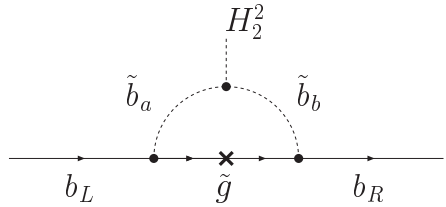}
\end{center}
\caption{One-loop SUSY-QCD diagram mediating an effective coupling between
the bottom quark and $H_2^2$. The cross in the gluino line here represents a mass
insertion, i.e.\ the $m_{\tilde g}$ term is chosen when computing the trace over 
the fermion line. The subscripts $a$ and $b$ of the sbottom particles take the 
values $1$ and $2$.}
\label{renorm:diag}
\end{figure}

At tree-level the bottom quark only couples to the first Higgs doublet $H_1$ as can be 
seen from the superpotential \eq{susy:superpotential}. A coupling to the second 
one $H_2$ is forbidden. Such a coupling can, however, be generated dynamically
at the one-loop level. Taking into account only SUSY-QCD corrections, i.e.\ corrections
with squarks and gluinos, this is done by the single diagram \fig{renorm:diag}.
Although this contribution is loop-suppressed, it can induce a potentially large 
shift in the tree-level relations, because it is enhanced by $t_\beta$.
By electroweak symmetry breaking the Higgs field $H_2^2$ acquires a 
vacuum expectation value $v_2$ and firstly we will consider only this part.
On tree-level the bottom-quark mass and its Yukawa coupling $\lambda_b$ are 
related via
\begin{equation}
m_b = \lambda_b v_1 .
\end{equation}
Adding the vacuum-expectation-value contribution 
from \fig{renorm:diag} changes this equation to
\begin{equation}
m_b = \lambda_b v_1 + \Delta\lambda_b v_2 
  = v_1 (\lambda_b + \Delta\lambda_b t_\beta)
  = \lambda_b v_1 ( 1 + \Delta m_b) \quad .
\end{equation}
As the numerical value of $m_b$ is fixed by experiments, this results in a change
of the effective Yukawa coupling of the bottom quark
\begin{equation}
\lambda_b = \frac{m_b}{v_1} \frac1{1 + \Delta m_b} \quad .
\label{renorm:lambdab}
\end{equation}

Computing the diagram in \fig{renorm:diag} in the limit of vanishing 
external momentum yields the following explicit form for $\Delta m_b$:
\begin{equation}
\Delta m_b = \frac{2 \alpha_s}{3 \pi} m_{\tilde g} \mu t_\beta
I\left(m_{{\tilde b}_1}, m_{{\tilde b}_2}, m_{\tilde g}\right)
\label{renorm:deltambeq}
\end{equation}
with
\begin{align}
I\left(m_{{\tilde b}_1}, m_{{\tilde b}_2}, m_{\tilde g}\right) =& 
- \left( 
  m_{{\tilde b}_1}^2 m_{{\tilde b}_2}^2 
    \ln\frac{m_{{\tilde b}_1}^2}{m_{{\tilde b}_2}^2} + 
  m_{{\tilde b}_2}^2 m_{\tilde g}^2
    \ln\frac{m_{{\tilde b}_2}^2}{m_{\tilde g}^2} + 
  m_{\tilde g}^2 m_{{\tilde b}_1}^2
    \ln\frac{m_{\tilde g}^2}{m_{{\tilde b}_1}^2}
\right) \nonumber\\
&\times \frac1{\left( m_{{\tilde b}_1}^2 - m_{{\tilde b}_2}^2 \right)
      \left( m_{{\tilde b}_2}^2 - m_{\tilde g}^2 \right)
      \left( m_{\tilde g}^2 - m_{{\tilde b}_1}^2 \right)
    } 
\label{renorm:integralI}
\end{align}
and $\mu$ denoting the MSSM parameter which couples the two Higgs doublets.
In the limit where the squark and gluino masses have approximately the same value,
denoted by a common SUSY mass $m_{\text{SUSY}}$, the last equation simplifies to
\begin{equation}
I\left(m_{\text{SUSY}},m_{\text{SUSY}},m_{\text{SUSY}}\right) =
  \frac{1}{2 m_{\text{SUSY}}^2} \quad .
\end{equation}
If additionally $\mu$ is of comparable size, this results in
\begin{equation}
\Delta m_b = \sign(\mu) \frac{\alpha_s(\mu_R=m_{\text{SUSY}})}{3 \pi} t_\beta \quad .
\end{equation}
So for large values of $t_\beta$ this effect can be of $\Order{1}$ and does not
vanish for heavy SUSY spectra.

For computations up to one-loop order \eq{renorm:lambdab} can be expanded
so that it contains only corrections up to $\Order{\alpha_s}$. The equation is then
modified and reads
\begin{equation}
\lambda_b = \frac{m_b}{v_1} \left( 1 - \Delta m_b \right) \quad .
\label{renorm:noresum}
\end{equation}
So for large absolute values of $\Delta m_b$, which are phenomenologically very
interesting, huge one-loop corrections appear. If $\Delta m_b$ exceeds one, the 
standard way of computing one-loop cross sections by adding the interference
term between tree-level and one-loop diagrams even yields negative total 
cross sections which are obviously wrong.
One might even question if perturbation theory is still valid in this regime, but 
definitely higher-order calculations would be needed to reduce the theoretical
uncertainty.

This problem is solved by the observation that these corrections do not 
appear at higher orders. In ref.~\cite{Carena:1999py} it was proven
that there are no contributions to $\Delta m_b$ of 
\begin{equation}
\Order{\left( \alpha_s \frac{\mu}{m_{\text{SUSY}}} t_\beta \right)^n}
\label{renorm:higherorder}
\end{equation}
for $n>1$. Higher-order corrections either lack the enhancement factor
$t_\beta$ or are suppressed by a mass ratio $\frac{m_b}{m_{\text{SUSY}}}$.
Therefore $\Delta m_b$ is a one-loop exact quantity and including it as in
\eq{renorm:lambdab} contains the corrections to all orders in $\alpha_s$ which have
the form given in \eq{renorm:higherorder}.

Using the resummed form \eq{renorm:lambdab} is only useful when computing total
cross sections. For a comparison with one-loop cross sections it is necessary
to use \eq{renorm:noresum} so that the same order in $\alpha_s$ is taken into account
in both calculations. This will be explained in more detail in \chap{bbWH}, where this
procedure is applied to a physical process. 

The $\Delta m_b$ corrections are universal. They occur in every coupling of 
the bottom quark to the different Higgs particles, both neutral and charged ones. 
They are also independent of the kinematic configuration. 

%%% Additional mod to b b h^0
When the bottom quark couples to the physical Higgs fields an additional term
occurs. It also originates from diagram \fig{renorm:diag}, but now not the coupling
to the vacuum expectation value 
but to the remaining neutral Higgs field $\phi_2^0$ is considered.
In addition to the tree-level coupling to the first Higgs doublet
\begin{equation}
\Gamma_{b\bar{b}\phi_1^0} = \lambda_b = \frac{m_b}{v_1}
\end{equation}
this induces another term
\begin{equation}
\Gamma_{b\bar{b}\phi_2^0} = \lambda_b \Delta \lambda_b 
  = \frac{m_b}{v_1} \frac{\Delta m_b}{t_\beta} \quad .
\end{equation}
After electroweak symmetry breaking the fields $\phi_1^0$ and 
$\phi_2^0$ must be rotated by the angle $\alpha$ to form the two CP-even mass
eigenstates $h^0$ and $H^0$. Combining everything this leads to
the following effective couplings of the bottom 
quark~\cite{Carena:1999py,Carena:1998gk,*Carena:1999bh}
\begin{align}
\Gamma_{b\bar{b}h^0} =& 
  \Gamma^0_{b\bar{b}h^0} \frac{1}{1+\Delta m_b} 
    \left( 1 - \frac{\Delta m_b}{t_\beta t_\alpha}\right) \\
\Gamma_{b\bar{b}H^0} =& 
  \Gamma^0_{b\bar{b}H^0} \frac{1}{1+\Delta m_b} 
    \left( 1 + \frac{\Delta m_b t_\alpha}{t_\beta}\right) \quad ,
\end{align}
where $\Gamma^0$ denotes the respective tree-level coupling.
Expanding these equations up to the one-loop order yields
\begin{align}
\Gamma^1_{b\bar{b}h^0} =& 
  \Gamma^0_{b\bar{b}h^0}
    \left( 1 - \Delta m_b \left(1 + \frac{1}{t_\beta t_\alpha} \right) \right) 
  \label{renorm:bbh}\\
\Gamma^1_{b\bar{b}H^0} =& 
  \Gamma^0_{b\bar{b}H^0}
    \left( 1 - \Delta m_b \left( 1 - \frac{t_\alpha}{t_\beta} \right) \right) \quad .
\end{align}

In the coupling of the top quark to the Higgs fields a similar effect occurs. 
On tree-level the top quark couples only to $H_2$ and a coupling to the 
second doublet $H_1$ is generated perturbatively. This results
in a modified Yukawa coupling which is given by
\begin{equation}
\lambda_t = \frac{m_t}{v_2}\frac{1}{1+\Delta m_t} \quad ,
\end{equation}
in complete analogy to \eq{renorm:lambdab}.
The correction term $\Delta m_t$ has the form~\cite{Dobado:2001mq}
\begin{equation}
\Delta m_t = \frac{2 \alpha_s}{3 \pi} m_{\tilde g} \mu \frac1{t_\beta} 
I\left(m_{{\tilde t}_1}, m_{{\tilde t}_2}, m_{\tilde g}\right) \quad .
\label{renorm:deltamt}
\end{equation}
In contrast to $\Delta m_b$ this equation has a suppression factor 
of $\frac1{t_\beta}$. Therefore its numerical impact is much smaller
than the $t_\beta$-enhanced bottom-quark correction and it is 
largest for small values of $t_\beta$. The contribution of this
correction is nevertheless significant and therefore it is justified
to include its effect in the same way as for the bottom-quark
correction.

%%% Additional mod to t t h^0; see above bbh-case
Also the coupling of the top quark to the physical Higgs particles 
gets an additional contribution from the coupling to the $H_1$ 
doublet. In this case the modified couplings are
\begin{align}
\Gamma_{t\bar{t}h^0} =& 
  \Gamma^0_{t\bar{t}h^0} \frac{1}{1+\Delta m_t} 
    \left( 1 - \Delta m_t t_\beta t_\alpha \right) \\
\Gamma_{t\bar{t}H^0} =& 
  \Gamma^0_{t\bar{t}H^0} \frac{1}{1+\Delta m_t} 
    \left( 1 + \frac{\Delta m_t t_\beta}{t_\alpha}\right) \quad ,
\end{align}
where $\Gamma^0$ denotes the respective tree-level coupling.
An expansion up to the one-loop order yields
\begin{align}
\Gamma^1_{t\bar{t}h^0} =& 
  \Gamma^0_{t\bar{t}h^0}
    \left( 1 - \Delta m_t \left(1 + t_\beta t_\alpha \right) \right) 
  \label{renorm:tth}\\
\Gamma^1_{t\bar{t}H^0} =& 
  \Gamma^0_{t\bar{t}H^0}
    \left( 1 - \Delta m_t \left( 1 - \frac{t_\beta}{t_\alpha} \right) \right) \quad .
\end{align}

\chapter{Hadronic Cross Sections}
\label{hadWQ}

The cross sections which are obtained by applying the Feynman rules
contain, amongst other particles, quarks and gluons. The leading
interaction between these particles is the strong interaction, which 
is described by quantum-chromo dynamics (QCD). This theory 
possesses two characteristic properties: asymptotic 
freedom~\cite{Gross:1973id,*Politzer:1973fx} and confinement.
Asymptotic freedom describes the behavior of the theory at small
distances. In this region the interaction is weak and the coupling constant
gets smaller with decreasing distance or, equivalently, with rising energy.
At large distances confinement 
appears, because the interaction becomes strong and binds the particles 
tightly together. If the space between them becomes even larger, it is energetically 
favorable to form new quark--anti-quark pairs. One consequence of this 
behavior is that quarks and gluons cannot be observed as free particles,
but only as constituents of hadrons, i.e. mesons, which are 
quark--anti-quark pairs, and baryons, which are states of three quarks or 
three anti-quarks. An example for these hadrons are protons, which are the 
colliding particles at the LHC.
To make theoretical predictions it is necessary to relate the interactions
at the parton level to the interactions at the hadron 
level~\cite{Brock:1993sz,*Brock:1994er}.
The basis for doing this is the parton 
model~\cite{Bjorken:1969ja,*Feynman:1969ej}, which will be described
in the next section.

\section{Parton Model}
The parton model describes the inner structure of hadrons in hard collisions.
It starts from the assumption that every observable hadron consists
of constituents, the so-called partons, which can be identified as 
quarks and gluons. Experimental evidence for this assumption comes
from the observation of 
scaling~\cite{Bloom:1969kc,*Breidenbach:1969kd,*Friedman:1972sy}
in deep inelastic electron-proton-scattering.
If the hadron carries some momentum ~$P^\mu$, the partons which 
take part in the partonic subprocess have momentum~$x P^\mu$ 
with $x \in [0,1]$. As normally the mass of the hadrons is small compared to
their kinetic energy one can assume $P^2 = 0$.

% high- and low-energy part
The interaction of an electron and a hadron or of two hadrons among each other
can be split into two parts. Because of Lorentz contraction and time dilation
the interaction time of the two incoming particles in the laboratory frame
is very short. Therefore effectively a static hadron is seen.
For the hard scattering process interactions between partons of the same hadron
need not be considered. Also the process of hadronization after the 
interaction happens on time scales which are much larger than the
interaction itself.

From this the theorem of factorization~\cite{Collins:1989gx} follows immediately.
It states that all diagrammatic contributions to the structure functions can be 
separated into a product of two functions $C$ and $f$, which depend on two
mass scales $\mu_R$ and $\mu_F$.
$\mu_R$ is the renormalization scale which was already defined in 
\chap{renorm}, $\mu_F$ is the so-called factorization scale and
separates the long-distance from the short-distance effects.
Slightly simplifying one can say that every parton propagator which 
is off-shell by $\mu_F$ or more contributes to $C$, while those 
which are below this value contribute to $f$.

\section{Integrated Hadronic Cross Sections}
% Hochenergie -> Feynmandiagramme
The hard scattering process $C$ therefore can be calculated in perturbation
theory by Feynman rules, using partons as incoming particles. 
It is independent of long-distance effects and especially from the type
of the colliding hadron.

% Niederenergie -> PDF, Altarelli-Parisi DGLAP
The parton distribution function (PDF) $f_{i/h}(x,\mu_F)$ 
contains the long-distance effects.
It is independent of the underlying scattering process, but depends
on $\mu_F$ and the type of hadron~$h$. It is normalized such that
it can also be interpreted as a probability density, namely the probability
of finding the parton $i$ in the hadron $h$ with a momentum $x P^\mu$.
Its behavior as a function of the parameters is determined by the
Altarelli-Parisi integro-differential 
equations~\cite{Altarelli:1977zs}. Its numerical value, however,
cannot be calculated a priori from the theory. At a single reference point it 
must be determined by experiments.

% Formel: Faltung mit PDFs
% -> integrierte WQ
Therewith one obtains the expression~\cite{Brock:1993sz,*Brock:1994er}
\begin{align}
 \sigma_{pp \rightarrow fin + X} =& 
  \sum_{\{m,n\}}
  \int_{\tau_0}^1 \di \tau \frac{\di \La}{\di \tau} 
  \hat{\sigma}_{m n \rightarrow fin}
   \left(\tau S,\alpha_s(\mu_R)\right) \label{inthadWQ}
\end{align}
for an integrated hadronic cross section with the parton luminosity
\begin{align}
 \frac{\di \La}{\di \tau} =& \int_\tau^1 \frac{\di x}{x} 
   \frac1{1+\delta_{mn}} \cdot\nonumber \\
  & \qquad \cdot \left( f_{m/p}\left(x,\mu_F\right) 
      f_{n/p}\left(\frac{\tau }{x},\mu_F\right) + 
    f_{n/p}\left(x,\mu_F\right) f_{m/p}\left(\frac{\tau}{x},\mu_F\right) 
   \right) \label{hadWQ:partonlumi}.
\end{align}
Here $\sqrt{S}$ denotes the hadronic center-of-mass energy, i.e.\ the 
one of the two colliding protons, and
$\hat{\sigma}_{m n \rightarrow fin}$ 
the partonic cross sections of the subprocesses, where the two incoming 
partons $m$ and $n$ produce some final state, labeled $fin$. The sum includes
all possible parton combinations $m$ and $n$ where the order 
of appearance is not taken into account. The integration variable 
$\tau$ relates the partonic and hadronic center-of-mass energies
with each other. More specifically, $\sqrt{\tau}$ can be interpreted
as the part of the hadronic center-of-mass energy which takes part
in the partonic subprocess, as the partonic center-of-mass energy
is given by $\sqrt{\hat{s}} = \sqrt{\tau S}$. The lower limit of the 
integral $\tau_0$ is determined by the kinematic configuration. $\sqrt{\tau_0
S}$ is the minimal energy which is necessary to produce the final state $fin$
and therefore denotes the production threshold.

% auch 2->1 moeglich
The formula given above is valid for processes with two or more
particles in the final state.
For hadronic cross sections it is also possible to calculate integrated
cross sections for $2\rightarrow1$ processes. One first obtains
for the partonic cross section of the process $mn \rightarrow f$
\begin{align}
\di \hat\sigma_{mn\rightarrow f} = \frac{\pi}
                       {4 p_f^0 \sqrt{\hat s} |\vec{p}_{m}|}
                  |\M_{fi}\left(mn\rightarrow f\right)|^2
                  \delta \left(p_m^0+p_n^0-p_f^0\right) \quad .
\end{align}
Again $m$ and $n$ specify the incoming partons, $f$ denotes
the outgoing particle, $m_f$ its mass, 
and $p_i^0$ the energy of the respective particle $i$.
$\vec{p}_{m}$ indicates the three-momentum of 
particle~$m$ in the partonic center-of-mass system and 
$\Mfi$ is the matrix element.

When convoluting with the parton distribution functions the single
remaining $\delta$-function in the partonic cross sections solves
the $\tau$ integral in \eq{inthadWQ} analytically.
Thus one obtains for the integrated hadronic cross section 
\begin{align}
\sigma_{pp\rightarrow f} = \sum_{\{m,n\}} 
  \left.\frac{\di \La}{\di \tau}\right|_{\tau =
    \frac{m_f^2}{S}}
         \frac{\pi}{2 m_f S |\vec{p}_{m}|} 
    |\M_{fi}\left(mn\rightarrow f\right)|^2 \quad .
\end{align}

\section{Differential Hadronic Cross Sections}
% differential hadronic cross sections
Additionally one can define hadronic cross sections that are differential
in one or more parameters. For these parameters it is useful to
take variables that are either invariant under Lorentz transformations
or at least have very simple transformation properties. In this thesis
three differential hadronic cross sections are presented which are also implemented in 
the \HadCalc{} program that is described below in section~\ref{hadWQ:hadcalc}. 
They are cross 
sections differential with respect to the invariant mass of the final-state particles, 
the rapidity of one final-state particle and, thirdly, the transverse momentum.

\subsection{Invariant Mass}

The first differential hadronic cross section is the one with respect to the invariant mass
of the final-state particles. The invariant mass of a process is equivalent to the partonic
center-of-mass energy $\sqrt{\hat s} \equiv \sqrt{\tau S}$ of the process or, in other
words, the sum of the final-state momenta of the outgoing particles.
The differential cross section takes the form
\begin{equation}
\frac{\di \sigma_{pp\rightarrow fin}}{\di \sqrt{\hat s}} = 
  4 \pi \frac{\sqrt{\hat s}}{S} 
  \sum_{\{m,n\}} \left. \frac{\di \La}{\di \tau} \right|_{\tau =
    \frac{\hat{s}}{S}}
  \hat\sigma_{mn \rightarrow fin} \quad ,
\end{equation}
where $fin$ again labeles a general final state.

\subsection{Rapidity}
% Two rapidities: normal and pseudo; transformation into each other

The rapidity $y$ of a particle is defined as
\begin{equation}
y = \artanh \frac{p_z}{p^0} \equiv \frac12 \ln \frac{p^0 + p_z}{p^0 - p_z}
\end{equation}
where $p_z = \vec p \cdot c_{\theta}$ denotes the fraction of the 
particle's three-momentum $\vec p$ that goes in the direction of 
the beam axis, labeled $z$. The mass of the particle will later be referred
to as $m$.
Using the rapidity instead of directly taking the angle 
$\theta$ between the particle and the beam axis 
possesses some advantages because the rapidity of a 
particle has a few useful properties. Under a boost in the $z$-direction
to a frame with a velocity $\beta$, the rapidity transforms as
$y \rightarrow y - \artanh \beta$. Thus the shape of the rapidity distribution
$\frac{\di \sigma}{\di y}$ stays unchanged. More generally, the sum of two
rapidities when the momenta point into the same direction is given by the
rapidity of the sum of the momenta, added via the formula for the relativistic
addition of velocities: 
$y\left(p_1\right) + y\left(p_2\right) = y \left(\frac{p_1+p_2}{1+ p_1 p_2}\right)$.
In experimental analyses often a slightly different measure, the pseudo-rapidity $\eta$,
is used. It is derived from the standard rapidity by taking the limit of a vanishing mass 
of the particle and is defined as
\begin{equation}
\eta = \frac12 \ln \frac{1+c_\theta}{1-c_\theta} \quad .
\end{equation}
In the \HadCalc{} program both normal rapidity and pseudo-rapidity are implemented.
As conversion between both variables can be performed by the simple transformation
\begin{equation}
y = \artanh\left(\sqrt{1-\frac{m^2}{{\vec p\,}^2+m^2}} \tanh \eta \right) \quad ,
\end{equation}
in the following only the shorter expressions for the standard rapidity are given. The
ones for pseudo-rapidity can then be deduced from them.

Using the above-mentioned definition of the rapidity the differential hadronic 
cross section with respect to the rapidity for $2 \rightarrow 2$ processes then reads
\begin{equation}
\frac{\di \sigma}{\di y} = \int_{\tau_0}^1 \di \tau 
  \frac{\di {\cal L}}{\di \tau} 
  \frac{\di  \hat\sigma}{\di  c_{\hat{\theta}}}
  \frac{\partial c_{\hat{\theta}}}{\partial y} \quad .
\label{hadWQ:haddiff_y}
\end{equation}
The momenta and masses given in the formulae always refer to the particle
for which the rapidity distribution is calculated.
The angle $c_{\hat\theta}$ between the particle and the beam axis in the
partonic center-of-mass system is fixed by the relation
\begin{equation}
c_{\hat{\theta}} = \sqrt{1+\frac{m^2}{\pvechat^2}}
 \tanh\left(y+\frac12\ln\frac{x^2}{\tau}\right)
\label{hadWQ:costheta_y}
\end{equation}
where the second term in the argument of $\tanh$ originates from the
boost from the hadronic center-of-mass system, which is the laboratory
frame, to the partonic one, in which the partonic subprocess is calculated.
This leads to
\begin{equation}
\frac{\partial c_{\hat{\theta}}}{\partial y} = 
 \sqrt{1+\frac{m^2}{\pvechat^2}}
 \frac1{\cosh^2\left(y+\frac12\ln\frac{x^2}{\tau}\right)} \quad .
\end{equation}

For processes with three or more particles in the final state the formula is
very similar. Additional phase-space integrals appear for the further
particles but otherwise \eq{hadWQ:haddiff_y} stays unchanged.
In the following equation the differential cross section for a $2\rightarrow3$ process
is given
\begin{equation}
\frac{\di \sigma}{\di y} = \int_{\tau_0}^1 \di \tau 
  \frac{\di {\cal L}}{\di \tau} 
   \int \di  k_3^0 \int\di  k_5^0 \int\di \hat\eta
  \frac{\di \hat\sigma}%
       {\di k_3^0 \di c_{\hat{\theta}}\di k_5^0 \di \hat\eta} 
  \frac{\partial c_{\hat{\theta}}}{\partial y} \quad .
\end{equation}
The parametrization of the three-particle phase space is described
in \app{ps:3}.

\subsection{Transverse Momentum}

The last implemented differential hadronic cross section is the one
with respect to the transverse momentum $p_T=\sqrt{p_x^2+p_y^2}$ 
of one of the final
state particles. For $2\rightarrow2$ processes it is defined as
\begin{equation}
\frac{\di \sigma}{\di p_T} = \int_{\tilde{\tau}_0}^1 \di \tau 
  \frac{\di{\cal L}}{\di \tau} 
  \frac{\di \hat\sigma}{\di c_{\hat{\theta}}}
  \frac{\partial c_{\hat{\theta}}}{\partial p_T} 
\end{equation}
with
\begin{equation}
\frac{\partial c_{\hat{\theta}}}{\partial p_T} = 
 \frac1{\sqrt{\frac{\pvechat^4}{p_T^2}-\pvechat^2}}
\end{equation}
which follows from
\begin{equation}
{c_{\hat{\theta}}}_\pm = 
 \pm \sqrt{1-\frac{p_T^2}{\pvechat^2}} \quad .
\label{hadWQ:costheta_pT}
\end{equation}
Here two possible solutions arise because of the sign ambiguity when
taking the square root. In principle both solutions have to be taken
into account and added up unless they are excluded by other constraints 
as shown below.
The lower limit of the $\tau$-integral $\tilde{\tau}_0$ must be
adjusted such that $s_{\hat{\theta}}$ is always inside its co-domain $[0;1]$
\begin{equation}
\tilde{\tau}_0 = \frac{\left( \sqrt{m_{f_1}^2 + p_T^2} 
                            + \sqrt{m_{f_2}^2 + p_T^2} \right)^2}{S} \quad ,
\end{equation}
$f_1$ and $f_2$ denoting the two final state particles.

For $2\rightarrow3$ processes the extension to include the third final-state particle is
straightforward. The lower limit for $\tau$ in these processes is
\begin{equation}
\tilde{\tau_0} = \frac{\left( \sqrt{m_{f_1}^2 + p_T^2}
  + \sqrt{\left( m_{f_2} + m_{f_3} \right)^2 + p_T^2}\right)^2}{S} \quad ,
\end{equation}
when the cross section is differential in the particle $f_1$.
Therefore the expression for the differential cross section reads
\begin{equation}
\frac{\di \sigma}{\di  p_T} = \int_{\tilde{\tau_0}}^1 \di \tau
  \frac{\di {\cal L}}{\di \tau}
   \int \di  k_3^0 \int\di k_5^0 \int\di \hat\eta
  \frac{\di \hat\sigma}%
       {\di k_3^0 \di c_{\hat{\theta}}\di k_5^0 \di \hat\eta}
  \frac{\partial c_{\hat{\theta}}}{\partial p_T} \quad .
\end{equation}

\section{Cuts}
\label{hadWQ:cuts}
%%% Cuts on rapidity and transverse momentum can be transformed into
%%%  limits on theta
%%% This then transformed into cut on integration variables
%%% Emphasize more this game with the regions: We can be sure so that there 
%%%  aren't any holes in the integration region

In order to improve the ratio of the signal-process cross section to that of 
the background processes it can be useful to place appropriate cuts on the final-state
particles. Also experimental techniques used in the reconstruction of events like
jet-clustering algorithms can mandate the use of cuts in theoretical predictions, so that
the behavior of these techniques is emulated there. 

In the \HadCalc{} program 
cuts on three different properties of the final-state particles are
implemented~\cite{Meier:2001da}.
The first two are cuts on the rapidity and the transverse momentum of a particle.
The definition of these two variables was already presented in the previous section.
The third one is a mutual property of two particles, the jet separation $\Delta R_{ij}$,
which is defined as
\begin{equation}
\Delta R_{ij} = \sqrt{\Delta y_{ij}^2 + \Delta \phi_{ij}^2} \quad .
\end{equation}
$\Delta y_{ij}$ denotes the rapidity difference between the two particles $i$ and $j$
and $\Delta \phi_{ij}$ the difference in the azimuthal angles of the two particles in 
the transverse plane. 
Its main use are exclusive hadronic cross sections where final-state jets shall
be observed explicitly. It mimics the behavior of jet-clustering algorithms. There two jets, which
are separated by a jet separation below a certain limit, are seen in the reconstruction 
as a single jet which has kinematic properties that are averaged over the two final-state
partons.

For the first two cut parameters, rapidity and transverse momentum, it is possible
to translate these cuts into a limit on the integration parameters of the phase space. The
most general case is assumed here that cuts on both the rapidity $y_{\text{cut}}$ 
and the transverse momentum ${p_T}_{\text{cut}}$ of a particle shall be applied. 
Using \eq{hadWQ:costheta_pT} the transverse-momentum cut can be
translated into a cut on $c_{\hat \theta}$ and one obtains
\begin{equation}
c_{{\hat{\theta}}_{p_T}}^{\text{min}} 
 \equiv - \sqrt{1-\frac{{p_T}_{\text{cut}}^2}{\pvechat^2}}
 < c_{\hat{\theta}}
 < \sqrt{1-\frac{{p_T}_{\text{cut}}^2}{\pvechat^2}}
 \equiv c_{{\hat{\theta}}_{p_T}}^{\text{max}} \quad .
\end{equation}
Likewise, the cut on the rapidity can also be turned into a cut on $c_{\hat \theta}$
via \eq{hadWQ:costheta_y}, yielding
\begin{align}
c_{\hat{\theta}} 
 &> \sqrt{1+\frac{m^2}{\pvechat^2}}
     \tanh \left(-y_{\text{cut}} + \frac12 \ln\frac{x^2}{\tau} \right)
 \equiv c_{{\hat{\theta}}_{y}}^{\text{min}} \nonumber \\
c_{\hat{\theta}} 
 &< \sqrt{1+\frac{m^2}{\pvechat^2}}
     \tanh \left( y_{\text{cut}} + \frac12 \ln\frac{x^2}{\tau} \right)
 \equiv c_{{\hat{\theta}}_{y}}^{\text{max}} \quad .
\end{align}
To shorten the notation the abbreviation
\begin{equation}
r = \sqrt{\frac{1-\frac{{p_T}_{\text{cut}}^2}{\pvechat^2}}%
               {1+\frac{m^2}{\pvechat^2}} }
\label{hadWQ:abbrr}
\end{equation}
is used in the following.
Again the momenta and mass used in the equations all refer to the particle
whose phase space should be constrained.

Applying both cuts requires that the conditions on $c_{\hat \theta}$ are all
fulfilled simultaneously. This also restricts the integral on $x$ which appears
in the parton luminosity given in \eq{hadWQ:partonlumi}. 
In total the $x$-interval divides into five different regions, which will be
labeled by roman numbers.
First the two cases where both cuts cannot be fulfilled
simultaneously, are considered, because the lower limit of one cut lies above the 
upper limit of the other one:
\begin{align} 
\text{Region I:}\qquad& c_{{\hat{\theta}}_{y}}^{\text{max}} 
                  \le c_{{\hat{\theta}}_{p_T}}^{\text{min}} &
\Rightarrow \qquad x &\le \sqrt{\tau} e^{-y_{\text{cut}}} \sqrt{\frac{1-r}{1+r}}
 \equiv x_{\text{I}} \\
\text{Region V:}\qquad& c_{{\hat{\theta}}_{y}}^{\text{min}} 
                  \ge c_{{\hat{\theta}}_{p_T}}^{\text{max}} &
\Rightarrow \qquad x &\ge \sqrt{\tau} e^{y_{\text{cut}}} \sqrt{\frac{1+r}{1-r}}
 \equiv x_{\text{V}} \quad .
\end{align} 
These two regions are excluded and the cross section vanishes there.

For specifying the other regions first two special cases are considered, where
the lower limits on $c_{\hat{\theta}}$ and the upper limits, respectively,
coincide. For these cases the according value of $x$ is determined
\begin{align}
c_{{\hat{\theta}}_{y}}^{\text{min}} &=
        c_{{\hat{\theta}}_{p_T}}^{\text{min}} & 
\Rightarrow \qquad x &= \sqrt{\tau} e^{y_{\text{cut}}} \sqrt{\frac{1-r}{1+r}}
  \equiv x_{\text{min}} \\
c_{{\hat{\theta}}_{y}}^{\text{max}} &=
        c_{{\hat{\theta}}_{p_T}}^{\text{max}} &
\Rightarrow \qquad x &= \sqrt{\tau} e^{-y_{\text{cut}}} \sqrt{\frac{1+r}{1-r}}
  \equiv x_{\text{max}} \quad .
\end{align}

Using these two definitions the other intermediate regions can be specified.
The ranges for $c_{\hat{\theta}}$ which are deduced from these 
following regions specify the allowed area where the cuts are fulfilled and 
therefore the cross section does not vanish.
The next two regions handle the cases where the limits on $c_{\hat \theta}$
from rapidity and transverse momentum overlap and one limit is given by 
the rapidity cut and the other one by the transverse-momentum cut:
\begin{align}
\text{Region II:}\qquad&
c_{{\hat{\theta}}_{y}}^{\text{min}} 
 \le c_{{\hat{\theta}}_{p_T}}^{\text{min}}
 < c_{\hat{\theta}}
 < c_{{\hat{\theta}}_{y}}^{\text{max}} 
 \le c_{{\hat{\theta}}_{p_T}}^{\text{max}} &
\Rightarrow \quad &x_{\text{I}}<x<\min(x_{\text{min}},x_{\text{max}}) \\
\text{Region IV:}\qquad& 
c_{{\hat{\theta}}_{p_T}}^{\text{min}}
 \le c_{{\hat{\theta}}_{y}}^{\text{min}} 
 < c_{\hat{\theta}}
 < c_{{\hat{\theta}}_{p_T}}^{\text{max}} 
 \le c_{{\hat{\theta}}_{y}}^{\text{max}} &
\Rightarrow \quad &\max(x_{\text{min}},x_{\text{max}})<x<x_{\text{V}} \quad .
\end{align}

Finally the definition of the last region is the case whether one
cut gives a range on $c_{\hat \theta}$ that completely lies inside
the other one. Depending on which cut this is, the limits on $x$ are
different:
\begin{align}
\text{Region III a):} \qquad&
 c_{{\hat{\theta}}_{p_T}}^{\text{min}}
 \le c_{{\hat{\theta}}_{y}}^{\text{min}} 
 < c_{\hat{\theta}}
 < c_{{\hat{\theta}}_{y}}^{\text{max}} 
 \le c_{{\hat{\theta}}_{p_T}}^{\text{max}} &
\Rightarrow \quad & x_{\text{min}}<x<x_{\text{max}} \quad \\
\text{Region III b):} \qquad&
c_{{\hat{\theta}}_{y}}^{\text{min}} 
 \le c_{{\hat{\theta}}_{p_T}}^{\text{min}}
 < c_{\hat{\theta}}
 < c_{{\hat{\theta}}_{p_T}}^{\text{max}} 
 \le c_{{\hat{\theta}}_{y}}^{\text{max}} &
\Rightarrow \quad &x_{\text{max}}<x<x_{\text{min}} \quad .
\end{align}

In addition to those regions the original constraint for $x$ for a hadronic cross
section without cuts applies:
\begin{equation}
\tau < x < 1 \quad .
\end{equation}
Combining the result of all regions one can see that no holes in the integration
over $x$ or $c_{\hat{\theta}}$ appear and the final borders of the integration
routine can be simplified to 
\begin{equation}
\max(\tau,x_{\text{I}}) < x < \min(x_{\text{V}},1)
\label{hadWQ:x_limit}
\end{equation}
and 
\begin{equation}
\max(c_{{\hat{\theta}}_{p_T}}^{\text{min}},
     c_{{\hat{\theta}}_{y}}^{\text{min}})
 < c_{\hat{\theta}}
 < \min(c_{{\hat{\theta}}_{p_T}}^{\text{max}},
        c_{{\hat{\theta}}_{y}}^{\text{max}}) \quad .
\label{hadWQ:costheta_limit}
\end{equation}

For a cross section which is differential with respect to the rapidity
of a final state particle the cut on the transverse momentum yields
a restriction on $c_{\hat{\theta}}$ in the same way as in 
\eq{hadWQ:costheta_limit}
\begin{equation}
c_{{\hat{\theta}}_{p_T}}^{\text{min}}
 < c_{\hat{\theta}}
 < c_{{\hat{\theta}}_{p_T}}^{\text{max}} \quad .
\end{equation}
The constraint on $x$ must then be adjusted such that 
$c_{\hat{\theta}}$ is always inside this allowed interval, yielding
\begin{equation}
\max(\tau,\sqrt{\tau} e^{-y} \sqrt{\frac{1-r}{1+r}}) < x 
  < \min(\sqrt{\tau} e^{-y} \sqrt{\frac{1+r}{1-r}},1) \quad , 
\end{equation}
which corresponds to \eq{hadWQ:x_limit} where the rapidity cut $y_{\text{cut}}$
is replaced by its value $y$ given as an argument to the cross section.

Similarly, for cross sections that are differential in the transverse momentum of a 
final-state particle a cut on the rapidity puts a further constraint on the allowed
interval for ${c_{\hat{\theta}}}_\pm$:
\begin{equation}
c_{{\hat{\theta}}_{y}}^{\text{min}}
 < {c_{\hat{\theta}}}_\pm
 < c_{{\hat{\theta}}_{y}}^{\text{max}}
\end{equation}
with
\begin{align}
c_{{\hat{\theta}}_{y}}^{\text{min}}
 &\equiv \sqrt{1+\frac{m_3^2}{\pvechat^2}}
     \tanh \left(-y_{\text{cut}} + \frac12 \ln\frac{x^2}{\tau} \right) \\
c_{{\hat{\theta}}_{y}}^{\text{max}}
 &\equiv \sqrt{1+\frac{m_3^2}{\pvechat^2}}
     \tanh \left(y_{\text{cut}} + \frac12 \ln\frac{x^2}{\tau} \right) \quad . 
\end{align}
Again this leads to a corresponding change in the limits of the $x$-integration 
which are given by
\begin{equation}
\max(\tau,\sqrt{\tau} e^{-y_{\text{cut}}} \sqrt{\frac{1-\tilde{r}}{1+\tilde{r}}}) < x 
  < \min(\sqrt{\tau} e^{y_{\text{cut}}} \sqrt{\frac{1+\tilde{r}}{1-\tilde{r}}},1)
\end{equation}
with
\begin{equation}
\tilde{r} = \sqrt{\frac{1-\frac{{p_T}^2}{\pvechat^2}}%
               {1+\frac{m_3^2}{\pvechat^2}} }\quad .
\end{equation}
This again corresponds to eqs.~(\ref{hadWQ:x_limit}) and (\ref{hadWQ:abbrr}) 
where instead of the cut on the
transverse momentum ${p_T}_{\text{cut}}$ its fixed value $p_T$, which is 
an argument to the differential cross section, is taken.

\section{HadCalc}
\label{hadWQ:hadcalc}

For the numerical evaluation of the cross sections presented in the following
chapters a program called \HadCalc{} was developed to 
facilitate this task. 
It is based on the established program packages 
\FeynArts~\cite{Kublbeck:1990xc,*Hahn:2000kx,*Hahn:2001rv} 
and \FormCalc~\cite{Hahn:1998yk,Hahn:2005vh}
which are used to generate the partonic cross sections. The main task
of \HadCalc{} then consists of the convolution with the PDFs that are
taken from the PDFlib~\cite{PDFlibmanual} or LHAPDF~\cite{LHAPDFmanual}
library packages that include PDF fits from various groups.

With this program it is possible to calculate both totally integrated and
differential hadronic cross sections of processes with up to three particles in
the final state.  The latter ones can be differential with respect to the
partonic center-of-mass energy, or the rapidity or the transverse momentum
of one of the outgoing particles. Several cuts can be applied to the
phase space. \HadCalc{} operates either in batch mode, where the 
parameters are read from a file and the cross sections are written back
to disk, allowing for easy post-processing with e.g.\ a tool that
generates plots. It can also be used in interactive mode where in- and
output are done via keyboard and screen and which allows the user for 
example to tune the parameters most easily.

A complete manual of \HadCalc{} can be found in appendix~\ref{hadcalc}.
The program code is available on request from the 
author\footnote{email: \texttt{mrauch@mppmu.mpg.de}}.

\chapter{Associated Production of \texorpdfstring{$W^\pm$ $H^\mp$}{W+- H-+}}
\label{bbWH}

The discovery of a charged Higgs boson would be a clear signal of an extended Higgs 
sector and therefore of physics beyond the Standard Model. For relatively light 
charged Higgs bosons with a mass $m_H \lesssim m_t - m_b$ the main production 
process is $t\bar{t}$-production via a subsequent decay sequence 
$t \rightarrow b H^+ \rightarrow b \tau^+ \nu_\tau$~\cite{Kunszt:1991qe}. Both
decay steps are enhanced by large Yukawa couplings.  The experimental signature is
an excess of $\tau \nu_\tau$ pairs in the detector. In the case of charged Higgs-boson
masses above the top-quark mass the dominant production process is 
$gb \rightarrow t H^-$~\cite{Gunion:1986pe,*Barnett:1987jw,*Olness:1987ep,Barger:1993th,Huang:1998vu,*Jin:1999tw,*Jin:2000vx}.
Afterwards the Higgs boson mainly decays into $b\bar{t}$ pairs with a branching ratio
of at least 90\%. 
The top-bottom-quark pairs lead to a detector signature which has 
a large QCD background at the LHC. 
The detection of a heavy charged Higgs boson is therefore much
more difficult. 
Later studies~\cite{Assamagan:2004mu,Branson:2001pj,Buscher:2005re} showed that
the cross section is large enough so that the main decay channel can be ignored.
It is sufficient to consider only the suppressed $H^- \rightarrow \tau \nu_\tau$ decay
channel which has a clear detector signal while still yielding enough events. 

In this chapter we investigate another production mechanism, the production in
association with a $W$ boson. The leptonic decay modes of the $W$ boson avoid large 
QCD backgrounds and can therefore provide an easier way of detecting a charged
Higgs boson.

\section{The \texorpdfstring{$H^+ W^-$}{H+W-} final state}

\begin{figure}
\begin{center}
\subfigure[s-channel contribution]{
\includegraphics[scale=0.4]{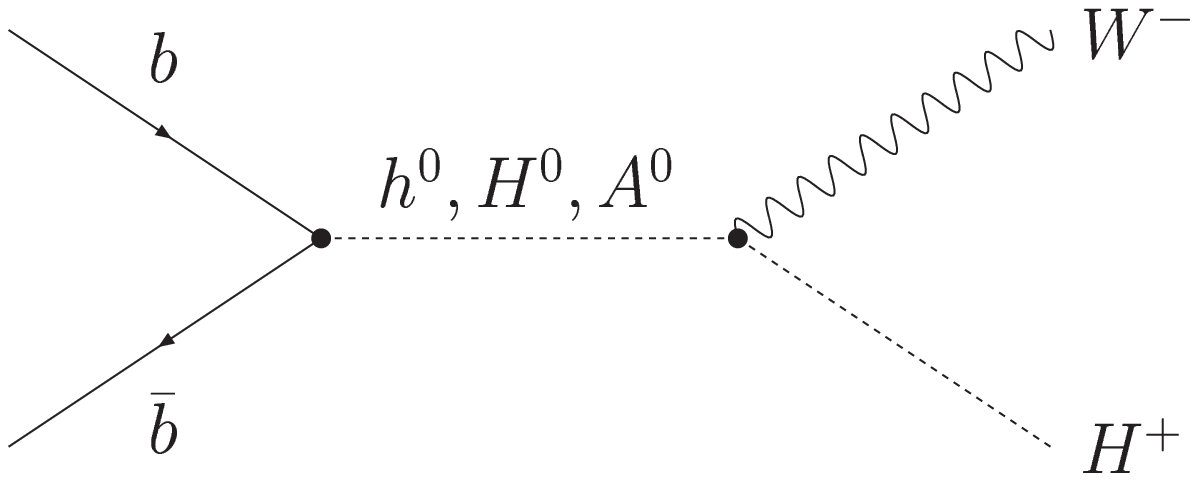}
\label{bbWH:bbWH-LO:s}
}
\subfigure[t-channel contribution]{
\hspace*{5ex}
\includegraphics[scale=0.4]{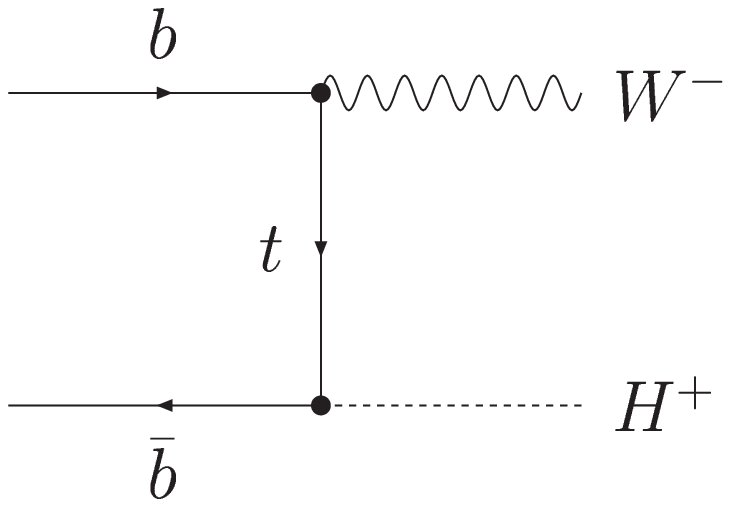}
\hspace*{5ex}
\label{bbWH:bbWH-LO:t}
}
\caption{Tree-level Feynman diagrams contributing to the 
  dominant subprocess $b\bar{b} \rightarrow H^+ W^-$}
\label{bbWH:bbWH-LO}
\end{center}
\end{figure}

\begin{figure}
\begin{center}
\subfigure[quark contribution]{
\includegraphics[scale=0.4]{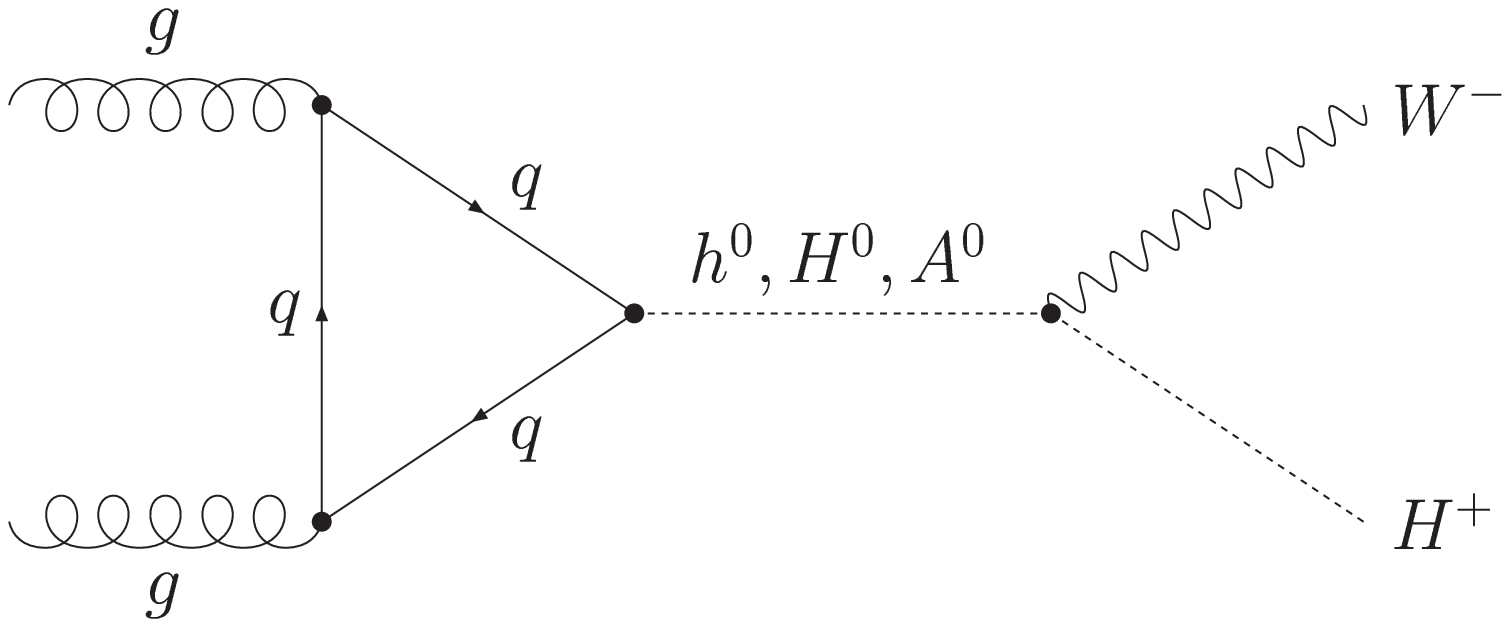}
\hspace*{1cm}
\includegraphics[scale=0.4]{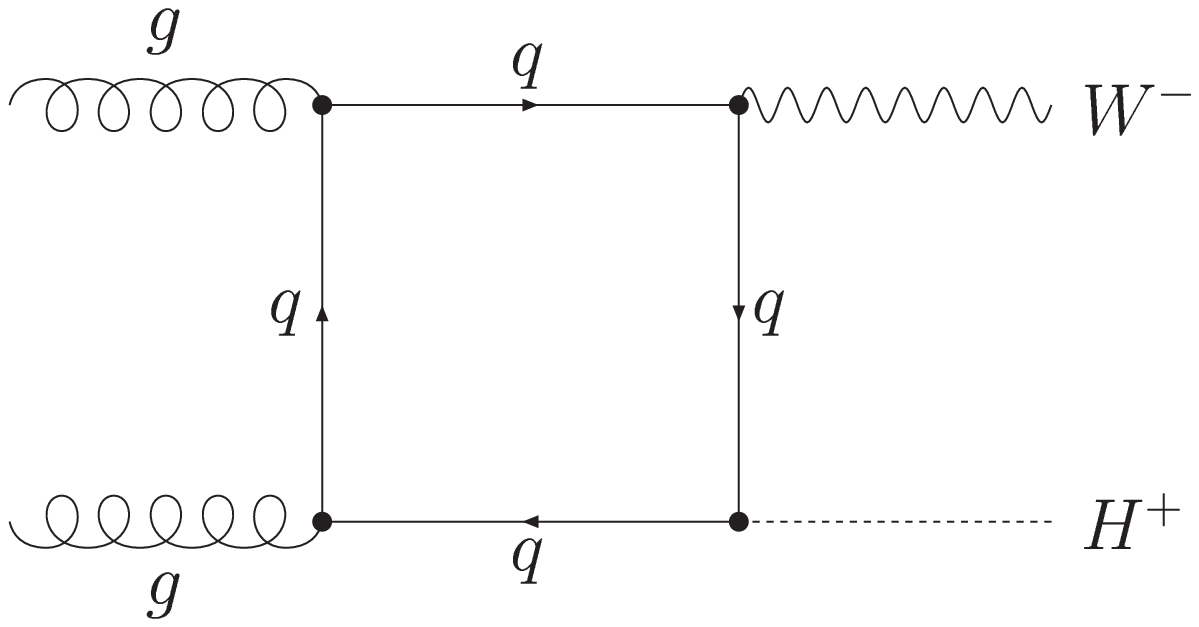}
}
\subfigure[squark contribution]{
\includegraphics[scale=0.4]{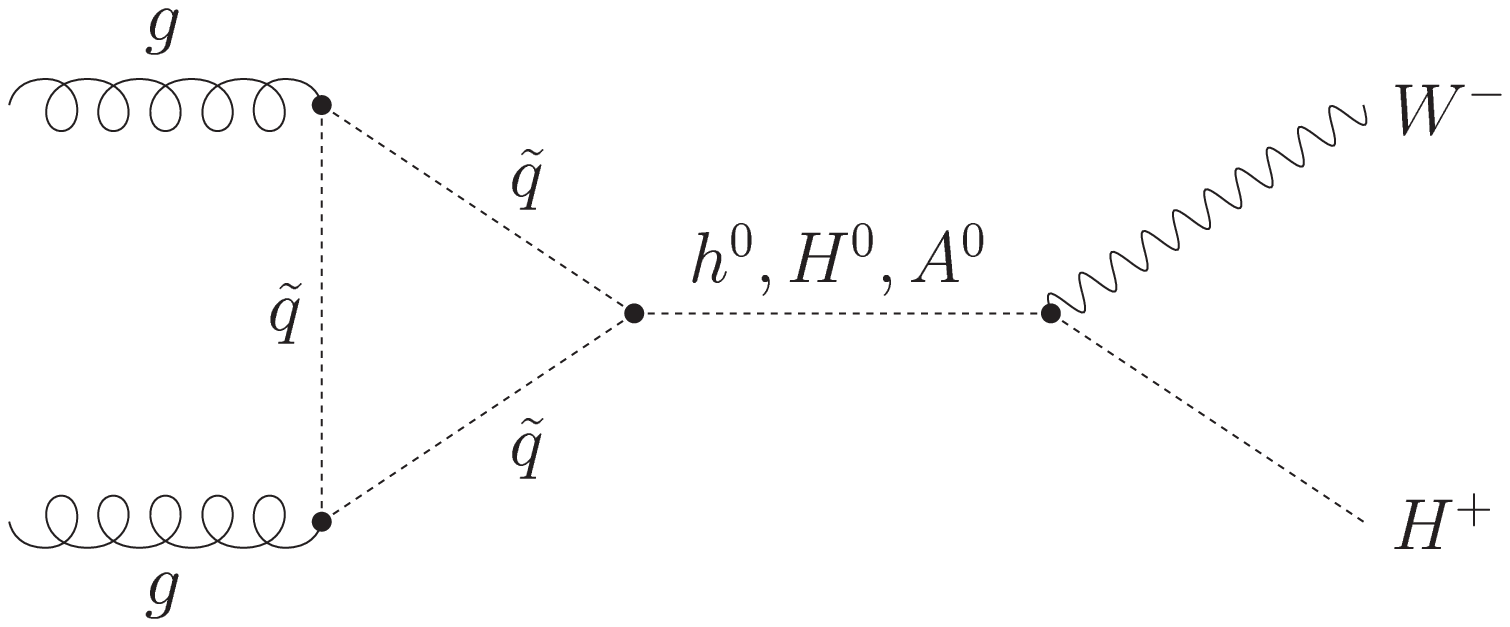}
\hspace*{1cm}
\includegraphics[scale=0.4]{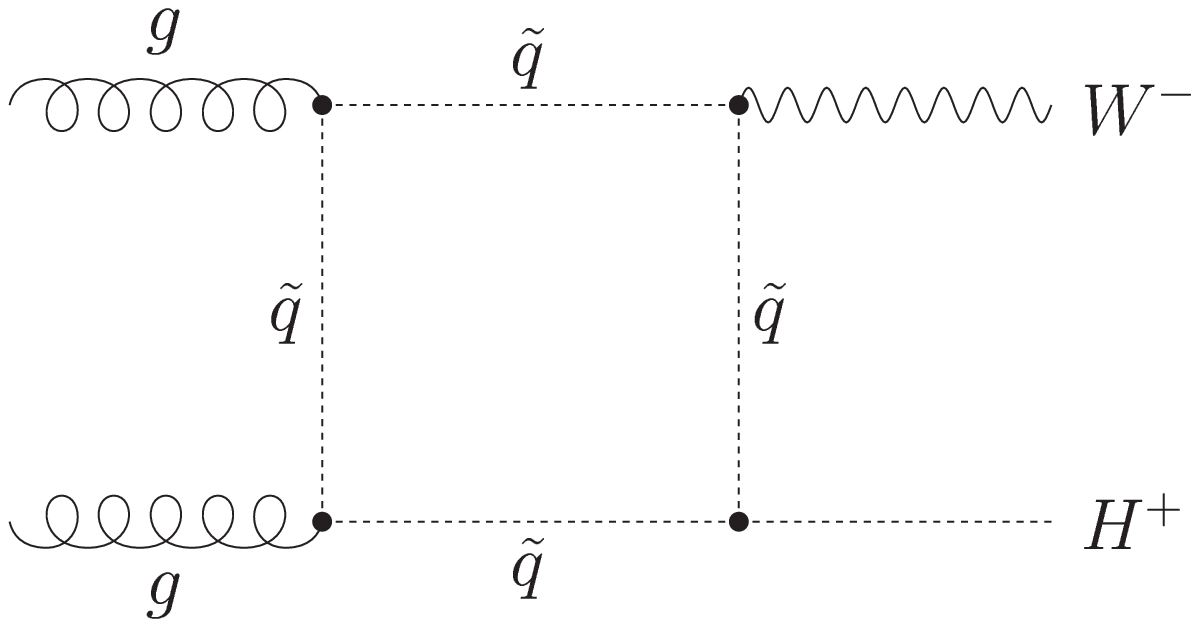}
}
\caption{Leading-order types of Feynman diagrams contributing to the 
  subprocess $gg \rightarrow H^+ W^-$}
\label{bbWH:ggWH}
\end{center}
\end{figure}

The production of a charged Higgs boson in association with a $W$ boson 
was first studied in ref.~\cite{Dicus:1989vf}. This process 
proceeds either via bottom quark--anti-quark annihilation (\fig{bbWH:bbWH-LO}) or 
via gluon fusion and an intermediate quark or squark loop (\fig{bbWH:ggWH}). 
The leptonic
decays of the $W$ boson could be used as a trigger for this process, thereby making
charged Higgs boson detection easier. The calculation was updated 
in ref.~\cite{BarrientosBendezu:1998gd,*BarrientosBendezu:1999vd} and triggered a 
detailed analysis~\cite{Moretti:1998xq} of the discovery potential at the LHC 
using this process. This paper concluded that an efficient separation of the signal 
process from the background processes such as top-quark pair production is 
difficult for semileptonic $W$ boson decays including both low and high values of
$t_\beta$. The cross sections were evaluated at leading order for both 
production processes. 

The later studies of refs.~\cite{Assamagan:2004mu,Branson:2001pj,Buscher:2005re} 
showed that a discovery is more likely in the main production channel 
$gb \rightarrow t H^-$ where only the rare decay into a $\tau\nu_\tau$ pair
is taken into account.
Nevertheless the associated production of a charged Higgs with a $W$ boson is 
an interesting process, especially when the existence of a charged Higgs boson
has already been established before. If at that point no supersymmetric particles were
detected, the question whether the $H^\pm$ originates from a Standard Model-like
theory with an extended Higgs sector, like the Two-Higgs-Doublet Model (THDM), or
from the MSSM remains open. In the latter case the cross section receives an additional
contribution from virtual superpartners running in the loop. This can be used to tell 
the two models apart.

As already mentioned earlier there are two important production processes for this 
final state in proton-proton
collisions. The dominant one is the tree-level production (see \fig{bbWH:bbWH-LO})
via bottom quark--anti-quark annihilation. 
The s-channel diagrams shown in \fig{bbWH:bbWH-LO:s} are mediated by
a virtual Higgs boson where all three neutral Higgs bosons of the MSSM 
($h^0$, $H^0$ and $A^0$) can appear in the intermediate state. The appearance of 
the massive particles in the s-channel leads to a propagator suppression of this
diagram type. In the t-channel diagrams the exchange of a top quark occurs and yields
the leading contribution to the bottom quark--anti-quark annihilation process.
As this class of diagrams contributes the most to the total $H^+ W^-$ production rate
one-loop corrections to this process are also important as they can modify
the cross section significantly.
Standard-QCD corrections of $\Order{\alpha_s}$ to this process were calculated
in ref.~\cite{Hollik:2001hy}. Both the \DRbar{} and the OS renormalization 
scheme were
considered and good agreement between the two schemes could be found. The
corrections are typically of $\Order{15\%}$ and can reach up to $30\%$ in the
small $t_\beta$ regime. The supersymmetric electroweak corrections, i.e.\ corrections
where squarks together with charginos and neutralinos appear in the loop, of 
$\Order{\alpha m_{t(b)}^2/m_W^2}$ and $\Order{\alpha m_{t(b)}^4/m_W^4}$
were calculated in ref.~\cite{Yang:2000yt}. 

The second parton process contributing to the $H^+ W^-$ final state proceeds via 
gluon fusion and an intermediate loop as shown in \fig{bbWH:ggWH}. 
Since there is no tree-level process and 
the leading order contains a quark or squark loop it is suppressed by a 
factor $\alpha_s^2$ with respect to the
bottom-quark annihilation process. The higher density of gluons in the proton
partly compensates this effect, making both processes comparable in size.
The contribution of quark loops was already included in the
calculation of ref.~\cite{BarrientosBendezu:1998gd,*BarrientosBendezu:1999vd}. The 
contribution of supersymmetric particles was calculated later on~\cite{Brein:2000cv} 
and it was shown that they can reach up to 40\%.
Together with the QCD corrections this can raise the cross section for small $t_\beta$
so that it becomes comparable in size to the bottom-quark annihilation process 
which is not loop-suppressed.

In this thesis the missing supersymmetric QCD corrections, 
i.e.\ corrections with squarks and
gluinos running in the loop, to the leading bottom-quark annihilation process
are considered. They are the last contribution of $\Order{\alpha^2 \alpha_s}$ to the
associated production of a charged Higgs boson with a W boson in the MSSM which has
not been calculated so far.

\section{SUSY-QCD corrections to 
  \texorpdfstring{$b\bar{b}\rightarrow H^+ W^-$}{bb -> H+ W-}}

In this chapter the supersymmetric QCD corrections to the main production process
$b\bar{b}\rightarrow H^+ W^-$ are calculated.
As already shown in \chap{renorm:deltamb} it is known that the coupling
of the bottom quark to the Higgs bosons receives large one-loop corrections. These
can be parametrized by introducing a correction term $\Delta m_b$
to the bottom-quark Yukawa coupling. Yet other terms also give significant
contributions, as we will see later. So it is 
necessary to not only use the tree-level result with an effective bottom-quark Yukawa 
coupling but to perform a full one-loop calculation.

\begin{figure}
\begin{center}
\subfigure[self-energy correction]{
\hspace*{1cm}
\includegraphics[scale=0.4]{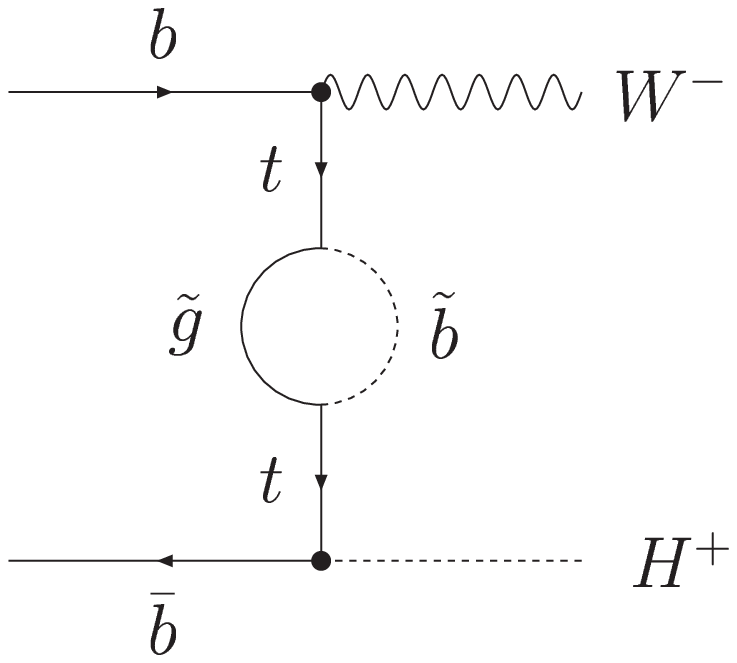}
\hspace*{1cm}
}
\subfigure[self-energy counter term]{
\hspace*{1cm}
\includegraphics[scale=0.4]{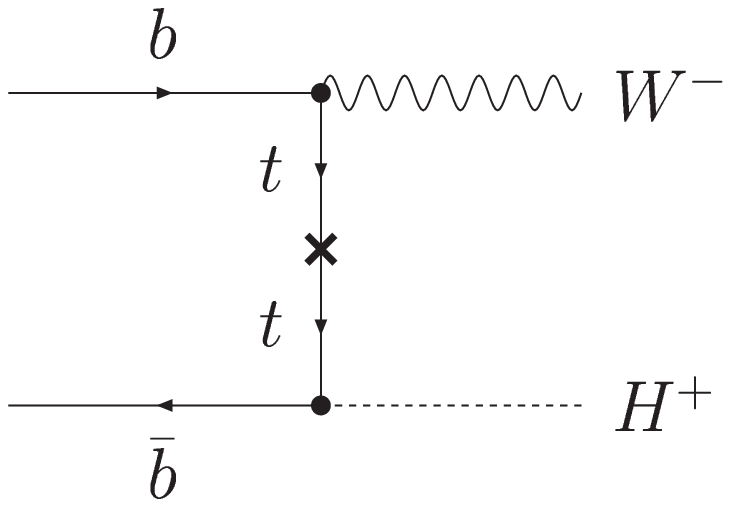}
\hspace*{1cm}
}
\subfigure[vertex corrections]{
\includegraphics[scale=0.4]{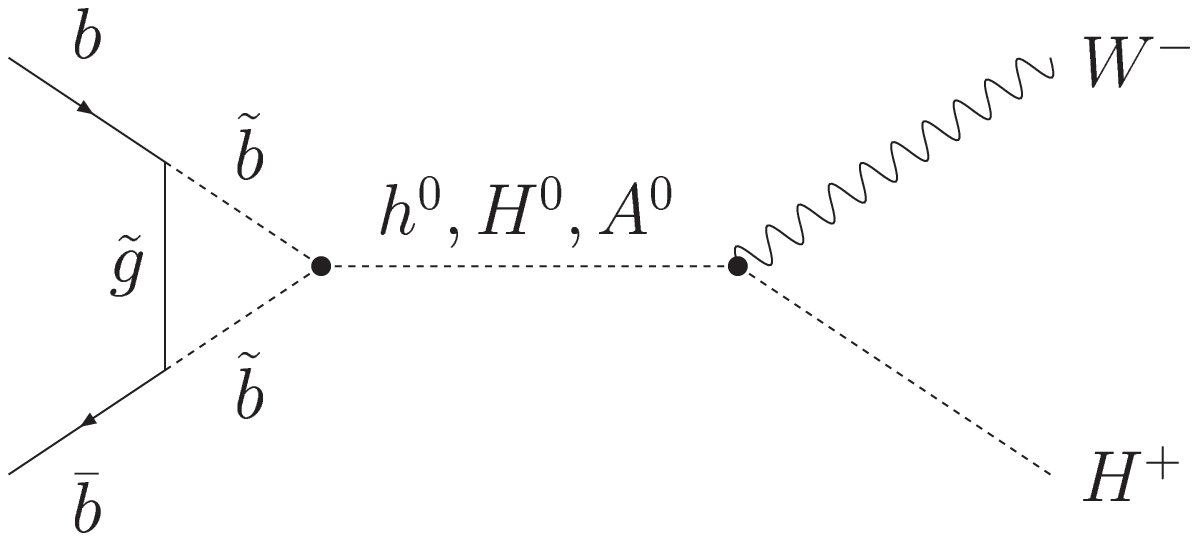}
\includegraphics[scale=0.4]{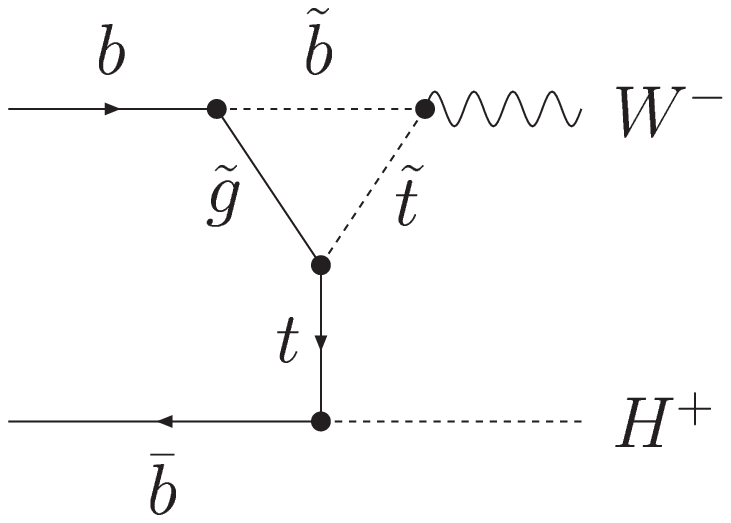}
\includegraphics[scale=0.4]{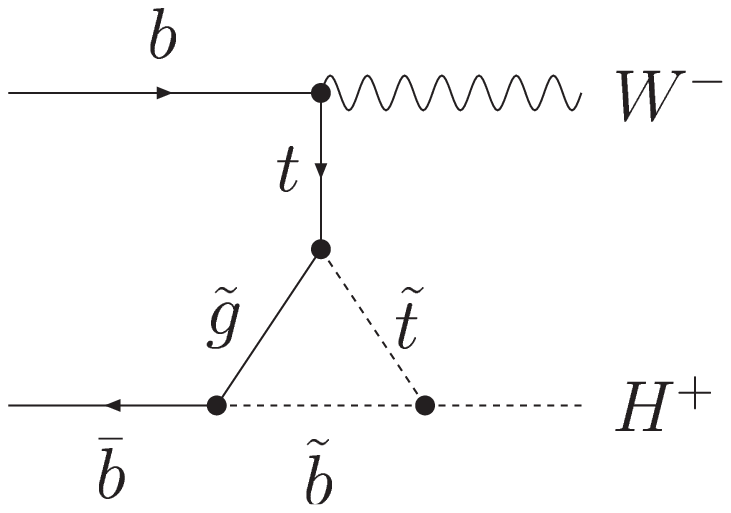}
}
\subfigure[vertex counter terms]{
\includegraphics[scale=0.4]{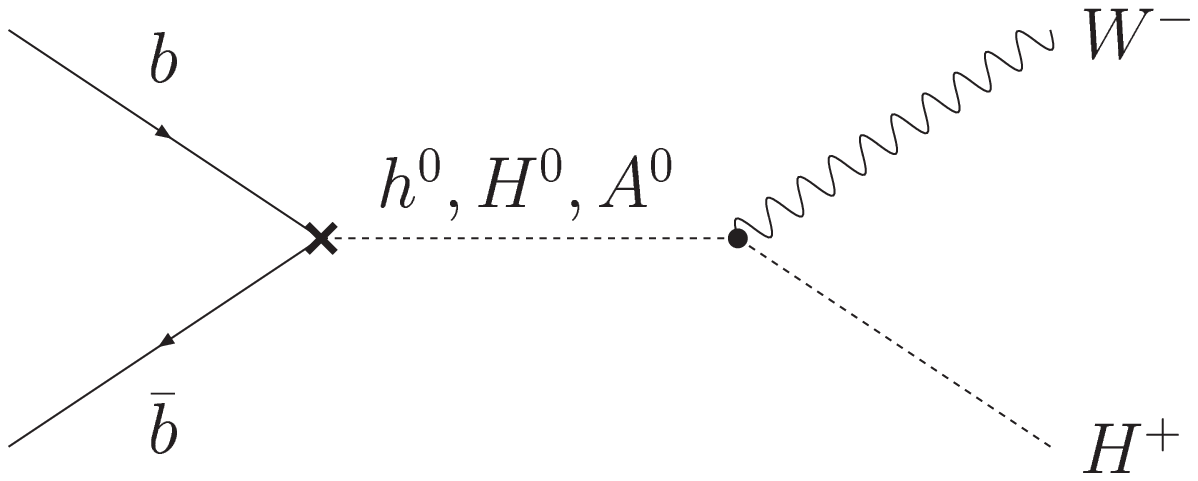}
\includegraphics[scale=0.4]{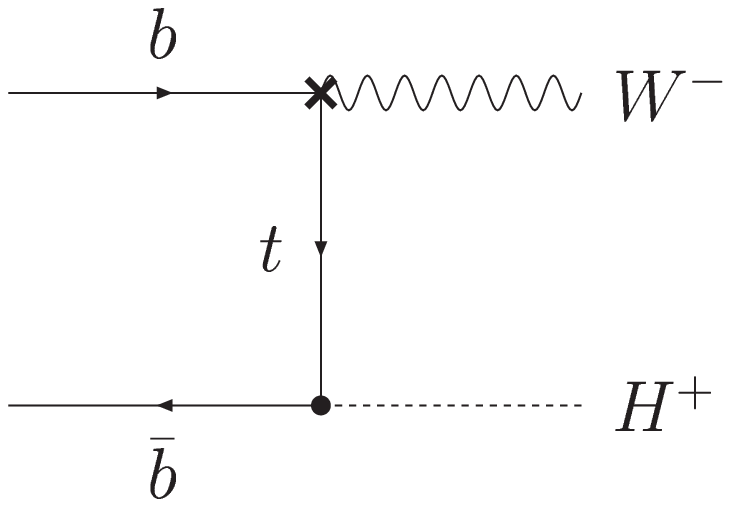}
\includegraphics[scale=0.4]{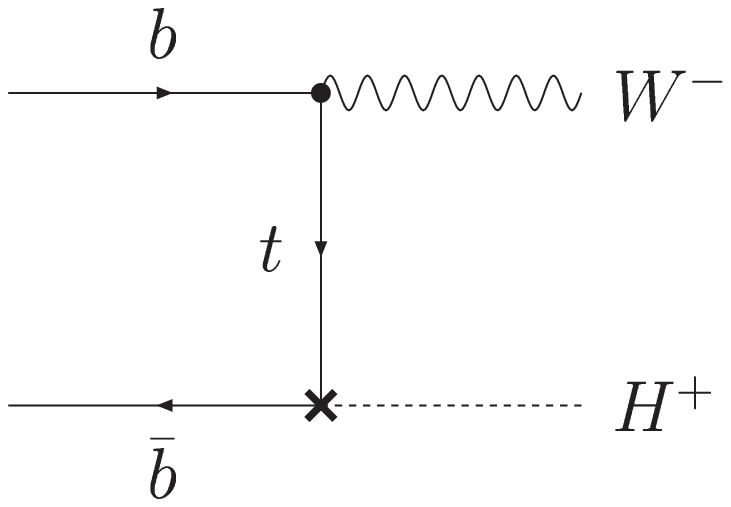}
}
\subfigure[box contribution]{
\includegraphics[scale=0.4]{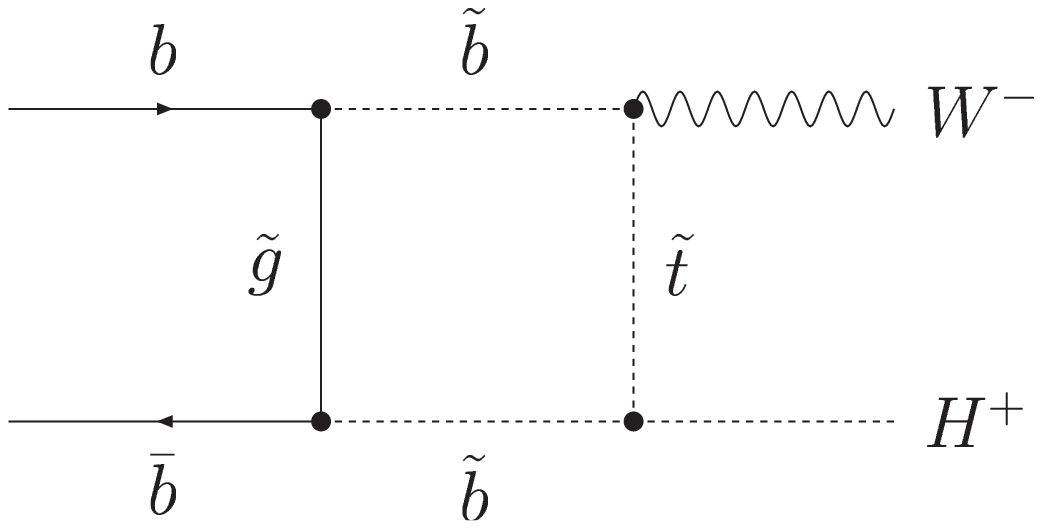}
}
\caption{Diagram types yielding SUSY-QCD corrections to the process 
  $b \bar{b} \rightarrow H^+ W^-$}
\label{bbWH:diags}
\end{center}
\end{figure}

The possible types of diagrams which appear in the calculation of 
SUSY-QCD corrections are depicted in \fig{bbWH:diags}.
A SUSY-QCD self-energy contribution to the bottom-quark propagator of
$\Order{\alpha_s}$ enters in the t-channel exchange diagram as shown in
\fig{bbWH:diags}(a). Vertex corrections (\fig{bbWH:diags}(c)) appear in the 
s-channel diagrams in the vertex where the incoming bottom quark and anti-quark 
couple to the intermediate Higgs boson. The t-channel diagram receives vertex
corrections at both the $btW$- and $btH$-vertices. Finally all four external particles 
can be connected via a box-shaped loop diagram (\fig{bbWH:diags}(e)). 
Additionally for the self-energy and vertex corrections appropriate counter-term
diagrams appear (\fig{bbWH:diags}(b), (d)).

The cross sections were calculated in both the OS and \DRbar{} renormalization
schemes. 
% \Delta m_b improvements
Additionally a $\Delta m_b$-corrected tree-level cross section was calculated.
As shown in \chap{renorm:deltamb} the bottom-quark mass and the bottom-quark
couplings to the Higgs fields receive large contributions from the one-loop SUSY-QCD
corrections which are parametrized in the variable $\Delta m_b$. To be able to 
compare the improved tree-level cross section with the full one-loop cross section it is
necessary to use the same order in perturbation theory for both calculations. This
means that one must use the non-resummed replacement \eq{renorm:noresum} 
\begin{equation}
m_b \rightarrow m_b \left( 1 - \Delta m_b \right) \nonumber
\end{equation}
and the non-resummed correction to the bottom-quark Yukawa coupling 
as in \eq{renorm:bbh}.
A matrix element with this replacement must be treated as a one-loop matrix element.
Let us recall that the standard way of computing a one-loop cross section is to add
the interference terms $(T^* L + T L^*)$ to the 
tree-level cross section $\left|T\right|^2$ and to discard the 
loop-squared term $\left|L\right|^2$ which is a two-loop quantity so that for the
squared matrix element
\begin{equation}
\left|\Mfi\right|^2 = \left|T\right|^2 + 2 \R \left(T^* L\right)
\end{equation}
is obtained.
In this equation $T$ denotes the tree-level amplitude and $L$ the amplitude
of the one-loop diagrams.
In complete analogy the squared matrix element of the $\Delta m_b$-corrected
cross section is defined as
\begin{equation}
\left|{\Mfi}_{\Delta m_b}\right|^2 = 
  \left|T\right|^2 + 2 \R \left(T^* L_{\Delta m_b}\right)
  \label{bbWH:deltamb}
\end{equation}
with
\begin{equation}
L_{\Delta m_b} = T_{\Delta m_b} - T \quad .
\end{equation}
$T_{\Delta m_b}$ denotes the tree-level cross section with the $\Delta m_b$ 
replacements of \eq{renorm:noresum} and \eq{renorm:bbh}. Therefore 
$L_{\Delta m_b}$ contains the additional contribution which originates from 
the correction terms. 
The corresponding cross section to $\left|{\Mfi}_{\Delta m_b}\right|^2$ is denoted 
by $\sigma_\Delta$ in the following.

% relative corrections
In order to present the numerical results several relative corrections $\Delta$ using
various contributions are defined.
Firstly there is the relative correction in the OS scheme,
\begin{equation}
\Delta_{OS} = \frac{\sigma_1^{OS} - \sigma_0^{OS}}{\sigma_0^{OS}} \quad .
\end{equation}
The relative correction in the \DRbar{} scheme is defined analogously as
\begin{equation}
\Delta_{DR} = 
\frac{\sigma_1^{\DRbar} - \sigma_0^{\DRbar}}{\sigma_0^{\DRbar}} \quad .
\end{equation}
The third relation consists of the difference between the one-loop result and 
the $\Delta m_b$-corrected tree-level result which is calculated according to
\eq{bbWH:deltamb}. Hence it signifies the true one-loop corrections.
It is defined as
\begin{equation}
\Delta_{\Delta m_b} 
  = \frac{\sigma_1^{OS} - \sigma_\Delta^{OS}}{\sigma_\Delta^{OS}} 
  \quad .
\end{equation}
The subscript of the cross section $\sigma$ always denotes the order of the 
respective loop contribution, i.e.\ $0$ for the tree-level result, $1$ for the one-loop
result including the SUSY-QCD corrections, and $\Delta$ for the 
$\Delta m_b$-corrected tree-level result. The superscript indicates the 
renormalization scheme in which the quantity is calculated. 

\section{Numerical Results}

\begin{figure}
\includegraphics{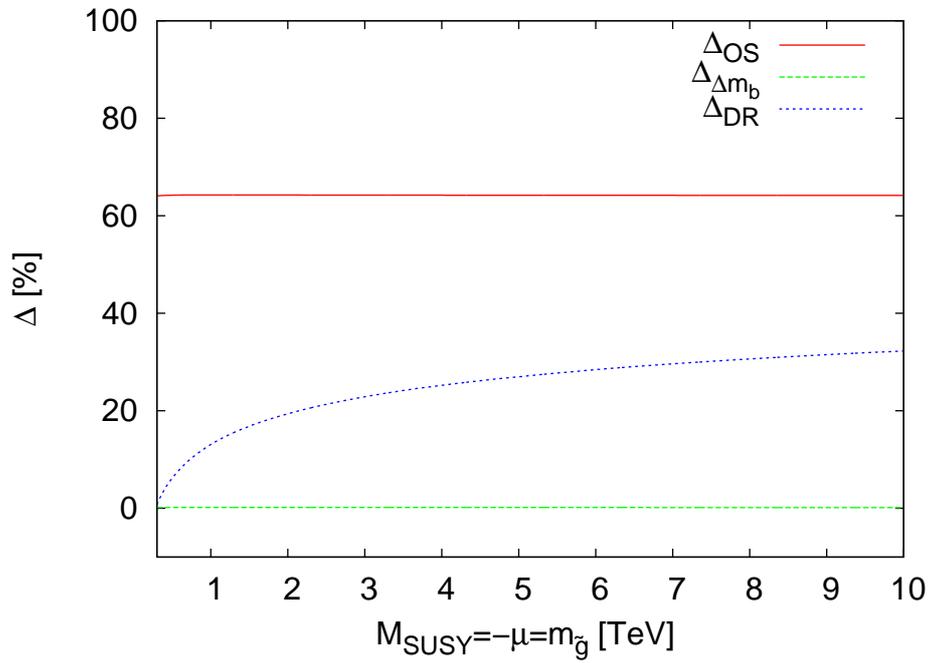}
\caption{Hadronic cross section differences for the process 
  $b \bar{b} \rightarrow H^+ W^-$ in the large $t_\beta$ regime. 
  The soft SUSY-breaking mass
  terms in the squark sector $M_{\text{SUSY}}$, 
  $- \mu$ and the gluino mass $m_{\tilde g}$ 
  are fixed to the same value and varied over a large mass range. The parameter
  set \eq{bbWH:param_SUSY} was used to obtain this plot.}
\label{bbWH:hadron_SUSY}
\end{figure}

In this section the numerical results for the production process 
$b \bar{b} \rightarrow H^+ W^-$ are presented. All quoted cross sections
are given for the LHC with a proton-proton center-of-mass energy of $14 \TeV$.
First of all we investigate the 
effect of varying the MSSM parameters on the SUSY-QCD corrections. To do so 
a parameter point is chosen for which the corrections of the $\Delta m_b$ term
are expected to have a large impact. To that effect the parameter point
\begin{align}
m_{H^+} &= 200 \GeV \nonumber\\
t_\beta &= 30 \nonumber\\
A_t  &= A_b = 0 &  \nonumber\\
M_{\tilde{Q}} &= M_{\tilde{U}} = M_{\tilde{D}} = - \mu = m_{\tilde{g}} 
  \in [0,10] \TeV
\label{bbWH:param_SUSY}
\end{align}
is used as input. $t_\beta$ of this point is fairly large to enhance the $\Delta m_b$
contribution. The soft SUSY-breaking mass terms in the squark sector, jointly denoted
 as  $M_{\text{SUSY}}=M_{\tilde{Q}} = M_{\tilde{U}} = M_{\tilde{D}}$, $-\mu$
and the gluino mass $m_{\tilde g}$ all take the same value 
which is varied over a large mass range.
Eq.~\ref{renorm:deltambeq} then predicts that the SUSY-QCD corrections in 
the OS scheme
should be large and independent of the varied mass scale as we are in the limit where
all these masses are equal and the mass dependence drops out. This is indeed the
case as one can see in \fig{bbWH:hadron_SUSY}. 
A naive calculation in the OS scheme
yields a correction of above $60 \%$. After subtracting the $\Delta m_b$ contribution
only the real one-loop corrections are left. Their size is of around $0.2 \%$. In the
\DRbar{} scheme the corrections are equally small for small mass values and 
show a logarithmic rise with growing mass. This is an artefact of the mismatch between
the renormalization scale and the masses of the squarks and gluinos appearing in 
the loop diagrams. The former one was fixed to the sum of the final-state particle
masses $\mu_R = m_{H^-} + m_W$. Terms of the order 
$\ln \frac{M_{SUSY}}{\mu_R}$ appear in the expression which originate from 
the dimensional regularization of the divergent loop integrals. Thus this 
logarithmic rise bears no physical meaning and will vanish if higher orders 
of perturbation theory are taken into account.

\begin{figure}
\includegraphics{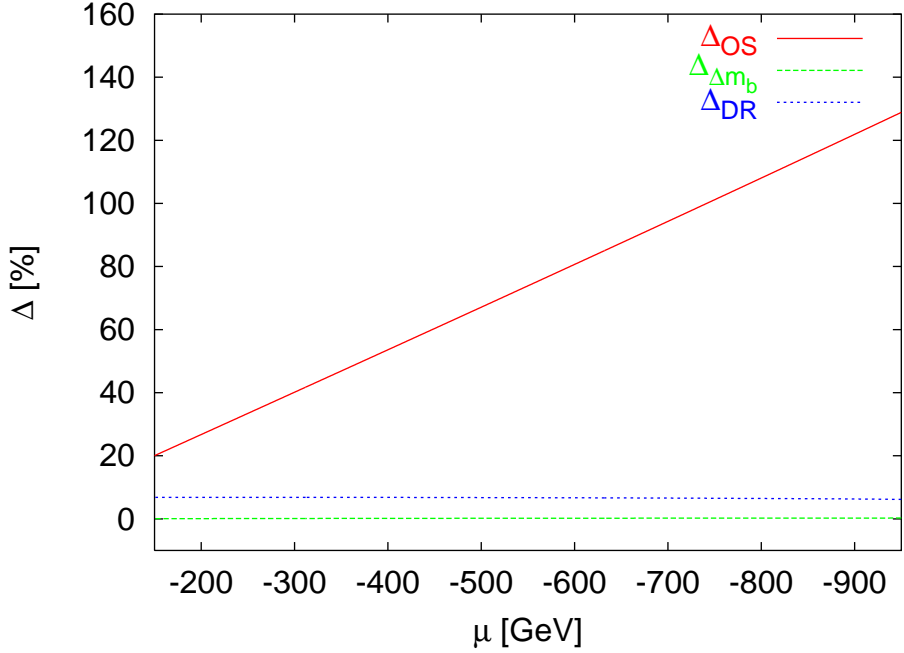}
\caption{Hadronic cross section differences for the process 
  $b \bar{b} \rightarrow H^+ W^-$ using the parameter set 
  \eq{bbWH:param_mue_ltb}. Only $\mu$ is varied in this plot. }
\label{bbWH:hadron_mue_ltb}
\end{figure}

In the next plot (\fig{bbWH:hadron_mue_ltb}) the squark and gluino masses are now 
fixed and only $\mu$ is left as a free parameter.
The parameter set for this plot is
\begin{align}
m_{H^+} &= 200 \GeV \nonumber\\
t_\beta &= 30 \nonumber\\
A_t  &= A_b = 0 \nonumber\\
M_{\tilde{Q}} &= M_{\tilde{U}} = M_{\tilde{D}} = 500 \GeV \nonumber\\
m_{\tilde{g}} &= 580 \GeV \quad .
\label{bbWH:param_mue_ltb}
\end{align}
As expected the cross section in the OS scheme grows linearly with $\mu$. For 
$\mu$ values less than about $-750 \GeV$ the corrections even exceed $100 \%$.
Again, when the $\Delta m_b$ corrections are subtracted and only the true 
one-loop SUSY-QCD corrections remain the order of the corrections is below $1\%$ 
and shows only a very low variation with $\mu$.
In the \DRbar{} scheme the corrections are also much smaller than in the OS scheme
and almost constant as a function of $\mu$, as is expected from the remaining
corrections appearing in this scheme.

\begin{figure}
\includegraphics{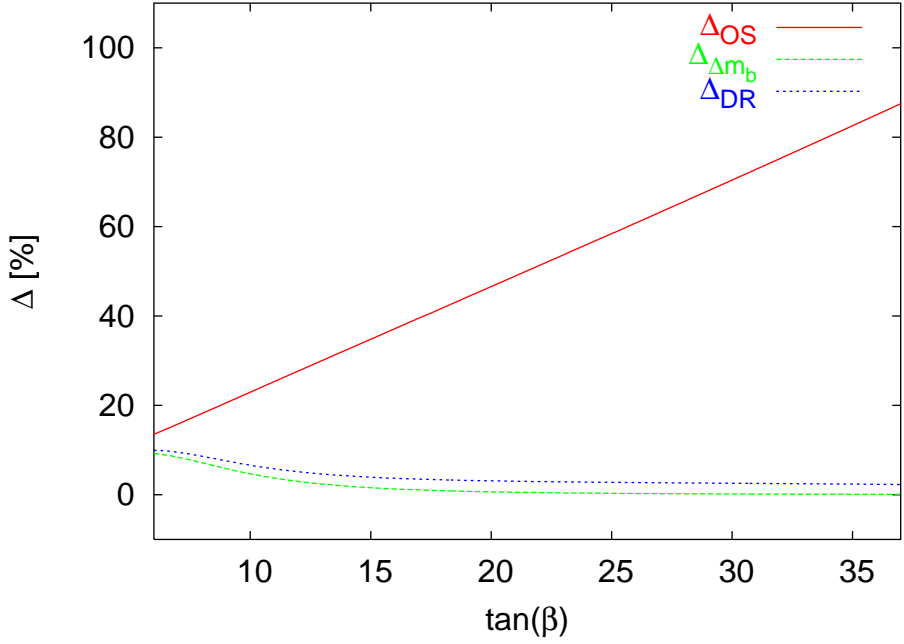}
\caption{Hadronic cross section differences for the process 
  $b \bar{b} \rightarrow H^+ W^-$ as a function of $t_\beta$. This plot was calculated
  with the parameter set \eq{bbWH:param_tanbeta}.}
\label{bbWH:hadron_tanbeta}
\end{figure}

\begin{figure}
\begin{center}
\includegraphics[scale=0.6]{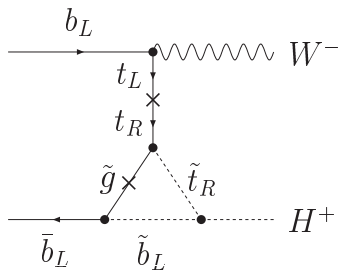}
\caption{Dominant one-loop contribution in the case of small $t_\beta$.
The crosses in this diagram denote a mass insertion, i.e.\ when calculating
the trace over the fermion line the mass term in the Dirac algebra
is chosen.}
\label{bbWH:t_largecontri}
\end{center}
\end{figure}

As next step the dependence of the hadronic cross section differences as a function of
$t_\beta$ is investigated in \fig{bbWH:hadron_tanbeta}.
Here a data point with a smaller soft-supersymmetry breaking mass is chosen to
emphasize the effect which will be discussed below, namely
\begin{align}
m_{H^+} &= 200 \GeV \nonumber\\
\mu &= -200 \GeV \nonumber\\
A_t  &= A_b = 0 \nonumber\\
M_{\tilde{Q}} &= M_{\tilde{U}} = M_{\tilde{D}} = 200 \GeV \nonumber\\
m_{\tilde{g}} &= 580 \GeV \quad .
\label{bbWH:param_tanbeta}
\end{align}
For large values of $t_\beta$ the respective corrections feature the behavior which was 
already observed in the previous plots.
The $\Delta m_b$ corrections are large, while the true
one-loop corrections almost vanish. The corrections in the OS scheme
rise linearly with $t_\beta$ as is expected from \eq{renorm:deltambeq}. For 
$t_\beta \lesssim 15$ however the behavior changes. There the difference between the
full one-loop computation and the $\Delta m_b$-corrected tree-level cross section can 
increase up to $10\%$. 
This contribution for small $t_\beta$ originates mainly from the diagram given in 
\fig{bbWH:t_largecontri}.
The Yukawa coupling of the charged Higgs to the stop and sbottom is 
proportional to the top-quark mass if a right-handed stop couples to a 
left-handed sbottom. Another factor $m_t$ appears in the trace over
the fermion line where for the internal top quark line the mass term in 
the Dirac algebra is chosen. These two factors cancel the top-quark
propagator which is dominated by the mass term and the top-quark mass
dependence drops out. To get this left-right mixing term in the Yukawa coupling
also the mass term for the gluino appears during the calculation of the
fermion trace, giving a factor $m_{\tilde{g}}$ in the nominator.
Accordingly this vertex correction to the $tbH^+$ vertex is proportional to 
\begin{equation}
\frac{\alpha_s}{3 \pi} \mu m_{\tilde{g}} 
I\left( m_{{\tilde b}}, m_{{\tilde t}}, m_{\tilde g} \right)
\label{bbWH:t_largecontri_prop}
\end{equation}
where $I$ was given in \eq{renorm:integralI} and is related to the three-point
integral in the limit of vanishing external momenta. The expression is independent 
of $t_\beta$. 
So for small values of $t_\beta$, where the bottom-quark Yukawa coupling is not
enhanced, this gives an important contribution. When $t_\beta$ takes larger values,
the bottom-quark terms dominate. These terms have a factor of $t_\beta$ appearing 
in the amplitude so the cross section increases 
quadratically with $t_\beta^2$. Hence the
relative contribution of \fig{bbWH:t_largecontri} is reduced. This plot underlines
the importance of performing
a full one-loop calculation because only in such a full calculation the non-universal
corrections are included and a tree-level calculation with effective couplings would give
wrong results in the low-$t_\beta$ regime.

\begin{figure}
\includegraphics{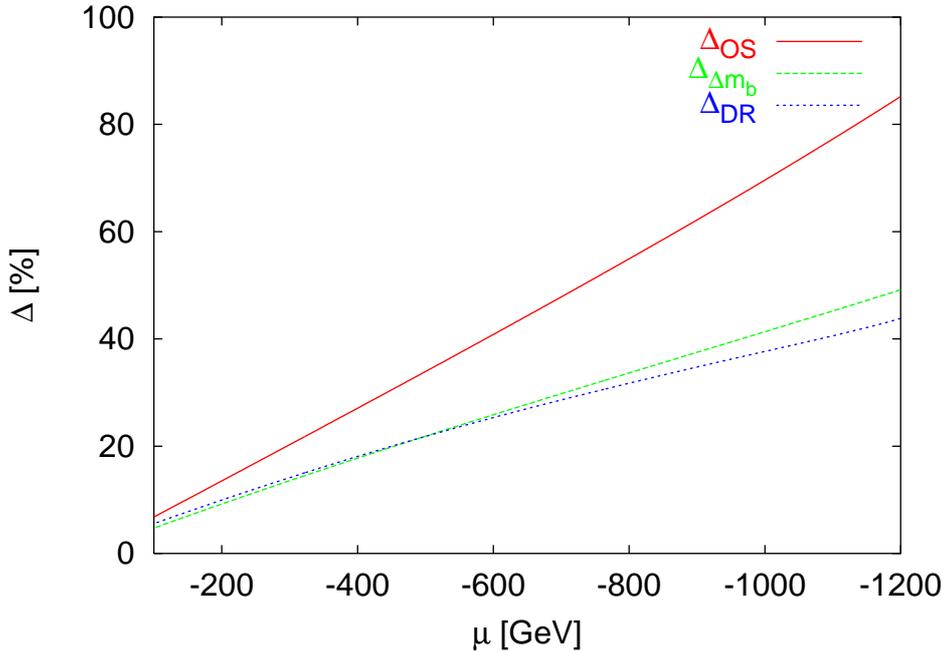}
\caption{Hadronic cross section differences for the process 
  $b \bar{b} \rightarrow H^+ W^-$ as a function of $\mu$. For 
  all other parameters the parameter set \eq{bbWH:param_tanbeta} 
  with $t_\beta = 6$ was used.}
\label{bbWH:hadron_mue}
\end{figure}

The variation of the cross section as a function of $\mu$ in the low-$t_\beta$ regime
is investigated in \fig{bbWH:hadron_mue}. For this plot $t_\beta = 6$ was chosen and
the other parameters were left unchanged from the last plot.
Again the rather small value of $200 \GeV$ is chosen for the soft-supersymmetry
breaking masses in the squark sector, so that the function $I$, which is proportional
to $\frac1{m_{SUSY}^2}$, has a small denominator.
One can clearly see two distinct properties. 
The one-loop corrections which cannot be absorbed into a redefinition
of the bottom-quark Yukawa coupling now give a significant contribution.
The $\Delta m_b$ corrections still yield an important contribution as can be 
seen from comparing the $\Delta_{OS}$ and $\Delta_{\Delta m_b}$ curves.
Yet after subtracting the universal corrections to the bottom-quark Yukawa
coupling the remaining non-universal corrections are large.
For large values of $\mu$ they can reach more than $50\%$.
They are equally present in the $\DRbar$ scheme where a numerical
contribution close to the $\Delta m_b$-corrected result is obtained.
Furthermore the correction to the cross section increases approximately linearly with
the absolute value of $\mu$ as is expected from \eq{bbWH:t_largecontri_prop}.

\begin{figure}
%% The long caption wouldn't fit on the pager otherwise.
\vspace*{-4mm}
\includegraphics{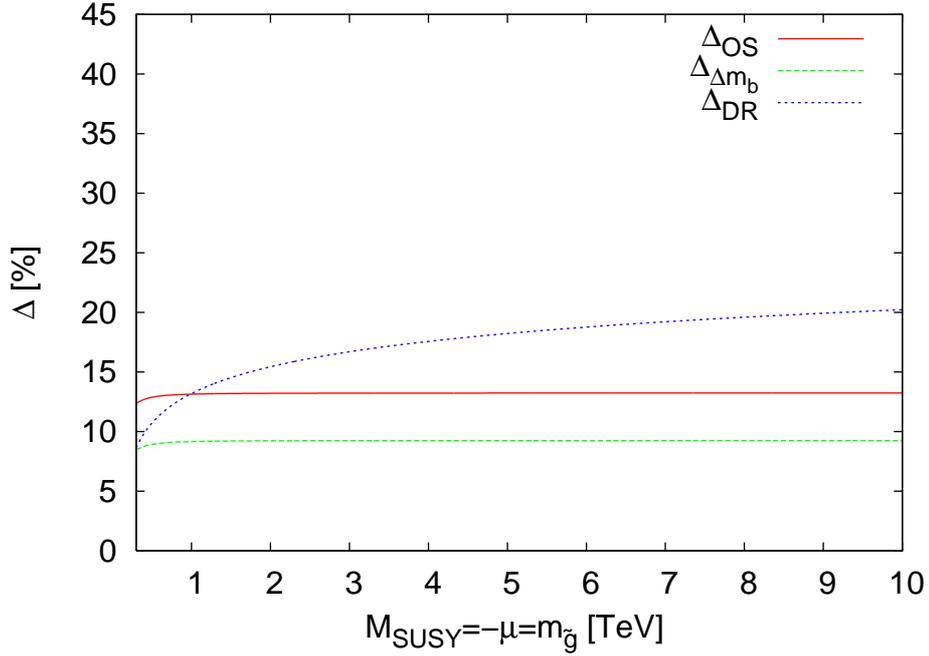}
\caption{Hadronic cross section differences for the process 
  $b \bar{b} \rightarrow H^+ W^-$ in the 
  low-$t_\beta$ regime, i.e.\ $t_\beta = 6$ is used. All other parameters
  take the values given in \eq{bbWH:param_SUSY}. The soft SUSY-breaking mass
  terms in the squark sector, $-\mu$ and the gluino mass are fixed
  to the same value and varied over a large mass range.}
\label{bbWH:hadron_SUSY_tb6}
\end{figure}

Furthermore, in \fig{bbWH:hadron_SUSY_tb6} a plot where the complete SUSY 
mass spectrum, i.e.\ soft-su\-per\-sym\-me\-try breaking mass terms for the 
squarks, $\mu$ and the gluino mass, is fixed to the same
value and run up to $10 \TeV$ is presented for the low-$t_\beta$ regime.
The behavior as a function of the mass scale is the same as before in the case
of large $t_\beta$. After having subtracted the $\Delta m_b$ corrections from the
complete one-loop contributions there is still a correction of the order of $10 \%$ left
which mainly originates from the diagram given in \fig{bbWH:t_largecontri}. For small
mass values the relative correction slightly drops because in this region other diagrams
also give a numerically significant contribution. 

\begin{figure}
\vspace*{-4mm}
\includegraphics{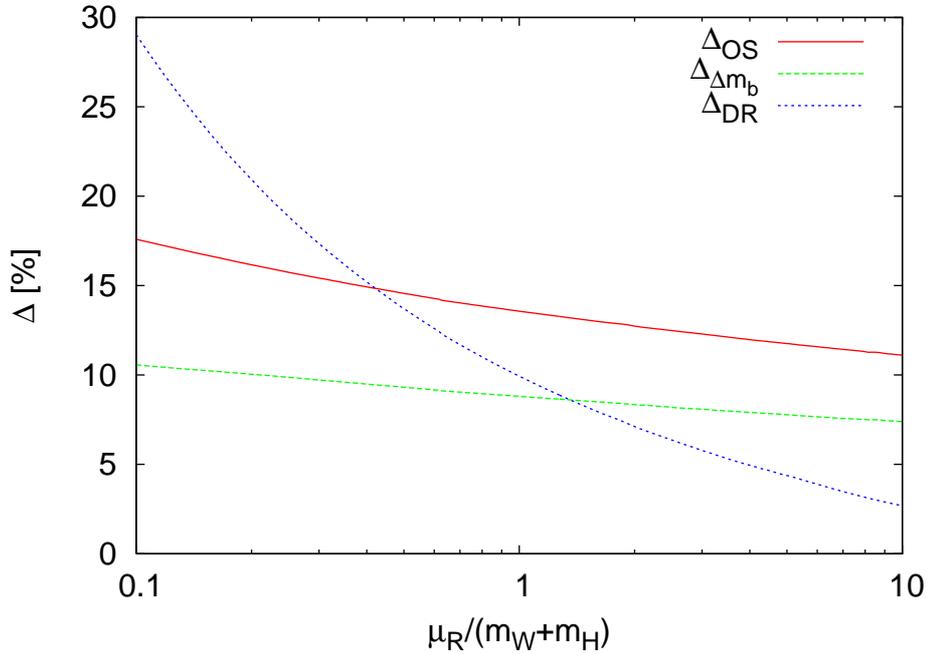}
\caption{Hadronic cross section differences for the process 
  $b \bar{b} \rightarrow H^+ W^-$ as a function of the renormalization
  and factorization scale. The plot was calculated using the parameter
  set \eq{bbWH:param_tanbeta} with additionally $t_\beta=6$.}
\label{bbWH:hadron_rscale}
\end{figure}

The scale dependence of the SUSY-QCD corrections is given in
\fig{bbWH:hadron_rscale}.
The factorization and renormalization scale of the process are fixed to the same
value and varied between 0.1 and 10 times their basic value 
$\mu_R=\mu_F=m_W + m_H$
which is used for the other plots. Even for this large scale variations there is only 
a mild dependence for the corrections in the OS scheme and the $\Delta m_b$
corrections. This is due to the fact that the only scale-dependent parameter
is the strong coupling constant $\alpha_s$ and the PDFs. 
On-shell conditions render all other paramaters independent of the scale. In the
\DRbar{} scheme also the quark masses are scale dependent resulting in a much
larger variation as a function of the scale. Including the Standard-QCD corrections 
which are
already known for this process~\cite{Hollik:2001hy} would reduce this variation, 
but implementing these additional contributions was beyond the scope of this dissertation.

\begin{figure}
\includegraphics{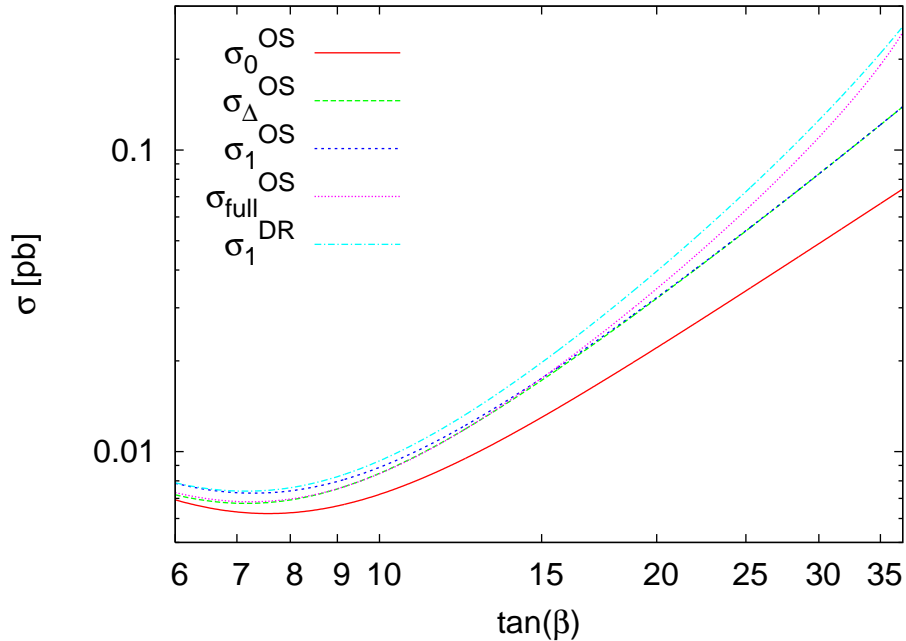}
\caption{Total hadronic cross sections for the process 
  $b \bar{b} \rightarrow H^+ W^-$ as a function of $t_\beta$ using the 
  parameter set \eq{bbWH:param_tanbeta}.}
\label{bbWH:hadron_tanbeta_v.absolut}
\end{figure}

The last plot in \fig{bbWH:hadron_tanbeta_v.absolut} shows the total hadronic 
cross section as a function of $t_\beta$ in both OS and \DRbar{} 
renormalization schemes. Various cross sections with different contributions 
taken into account are presented here.
The same parameter set as in \eq{bbWH:param_tanbeta} is used for this plot.
\begin{table}
\begin{center}
\setlength{\unitlength}{1ex}%
\begin{tabular}{|lc|l|}
\hline
${\sigma_0}^{\text{OS}}$ & \textcolor{red}{
\begin{picture}(6,1)
\put(0,1){\line(1,0){6}}
\end{picture}}
  & Tree-level cross section\\\hline
${\sigma_\Delta}^{\text{OS}}$ & \textcolor{green}{
\begin{picture}(6,1)
\put(0.0,1){\line(1,0){0.4}}
\put(0.6,1){\line(1,0){0.4}}
\put(1.2,1){\line(1,0){0.4}}
\put(1.8,1){\line(1,0){0.4}}
\put(2.4,1){\line(1,0){0.4}}
\put(3.0,1){\line(1,0){0.4}}
\put(3.6,1){\line(1,0){0.4}}
\put(4.2,1){\line(1,0){0.4}}
\put(4.8,1){\line(1,0){0.4}}
\put(5.4,1){\line(1,0){0.4}}
\end{picture}}
  & $\Delta m_b$-corrected tree-level result \\\hline
${\sigma_1}^{\text{OS}}$ & \textcolor{blue}{
\begin{picture}(6,1)
\put(0.0,1){\line(1,0){0.2}}
\put(0.5,1){\line(1,0){0.2}}
\put(1.0,1){\line(1,0){0.2}}
\put(1.5,1){\line(1,0){0.2}}
\put(2.0,1){\line(1,0){0.2}}
\put(2.5,1){\line(1,0){0.2}}
\put(3.0,1){\line(1,0){0.2}}
\put(3.5,1){\line(1,0){0.2}}
\put(4.0,1){\line(1,0){0.2}}
\put(4.5,1){\line(1,0){0.2}}
\put(5.0,1){\line(1,0){0.2}}
\put(5.5,1){\line(1,0){0.2}}
\end{picture}}
  & One-loop OS cross section \\\hline
${\sigma_{\text{full}}}^{\text{OS}}$ & \textcolor{magenta}{
\begin{picture}(6,1)
\put(0.00,1){\line(1,0){0.1}}
\put(0.25,1){\line(1,0){0.1}}
\put(0.50,1){\line(1,0){0.1}}
\put(0.75,1){\line(1,0){0.1}}
\put(1.00,1){\line(1,0){0.1}}
\put(1.25,1){\line(1,0){0.1}}
\put(1.50,1){\line(1,0){0.1}}
\put(1.75,1){\line(1,0){0.1}}
\put(2.00,1){\line(1,0){0.1}}
\put(2.25,1){\line(1,0){0.1}}
\put(2.50,1){\line(1,0){0.1}}
\put(2.75,1){\line(1,0){0.1}}
\put(3.00,1){\line(1,0){0.1}}
\put(3.25,1){\line(1,0){0.1}}
\put(3.50,1){\line(1,0){0.1}}
\put(3.75,1){\line(1,0){0.1}}
\put(4.00,1){\line(1,0){0.1}}
\put(4.25,1){\line(1,0){0.1}}
\put(4.50,1){\line(1,0){0.1}}
\put(4.75,1){\line(1,0){0.1}}
\put(5.00,1){\line(1,0){0.1}}
\put(5.25,1){\line(1,0){0.1}}
\put(5.50,1){\line(1,0){0.1}}
\put(5.75,1){\line(1,0){0.1}}
\end{picture}}
  & One-loop OS cross section including \\
  & & \quad resummed higher-order $\Delta m_b$ corrections \\\hline
${\sigma_1}^{\text{DR}}$ & \textcolor{cyan}{
\begin{picture}(6,1)
\put(0.0,1){\line(1,0){0.5}}
\put(0.7,1){\line(1,0){0.1}}
\put(1.0,1){\line(1,0){0.5}}
\put(1.7,1){\line(1,0){0.1}}
\put(2.0,1){\line(1,0){0.5}}
\put(2.7,1){\line(1,0){0.1}}
\put(3.0,1){\line(1,0){0.5}}
\put(3.7,1){\line(1,0){0.1}}
\put(4.0,1){\line(1,0){0.5}}
\put(4.7,1){\line(1,0){0.1}}
\put(5.0,1){\line(1,0){0.5}}
\put(5.7,1){\line(1,0){0.1}}
\end{picture}}
  & One-loop \DRbar{} cross section \\\hline
\end{tabular}
\end{center}
\caption{Key to the total hadronic cross sections of the process 
$b \bar{b} \rightarrow H^+ W^-$ plotted in
\fig{bbWH:hadron_tanbeta_v.absolut}.}
\label{bbWH:hadron_tanbeta_v.absolut_key}
\end{table}
In all cases the total cross section rises quadratically with $t_\beta$ in the region where
$t_\beta$ is larger than about 15. This is the parameter space where the
Yukawa coupling to the charged Higgs boson is dominated by the term proportional 
to the bottom-quark mass, which scales with $t_\beta$, and gives the leading
contribution to the cross section. On the left-hand side of the plot, where $t_\beta$ 
is small, in contrast the top-quark mass part is responsible for 
the overall behavior of the 
cross section and leads to a decrease with $t_\beta$. In the intermediate region 
both terms contribute equally much, leading to a minimum of the cross section 
for $t_\beta \approx 8$.

In total five different cross-section types including miscellaneous contributions 
are depicted. An overview is given in table~\ref{bbWH:hadron_tanbeta_v.absolut_key}.
${\sigma_0}^{\text{OS}}$, the straight red line, denotes the tree-level 
contribution in 
the OS renormalization scheme. The short-dashed blue line,
${\sigma_0}^{\text{OS}}$, is the 
one-loop cross section in the OS scheme without having 
used any further effective couplings. ${\sigma_0}^{\text{OS}}$, 
the long-dashed green line contains the
$\Delta m_b$-corrected tree-level result. As seen before in the plots of 
the relative corrections, this line agrees with the complete one-loop result 
in the case of large $t_\beta$ as in this region only the universal 
$\Delta m_b$ corrections are relevant. In the small-$t_\beta$ regime these
terms can only account for a part of the total corrections. There are also non-universal
terms, mainly coming from the sub-diagram given in \fig{bbWH:t_largecontri_prop},
which cannot be absorbed into an effective coupling. These ones do not, as 
observed before, contain any factors of $t_\beta$ and hence their effects
diminish for higher $t_\beta$ values as the 
total cross section scales with $t_\beta^2$. 
For ${\sigma_{\text{full}}}^{\text{OS}}$, the dotted pink line,
the $\Delta m_b$
corrections are included in the resummed version of \eq{renorm:lambdab}.
Additionally the non-universal one-loop corrections are added. To avoid
double-counting the non-resummed $\Delta m_b$ contribution must then be 
subtracted again, resulting in the following formula
\begin{equation}
\sigma_{\text{full}}^{OS} = \sigma_{\Delta \text{resum}}^{OS} 
  + \left( \sigma_1^{OS} - \sigma_{\Delta}^{OS} \right) \quad .
\end{equation}
In this line also phase-space effects from the reduced bottom-quark mass are 
taken into account, leading to an additional shift with respect to the tree-level
cross section. Nevertheless the curve again shows the expected behavior which 
can be deduced from the results given above with rising corrections for 
increasing $t_\beta$. For large values of $t_\beta$ they largely exceed the
one-loop result because of the $\Delta m_b$ resummation where contributions
are added that in a naive calculation would appear in higher-order diagrams
of perturbation theory. This curve presents the current best estimate for the total
cross section in the OS scheme where higher-order contributions are included as 
much as possible.
Lastly the one-loop expression in the \DRbar{} scheme is plotted as dash-dotted 
light-blue line and labeled
${\sigma_1}^{\text{DR}}$. It has a shape very similar to the previous curve because
the $\Delta m_b$ corrections appear in the self-energy contributions to the 
bottom-quark mass. In this renormalization scheme they enter completely at 
one-loop order via the bottom-quark propagators and no further contributions
at higher orders appear. Hence this corresponds to the resummed result in the
OS scheme. The remaining difference between the two curves is a measure 
for the theoretical uncertainty of the calculation because of missing non-leading
higher-order contributions from perturbation theory.

% SPA1a'
Finally the numerical results for the parameter point $\spa$~\cite{spa} are given in
table~\ref{bbWH:spa}. This parameter point is described in \app{param:spa},
it was chosen as a reference point for MSSM calculations.
\begin{table}
\begin{center}
\begin{tabular}{|c|c@{$=$}c|}
\hline
Tree-level cross section & $\sigma_0$  & $2.684$ fb \\\hline
$\Delta m_b$-corrected tree-level cross section & 
  $\sigma_{\Delta m_b}$ & $2.266$ fb \\\hline
One-loop OS cross section & $\sigma_1$  & 2.176 fb \\\hline
Relative one-loop OS correction & $\Delta_{OS}$ & $-15.6$ \% \\\hline
Relative true one-loop OS correction & $\Delta_{\Delta m_b}$ & $-4.0$ \% \\\hline
\end{tabular}
\end{center}
\caption{Hadronic cross sections for the reference point $\spa$, which is
described in \app{param:spa}.}
\label{bbWH:spa}
\end{table}
Because of the positive sign of $\mu$ the one-loop cross section is now reduced
with respect to the tree-level result. The $t_\beta$ value of $10$ is in a region
where the $\Delta m_b$ corrections are already the dominant ones, but the
non-universal corrections still yield a numerically significant contribution.

\chapter{Higgs-Boson Production via Vector Boson Fusion}
\label{vbf}

\begin{figure}
\begin{center}
\includegraphics[scale=0.6]{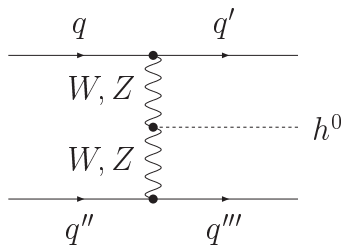}
\caption{Generic tree-level Feynman diagram of the vector-boson-fusion process
 $qq\rightarrow qqh^0$}
\label{vbf:vbf_born}
\end{center}
\end{figure}

\begin{figure}
\begin{center}
\subfigure[quark loop]{
\includegraphics[scale=0.6]{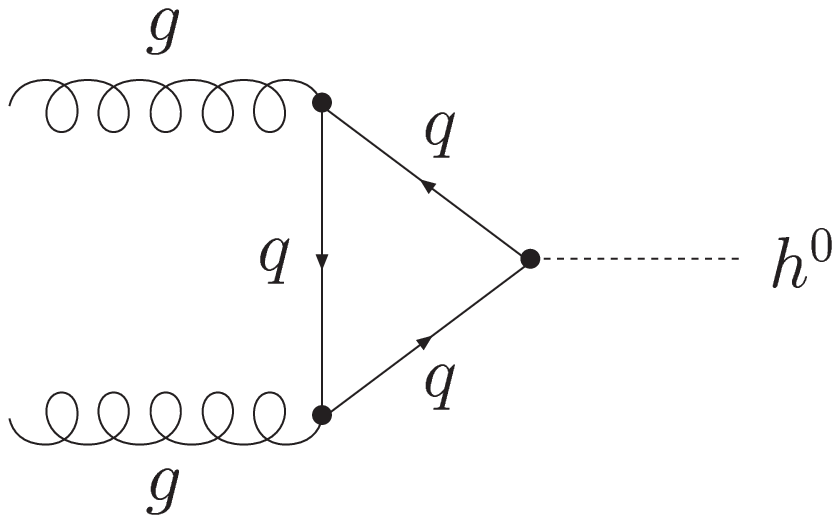}
}
\subfigure[squark loop]{
\includegraphics[scale=0.6]{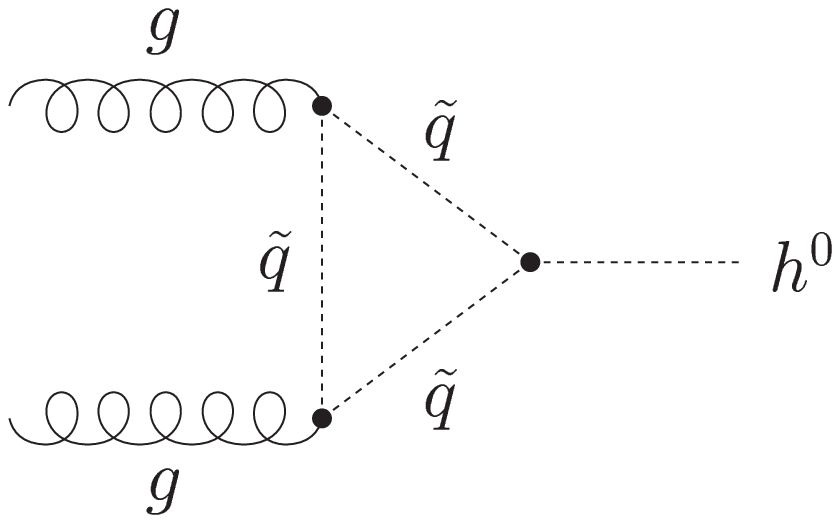}
}
\caption{Leading-order Feynman diagrams of the gluon-fusion process
 $gg\rightarrow h^0$}
\label{vbf:gluon_fusion}
\end{center}
\end{figure}

Proving the existence of a neutral Higgs boson is one of the main tasks of the LHC.
Its main production processes for both SM and MSSM Higgs bosons 
include Higgs-boson production via vector-boson 
fusion
(\fig{vbf:vbf_born})~\cite{Cahn:1983ip,*Cahn:1983ip2,Dicus:1985zg,Altarelli:1987ue}.
Its rate is surpassed 
only by the gluon-fusion process $gg \rightarrow h^0$~\cite{Georgi:1977gs} shown in 
\fig{vbf:gluon_fusion}.
This process has large NLO-QCD corrections with K-factors larger 
than $2$~\cite{Graudenz:1992pv,*Spira:1995rr,*Spira:1997dg}. Even after including 
the NNLO-QCD
corrections~\cite{Catani:2001ic,*Harlander:2001is,*Harlander:2002wh,*Anastasiou:2002yz,*Ravindran:2003um} 
theoretical uncertainties of $\Order{10-20\%}$ remain. They make the extraction of
coupling constants from the gluon-fusion process difficult and lead to large
errors. 

The Standard-QCD corrections to the vector-boson-fusion process were 
already calculated before~\cite{Han:1991ia,Figy:2003nv}.
At tree-level the process only consists of a t-channel exchange of a colorless
gauge boson, which is why the contribution where an additional gluon 
connects the two quark lines
has no interference term with the tree-level diagram. Only the 
quark--anti-quark--vector-boson vertex receives corrections 
of $\Order{\alpha_s}$ and
hence the overall QCD corrections are relatively small and typically between 
$5$ and $10 \%$. 
Therefore, the extraction of Higgs coupling constants leads to much smaller
errors than in the gluon-fusion case,
and the vector-boson-fusion production process is, despite its
lower cross section, an ideal instrument to study the Higgs boson. 

This process possesses a clear experimental signature of two jets in the forward
region and thus can be easily separated from background processes by 
applying appropriate cuts~\cite{Berger:2004pc}. 
In this chapter the production of the lightest MSSM
Higgs boson $h^0$ via vector-boson fusion is investigated and the SUSY-QCD
corrections to this process are calculated. 

\section{The Partonic Process}

The partonic processes which contribute to the production of a Higgs boson via 
vector-boson fusion can be summarized in a single general Feynman diagram
which is depicted in \fig{vbf:vbf_born}.
It can be seen as scattering of two quarks which is mediated via 
a vector boson in the t-channel with a Higgs boson being radiated off the 
intermediate vector boson. This is why the process has a clear experimental
signature of two jets in the forward region of the detector which allows one to 
easily distinguish the signal from background processes by using appropriate cuts.

In a strict sense, the general diagram given in \fig{vbf:vbf_born} is not the only one
which contributes to this final state. When a quark--anti-quark pair of the same 
flavor appears
in the initial state they can form a $Z h^0$ pair via an intermediate virtual $Z$ boson 
and the $Z$ subsequently decays again into a quark--anti-quark pair. Hence these 
diagrams have exactly the same particle content in both the external and internal lines.
There are also similar processes where an intermediate $W$ boson 
can appear in the same way. However,
the vector-boson-fusion process has a very distinct signature of two jets in 
the forward region. Using only this particular phase-space region 
the interference between the 
two diagram types is strongly suppressed by the large momentum transfer appearing
in the intermediate gauge bosons. Additionally a color suppression factor appears
in the interference term~\cite{Figy:2003nv}.
Hence the effects from these additional diagrams can be safely 
neglected if appropriate
cuts~\cite{Berger:2004pc} to the phase space are applied.

An analysis of the statistical accuracies of the cross section which are 
achievable at the LHC was done in
refs.~\cite{Zeppenfeld:2000td,*Zeppenfeld:2002ng,ATLAS-TDR}.
It could be shown that a measurement with an accuracy of $5$ to $10 \%$
can be performed, also taking uncertainties in the decay branching ratios of the
Higgs boson into account. So the size of the NLO-QCD corrections already matches
the accuracy which is achievable in experiment. Theoretical uncertainties
are not expected to limit the precision with which the cross sections can be measured.

In the MSSM besides the Standard-QCD corrections also SUSY-QCD corrections 
appear which are of the same order $\Order{\alpha_s}$ or, in case of the pentagon 
diagrams, even enhanced and of $\Order{\frac{\alpha_s^2}{\alpha}}$ 
in the coupling constant. A full one-loop calculation of
these corrections has not been done before and is presented in this thesis.

\section{SUSY-QCD Corrections}

In this chapter the SUSY-QCD corrections to $h^0$-production via vector-boson fusion
are studied. If their size is larger than
the experimental uncertainties one might be able to use this to tell the SM
and the MSSM apart. In the limit of large $m_A$ the couplings of the $h^0$ 
become SM-like. So if at the LHC only one Higgs boson with Standard-Model couplings
and a mass below $140 \GeV$ is found, the question arises whether a 
SM or a MSSM Higgs boson was seen in the detector. The SUSY-QCD corrections,
which exist only in the case of a MSSM Higgs boson, could modify the Higgs boson
coupling by an amount large enough 
and therefore make the distinction between the two
models possible. Also if supersymmetry could be established by other means 
beforehand, these corrections give an indirect contribution to the coupling between the
Higgs boson and two gauge bosons. To be able to extract the value from the 
experiment as precisely as possible it is necessary to include these higher-order
corrections.

\begin{figure}
\begin{center}
\subfigure[vertex corrections]{
\includegraphics[scale=0.5]{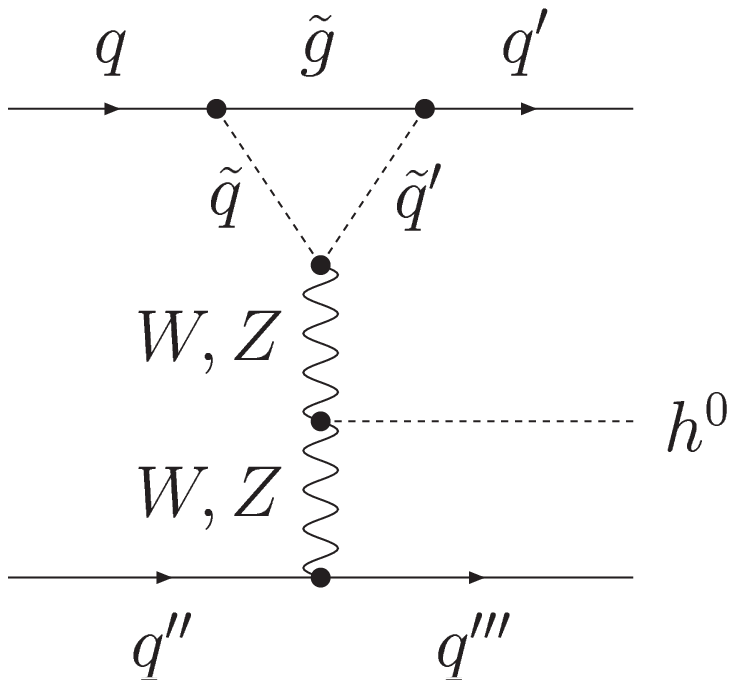}
\hspace*{1cm}
\includegraphics[scale=0.5]{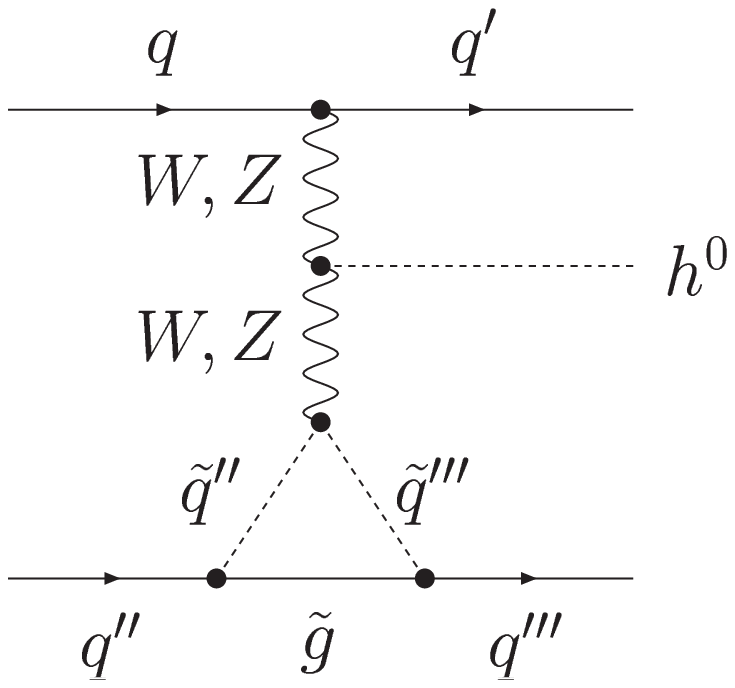}
}
\subfigure[vertex counter terms]{
\includegraphics[scale=0.5]{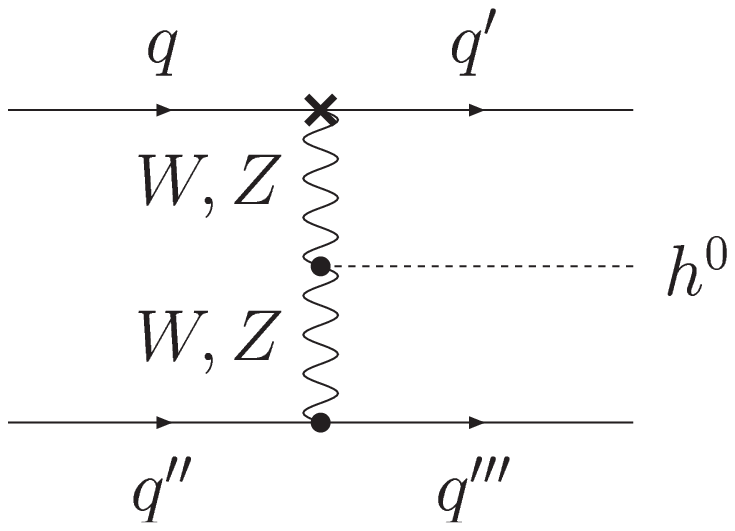}
\hspace*{1cm}
\includegraphics[scale=0.5]{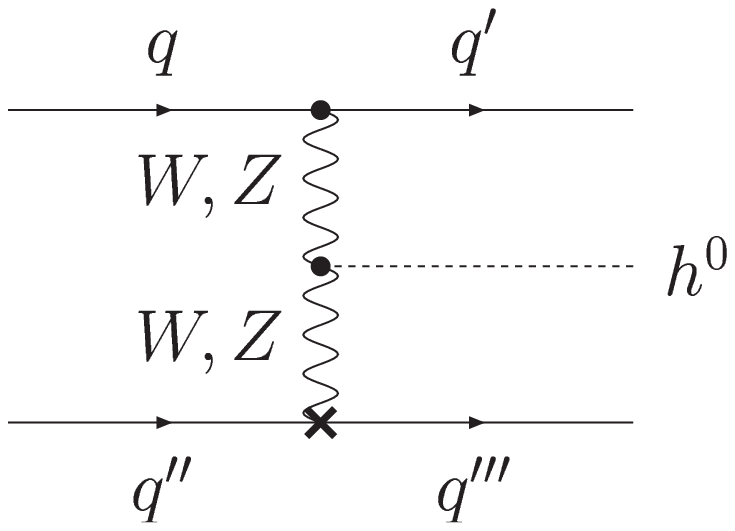}
}
\subfigure[box contribution]{
\parbox{\textwidth}{
\centering
\includegraphics[scale=0.5]{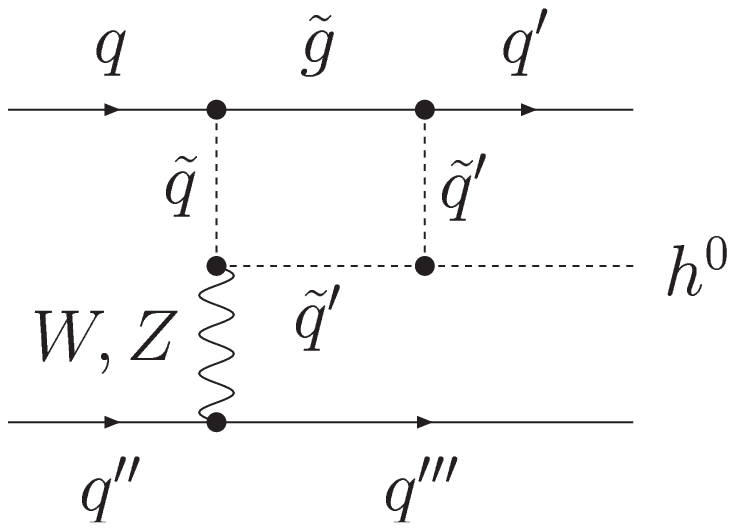}
\hspace*{1cm}
\includegraphics[scale=0.5]{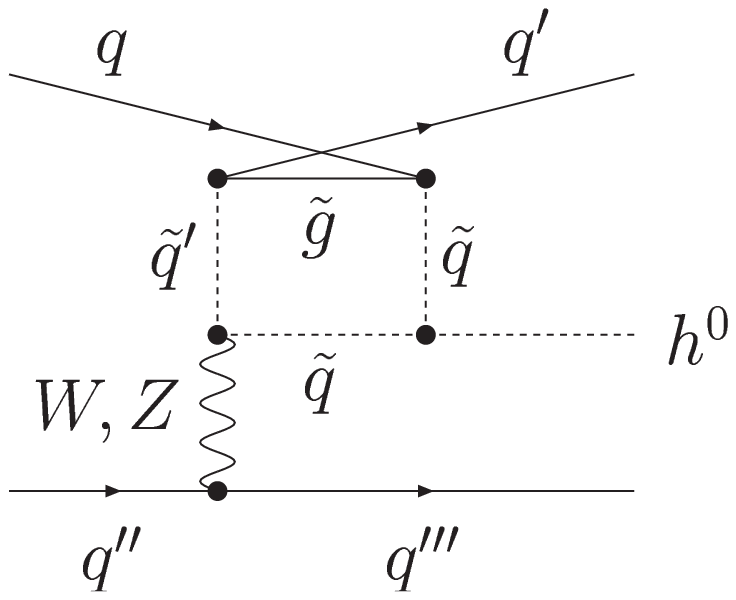}
\\[\baselineskip]
\includegraphics[scale=0.5]{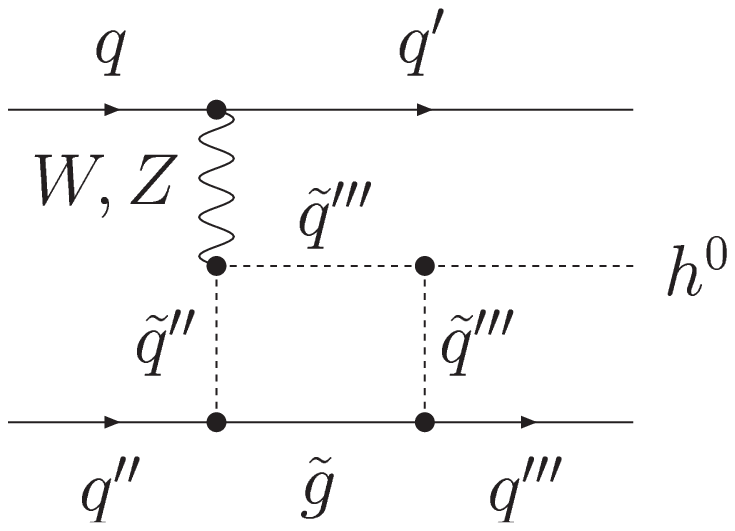}
\hspace*{1cm}
\includegraphics[scale=0.5]{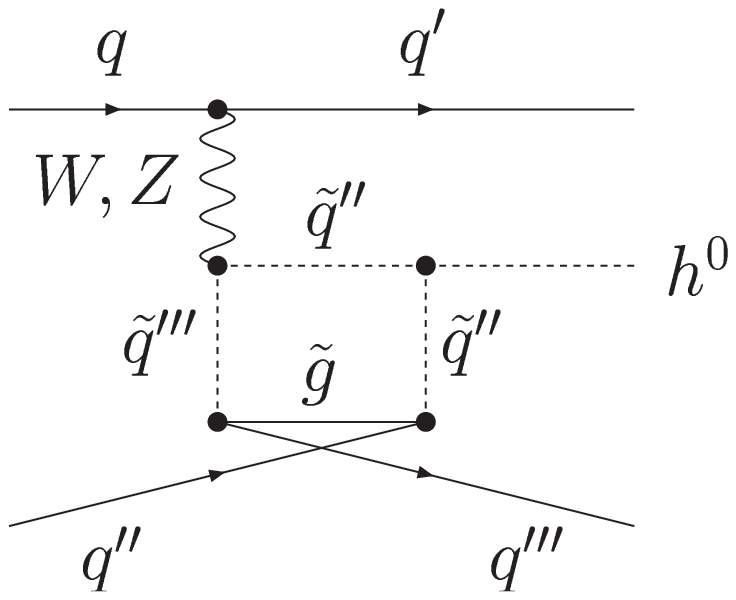}}
}
\subfigure[five-point contribution]{
\includegraphics[scale=0.5]{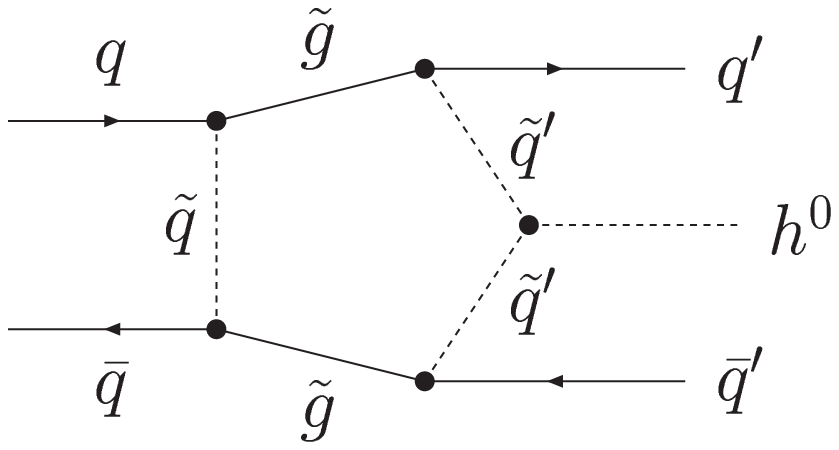}
\hspace*{1cm}
\includegraphics[scale=0.5]{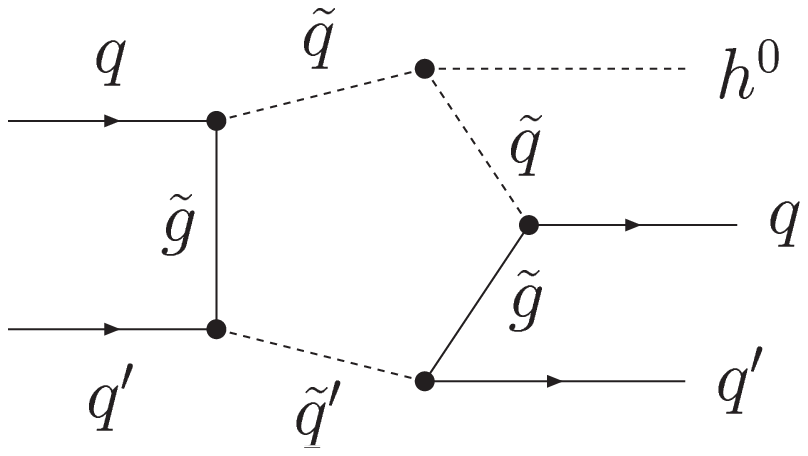}
}
\caption{Diagram types contributing to SUSY-QCD corrections to 
  $h^0$ production via vector-boson fusion}
\label{vbf:diags}
\end{center}
\end{figure}

The possible types of diagrams which appear in the 
SUSY-QCD corrections are depicted in \fig{vbf:diags}.
The quark--quark--gauge boson vertices receive corrections which are depicted
in \fig{vbf:diags}(a). Their divergencies are cancelled by appropriate counter 
terms which are shown in \fig{vbf:diags}(b). The contribution of these diagrams
was already investigated before~\cite{Djouadi:1999ht} for the special case 
where all squarks have equal mass. Additionally one of the gauge bosons
can be replaced by a box-shaped sparticle loop as shown in \fig{vbf:diags}(c).
Finally, all external particles can be coupled via a pentagon-type loop as in
\fig{vbf:diags}(d). Because of the Majorana nature of the gluinos also the diagram
displayed on the right-hand side of the figure, where two quark lines are connected,
exists. 

All cross sections are calculated in the OS renormalization scheme. For the 
tree-level diagrams the leading-order parton distribution functions of 
ref.~\cite{Martin:2002dr} were used. The one-loop cross sections were 
convoluted with the NLO-PDFs of the same group given in ref.~\cite{Martin:2002aw}.
In both cases the implementation from the program package
\LHAPDF{}~\cite{LHAPDFmanual} was used to
obtain the numerical results which are presented in the next section.

\section{Numerical Results}

In this section numerical results for the SUSY-QCD corrections to $h^0$-production
via vector-boson fusion are presented. The hadronic cross sections for each 
individual process for the MSSM 
reference point $\spa$, which is described in detail in appendix~\ref{param:spa}, are 
listed in the following table. The one-loop corrections are separated
according to the loop type, where $\sigma_{\text{vertex}}$ includes the 
contributions from the diagrams shown in \fig{vbf:diags}(a) and (b), 
$\sigma_{\text{box}}$ from the ones of \fig{vbf:diags}(c) and 
$\sigma_{\text{five-pt.}}$ from the pentagon diagrams depicted in 
\fig{vbf:diags}(d).
\begin{longtable}{l|r|r|r|r}
Partonic subprocess &\multicolumn{1}{c|}{$ \sigma_{\text{tree}} $ [pb] }&\multicolumn{1}{c|}{$ \frac{\sigma_{\text{vertex}}}{\sigma_{\text{tree}}}$} &\multicolumn{1}{c|}{ $ \frac{\sigma_{\text{box}}}{\sigma_{\text{tree}}}$} &\multicolumn{1}{c}{ $\frac{\sigma_{\text{five-pt.}}}{\sigma_{\text{tree}}}$} \\\hline\endhead
$      d      d  \rightarrow      d      d  h^0 $ & $ 1.76 \cdot 10^{-2} $ & $ -1.72 \cdot 10^{-4} $ & $ 8.34 \cdot 10^{-4} $ & $ 1.24 \cdot 10^{-5} $ \\
$      d      u  \rightarrow      d      u  h^0 $ & $ 3.46 \cdot 10^{-1} $ & $ -1.60 \cdot 10^{-4} $ & $ 6.68 \cdot 10^{-5} $ & $ 1.46 \cdot 10^{-6} $ \\
$      d      s  \rightarrow      d      s  h^0 $ & $ 1.21 \cdot 10^{-2} $ & $ -1.58 \cdot 10^{-4} $ & $ 7.89 \cdot 10^{-4} $ & $ -1.67 \cdot 10^{-5} $ \\
$      d      c  \rightarrow      d      c  h^0 $ & $ 6.39 \cdot 10^{-3} $ & $ -1.58 \cdot 10^{-4} $ & $ 2.69 \cdot 10^{-5} $ & $ 1.83 \cdot 10^{-6} $ \\
$      d      c  \rightarrow      u      s  h^0 $ & $ 3.34 \cdot 10^{-2} $ & $ -1.43 \cdot 10^{-4} $ & $ 9.07 \cdot 10^{-5} $ & $ 0 $ \\
$      d \bar{d} \rightarrow      d \bar{d} h^0 $ & $ 1.58 \cdot 10^{-2} $ & $ -1.55 \cdot 10^{-4} $ & $ 7.77 \cdot 10^{-4} $ & $ 2.48 \cdot 10^{-6} $ \\
$      d \bar{d} \rightarrow      u \bar{u} h^0 $ & $ 6.33 \cdot 10^{-2} $ & $ -1.40 \cdot 10^{-4} $ & $ 1.02 \cdot 10^{-5} $ & $ 1.52 \cdot 10^{-6} $ \\
$      d \bar{u} \rightarrow      d \bar{u} h^0 $ & $ 1.09 \cdot 10^{-2} $ & $ -1.53 \cdot 10^{-4} $ & $ 2.16 \cdot 10^{-5} $ & $ 1.66 \cdot 10^{-6} $ \\
$      d \bar{s} \rightarrow      d \bar{s} h^0 $ & $ 1.18 \cdot 10^{-2} $ & $ -1.56 \cdot 10^{-4} $ & $ 7.83 \cdot 10^{-4} $ & $ -8.90 \cdot 10^{-6} $ \\
$      d \bar{s} \rightarrow      u \bar{c} h^0 $ & $ 4.75 \cdot 10^{-2} $ & $ -1.41 \cdot 10^{-4} $ & $ 1.52 \cdot 10^{-5} $ & $ 0 $ \\
$      d \bar{c} \rightarrow      d \bar{c} h^0 $ & $ 6.26 \cdot 10^{-3} $ & $ -1.56 \cdot 10^{-4} $ & $ 2.86 \cdot 10^{-5} $ & $ 1.96 \cdot 10^{-6} $ \\
$      u      u  \rightarrow      u      u  h^0 $ & $ 3.68 \cdot 10^{-2} $ & $ -1.86 \cdot 10^{-4} $ & $ -6.65 \cdot 10^{-4} $ & $ -1.26 \cdot 10^{-5} $ \\
$      u      s  \rightarrow      d      c  h^0 $ & $ 1.10 \cdot 10^{-1} $ & $ -1.49 \cdot 10^{-4} $ & $ 8.16 \cdot 10^{-5} $ & $ 0 $ \\
$      u      s  \rightarrow      u      s  h^0 $ & $ 2.12 \cdot 10^{-2} $ & $ -1.65 \cdot 10^{-4} $ & $ 1.53 \cdot 10^{-4} $ & $ 1.84 \cdot 10^{-6} $ \\
$      u      c  \rightarrow      u      c  h^0 $ & $ 1.14 \cdot 10^{-2} $ & $ -1.64 \cdot 10^{-4} $ & $ -6.23 \cdot 10^{-4} $ & $ 2.20 \cdot 10^{-5} $ \\
$      u \bar{d} \rightarrow      u \bar{d} h^0 $ & $ 2.75 \cdot 10^{-2} $ & $ -1.63 \cdot 10^{-4} $ & $ 1.56 \cdot 10^{-4} $ & $ 1.59 \cdot 10^{-6} $ \\
$      u \bar{u} \rightarrow      d \bar{d} h^0 $ & $ 1.27 \cdot 10^{-1} $ & $ -1.44 \cdot 10^{-4} $ & $ 1.60 \cdot 10^{-4} $ & $ 2.00 \cdot 10^{-6} $ \\
$      u \bar{u} \rightarrow      u \bar{u} h^0 $ & $ 1.92 \cdot 10^{-2} $ & $ -1.61 \cdot 10^{-4} $ & $ -6.05 \cdot 10^{-4} $ & $ -3.15 \cdot 10^{-5} $ \\
$      u \bar{s} \rightarrow      u \bar{s} h^0 $ & $ 2.08 \cdot 10^{-2} $ & $ -1.64 \cdot 10^{-4} $ & $ 1.52 \cdot 10^{-4} $ & $ 1.82 \cdot 10^{-6} $ \\
$      u \bar{c} \rightarrow      d \bar{s} h^0 $ & $ 7.45 \cdot 10^{-2} $ & $ -1.46 \cdot 10^{-4} $ & $ 1.56 \cdot 10^{-4} $ & $ 0 $ \\
$      u \bar{c} \rightarrow      u \bar{c} h^0 $ & $ 1.12 \cdot 10^{-2} $ & $ -1.63 \cdot 10^{-4} $ & $ -6.17 \cdot 10^{-4} $ & $ 1.77 \cdot 10^{-5} $ \\
$      s      s  \rightarrow      s      s  h^0 $ & $ 8.40 \cdot 10^{-4} $ & $ -1.48 \cdot 10^{-4} $ & $ 7.62 \cdot 10^{-4} $ & $ 7.40 \cdot 10^{-6} $ \\
$      s      c  \rightarrow      s      c  h^0 $ & $ 4.37 \cdot 10^{-3} $ & $ -1.31 \cdot 10^{-4} $ & $ 6.04 \cdot 10^{-5} $ & $ 1.81 \cdot 10^{-6} $ \\
$      s \bar{d} \rightarrow      s \bar{d} h^0 $ & $ 2.22 \cdot 10^{-3} $ & $ -1.41 \cdot 10^{-4} $ & $ 7.34 \cdot 10^{-4} $ & $ -7.60 \cdot 10^{-6} $ \\
$      s \bar{d} \rightarrow      c \bar{u} h^0 $ & $ 9.10 \cdot 10^{-3} $ & $ -1.27 \cdot 10^{-4} $ & $ 1.93 \cdot 10^{-5} $ & $ 0 $ \\
$      s \bar{u} \rightarrow      s \bar{u} h^0 $ & $ 1.50 \cdot 10^{-3} $ & $ -1.40 \cdot 10^{-4} $ & $ -5.07 \cdot 10^{-6} $ & $ 1.72 \cdot 10^{-6} $ \\
$      s \bar{s} \rightarrow      s \bar{s} h^0 $ & $ 1.62 \cdot 10^{-3} $ & $ -1.42 \cdot 10^{-4} $ & $ 7.42 \cdot 10^{-4} $ & $ 1.58 \cdot 10^{-6} $ \\
$      s \bar{s} \rightarrow      c \bar{c} h^0 $ & $ 6.69 \cdot 10^{-3} $ & $ -1.28 \cdot 10^{-4} $ & $ 2.33 \cdot 10^{-5} $ & $ 3.29 \cdot 10^{-8} $ \\
$      s \bar{c} \rightarrow      s \bar{c} h^0 $ & $ 8.27 \cdot 10^{-4} $ & $ -1.43 \cdot 10^{-4} $ & $ 1.79 \cdot 10^{-6} $ & $ 1.96 \cdot 10^{-6} $ \\
$      c      c  \rightarrow      c      c  h^0 $ & $ 1.70 \cdot 10^{-4} $ & $ -1.49 \cdot 10^{-4} $ & $ -5.99 \cdot 10^{-4} $ & $ -6.09 \cdot 10^{-6} $ \\
$      c \bar{d} \rightarrow      c \bar{d} h^0 $ & $ 9.00 \cdot 10^{-4} $ & $ -1.41 \cdot 10^{-4} $ & $ 1.88 \cdot 10^{-4} $ & $ 1.75 \cdot 10^{-6} $ \\
$      c \bar{u} \rightarrow      s \bar{d} h^0 $ & $ 4.17 \cdot 10^{-3} $ & $ -1.27 \cdot 10^{-4} $ & $ 1.35 \cdot 10^{-4} $ & $ 0 $ \\
$      c \bar{u} \rightarrow      c \bar{u} h^0 $ & $ 6.09 \cdot 10^{-4} $ & $ -1.40 \cdot 10^{-4} $ & $ -5.66 \cdot 10^{-4} $ & $ 1.54 \cdot 10^{-5} $ \\
$      c \bar{s} \rightarrow      c \bar{s} h^0 $ & $ 6.56 \cdot 10^{-4} $ & $ -1.43 \cdot 10^{-4} $ & $ 1.85 \cdot 10^{-4} $ & $ 1.91 \cdot 10^{-6} $ \\
$      c \bar{c} \rightarrow      s \bar{s} h^0 $ & $ 2.27 \cdot 10^{-3} $ & $ -1.31 \cdot 10^{-4} $ & $ 1.33 \cdot 10^{-4} $ & $ 4.08 \cdot 10^{-6} $ \\
$      c \bar{c} \rightarrow      c \bar{c} h^0 $ & $ 3.31 \cdot 10^{-4} $ & $ -1.44 \cdot 10^{-4} $ & $ -5.84 \cdot 10^{-4} $ & $ -3.33 \cdot 10^{-5} $ \\
$ \bar{d}\bar{d} \rightarrow \bar{d}\bar{d} h^0 $ & $ 1.81 \cdot 10^{-3} $ & $ -1.43 \cdot 10^{-4} $ & $ 7.32 \cdot 10^{-4} $ & $ 7.41 \cdot 10^{-6} $ \\
$ \bar{d}\bar{u} \rightarrow \bar{d}\bar{u} h^0 $ & $ 1.36 \cdot 10^{-2} $ & $ -1.28 \cdot 10^{-4} $ & $ 5.79 \cdot 10^{-5} $ & $ 1.69 \cdot 10^{-6} $ \\
$ \bar{d}\bar{s} \rightarrow \bar{d}\bar{s} h^0 $ & $ 2.64 \cdot 10^{-3} $ & $ -1.40 \cdot 10^{-4} $ & $ 7.29 \cdot 10^{-4} $ & $ -1.72 \cdot 10^{-5} $ \\
$ \bar{d}\bar{c} \rightarrow \bar{d}\bar{c} h^0 $ & $ 1.35 \cdot 10^{-3} $ & $ -1.41 \cdot 10^{-4} $ & $ 4.16 \cdot 10^{-7} $ & $ 1.73 \cdot 10^{-6} $ \\
$ \bar{d}\bar{c} \rightarrow \bar{u}\bar{s} h^0 $ & $ 7.16 \cdot 10^{-3} $ & $ -1.28 \cdot 10^{-4} $ & $ 8.44 \cdot 10^{-5} $ & $ 0 $ \\
$ \bar{u}\bar{u} \rightarrow \bar{u}\bar{u} h^0 $ & $ 7.23 \cdot 10^{-4} $ & $ -1.44 \cdot 10^{-4} $ & $ -5.70 \cdot 10^{-4} $ & $ -6.54 \cdot 10^{-6} $ \\
$ \bar{u}\bar{s} \rightarrow \bar{d}\bar{c} h^0 $ & $ 8.30 \cdot 10^{-3} $ & $ -1.31 \cdot 10^{-4} $ & $ 7.43 \cdot 10^{-5} $ & $ 0 $ \\
$ \bar{u}\bar{s} \rightarrow \bar{u}\bar{s} h^0 $ & $ 1.56 \cdot 10^{-3} $ & $ -1.44 \cdot 10^{-4} $ & $ 1.76 \cdot 10^{-4} $ & $ 1.90 \cdot 10^{-6} $ \\
$ \bar{u}\bar{c} \rightarrow \bar{u}\bar{c} h^0 $ & $ 7.90 \cdot 10^{-4} $ & $ -1.45 \cdot 10^{-4} $ & $ -5.81 \cdot 10^{-4} $ & $ 2.19 \cdot 10^{-5} $ \\
$ \bar{s}\bar{s} \rightarrow \bar{s}\bar{s} h^0 $ & $ 8.40 \cdot 10^{-4} $ & $ -1.49 \cdot 10^{-4} $ & $ 7.63 \cdot 10^{-4} $ & $ 7.40 \cdot 10^{-6} $ \\
$ \bar{s}\bar{c} \rightarrow \bar{s}\bar{c} h^0 $ & $ 4.37 \cdot 10^{-3} $ & $ -1.31 \cdot 10^{-4} $ & $ 6.02 \cdot 10^{-5} $ & $ 1.81 \cdot 10^{-6} $ \\
$ \bar{c}\bar{c} \rightarrow \bar{c}\bar{c} h^0 $ & $ 1.70 \cdot 10^{-4} $ & $ -1.49 \cdot 10^{-4} $ & $ -5.98 \cdot 10^{-4} $ & $ -6.10 \cdot 10^{-6} $ \\\hline
$ \sum\left(h^0 \text{ via VBF}\right)$ & $ 1.11 \hspace{32pt}$ & $ -1.53 \cdot 10^{-4} $ & $ 7.69 \cdot 10^{-5} $ & $ 3.44 \cdot 10^{-7} $ 
\end{longtable}
The quoted cross sections are hadronic ones, where the convolution with the PDFs
has already been performed.
They are given separately for each partonic subprocess
to facilitate an easier 
analysis of the characteristic effects appearing in this process. Additionally
the total hadronic cross section is stated, which is the sum of all subprocesses.
For some partonic subprocesses no five-point loop diagrams exist at all. In this
case the entry in the last column of the table is exactly zero.
To exploit the unique characteristics of the final state of this process cuts 
were used to obtain the cross sections. A lower limit was placed on the 
transverse momentum $p_T$ and the pseudo-rapidity $\eta$ of the outgoing quarks and 
anti-quarks, so that the final-state jets are clearly separated from the beam pipe,
but still in the forward region of the detector. Also a cut on the 
jet separation $\Delta R$ between each combination of outgoing particles 
was set to emulate the behavior of the jet-clustering algorithms
used in experimental analyses 
and to be able to resolve the particles in the detector separately.
Thus the applied cuts were
\begin{align}
p_T(q,\bar{q}) &\ge 40 \GeV & \eta(q,\bar{q}) &\ge 2 & 
\Delta R_{qq, q\bar{q}, \bar{q}\bar{q}, qh^0, \bar{q}h^0} &\ge 0.4 \quad .
\label{vbf:cuts}
\end{align}
The formal definition of these quantities was given in \chap{hadWQ:cuts}.

There is an interesting observation already at tree-level. The partonic subprocesses
which enter via a $Z$-boson exchange are suppressed with respect to the ones
with a $W$ boson as intermediate vector boson. Firstly, the coupling of the $W$ boson
to the quarks is enhanced by a factor of $\frac1{c_W}$, the inverse of the 
cosine of the electroweak
mixing angle, with respect to the leading term of the $qqZ$-coupling. Secondly, 
the $Z$ boson is heavier than the $W$ boson, and the ratio of the two masses is also 
equal to $\frac1{c_W}$. As the gauge-boson propagators are dominated by their 
mass terms this leads to an additional enhancement of $\frac1{c_W^2}$ for each 
$W$ boson propagator. So in total the amplitude of a $W$ boson-exchange diagram
is enhanced by about $\frac1{c_W^6}$ over one with a $Z$ boson exchange.
Accordingly, this amounts to a factor 
$\frac1{c_W^{12}} = \frac1{{(0.877)}^{12}}\simeq 4.8$ for the tree-level 
cross section, which corresponds to the observed partonic cross-section ratios.
Because of this effect and the large
valence-quark densities of the up- and down-quarks in the proton, the partonic
subprocess $ud\rightarrow udh^0$ gives the leading contribution to the hadronic
process.

The vertex corrections all have the same size relative to the respective tree-level
cross section and correspond to the results obtained in ref.~\cite{Djouadi:1999ht}. 
Since the
coupling of the $W$ boson to the quarks is purely left-handed, they are largest if 
the off-diagonal elements in the squark mixing matrix are small and therefore 
left- and right-handed squarks have almost equal masses. Also for the intermediate
gluino propagator only the momentum term survives when calculating the trace over
the fermion line. It is proportional to the momentum transfer in the t-channel
and thus small.
For diagrams with $Z$-boson exchange the situation is more
complicated, because the $qqZ$-coupling contains both left- and right-handed parts.
Nevertheless, any subdiagrams involving a mixing of left- and right-handed squarks
are proportional to the off-diagonal terms in the squark mixing matrix, which contain 
the Yukawa coupling of the corresponding quark and are hence small. Hence, the
same effects as in the $W$ boson case appear and lead to a similar relative
size of the vertex corrections.

In the case of box diagrams the relative size of the corrections shows a much wider
range. For some partonic subprocesses they exceed the vertex corrections
significantly, which underlines the importance of performing a full 
one-loop calculation to take 
all effects into account. Yet for other subprocesses they are much smaller.
This is due to the fact that for $W$-boson exchange large cancellation is manifest.
It occurs between the diagrams on the left-hand side of each row of \fig{vbf:diags}(c), 
where the $h^0$ couples 
to the squark with the flavor of the outgoing quark, and the ones on the right-hand side
which have an $h^0$-coupling to the squark with the incoming-quark flavor.
Because of the $W$ boson one of the squarks is always up-type and the other 
one down-type.
Their couplings to the $h^0$ have a minus sign relative to each other. 
Any effects from CKM-mixing are neglected, therefore only squarks of the same
generation can appear in a single diagram. This is also why only the superpartners 
of the four light quarks contribute at all. They all have 
very similar masses, so the absolute value of the $\tilde{q}\tilde{q}W$-coupling
is roughly the same everywhere. Hence the diagrams on the left- and right-hand side of 
\fig{vbf:diags}(c) almost exactly cancel for $W$-boson exchange, leading to a strong
suppression of this contribution.
For $Z$-boson exchange no such effect occurs and the relative box-diagram
contribution is larger by an order of magnitude. Since, as mentioned above,
already the tree-level amplitude is smaller for this diagram type, 
the absolute value of the correction is small as well and these corrections
cannot give a significant contribution to the total cross section.
Again the correction is maximized in the case where the off-diagonal elements in the
squarx mixing matrix are small. Also the change of the coupling constant between
tree-level, which is proportional to $s_{\alpha+\beta}$, to the one-loop one of 
$s_{\beta-\alpha}$ cannot give an important effect, because the ratio of the two
is always above $0.9$ and approaches $1$ in the decoupling scenario, where the
additional Higgs bosons of the MSSM are heavy and the $h^0$-coupling becomes
SM-like.

For the five-point diagrams a cancellation similar to the box diagrams occurs
when the corresponding tree-level process is mediated by a $W$-boson exchange.
Additionally at least twice a left-right mixing term in the squark sector 
appears in the amplitude. 
For this reason a term proportional to the Yukawa coupling of the four light quarks
enters the expression and leads to a suppression of the one-loop correction.
The choice of parameters which yield the biggest one-loop corrections
are in this case large terms in the off-diagonal
entries of the squark mixing matrices and therefore a larger mass splitting
in the squark sector.
Additionally for both types the larger masses of the gluinos and squarks,
where experimental limits require that they are heavier than the $W$  or $Z$ boson,
lead to a further reduction of the cross section. Yet, there is also an enhancement
factor. Except for the Higgs coupling, all other four couplings of the pentagon 
diagrams are proportional to the strong coupling constant, while for the triangular
and box-type diagrams two of the couplings are strong and two of them are
electroweak. Hence this type of diagram contains an enhancement factor of 
$\frac{\alpha_s}{\alpha} \simeq 14$. This is however not sufficient to give a 
significant contribution to the cross section.

\section{\texorpdfstring{$h^0$}{h0}-Production with External Gluons}

\begin{figure}
\begin{center}
\subfigure[vertex corrections]{
\parbox{\textwidth}{
\begin{center}
\includegraphics[scale=0.5]{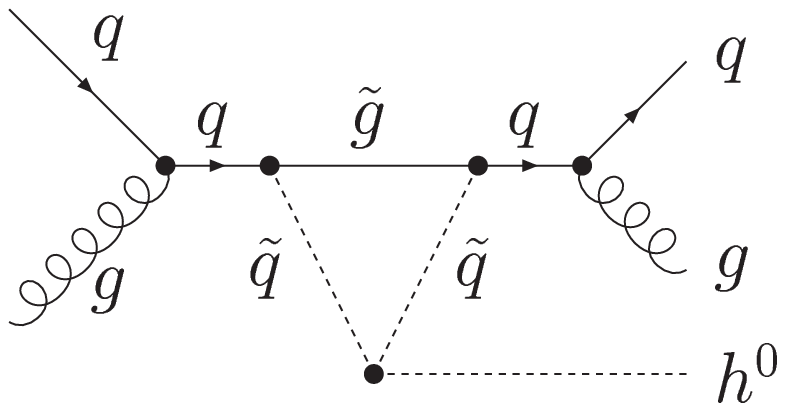}
\hspace*{1cm}
\includegraphics[scale=0.5]{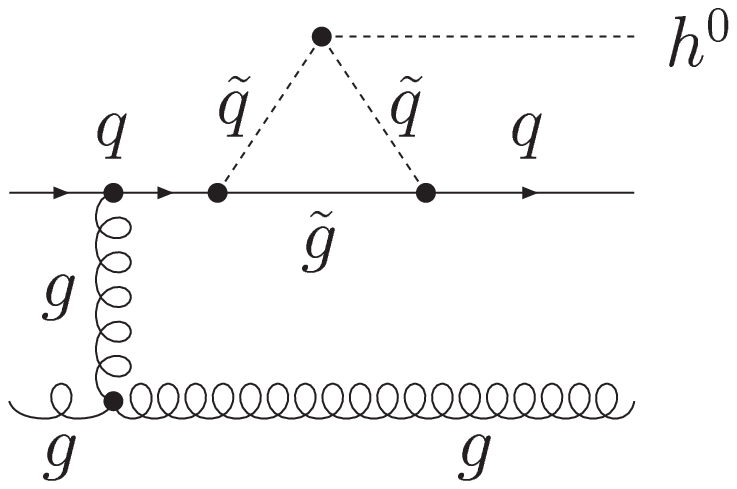} \\
\includegraphics[scale=0.5]{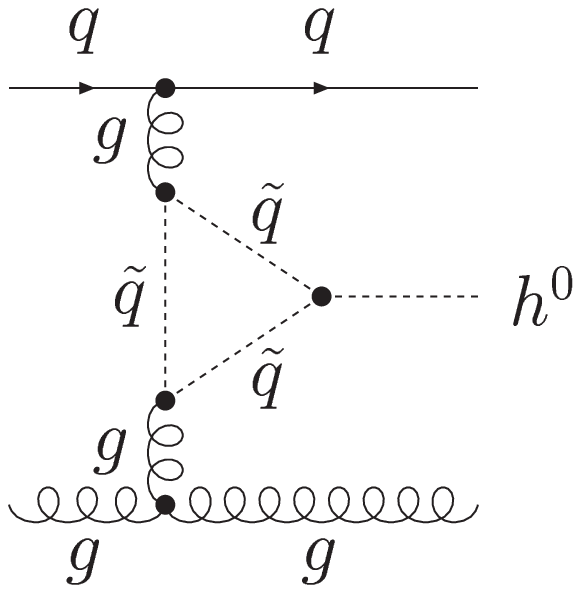}
\hspace*{1cm}
\includegraphics[scale=0.5]{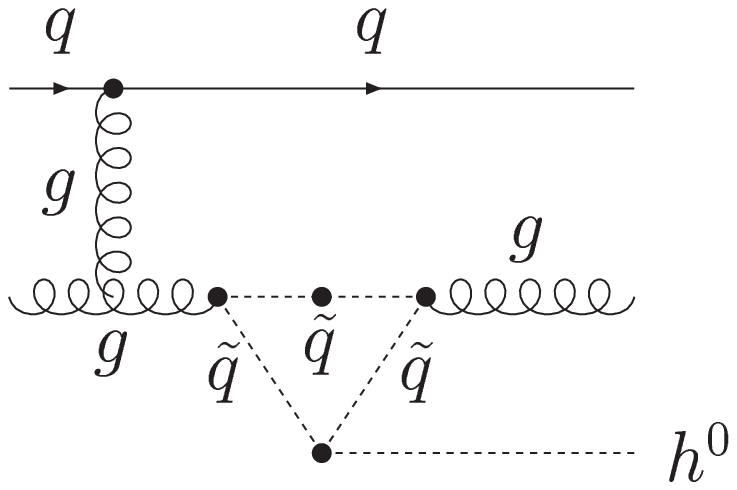}
\end{center}
}
}
\subfigure[box diagrams]{
\parbox{\textwidth}{
\begin{center}
\includegraphics[scale=0.5]{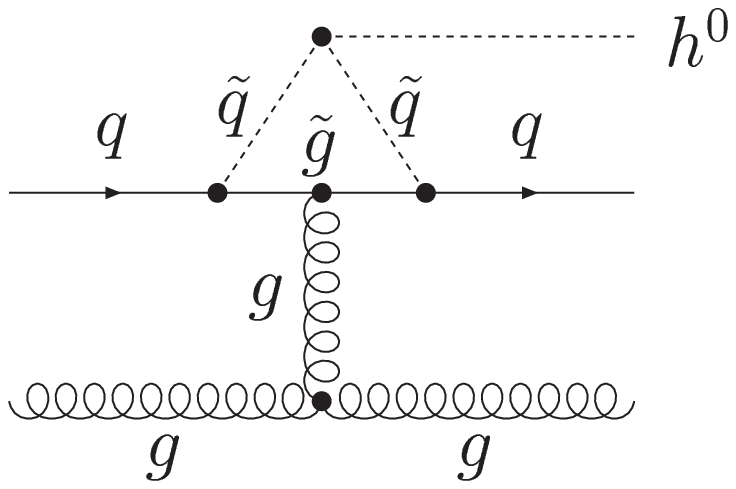}
\hspace*{1cm}
\includegraphics[scale=0.5]{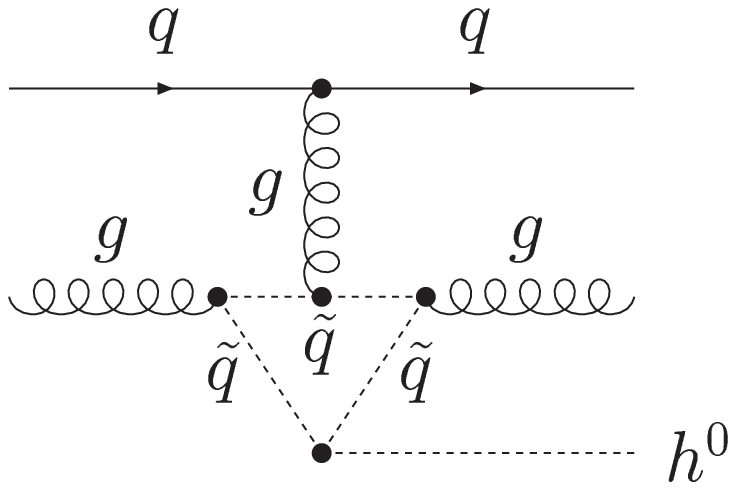} 
\hspace*{1cm}
\includegraphics[scale=0.5]{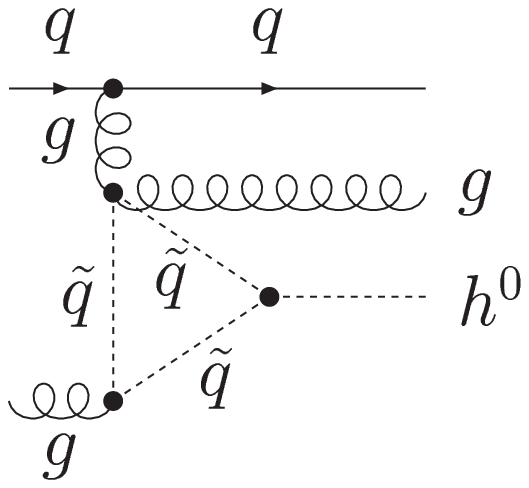} \\
\includegraphics[scale=0.5]{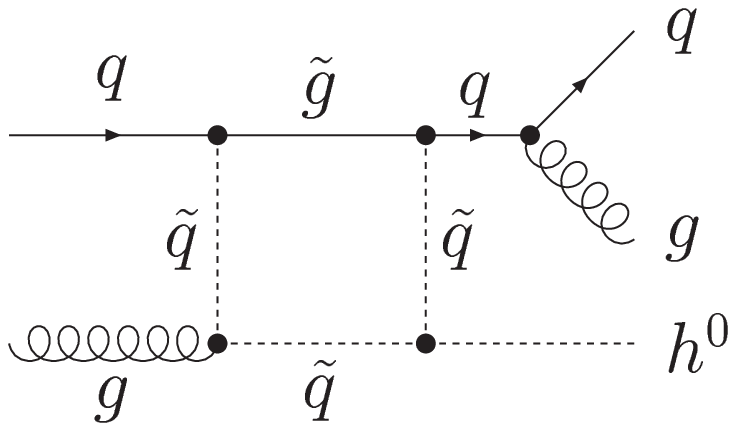}
\hspace*{1cm}
\includegraphics[scale=0.5]{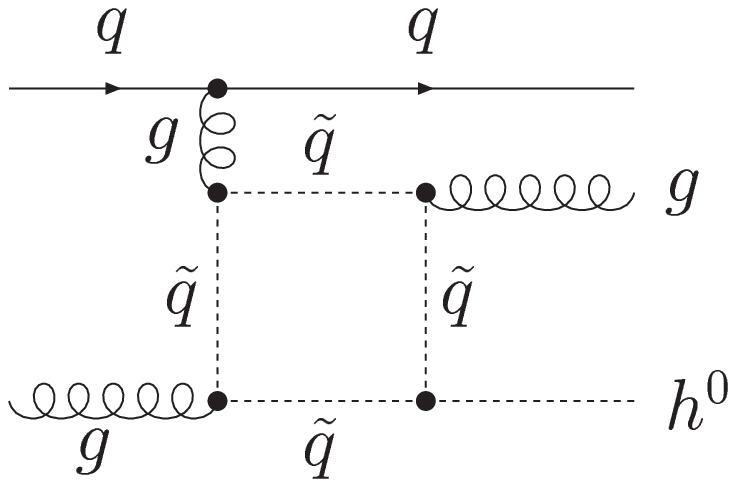}
\end{center}
}
}
\subfigure[pentagon diagrams]{
\includegraphics[scale=0.5]{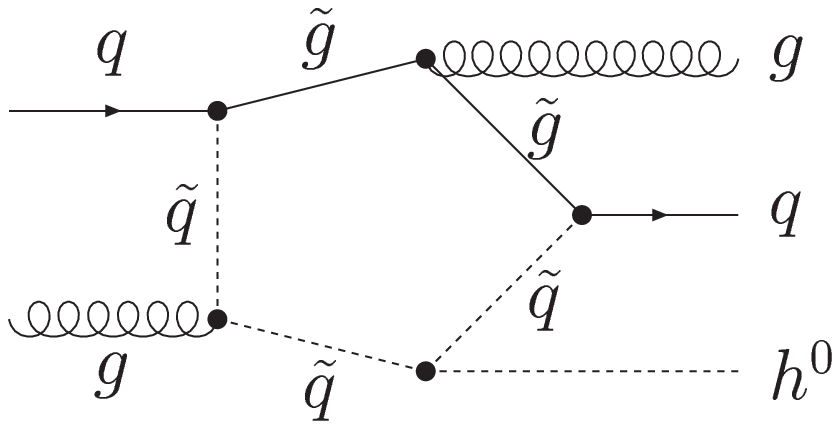}
\hspace*{1cm}
\includegraphics[scale=0.5]{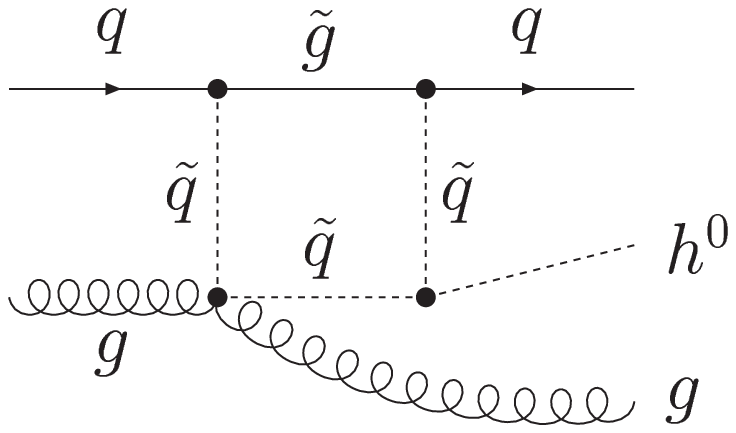} 
}
\caption{Leading-order diagram types for $h^0$-production with one external gluon
in the initial and final state. The Feynman diagrams with two gluons in the initial state
have the same topology. They are obtained from these ones by taking both gluons
as incoming particles and changing the incoming quark to an outgoing anti-quark.}
\label{vbf:diag_gluon}
\end{center}
\end{figure}

Additionally the production of an $h^0$ with one or two gluons in the initial state
was considered. The corresponding Feynman diagrams are shown in
\fig{vbf:diag_gluon}.
Since there is no tree-level coupling of the gluons to the Higgs bosons, these
processes occur at the one-loop level in leading order. This leads to an additional
factor of $\alpha_s$ in the cross section so the total amplitude is of
$\Order{\alpha_s^4 \alpha}$, 
while the vector-boson-fusion diagram has a factor $\alpha^3$.
Given that $\alpha_s^2$ is of the same numerical size as $\alpha$ these contributions
might prove important. Additionally, the gluon densities in the proton are much higher
than those of the quarks at typical LHC energies and this further enhances this 
type of diagram.

The numerical results for these processes with external gluons for the
reference point $\spa$ are given in table~\ref{vbf:sigma_hgluon}.
\begin{table}
\begin{center}
\begin{tabular}{l|r}
Partonic subprocess &\multicolumn{1}{c}{$ \sigma_{\text{one-loop}} $ [pb] }\\\hline
$ g      d  \rightarrow      g      d  h^0 $  & $ 2.2 \cdot 10^{-5} $ \\
$ g      u  \rightarrow      g      u  h^0 $  & $ 1.7 \cdot 10^{-5} $ \\
$ g      s  \rightarrow      g      s  h^0 $  & $ 2.3 \cdot 10^{-6} $ \\
$ g      c  \rightarrow      g      c  h^0 $  & $ 9.3 \cdot 10^{-7} $ \\
$ g \bar{d} \rightarrow      g \bar{d} h^0 $  & $ 3.2 \cdot 10^{-6} $ \\
$ g \bar{u} \rightarrow      g \bar{u} h^0 $  & $ 2.5 \cdot 10^{-6} $ \\
$ g \bar{s} \rightarrow      g \bar{s} h^0 $  & $ 2.1 \cdot 10^{-6} $ \\
$ g \bar{c} \rightarrow      g \bar{c} h^0 $  & $ 8.2 \cdot 10^{-7} $ \\
$ g      g  \rightarrow      d \bar{d} h^0 $  & $ 1.8 \cdot 10^{-7} $ \\
$ g      g  \rightarrow      u \bar{u} h^0 $  & $ 2.3 \cdot 10^{-7} $ \\
$ g      g  \rightarrow      s \bar{s} h^0 $  & $ 2.6 \cdot 10^{-7} $ \\
$ g      g  \rightarrow      c \bar{c} h^0 $  & $ 2.5 \cdot 10^{-7} $ 
\end{tabular}
\caption{One-loop hadronic cross sections for the subprocesses with one or 
  two gluons in the initial state for the MSSM reference point $\spa$.}
\label{vbf:sigma_hgluon}
\end{center}
\end{table}
In this case the convolution with the PDFs is also already included in 
the numbers for the cross section.
The same cuts as for the vector-boson-fusion process, given in \eq{vbf:cuts},
were used. Also the same cuts were applied to the final-state gluons
as to the quarks and anti-quarks.

Formally this process type constitutes a background to the previously 
considered process of $h^0$-production via vector-boson fusion.
Therefore one wants to pursue the question, how large the total contribution
of these diagrams is and if there are cuts which reduce their size with 
respect to the signal process.

The processes with one gluon in the initial and final state have a momentum 
distribution similar to the vector-boson-fusion one. The gluon densities
in the proton are very large for small $x$, but rapidly diminish for larger $x$.
In contrast the sea-quark densities fall off much slower and the valence-quark
densities have their maximum at about $\frac16$. So the most favorable 
configuration is the one where the energy to produce the final state mostly 
originates from the quark.
Thus the hadronic center-of-mass frame is strongly boosted with respect 
to the partonic one, which leads to jets in the forward region of the detector.
This would mimic the signature of a vector-boson-fusion process 
and produce events which cannot be eliminated by cuts.
In contrast for processes with two gluons in the initial state the 
momentum configuration which maximizes the hadronic
cross section is the one where both gluons have similar values of $x$.
This leads to more central jets which are suppressed by the applied cuts.

As one can see from the cross sections in table~\ref{vbf:sigma_hgluon}, 
the total contribution of these diagrams is of $\Order{10^{-4}}$ 
and therefore well below the experimental uncertainties,
which are in the range of $5$ to $10\%$.
As the total contribution of these background processes is below the 
statistical uncertainties which can be reached in a measurement of the
vector-boson-fusion cross section, these background processes
do not affect the experimental determination of the $h^0$-production
rate via vector-boson fusion.

\chapter{Higgs-Boson Production in Association with Heavy Quarks}
\label{hq}

The coupling of Higgs bosons to fermions is of the Yukawa type and therefore
proportional to the mass of the fermion. The four light quarks, 
$u$, $d$, $s$ and $c$, all have a mass below or of about $1 \GeV$. 
This mass should be compared to the Higgs vacuum expectation value $v$, the scale of electroweak 
symmetry breaking, to obtain the strength of their respective Yukawa couplings, 
which are therefore small. In contrast the top-quark 
mass is of the same order as the electroweak symmetry-breaking scale, 
making the top-Higgs coupling numerically sizable. The bottom-quark mass
of a few$\GeV$ also leads to a rather weak coupling to the Higgs boson 
in the Standard Model. In the MSSM, the coupling to the $h^0$ is enhanced
by a factor $t_\beta$, so for large values of $t_\beta$ its size can become
comparable to the top-quark Yukawa coupling.
These large Yukawa couplings make Higgs-boson production in association with heavy 
quarks~\cite{Raitio:1978pt,*Bagdasaryan:1986fj,*Ng:1983jm,Kunszt:1984ri} 
a phenomenologically interesting process.

In this chapter the production of the lightest CP-even neutral MSSM-Higgs boson
$h^0$ in association with a bottom or top quark--anti-quark pair is studied.
The top quarks decay rapidly into mainly $bW$ and the outgoing bottom 
quarks can be identified in the detector via $b$-quark tagging. 
Therefore both processes form distinct final states which 
neither interfere with each other nor with the Higgs-boson production
via vector-boson fusion presented in \chap{vbf}.
First the peculiarities of each of the two processes are discussed separately.
Then the one-loop SUSY-QCD corrections for both processes are described. Since
the same basic Feynman diagrams appear in both cases, this task is done jointly.
The Standard-QCD corrections have already been calculated in 
refs.~\cite{Balazs:1998sb,*Dicus:1998hs,Dittmaier:2003ej,*Dawson:2003kb,*Dawson:2005vi}
for $b\bar{b}h^0$-production and 
refs.~\cite{Beenakker:2001rj,Reina:2001sf,*Dawson:2002tg,Beenakker:2002nc}
for $t\bar{t}h^0$-production.
Finally in the last two sections the numerical results for both processes are shown.

\section{The \texorpdfstring{$b\bar{b}h^0$}{b b h0} Final State}

The production of a Higgs boson in association with bottom quarks in the Standard
Model was intensively studied in the 
literature~\cite{ATLAS-TDR,Dicus:1988cx,Campbell:2004pu}. 
\begin{figure}
\begin{center}
\subfigure[s-channel contribution]{
\includegraphics[scale=0.5]{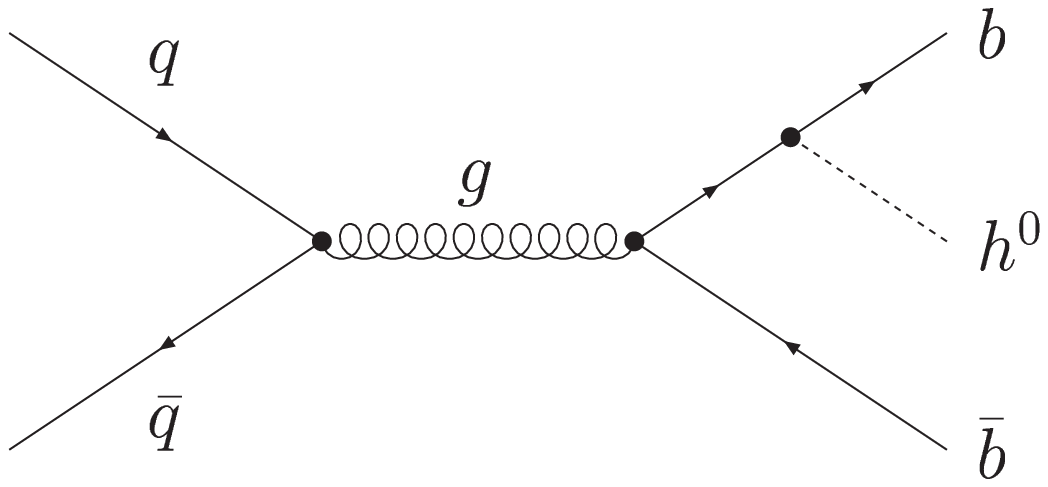}
\hspace*{1cm}
\includegraphics[scale=0.5]{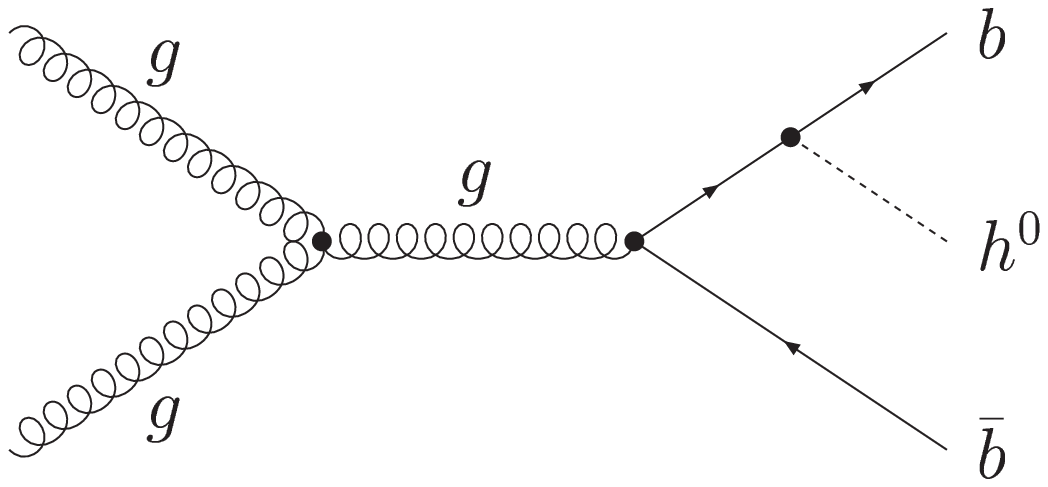}
}
\subfigure[t-channel contribution]{
\hspace*{5ex}
\includegraphics[scale=0.5]{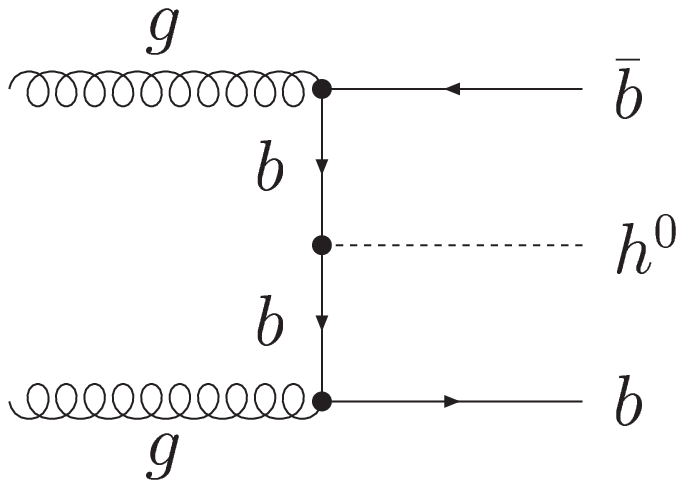}
\hspace*{5ex}
}
\caption{Tree-level Feynman diagrams of the 
  process $pp \rightarrow b\bar{b}h^0$}
\label{hq:tree}
\end{center}
\end{figure}
At tree-level it originates from the annihilation of a quark--anti-quark pair
or from a gluon fusion process, where the final-state
$b\bar{b}$-pair is produced via an intermediate gluon, and the Higgs boson
radiates off from one of the bottom quarks (\fig{hq:tree}(a)). 
Besides these s-channel
diagrams the partonic gluon-fusion process also proceeds via a 
t-channel diagram shown in \fig{hq:tree}(b), where the Higgs boson can 
be emitted from both the internal and external bottom-quark lines.
The analysis was soon
extended~\cite{Carena:1998gk,*Diaz-Cruz:1998qc,*Balazs:1998nt,
  Kunszt:1991qe,Dai:1994vu,*Dai:1996rn,*Richter-Was:1997gi}
to include the lightest MSSM-Higgs boson $h^0$. The diagram types are exactly 
the same as in the Standard Model case. 
Only the bottom-quark--Higgs coupling is changed
to its supersymmetric counterpart, resulting in 
\begin{equation}
\sigma_{\text{MSSM}}\left(pp\rightarrow b\bar{b}h^0\right)
 = \left(- \frac{s_\alpha}{c_\beta} \right)^2
  \sigma_{\text{SM}}\left(pp\rightarrow b\bar{b}H\right) \quad ,
\end{equation}
where $- \frac{s_\alpha}{c_\beta}$
is the ratio of the bottom-quark coupling to the MSSM $h^0$ boson and 
to the one of the SM Higgs boson $H$.

The Standard-QCD
corrections~\cite{Balazs:1998sb,*Dicus:1998hs,Dittmaier:2003ej,*Dawson:2003kb,*Dawson:2005vi}
to this process are also known and reduce the dependence of the 
cross section on the factorization and renormalization scales.

However, there are subtleties when making a theoretical prediction for
total integrated $h^0$-production via this process, 
i.e.\ when the final-state bottom quarks are not explicitly detected. 
In a four-flavor-number scheme, where only gluons and 
the four light quarks, but no bottom quarks appear in the initial state, large logarithms of 
$\Order{\ln{\frac{Q^2}{m_b^2}}}$ in the total cross section emerge, where $Q$ 
is of the order of the Higgs-boson mass. They arise
from the kinematical configuration where a gluon splits into a $b\bar{b}$-pair and the
bottom quarks are collinear to the gluon. These logarithms can be resummed
using bottom-quark parton densities, thereby using a five-flavor-number
scheme. The bottom-quark densities in the proton originate purely from such splitting
gluons. So for every bottom quark which appears in a partonic process another 
bottom (anti-)quark exists in the hadronic final state.
This scheme uses the approximation that the outgoing bottom quarks have
small transverse momentum and they are assigned zero transverse momentum 
at leading order. 
In the five-flavor-number scheme the leading-order partonic process is then
$b\bar{b} \rightarrow h^0$. $gg \rightarrow b\bar{b}h^0$ only appears
at NNLO together with the two-loop corrections to this
process~\cite{Harlander:2003ai}.
% Using this scheme care must be taken not to double-count any 
% contribution~\cite{Barnett:1987jw,Olness:1987ep}.

In our case though these large logarithms are avoided by requiring
bottom-quark jets with high transverse momenta and a tagging of
the final-state bottom quarks in the detector. The additional cuts 
reduce the cross section by one or two orders of magnitude, but also
greatly reduce the background and make this approach more interesting.
The existence of bottom-quark jets with large transverse momenta also
guarantees that the Higgs boson was emitted from a bottom quark and
is therefore proportional to the bottom-quark Yukawa coupling, allowing its
precise measurement.

The SUSY-QCD corrections to this process were partly calculated in
ref.~\cite{Gao:2004wg}. There an effective $b\bar{b}h^0$-coupling 
was used which includes the one-loop squark and gluino contributions,
but no box-type or pentagon diagrams were added in their analysis.
In this dissertation a full one-loop calculation of the SUSY-QCD corrections
is performed.

\section{The \texorpdfstring{$t\bar{t}h^0$}{t t h0} Final State}

The production of a Higgs boson in association with a top quark--anti-quark
pair~\cite{Kunszt:1984ri,Marciano:1991qq,*Gunion:1991kg,Goldstein:2000bp}
proceeds in the same way as the one with a bottom quark--anti-quark pair discussed 
in the previous section and the same diagrams as in \fig{hq:tree} appear.
Since the mass of the top quark is of the same order as the electroweak
symmetry-breaking scale, its Yukawa coupling gives a sizable contribution and 
the process is an important channel for Higgs-boson production in the
mass region below $125 \GeV$~\cite{Richter-Was:1999sa,*Drollinger:2001ym}.
Furthermore, this process can be used to measure the top-quark Yukawa
coupling precisely~\cite{Maltoni:2002jr,*Belyaev:2002ua}.
The extension of the SM tree-level calculations to the MSSM, where the Higgs boson 
is an $h^0$, is again straightforward and amounts to a replacement of the Yukawa
coupling such that
\begin{equation}
\sigma_{\text{MSSM}}\left(pp\rightarrow t\bar{t}h^0\right)
 = \left(\frac{c_\alpha}{s_\beta}\right)^2
  \sigma_{\text{SM}}\left(pp\rightarrow t\bar{t}H\right) \quad , 
\end{equation}
where $\frac{c_\alpha}{s_\beta}$ 
is the ratio of the top-quark coupling to the MSSM $h^0$ boson and 
to the Higgs boson $H$ of the SM.
Thus the total cross section in the MSSM is reduced with respect to the SM one
by approximately a factor of $\frac1{t_\beta^2}$.

The Standard-QCD corrections for this process are available in the 
literature~\cite{Beenakker:2001rj,Reina:2001sf,*Dawson:2002tg,Beenakker:2002nc}.
Their numerical size is of $\Order{20\%-40\%}$ and leads to a stable prediction
of total and differential cross sections with respect to variation of the 
renormalization and factorization scales.

The large mass of the top quark also reduces the size of the collinear logarithms to
$\Order{\ln\frac{Q^2}{m_t^2}}$. Now the argument of the logarithm is close
to $1$. So the higher-order corrections are small and no resummation of these terms
needs to be performed. Hence one can safely use the four-flavor-number scheme
for this process and need not apply any additional cuts to the final-state top quarks
in this case.

A calculation of the SUSY-QCD corrections was performed quite recently
in ref.~\cite{Peng:2005ti}. As the figures of this article include both the Standard-QCD 
and the SUSY-QCD one-loop contributions a direct comparison of the numerical 
results is difficult. As far as the principal behavior with respect
to a variation of the MSSM parameters is concerned, agreement could be found.

\section{SUSY-QCD Corrections}

In this section the SUSY-QCD corrections to $h^0$-production in association
with a heavy quark--anti-quark pair are described. In the Feynman diagrams
the heavy quark is denoted by a $Q$, which represents a $b$ for the bottom-quark
and a $t$ for the top-quark final state. Correspondingly, in part of the 
diagrams the supersymmetric partners to the heavy quark appears, which is 
marked by $\tilde{Q}$, specifying $\tilde{b}$ and $\tilde{t}$, respectively.
A small $\tilde{q}$ on the other hand signifies that all squarks can be inserted
in the propagator. The $\tilde{q}^\prime$ in the $q\bar{q}$ diagrams denotes
the superpartner to the initial-state quark.
\begin{figure}[p]
\begin{center}
\vspace*{-12mm}
\subfigure[self-energy corrections]{
\parbox{\textwidth}{
\begin{center}
\includegraphics[scale=0.5]{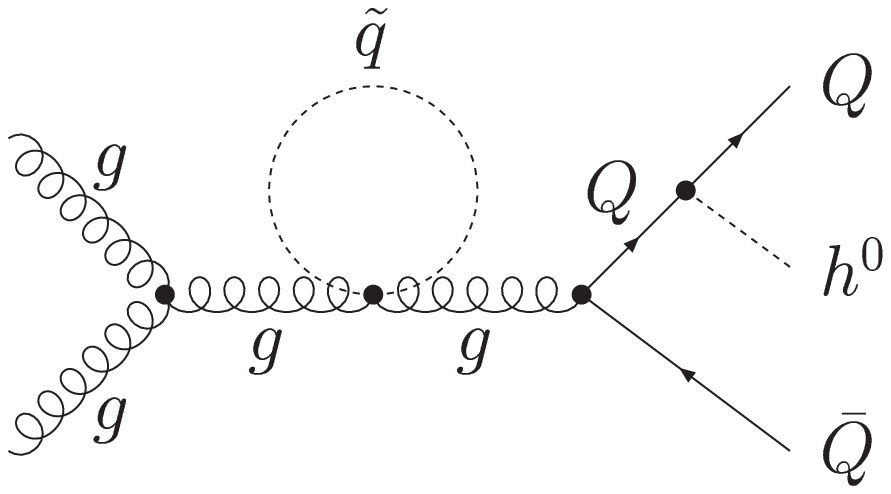}%
\includegraphics[scale=0.5]{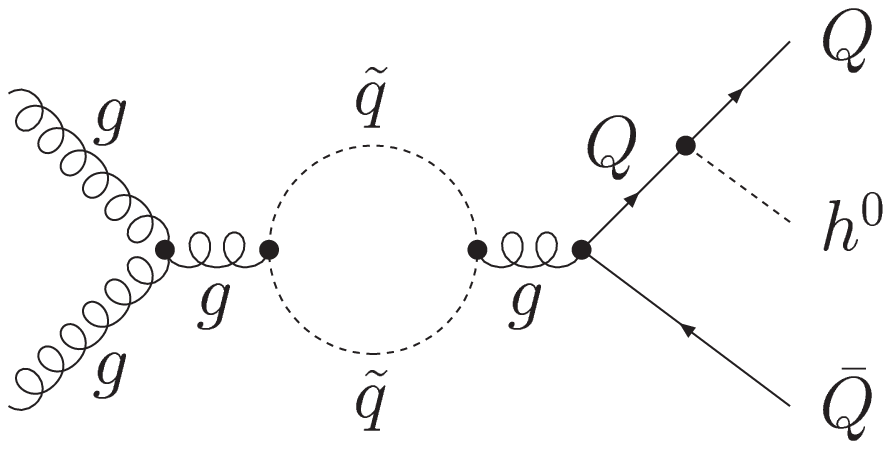}%
\includegraphics[scale=0.5]{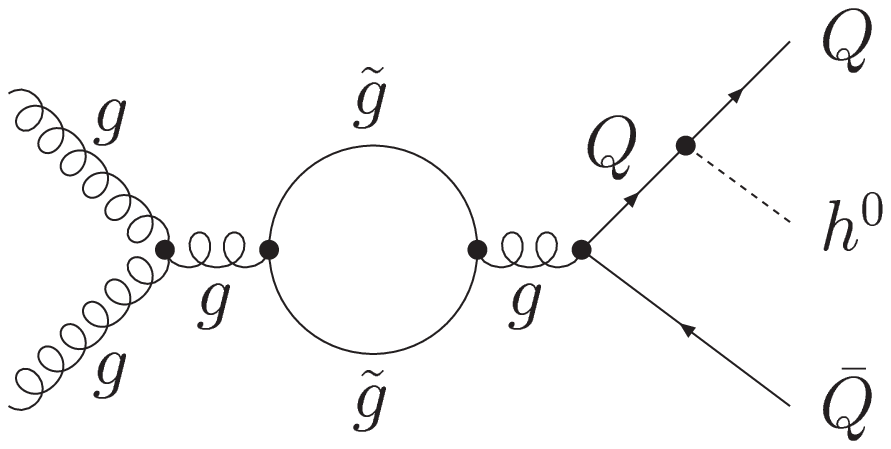}\\
\includegraphics[scale=0.5]{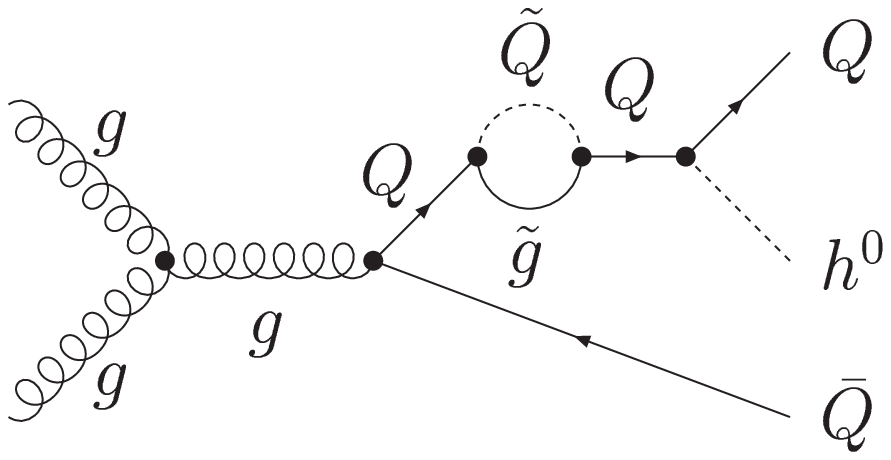}%
\includegraphics[scale=0.5]{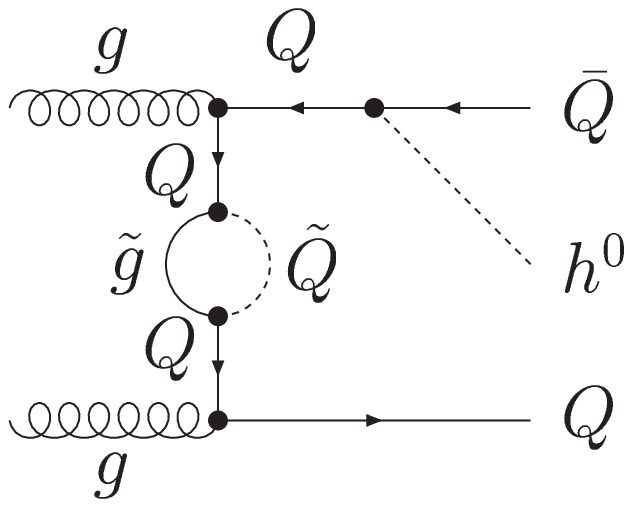}
\end{center}
}
}
\subfigure[vertex corrections]{
\parbox{\textwidth}{
\begin{center}
\includegraphics[scale=0.5]{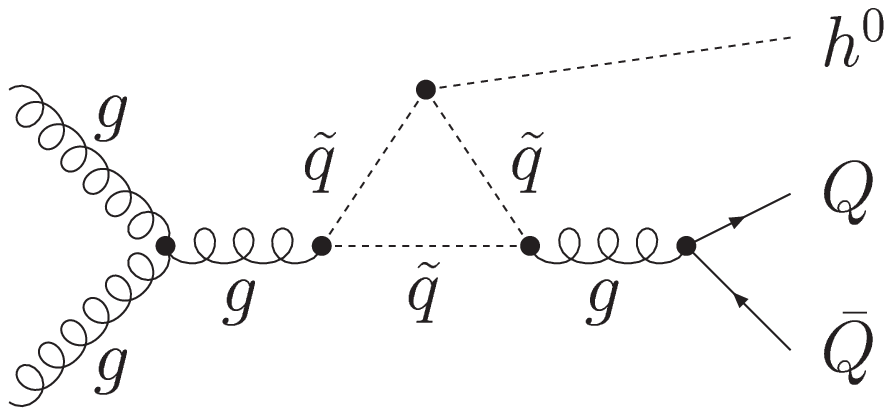}%
\includegraphics[scale=0.5]{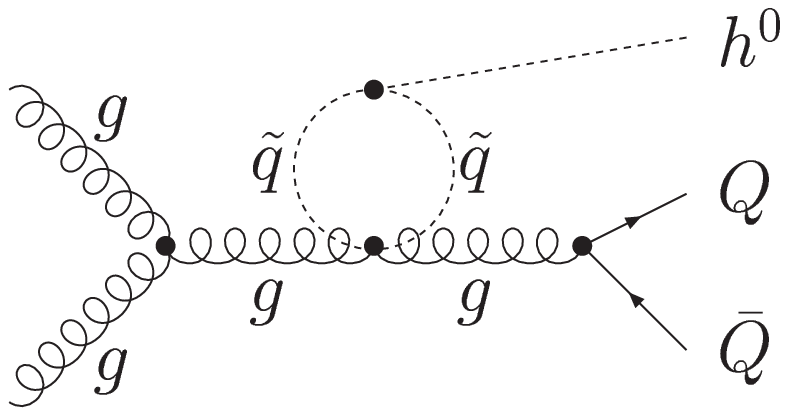}%
\includegraphics[scale=0.5]{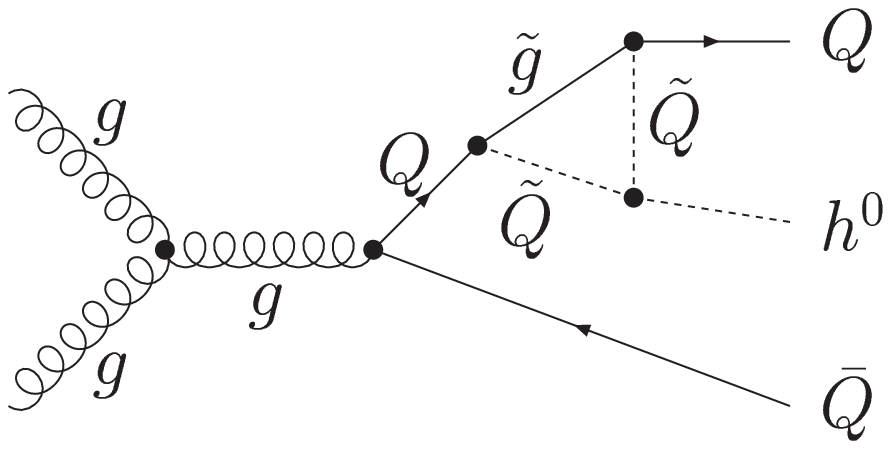}\\
\includegraphics[scale=0.5]{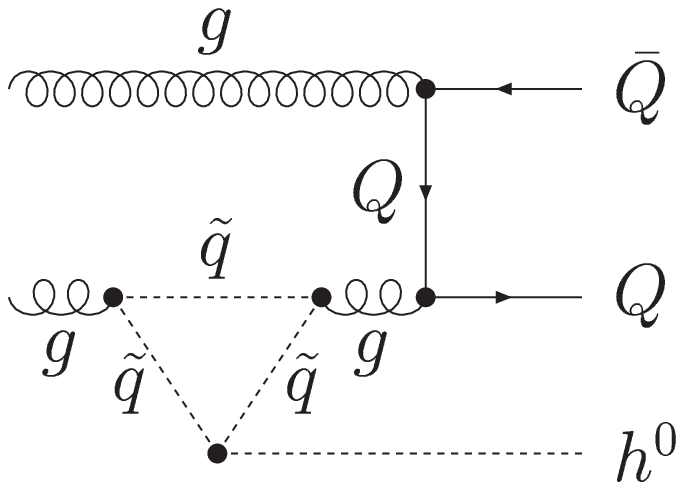}%
\includegraphics[scale=0.5]{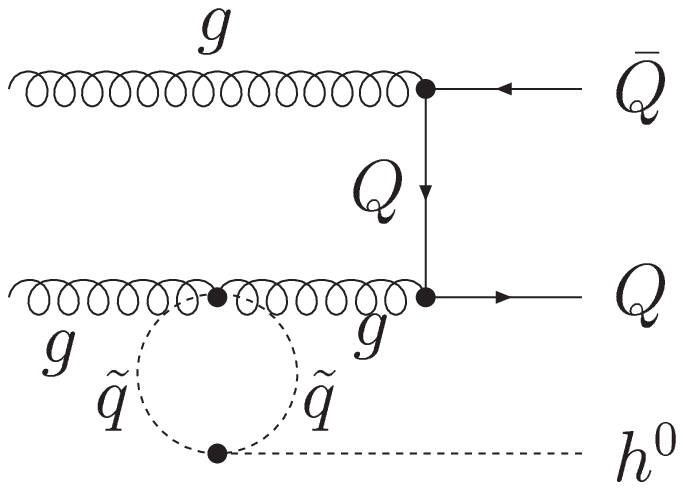}%
\includegraphics[scale=0.5]{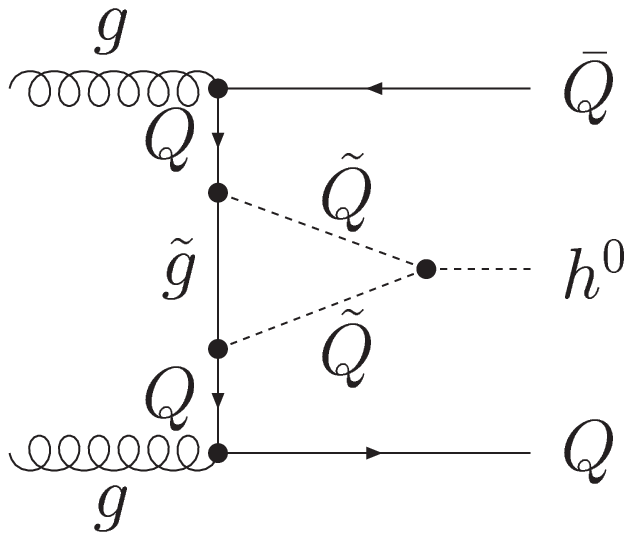}\\
\includegraphics[scale=0.5]{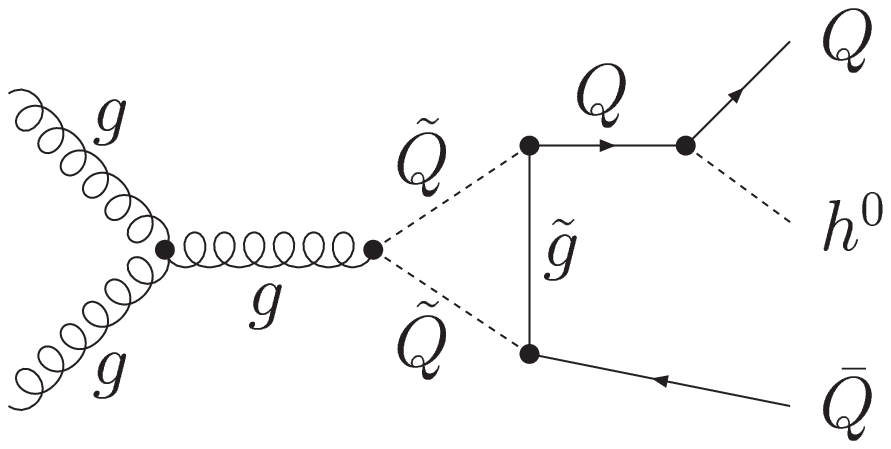}%
\includegraphics[scale=0.5]{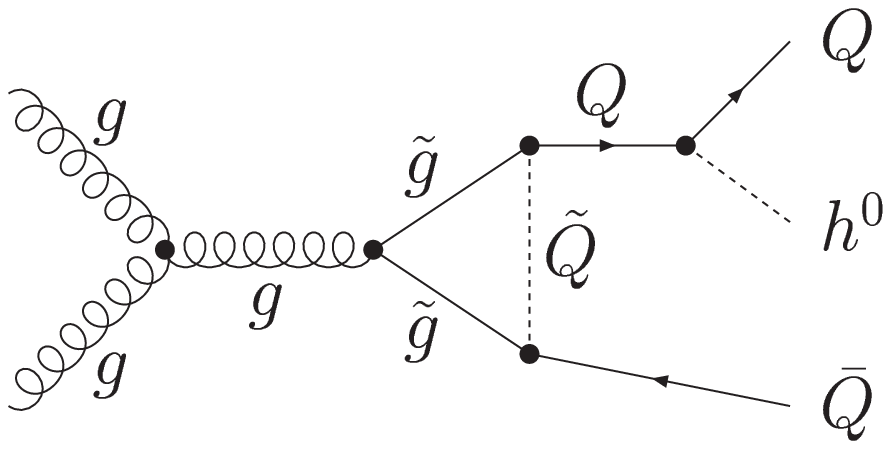}\\
\includegraphics[scale=0.5]{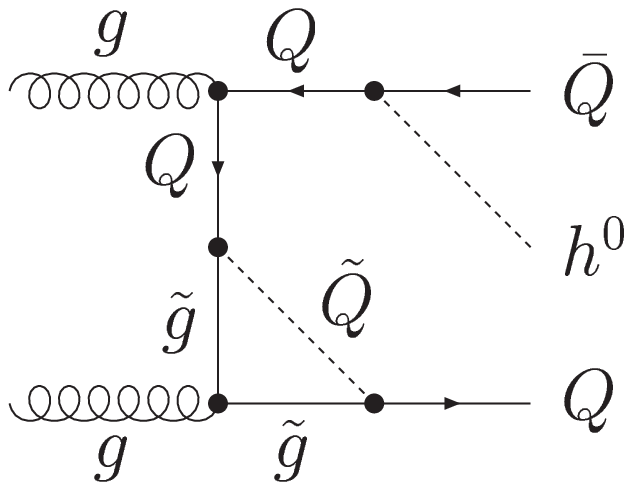}%
\includegraphics[scale=0.5]{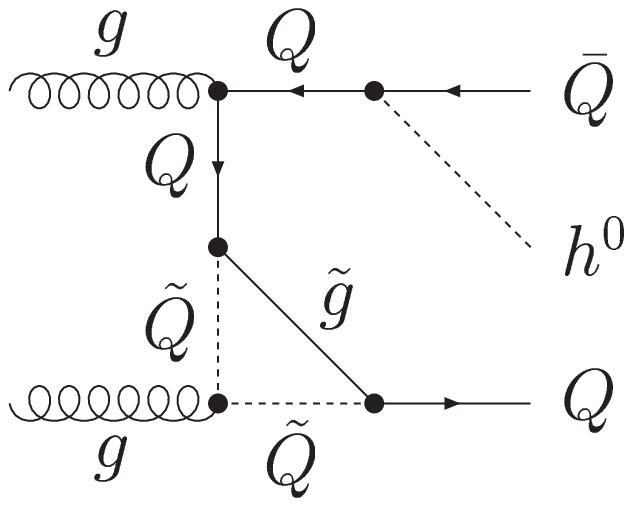}\\
\includegraphics[scale=0.5]{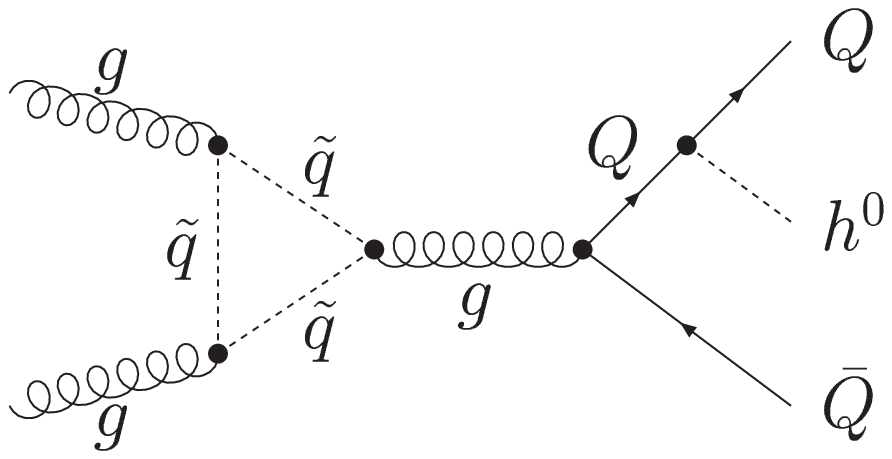}%
\includegraphics[scale=0.5]{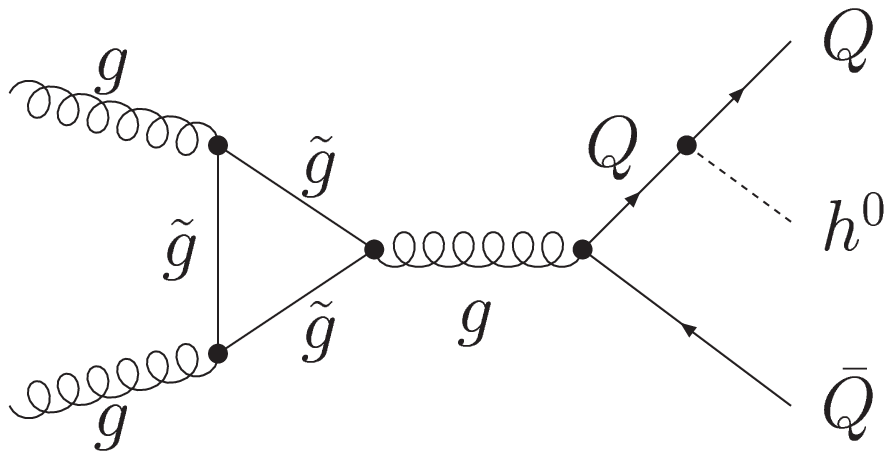}%
\includegraphics[scale=0.5]{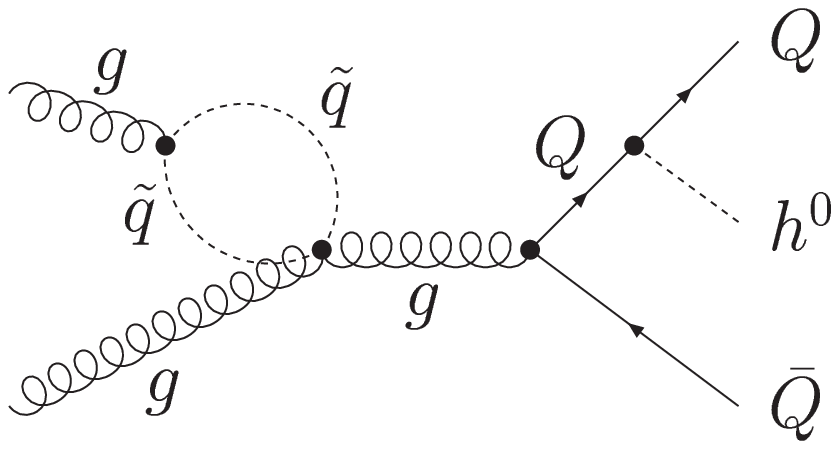}\\
\end{center}
}
}
\caption{Types of Feynman diagrams contributing to SUSY-QCD corrections 
  to $h^0$-production in association with heavy quarks}
\label{hq:susydiag}
\end{center}
\end{figure}%
\addtocounter{figure}{-1}%
\setcounter{subfigure}{2}%
\begin{figure}[p]
\begin{center}
\subfigure[box diagrams]{
\parbox{\textwidth}{
\begin{center}
\includegraphics[scale=0.5]{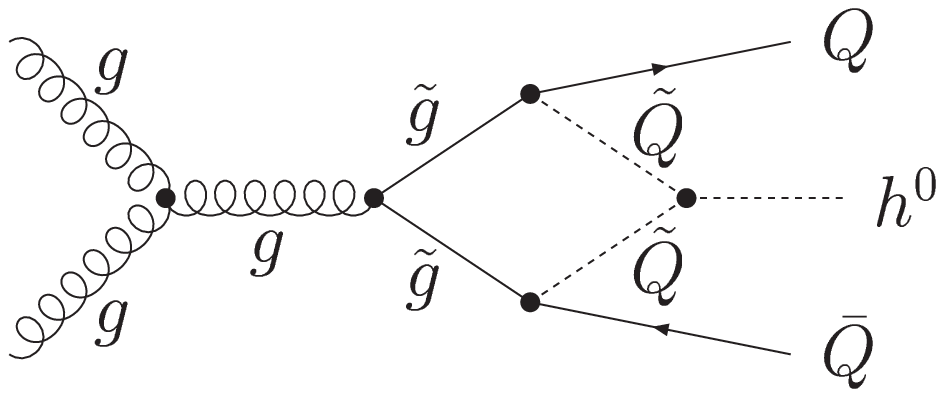}%
\includegraphics[scale=0.5]{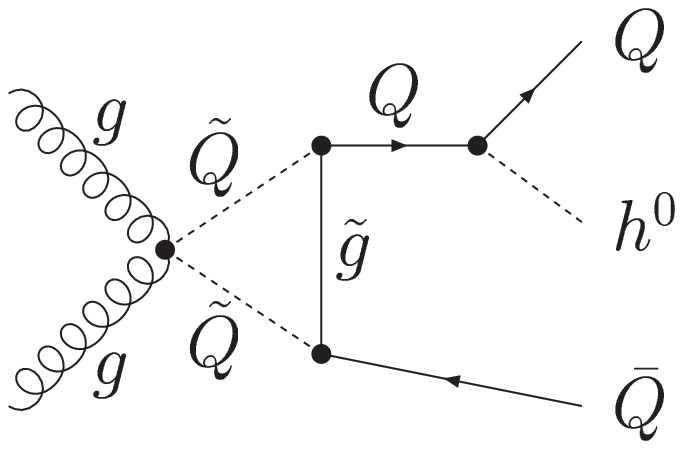}\\
\includegraphics[scale=0.5]{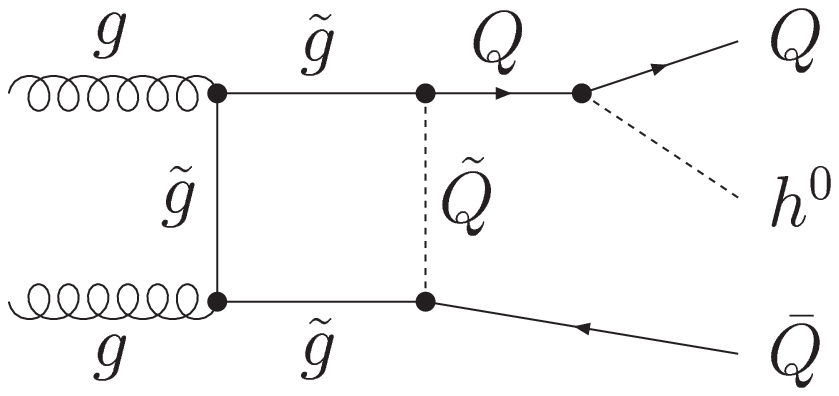}%
\includegraphics[scale=0.5]{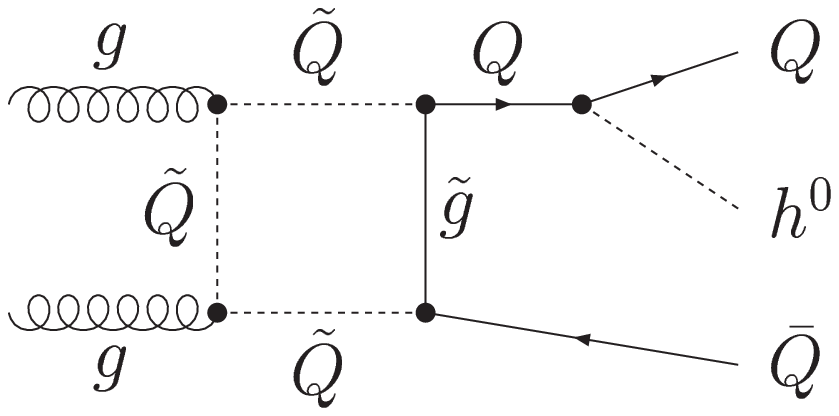}%
\includegraphics[scale=0.5]{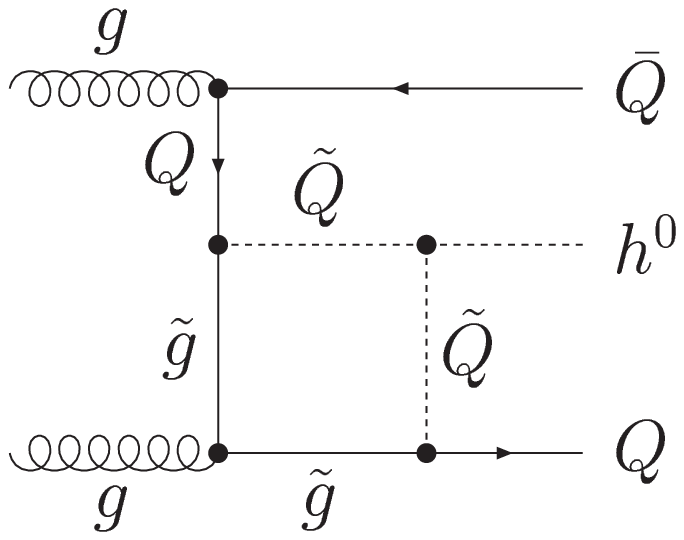}\\
\includegraphics[scale=0.5]{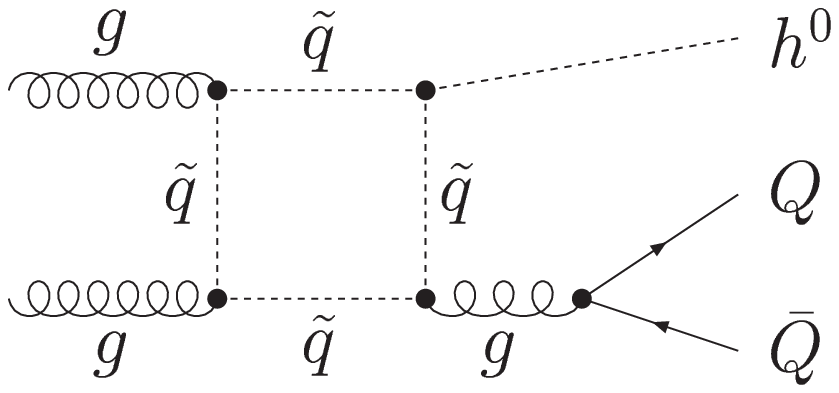}%
\includegraphics[scale=0.5]{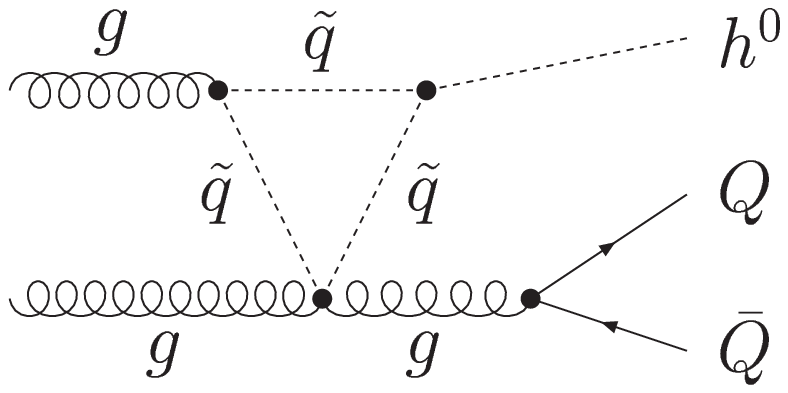}%
\end{center}
}
}
\subfigure[pentagon diagrams]{
\parbox{\textwidth}{
\begin{center}
\includegraphics[scale=0.5]{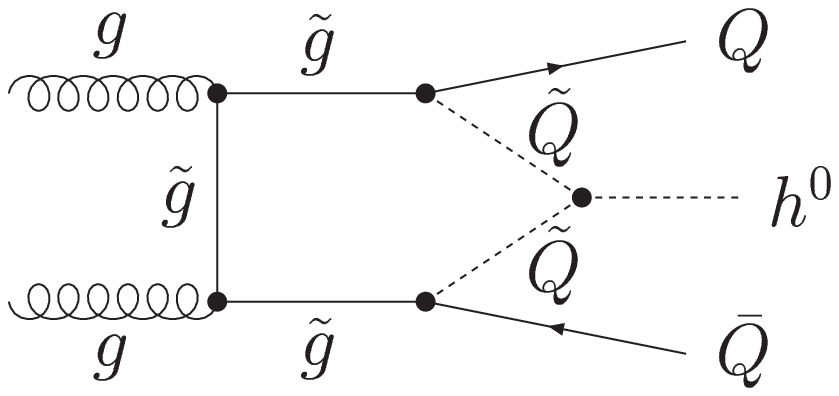}%
\includegraphics[scale=0.5]{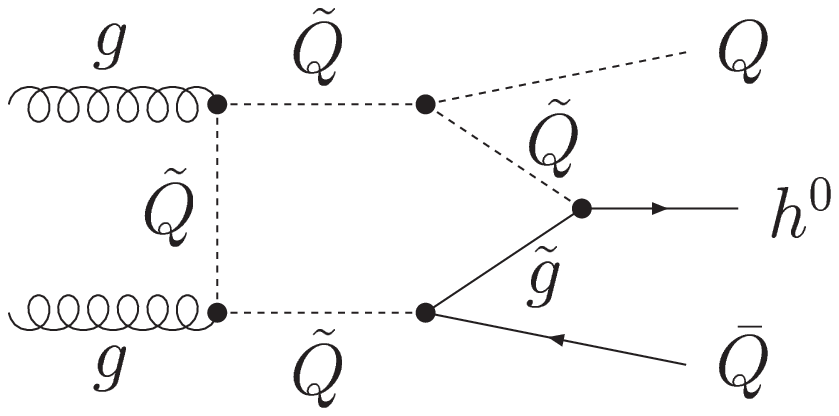}\\
\includegraphics[scale=0.5]{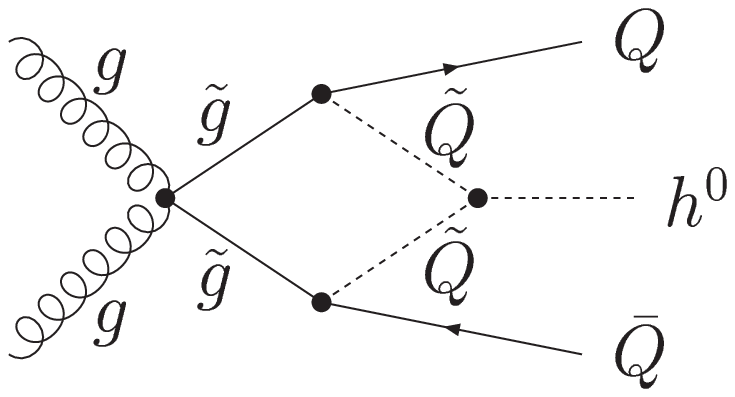}%
\includegraphics[scale=0.5]{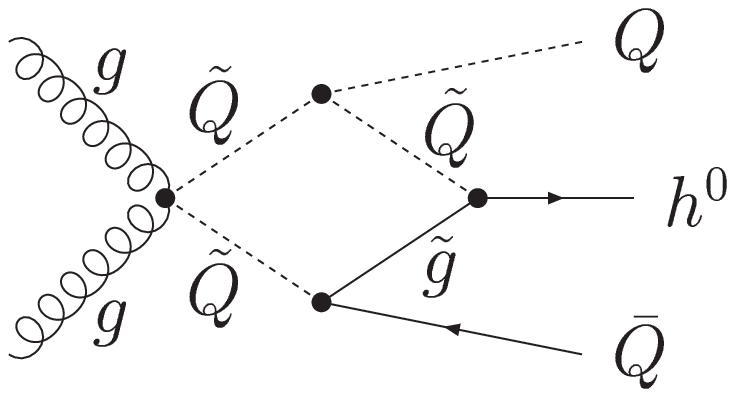}
\end{center}
}
}
\subfigure[self-energy counter terms]{
\includegraphics[scale=0.5]{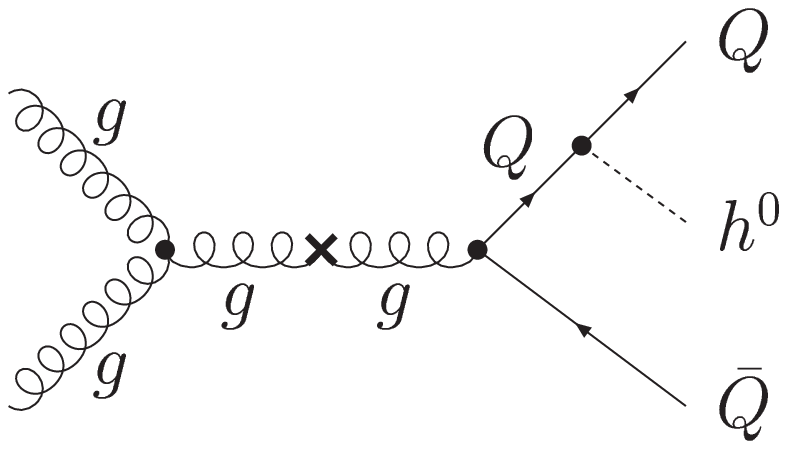}
\includegraphics[scale=0.5]{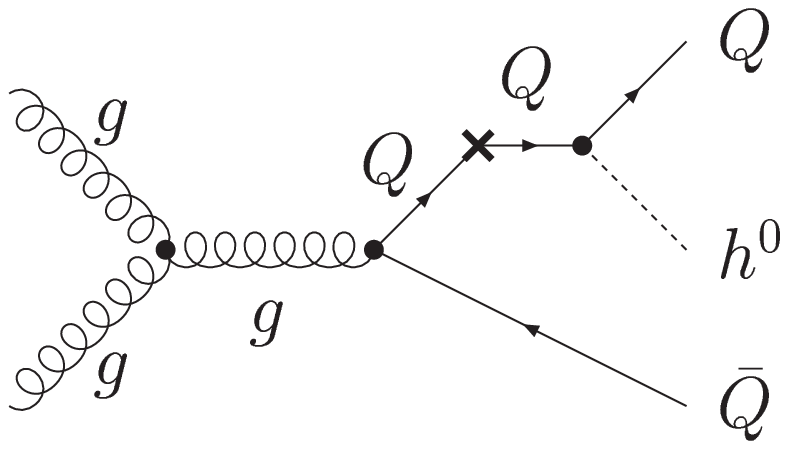}
\includegraphics[scale=0.5]{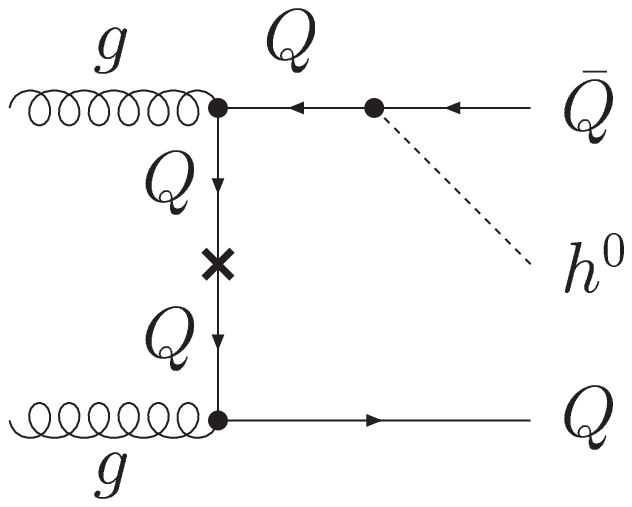}
}
\caption{(continued)}
\end{center}
\end{figure}%
\addtocounter{figure}{-1}%
\setcounter{subfigure}{5}%
\begin{figure}
\begin{center}
\subfigure[vertex counter terms]{
\parbox{\textwidth}{
\begin{center}
\includegraphics[scale=0.5]{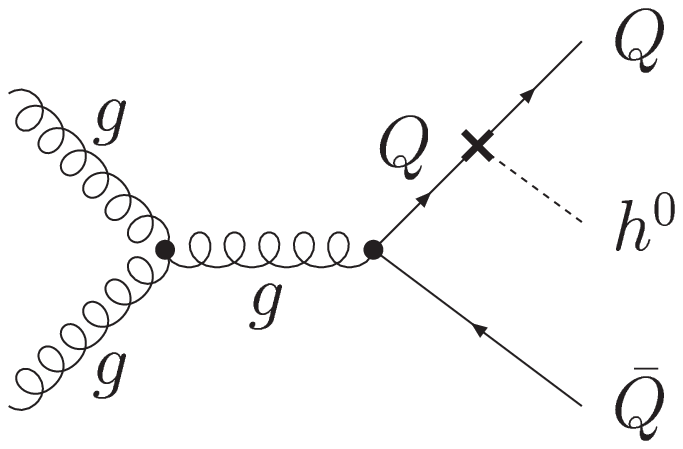}%
\includegraphics[scale=0.5]{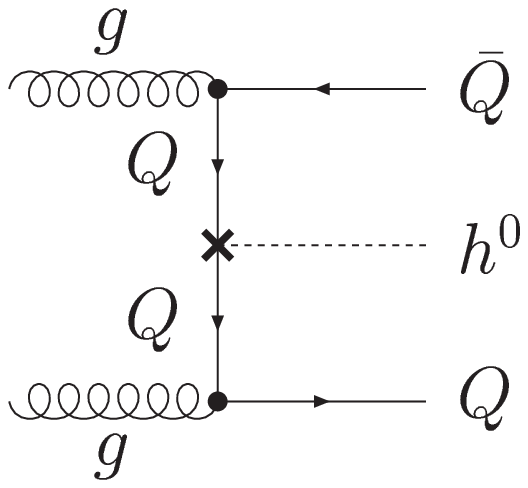}\\
\includegraphics[scale=0.5]{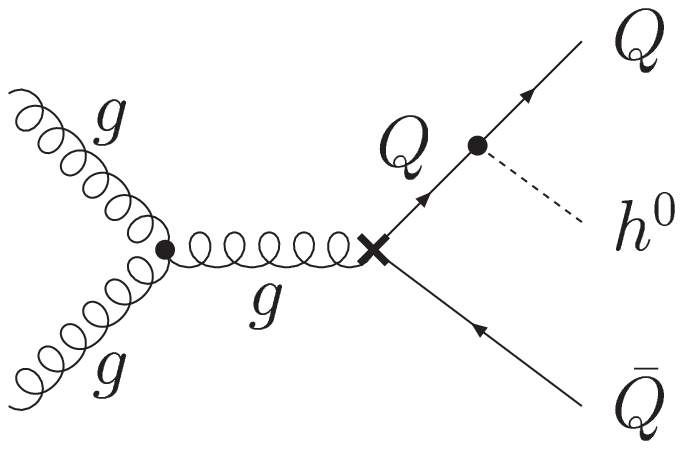}%
\includegraphics[scale=0.5]{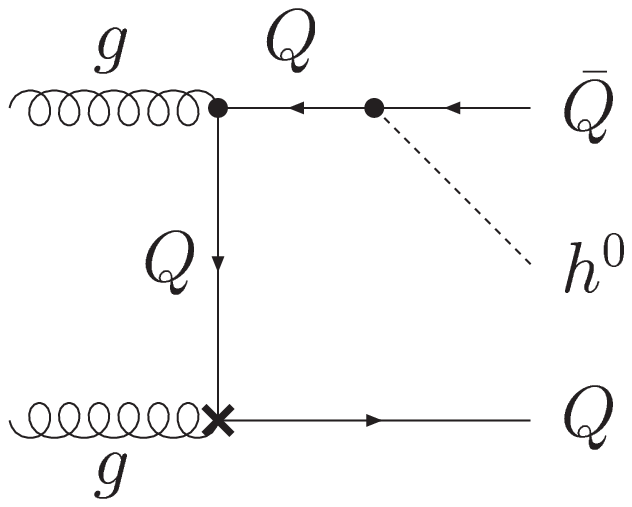}%
\includegraphics[scale=0.5]{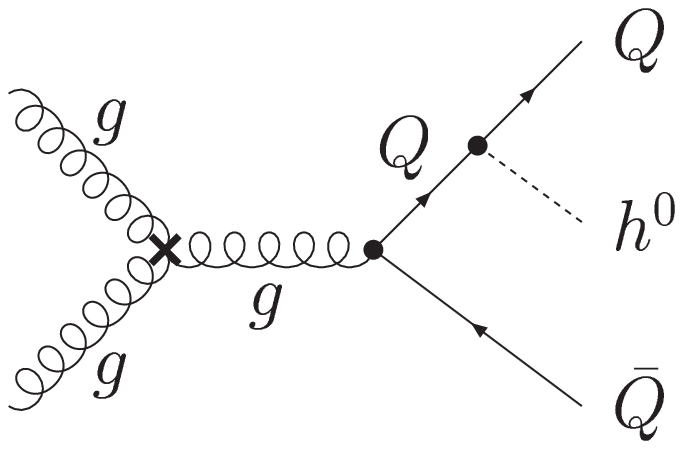}\\
\end{center}
}
}
\subfigure[additional diagrams for quark--anti-quark annihilation]{
\parbox{\textwidth}{
\begin{center}
\includegraphics[scale=0.5]{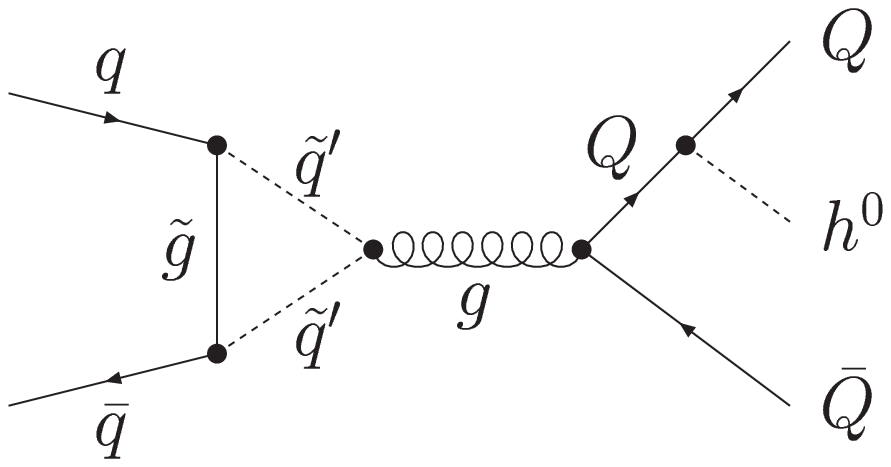}%
\includegraphics[scale=0.5]{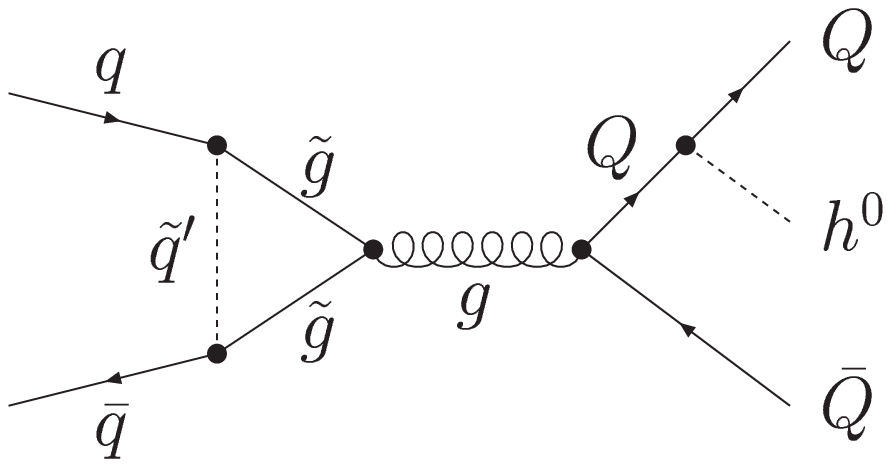}%
\includegraphics[scale=0.5]{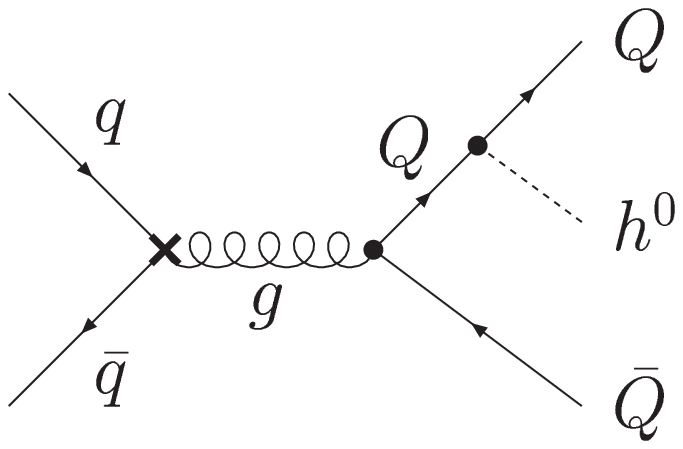}\\
\includegraphics[scale=0.5]{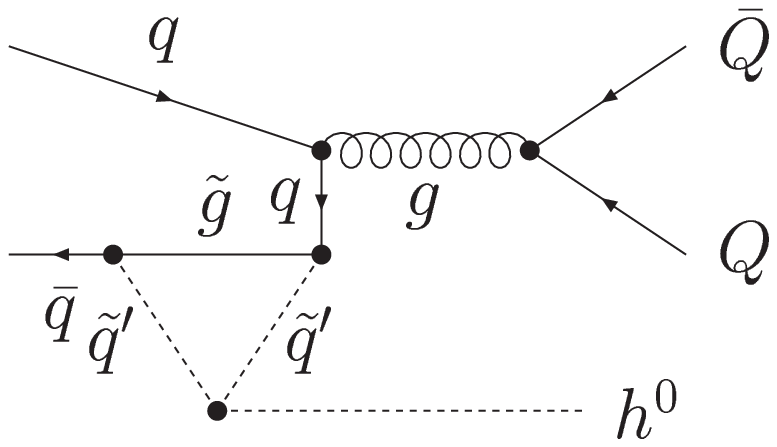}\\
\includegraphics[scale=0.5]{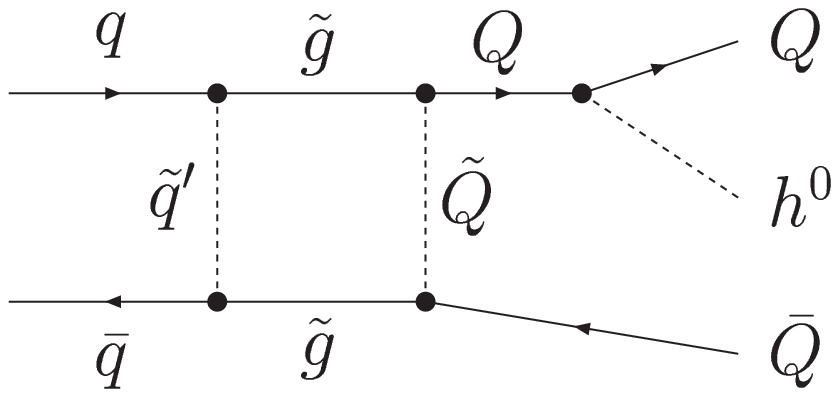}%
\includegraphics[scale=0.5]{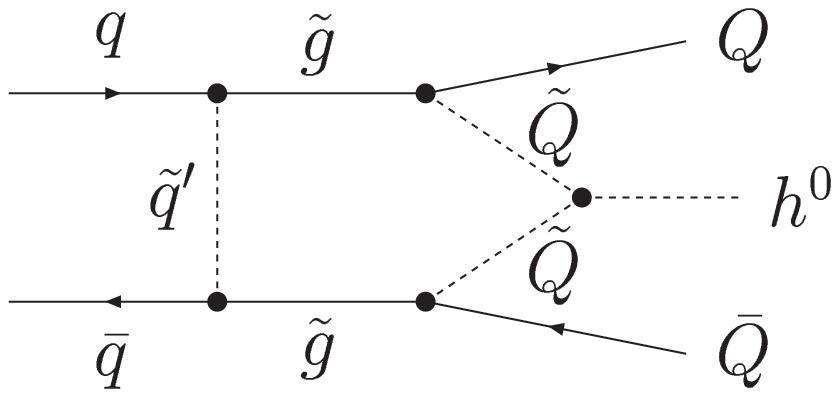}%
\end{center}
}
}
\caption{(continued)}
\end{center}
\end{figure}%
%%% description of diagrams 
In \fig{hq:susydiag}(a)-(f) the basic types of Feynman diagrams
which appear as one-loop SUSY-QCD corrections to $h^0$-production
via gluon fusion are depicted.
Self-energy corrections (\fig{hq:susydiag}(a)) enter either via a squark
or gluino loop which is inserted into the intermediate gluon propagator, 
or a combined squark-gluino two-point loop inserted into the heavy-quark line.
In \fig{hq:susydiag}(b) the possible vertex corrections are displayed.
Squark loops induce an effective gluon-Higgs coupling appearing in the
two diagrams on the left-hand side of the first two rows. The diagrams on the
right-hand side of the first two rows contain a correction to the coupling between
$h^0$ and the heavy quark. The third and fourth row feature corrections
to the gluon-quark interaction and the last row an additional contribution to the
triple-gluon vertex.
Diagrams where four particles are connected via a sparticle loop are presented
in \fig{hq:susydiag}(c). 
Finally, all five external particles can be joined by a squark-gluino loop as shown in
\fig{hq:susydiag}(d). 
The emerging divergences are cancelled by counter-term diagrams shown in
\fig{hq:susydiag}(e) for the self-energy contributions and (f) for the vertex corrections.
All gluon-fusion s-channel diagrams, where the two initial-state gluons 
couple to a further, intermediate gluon, also exist in the 
quark--anti-quark--annihilation subprocesses. The only change is the replacement 
of the two incoming gluons by a quark--anti-quark pair. The additional diagrams
which appear for this type of subprocess are depicted in \fig{hq:susydiag}(g). 
They are corrections to the quark--anti-quark--gluon coupling together with the 
associated counter-term diagram as shown in the first row. Secondly, the coupling of 
the incoming light quark to the Higgs boson, which is neglected at tree-level, appears
at one-loop order as indicated by the diagram in the second row. Finally in the last row
the additional box and pentagon diagrams are shown.

In the remaining sections of this chapter the numerical results of 
$h^0$-production in association with a heavy quark--anti-quark pair
are presented. In analogy to \chap{bbWH} several cross-section
differences are defined to illustrate the results.
All calculations in this chapter were performed in the 
OS renormalization scheme, so the label indicating the 
renormalization scheme will in the following be dropped for all items.
The relative one-loop correction is defined as
\begin{equation}
\Delta_1 = \frac{\sigma_1 - \sigma_{0}}{\sigma_0} \quad ,
\end{equation}
where $\sigma_0$ denotes the tree-level and $\sigma_1$ the one-loop
cross section.
For the calculation of hadronic cross sections the PDF set of 
ref.~\cite{Martin:2002aw} was used.
Additionally, a $\Delta m_{b,t}$-corrected tree-level cross section
$\sigma_{\Delta}$ 
was calculated in a similar way as already described in \chap{bbWH}, 
i.e.\ using the non-resummed version of \eq{renorm:noresum} and treating the 
$\Delta m_{b,t}$ term as a one-loop contribution. Additionally, the contribution to the
vertex from the term proportional to the second mixing angle in the
MSSM-Higgs sector, $\alpha$, was included in $\sigma_{\Delta}$ according to 
eqs.~(\ref{renorm:bbh}) and~(\ref{renorm:tth}).
The relative correction using only these contributions is defined as
\begin{equation}
\Delta_{\tilde{0}} = 
  \frac{\sigma_{\Delta} - \sigma_{0}}{\sigma_0} \quad .
\end{equation}
Finally, the difference between the $\Delta m_{b,t}$-corrected tree-level cross section
and the full one-loop result, which denotes the true one-loop corrections, is given by
\begin{equation}
\Delta_{\Delta m_{b,t}} = \frac{\sigma_1 - \sigma_\Delta}{\sigma_0} \quad .
\end{equation}
The renormalization of the strong coupling constant 
$\alpha_s$ was performed as described in \chap{renorm:alphas}.

\section{Numerical Results for \texorpdfstring{$b\bar{b}h^0$}{b b h0}}

In this section the numerical results for the process $ p p \rightarrow b\bar{b}h^0$ 
are presented. 
First the total hadronic cross section for the MSSM reference point $\spa$ is given
in table~\ref{hq:bspa}.
\begin{table}
\begin{center}
\begin{tabular}{l|r|r|r|r}
Partonic subprocess & $\sigma_0$ [fb]
& $\sigma_1$ [fb] 
&  $ \Delta_1 $ [\%] & $ \Delta_{\tilde{0}} $ [\%] \\\hline
$d\bar{d}\rightarrow b\bar{b}h^0$ & $0.107$ & $0.104$ & $-2.48$ & $-1.95$ \\
$u\bar{u}\rightarrow b\bar{b}h^0$ & $0.168$ & $0.164$ & $-2.56$ & $-1.95$ \\
$s\bar{s}\rightarrow b\bar{b}h^0$ & $0.028$ & $0.028$ & $-2.26$ & $-1.95$ \\
$c\bar{c}\rightarrow b\bar{b}h^0$ & $0.013$ & $0.012$ & $-2.20$ & $-1.95$ \\
$gg\rightarrow b\bar{b}h^0$ & $35.647$ & $33.734$ & $-5.37$ & $-1.95$ \\\hline
$\sum\left(pp\rightarrow b\bar{b}h^0\right)$ & $35.963$ & $34.042$ & $-5.34$ & $-1.95$ 
\end{tabular}
\caption{Hadronic cross sections for $b\bar{b}h^0$-production at the 
  parameter point $\spa$ (see \app{param:spa}).}
\label{hq:bspa}
\end{center}
\end{table}
It is also given separately for each partonic subprocess. As described before,
the outgoing bottom-quark jets are required to have a high transverse momentum,
so that large logarithms are avoided and the background processes, where the 
Higgs boson does not radiate off a bottom quark, are reduced.
To this end a cut on the bottom quarks,
\begin{equation}
p_T(b,\bar b) \ge 20 \GeV \quad ,
\end{equation}
was applied to obtain these results. The same cut will also be used for all other
cross sections of this section.

As one can see in
the table, the dominant contribution originates from the gluon-fusion process and
the quark--anti-quark--annihilation processes are suppressed by two
orders of magnitude. Hence their contribution is negligible and in the following
analysis only the gluon-fusion subprocess is considered. This large difference
in the cross sections is due to the fact that the quark--anti-quark--annihilation
process can only proceed via the s-channel diagram shown on the left-hand
side of \fig{hq:tree}(a). It contains a propagator suppression from the 
intermediate gluon which must carry at least the energy to procuce the final-state
Higgs boson and the two bottom quarks. In contrast the gluon-fusion subprocess
also contains a t-channel diagram (\fig{hq:tree}(b)) which does not suffer from such 
a suppression. In fact, if one takes only the s-channel diagram on the right-hand side of
\fig{hq:tree}(a) into account, the cross section of the gluon-fusion contribution is of
comparable size $(\sigma_1 = 1.103 \text{ fb})$ to the one of 
quark--anti-quark annihilation. 

In the following plots the effect of varying MSSM parameters on the SUSY-QCD
contributions is investigated. To that end a parameter point with a fairly light 
SUSY spectrum was chosen, namely
\begin{align}
m_A &= 200 \GeV \nonumber\\
\mu &= 300 \GeV \nonumber\\
A_t  &= A_b = 0 \nonumber\\
M_{\text{SUSY}} &\equiv M_{\tilde{Q}} = M_{\tilde{U}} = M_{\tilde{D}} = 250 \GeV \nonumber\\
m_{\tilde{g}} &= 400 \GeV \quad .
\label{hq:b_param}
\end{align}
The MSSM parameters were then varied around this point. 
The renormalization scale, which appears in $\alpha_s$ (see \eq{renorm:alphasrun}), 
was set to $\mu_R =2 m_b + m_{h^0}$. 
As the contribution of the 
quark--anti-quark annihilation diagrams is negligible compared to the 
gluon-fusion subprocess, only the latter one is considered in the following.
Also for simplicity the quoted cross section differences are partonic ones
with a center-of-mass energy of $\sqrt{\hat s} = 500 \GeV$.

\begin{figure}
\includegraphics{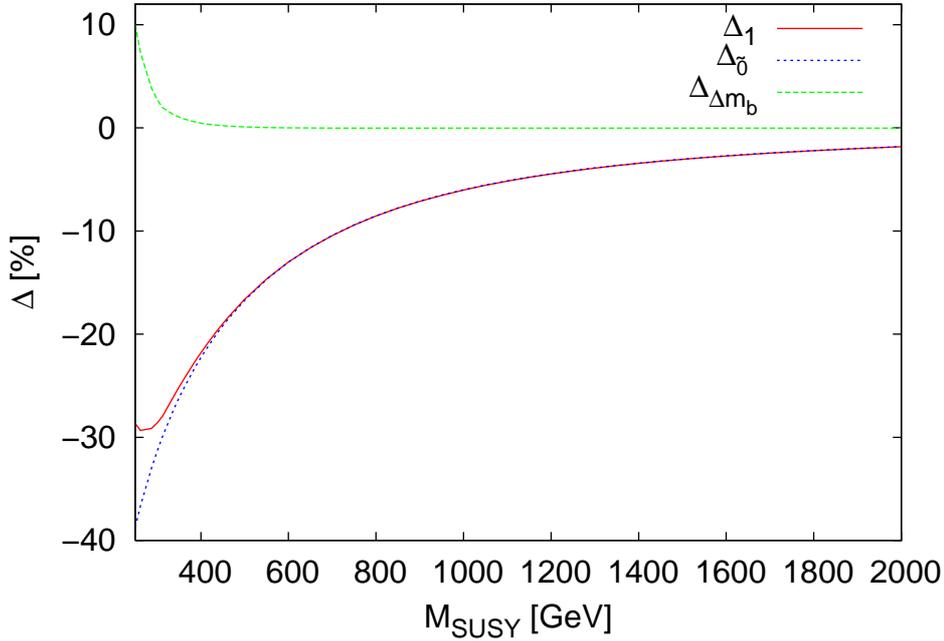}
\caption{Partonic cross section differences for the process 
  $g g \rightarrow b\bar{b} h^0$, using $t_\beta=30$, as a function of 
  $M_{\text{SUSY}} \equiv M_{\tilde{Q}} = M_{\tilde{U}} = M_{\tilde{D}}$. 
  All other parameters take the values given in \eq{hq:b_param}.}
\label{hq:b_parton_MSUSY_ltb}
\end{figure}

For the first plots $t_\beta$ is set to the large value $30$. 
In the plot given in \fig{hq:b_parton_MSUSY_ltb} a common 
mass scale $M_{\text{SUSY}}$, where 
all soft-supersymmetry breaking masses in the squark sector take the 
same value, is chosen. 
According to \chap{renorm:deltamb}, where the bottom-quark Yukawa coupling 
was studied, one would expect the universal corrections, which are parametrized
in $\Delta m_b$, to give the dominant contribution. This is indeed the case for almost all
$M_{\text{SUSY}}$-values. Also the decrease of the corrections with growing 
SUSY mass scale, which is predicted by \eq{renorm:deltambeq} to fall off 
as $\frac1{M_{\text{SUSY}}^2}$, can be seen in the plot. Only for rather small values 
of $M_{\text{SUSY}}$ a deviation from this behavior occurs. Other terms contribute
significantly in this region and lead to smaller cross-section differences than one would
expect from the $\Delta m_b$ terms alone.

\begin{figure}
\includegraphics{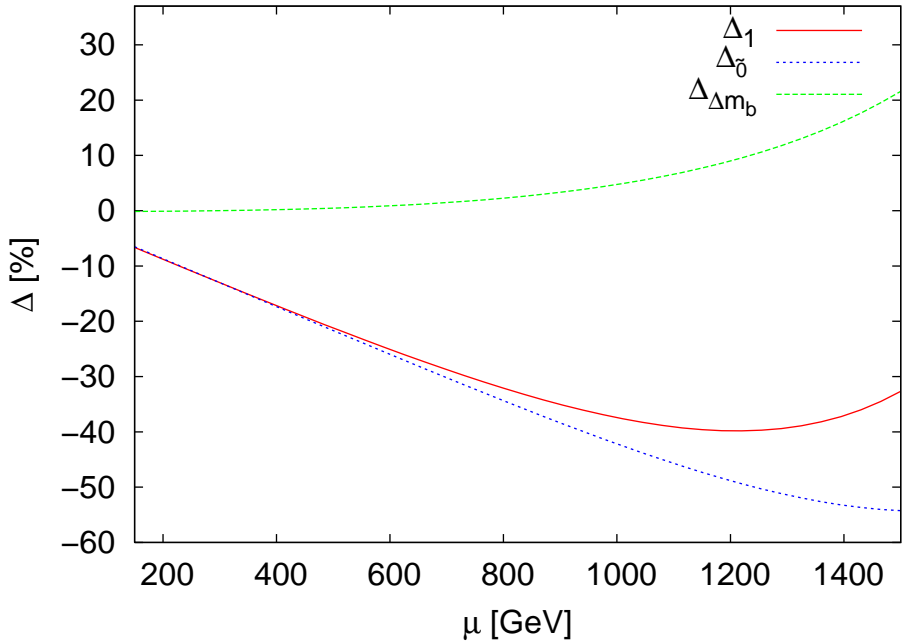}
\caption{Partonic cross section differences for
  $b\bar{b} h^0$-production via gluon fusion in the large $t_\beta$-regime   
  ($t_\beta=30$) as a function of $\mu$. 
  For the value of the other parameters see \eq{hq:b_param}.}
\label{hq:b_parton_mue_ltb}
\end{figure}

The numbers for the second plot (\fig{hq:b_parton_mue_ltb}) are also calculated in the
regime of large $t_\beta$, but now $\mu$ is varied.
For small values of $\mu$, the $\Delta m_b$-corrected tree-level result and the 
full one-loop cross section coincide and show the expected linear rise with $\mu$.
When $\mu$ becomes large, and thus the off-diagonal elements in the sbottom
mixing matrix lead to a larger split between the lighter and heavier sbottom, 
this behavior changes and leads to a decelerated increase with $\mu$. Also 
other terms begin to contribute to the cross section in a significant way and 
induce a deviation of the full one-loop result from the $\Delta m_b$ corrections
by up to $20 \%$. 

\begin{figure}
\includegraphics{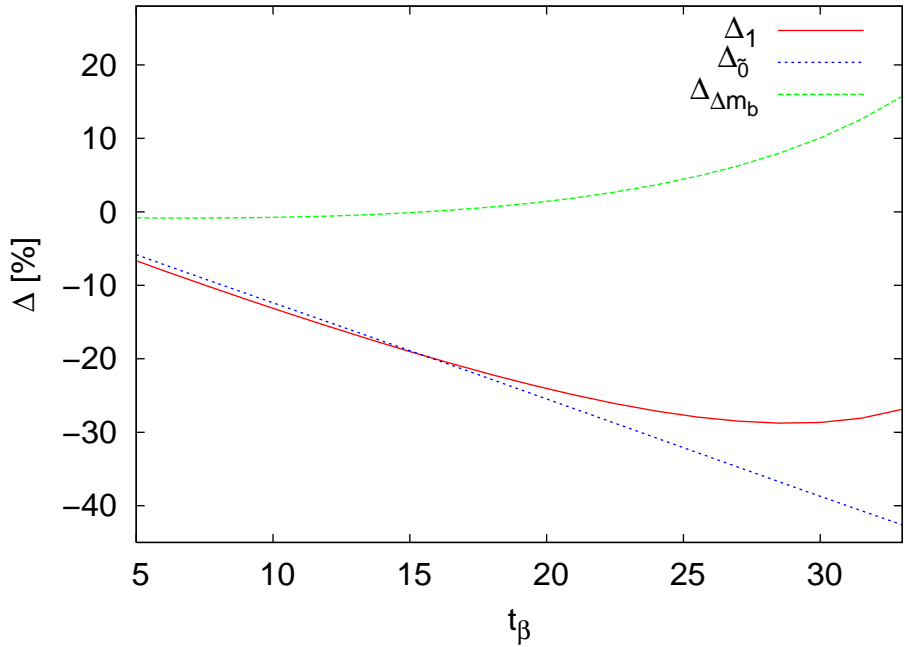}
\caption{$t_\beta$-dependence of the partonic cross section differences for
  $gg \rightarrow b\bar{b} h^0$.
  The values of all other parameters are given in \eq{hq:b_param}.}
\label{hq:b_parton_tb}
\end{figure}

\begin{figure}
\includegraphics{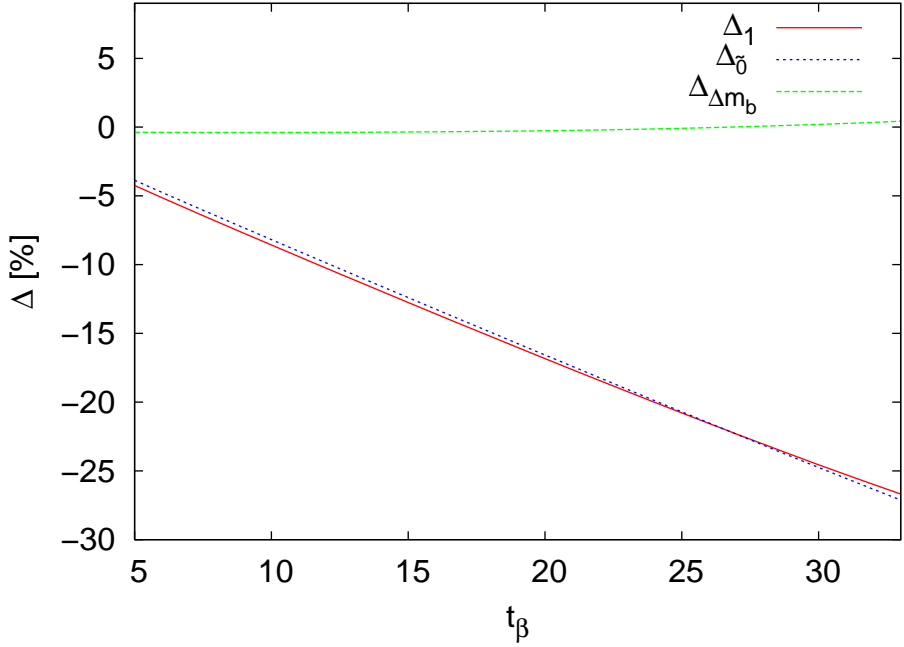}
\caption{$t_\beta$-dependence of the partonic cross section differences for
  $gg \rightarrow b\bar{b} h^0$. For this plot a slightly higher 
  $M_{\text{SUSY}} = 400 \GeV$ and $m_{\tilde g} = 640 \GeV$ was used
  than the one of \eq{hq:b_param}.}
\label{hq:b_parton_tb2}
\end{figure}

The effect of varying $t_\beta$ is studied in \fig{hq:b_parton_tb}.
The $\Delta m_b$-corrected tree-level result grows linearly with $t_\beta$ 
as predicted from \eq{renorm:deltambeq}. In the small $t_\beta$-regime
it approximates the full one-loop result rather well with a deviation of only about
one percent. For larger values of $t_\beta$ the complete one-loop corrections
begin to deviate and additional contributions lead to a slower rise. Finally,
the absolute value of the full corrections slightly decreases again. This is the same
effect which was already observed on the left-hand side 
of \fig{hq:b_parton_MSUSY_ltb}. As the common value of the 
soft supersymmetry-breaking masses $M_{\text{SUSY}}$ was chosen to be 
$250 \GeV$ we are exactly in this regime. A higher value for $M_{\text{SUSY}}$
leads to a one-loop cross-section difference which coincides with the 
$\Delta m_b$-corrected tree-level one over the whole range of $t_\beta$. 
To verify this a value of $M_{\text{SUSY}} = 400 \GeV$ was chosen 
to obtain \fig{hq:b_parton_tb2}.
Additionally, the gluino mass was set to $m_{\tilde g} = 640 \GeV$ such that 
the ratio of the two masses is the same as in the parameter set \eq{hq:b_param}.
In this case the discrepancy $\Delta_{\Delta m_b}$ between 
$\Delta_1$ and $\Delta_{\tilde 0}$ stays below one percent for all $t_\beta$-values.

\begin{figure}
\vspace*{-4mm}
\includegraphics{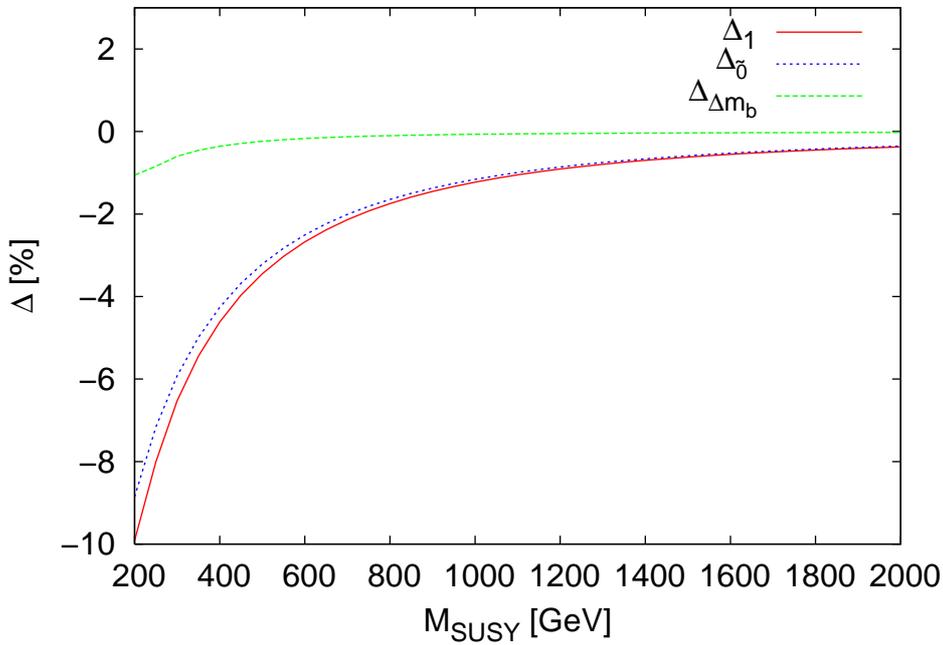}
\caption{Partonic cross section differences for
  $b\bar{b} h^0$-production via gluon fusion for $t_\beta=6$ as 
  function of a common mass $M_{\text{SUSY}}$ for the soft 
  supersymmetry-breaking terms.
  All other parameters take the values given in \eq{hq:b_param}.}
\label{hq:b_parton_SUSY}
\end{figure}

In the final two plots the behavior of the cross-section differences in the small
$t_\beta$-regime, namely for $t_\beta=6$, is studied. Firstly, in \fig{hq:b_parton_SUSY}
the common mass scale $M_{\text{SUSY}}$ of the 
soft supersymmetry-breaking masses appearing in the squark sector is varied.
The $\Delta m_b$-corrected tree-level result is a good approximation of the 
full one-loop result over the whole mass range. The difference is about one 
percent for small $M_{\text{SUSY}}$ and rapidly vanishes for larger
values.

\begin{figure}
\vspace*{-4mm}
\includegraphics{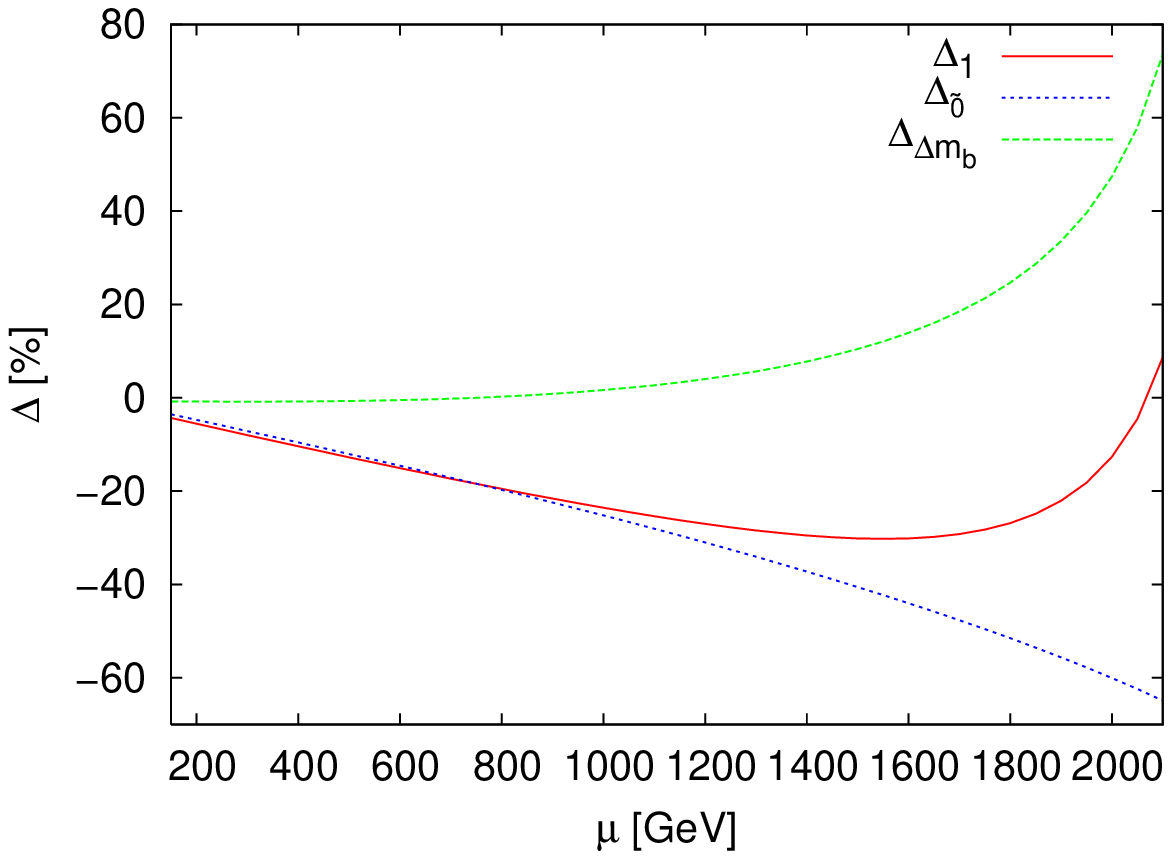}
\caption{$\mu$-dependence of the partonic cross section differences for
  $b\bar{b} h^0$-production via gluon fusion in the small $t_\beta$-regime 
  ($t_\beta = 6$).
  The values of all other parameters is given in \eq{hq:b_param}.}
\label{hq:b_parton_mue}
\end{figure}

The last plot (\fig{hq:b_parton_mue}) depicts the dependence of the 
cross-section differences on $\mu$ for $t_\beta = 6$.
It shows a similar behavior as the same plot for large $t_\beta$
(\fig{hq:b_parton_mue_ltb}). For small values of $\mu$ 
the $\Delta m_b$-corrected tree-level result and the full one-loop
calculation coincide, while for larger ones significant deviations occur.
In the small $t_\beta$-regime the absolute value of the $\Delta_{\tilde 0}$
corrections even grows slightly larger than the linear behavior of
\eq{renorm:deltambeq}, which is due to additional effects originating from the
growing mass splitting between the two sbottoms. The one-loop
corrections in contrast decrease again, once the absolute value has 
reached a maximum at about $1500 \GeV$, and even change sign and 
become positive. So in this parameter region true one-loop corrections 
contribute significantly.

\section{Numerical Results for \texorpdfstring{$t\bar{t}h^0$}{t t h0}}

The numerical results for the second process of Higgs production in association with
heavy quarks, $ p p \rightarrow t\bar{t}h^0$, are presented in this section.
In table~\ref{hq:tspa} the hadronic cross section for the MSSM 
reference point $\spa$ is denoted. It is given separately for each partonic subprocess
which contributes to the $t\bar{t}h^0$-final state.
\begin{table}
\begin{center}
\begin{tabular}{l|r|r|r}
Partonic subprocess & $\sigma_0$ [fb]
& $\sigma_1$ [fb] 
&  $ \Delta_1 $ [\%] \\\hline
$d\bar{d}\rightarrow t\bar{t}h^0$ & $42.7$ & $37.6$ & $-11.77$ \\
$u\bar{u}\rightarrow t\bar{t}h^0$ & $71.9$ & $63.4$ & $-11.81$ \\
$s\bar{s}\rightarrow t\bar{t}h^0$ & $7.5$ & $6.6$ & $-11.58$ \\
$c\bar{c}\rightarrow t\bar{t}h^0$ & $2.8$ & $2.5$ & $-11.53$ \\
$gg\rightarrow t\bar{t}h^0$ & $273.7$ & $264.7$ & $-3.30$ \\\hline
$\sum\left(pp\rightarrow t\bar{t}h^0\right)$ & $399.0$ & $374.8$ & $-5.96$
\end{tabular}
\caption{Hadronic cross sections for $t\bar{t}h^0$-production at the 
  parameter point $\spa$, which is defined in \app{param:spa}.}
\label{hq:tspa}
\end{center}
\end{table}

In this case the quark--anti-quark annihilation diagrams give a 
contribution which is of comparable size to the gluon-fusion ones. On the
partonic level the same analysis as in the previous section for 
$h^0$-production in association with a bottom quark--anti-quark pair
holds. The quark--anti-quark--annihilation processes are suppressed
because there only a propagator-suppressed s-channel diagram
exists, while the gluon-fusion subprocess also proceeds via a t-channel
diagram which does not suffer from such a suppression. After the 
convolution with the parton distribution functions the situation however
changes. The gluon densities in the proton show a much steeper fall with 
growing parton-momentum fraction $x$ than the sea-quark ones. 
As the top quarks are much heavier 
than the bottom quarks, also the energy and thus the $x$ of the incoming 
partons must be larger to be above the threshold for $t\bar{t}h^0$-production.
For this final state it approximately compensates the effect from the propagator 
suppression. The gluon-fusion process is still the dominant production mode, but all 
processes need to be taken into account for a complete analysis.

In the following plots the effect of varying MSSM parameters on the SUSY-QCD
contributions is investigated. To that end the same parameter point as in the 
previous section with a fairly light SUSY spectrum was chosen, namely 
\begin{align}
m_A &= 200 \GeV \nonumber\\
t_\beta &= 6 \nonumber\\
\mu &= 300 \GeV \nonumber\\
A_t  &= A_b = 0 \nonumber\\
M_{\tilde{Q}} &= M_{\tilde{U}} = M_{\tilde{D}} = 250 \GeV \nonumber\\
m_{\tilde{g}} &= 400 \GeV  \quad .
\label{hq:t_param}
\end{align}
The MSSM parameters were then varied around this point. $t_\beta = 6$ was
kept fixed for all plots of this section.
The renormalization scale, which enters $\alpha_s$ via \eq{renorm:alphasrun}, 
and the factorization scale 
were set to $\mu_R = \mu_F = 2 m_t + m_{h^0}$. 
The hadronic cross-section calculations were performed for 
the LHC with a proton-proton center-of-mass energy of $14 \TeV$.

\begin{figure}
\includegraphics{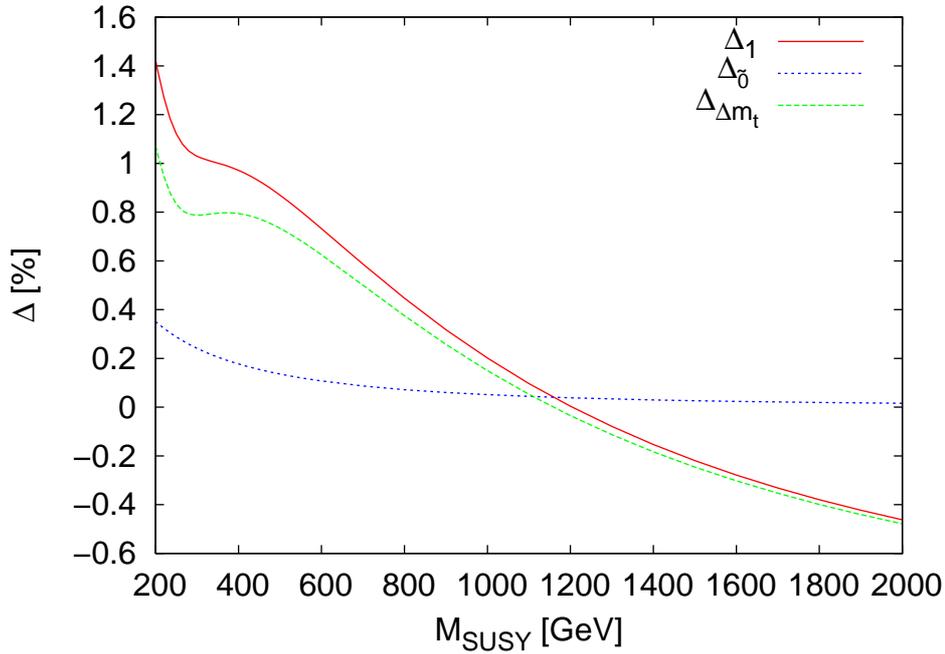}
\caption{Hadronic cross section differences for the process 
  $p p \rightarrow t\bar{t} h^0$ as a function of 
  $M_{\text{SUSY}} \equiv M_{\tilde{Q}} = M_{\tilde{U}} = M_{\tilde{D}}$.
  For the value of the other parameters see \eq{hq:t_param}.}
\label{hq:t_hadron_MSUSY}
\end{figure}

First a common mass scale $M_{\text{SUSY}}$, where the 
soft supersymmetry-breaking squark mass terms all take the same value, 
is introduced and varied between $200$ and $2000 \GeV$, as shown 
in \fig{hq:t_hadron_MSUSY}. 
The difference between the tree-level cross section and the $\Delta m_t$-corrected 
one falls off as $\frac1{M_{\text{SUSY}}^2}$, as expected from the form of the
$\Delta m_t$ term given in \eq{renorm:deltamt}. 
The total one-loop contributions show a similar decrease, but with a larger 
coefficient which leads to a much steeper descent. This originates from the fact that 
the $\Delta m_t$ term only includes vertex corrections to the $t\bar{t}h^0$ vertex.
Yet there are many other one-loop diagrams which also contribute and lead to the
modified behavior. In contrast to $\Delta m_b$, which is enhanced by a factor $t_\beta$,
the $\Delta m_t$ corrections are suppressed by $\frac1{t_\beta}$ and their numerical
effect is expected to be smaller. 
For small SUSY masses threshold effects of the squark 
masses induce a deviation from the scaling with $\frac1{M_{\text{SUSY}}^2}$.
% In the region of large SUSY masses a finite negative correction remains.

\begin{figure}
\includegraphics{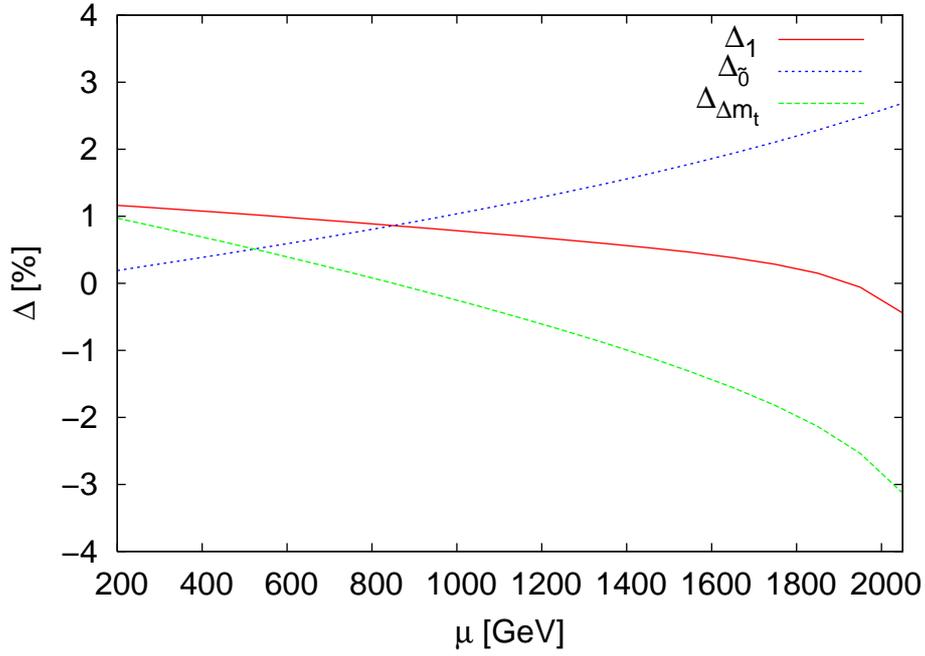}
\caption{Hadronic cross section differences for
  $t\bar{t} h^0$-production as a function of $\mu$.
  The values of all other parameters are given in \eq{hq:t_param}.}
\label{hq:t_hadron_mue}
\end{figure}

In the second plot (\fig{hq:t_hadron_mue}) the dependence of the relative corrections
on $\mu$, the mass parameter mixing the two Higgs doublets, is presented.
Also in this case the $\Delta m_t$-corrected tree-level cross section is not a good 
approximation. Whereas the term rises with growing $\mu$, the full one-loop correction
decreases in this case. The slope is constant over a large range of $\mu$. 
Only for bigger values, when the off-diagonal entries in the squark mixing
matrices become very large and yield an additional contribution, 
also the gradient increases.

\begin{figure}
\includegraphics{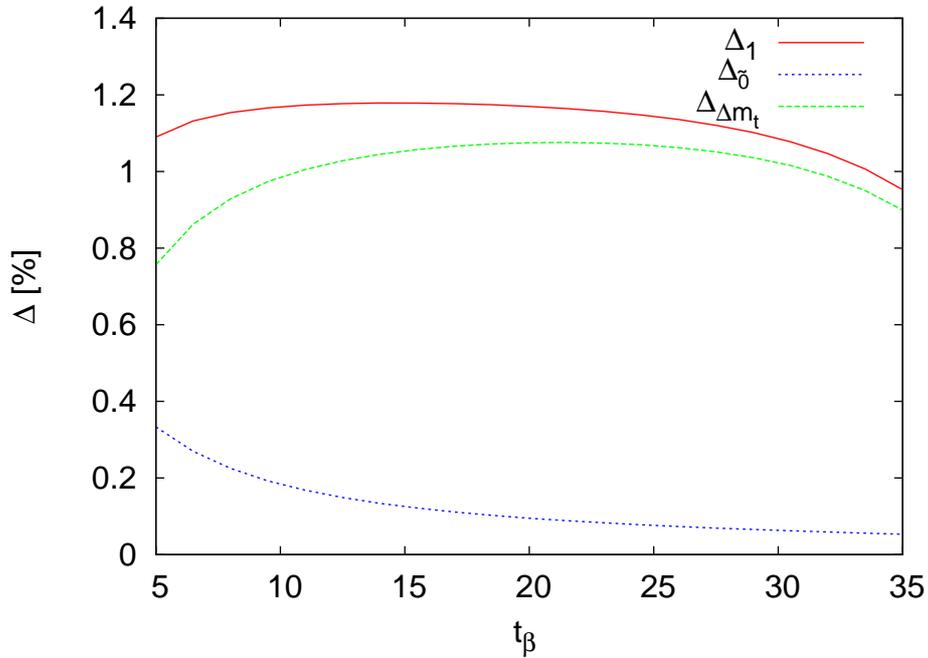}
\caption{$t_\beta$-dependence of the hadronic cross section differences 
  for $p p \rightarrow t\bar{t} h^0$.
  All other parameters take the values given in \eq{hq:t_param}.}
\label{hq:t_hadron_tb}
\end{figure}

Finally, in \fig{hq:t_hadron_tb} $t_\beta$ is varied.
The corrections to the $\Delta m_t$-corrected tree-level result fall off 
with growing $t_\beta$. This is again the expected behavior of~\eq{renorm:deltamt}.
The full one-loop corrections are significantly larger in size.
They show a mild dependence on this parameter, with a maximum at around
$t_\beta = 15$.

%Finally, the total cross section of $h^0$-production in association
%with a top quark--anti-quark pair is shown in \fig{hq:t_hadron_MsusyMGl}.
%\begin{figure}
%\includegraphics{plots/hq/gnuplot.t_hadron_MsusyMGl.eps}
%\caption{Total hadronic cross section for the process 
%  $p p \rightarrow t\bar{t} h^0$. The common SUSY mass scale $M_{\text{SUSY}}$
%  is altered together with the gluino mass $m_{\tilde{g}}$, where the ratio of the
%  two is kept fixed at $\frac{250}{400}$.}
%\label{hq:t_hadron_MsusyMGl}
%\end{figure}
%In this plot both the common SUSY mass scale and the gluino mass are 
%changed simultaneously and the mass ratio of the parameter set \eq{hq:t_param}
%of $\frac{250 \GeV}{400 \GeV}$ is kept fixed. Both tree-level and one-loop
%cross sections show a quadratic decrease with growing mass values.
%Most of the difference between those two comes from the use of different
%PDFs. Therefore also a curve is included, where the partonic tree-level 
%cross section was convoluted with the NLO-PDFs.

\chapter{Quartic Higgs Coupling at Hadron Colliders}
\label{quartic}

In this chapter another possibility is investigated to test the means of electroweak 
symmetry breaking. This is achieved by measuring the quartic Higgs coupling
and hence fully
determining the Higgs potential~\cite{Plehn:2005nk}.

After the discovery of a light Higgs boson the next step will be to study its properties,
including its couplings to other particles.
At the planned International Linear Collider (ILC) measuring these couplings 
with high precision will be possible for all 
Standard Model bosons and fermions~\cite{Desch:2001xh}.
Furthermore, if supersymmetric particles are found, the coupling of the Higgs to
charginos and neutralinos can be measured precisely~\cite{Kilian:2004uj}.
To fully understand electroweak symmetry breaking it is important to measure the
Higgs self-couplings and to thereby determine the parameters of the Higgs potential. 

\section{Higgs potential}

The Higgs potential of the Standard Model was already given in 
\eq{sm:higgspot}. In this model 
the trilinear ($\lambda_3$) and quartic ($\lambda_4$) Higgs self-coupling 
are related to the Higgs mass via
\begin{align}
\lambda_3 =& \frac{-3i m_H^2}{v} & 
\lambda_4 =& \frac{-3i m_H^2}{v^2} = \frac{\lambda_3}{v} ,
\end{align}
where $v$ is the vacuum expectation value of the Higgs field.

In models with more than one Higgs field, like the MSSM with its two fundamental
Higgs doublets, the relations between the trilinear and quartic Higgs coupling 
can change significantly.
For the lighter CP-even MSSM Higgs boson $h^0$ the ratio of the self-couplings is
\begin{equation}
\frac{\lambda_{3h^0}}{\lambda_{4h^0}} = 
  v \frac{s_{\beta+\alpha}}{c_{2\alpha}} ,
\end{equation}
where $v=\sqrt{v_1^2 + v_2^2}$ and $v_1$ and $v_2$ are the 
vacuum expectation values of the 
two Higgs fields.
If the parameter $m_A$ is sufficiently large there is a mass splitting between $h^0$
and the remaining Higgs sector. Additionally, the angles $\alpha$ and $\beta$ 
are related via $s_\alpha \simeq -c_\beta$ in this limit.
Therefore $\frac{s_{\beta+\alpha}}{c_{2\alpha}}$ approaches $1$ and the 
$h^0$-coupling becomes Standard Model-like.

In this chapter we will not refer to the MSSM as our underlying theory. Instead we use 
an effective theory whose particle content is the same as the one of the 
Standard Model.
Its Higgs sector also contains one doublet but the trilinear and quartic couplings 
are left as free parameters of the theory. In this way we are not restricted on a 
specific model but can accommodate for many different ones. Such deviations from 
the Standard Model couplings can for example be generated when 
higher-dimensional powers of the Higgs doublet are added to the potential as 
shown in \chap{sm:higherdimhiggs}.
%%% ??? Potential nochmal anschreiben?
%%% ??? Wieviel erklären, sodass man nicht zuviel blättern muss?
Taking the first two higher-order terms into account the Higgs self-couplings become
\begin{align}
\lambda_3 =& \lambda_3 \left( 1 
  + \frac{{\tilde\lambda}_1 v^2}{{\tilde\lambda}_0 \Lambda^2} \right) \nonumber\\
\lambda_4 =& \lambda_4 \left( 1 
  + \frac{6{\tilde\lambda}_1 v^2}{{\tilde\lambda}_0 \Lambda^2} 
  + \frac{4{\tilde\lambda}_2 v^4}{{\tilde\lambda}_0 \Lambda^4} \right) \quad .
\end{align}
The additional terms are suppressed by $\Lambda$ which is the scale where new
physics sets in. Both self-couplings receive different contributions from the additional
terms. In general, the self-couplings may even become negative. The stability of the 
Higgs potential is guaranteed if the highest non-vanishing term in the potential has 
a positive sign. All other terms can have arbitrary values as long as the ground state
has a non-vanishing vacuum expectation value to break the electroweak symmetry.

\section{Trilinear Higgs coupling}

\begin{figure}
\begin{center}
\subfigure[]{
\includegraphics[scale=0.4]{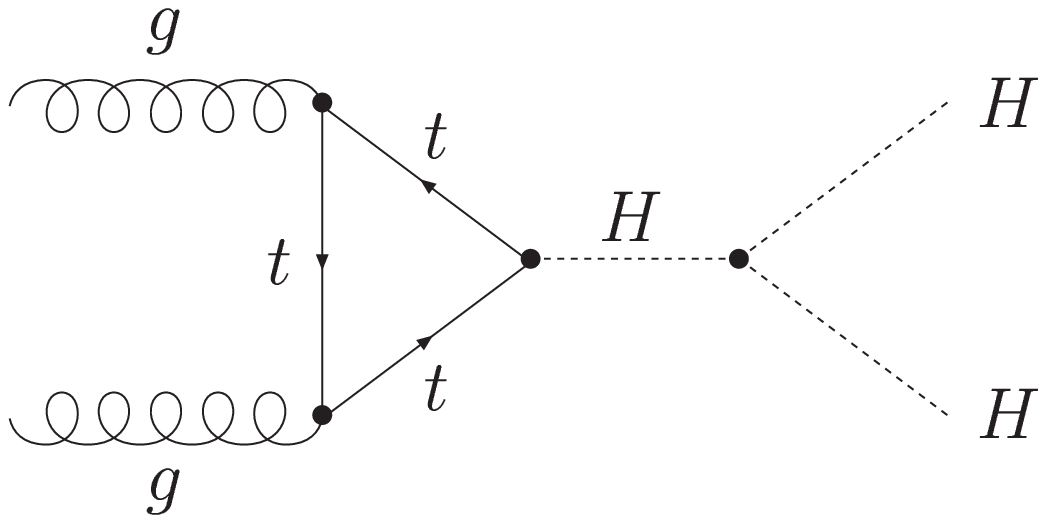}
}
\subfigure[]{
\includegraphics[scale=0.4]{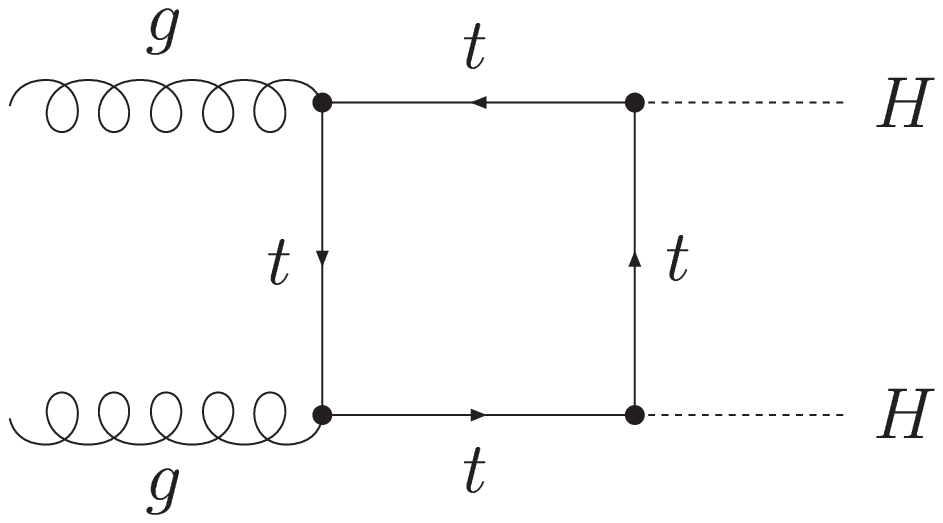}
}
\caption{Leading-order types of Feynman diagrams contributing to the 
  process $gg \rightarrow HH$}
\label{quartic:feynman_two}
\end{center}
\end{figure}

As we will see below it is essential for the measurement of the quartic Higgs coupling
to know the value of the trilinear Higgs coupling as precisely as possible. For a Higgs
boson with a mass larger than $150\GeV$ this coupling can be extracted at the 
LHC~\cite{Plehn:1996wb,*Plehn:1996wberr, 
Dicus:1987ic,*Glover:1987nx,*Djouadi:1999rc,*Dawson:1998py,
Baur:2002rb,*Baur:2002qd,*Dahlhoff:2005sz,*Gianotti:2002xx}.
At least two Higgs bosons must be produced to measure the three-Higgs coupling.
At hadron colliders this is performed via a gluon fusion process. In this process two
distinct types of diagrams appear, as shown in \fig{quartic:feynman_two}.
Either (a) an intermediate Higgs boson is produced via a 
three-point top-quark loop diagram which couples to the two final-state Higgs bosons and 
contains the required trilinear coupling, or (b) the particles couple via a four-point 
box-type top-quark loop. For the detection of the Higgs bosons the decay channel 
into two $W^+ W^-$ boson pairs is
analyzed~\cite{Baur:2002rb,*Baur:2002qd,*Dahlhoff:2005sz,*Gianotti:2002xx}. 
Two or three of the four $W$ bosons
are required to decay leptonically into a lepton and a neutrino to have a clear 
detector signal and the other two or one, respectively, decay hadronically into two jets.

% dsigma/dMinv
Additional information can be obtained from the kinematic distributions of the 
Higgs bosons. Not only the total cross section carries information on the trilinear 
Higgs coupling but also the differential hadronic distribution with respect to the
invariant mass of the final state. This can only be calculated correctly if the 
top-quark loop is fully taken into account. Using the infinite top-quark mass limit
and an effective gluon-gluon-Higgs coupling 
will yield completely incorrect results, as was shown in ref.~\cite{Baur:2002rb}.

Using this information it was 
evaluated~\cite{Baur:2002rb,*Baur:2002qd,*Dahlhoff:2005sz,*Gianotti:2002xx} with
which precision the trilinear Higgs coupling for Higgs bosons heavier than $150 \GeV$
can be measured.
In the beginning of LHC a non-zero value can be established with a confidence level 
of $95 \%$ after having accumulated a luminosity of $300~\text{fb}^{-1}$. After a 
luminosity upgrade the self-coupling can be measured with a precision of up to
$20 \%$ at the $95 \%$ confidence level using an integrated luminosity 
of $3~\text{ab}^{-1}$. At a future high-energy Very Large Hadron Collider (VLHC)
with a hadronic center-of-mass energy of $200 \GeV$, the measurement 
of $\lambda_3$ can be performed with an uncertainty of about $10 \%$ at 
a confidence level of $95 \%$ after having accumulated a luminosity of 
$1~\text{ab}^{-1}$. 

\begin{figure}[!htb]
\begin{center}
\subfigure[]{
\includegraphics[scale=0.5]{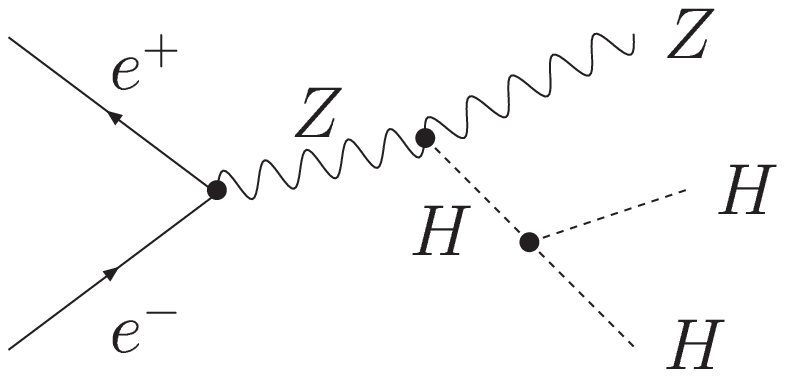}
}
\subfigure[]{
\includegraphics[scale=0.5]{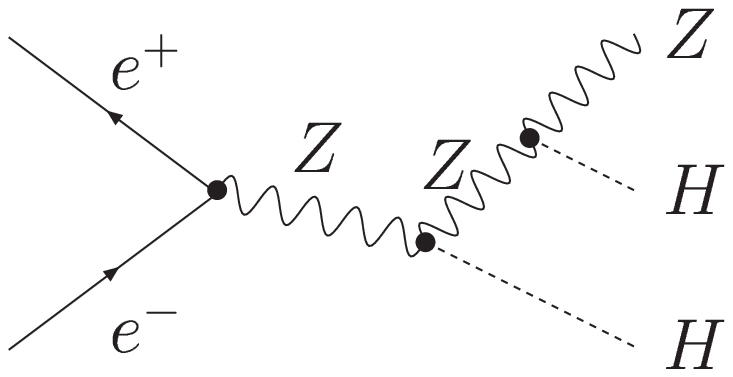}
}
\subfigure[]{
\includegraphics[scale=0.5]{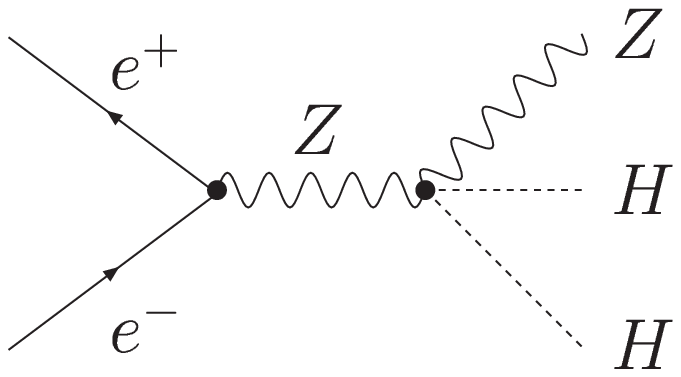}
}
\caption{Leading-order types of Feynman diagrams contributing to the 
  process $e^+ e^- \rightarrow ZHH$}
\label{quartic:feynman_el}
\end{center}
\end{figure}

\begin{figure}[!htb]
\begin{center}
\subfigure[]{
\includegraphics[scale=0.5]{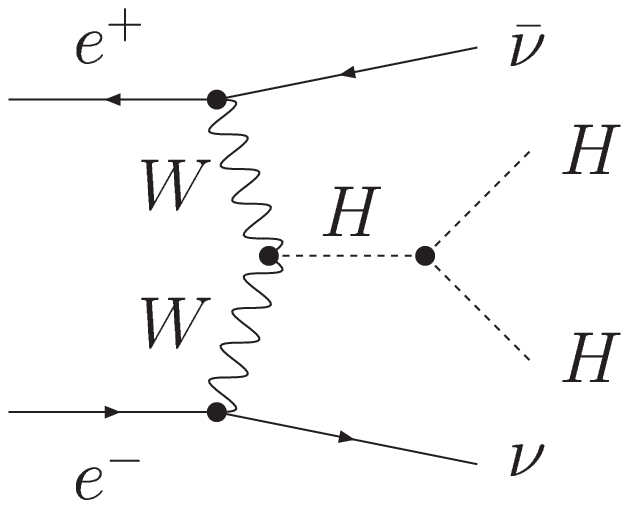}
}
\subfigure[]{
\includegraphics[scale=0.5]{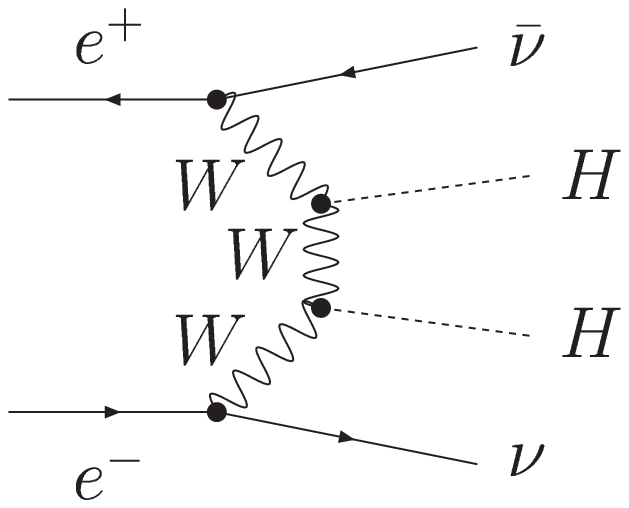}
}
\subfigure[]{
\includegraphics[scale=0.5]{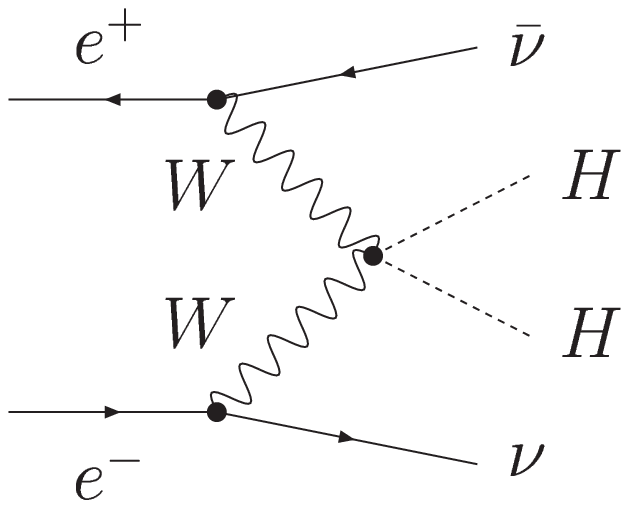}
}
\caption{Leading-order types of Feynman diagrams contributing to the 
  process $e^+ e^- \rightarrow \bar{\nu}\nu HH$}
\label{quartic:feynman_dhf}
\end{center}
\end{figure}

In contrast a linear collider can measure the three-Higgs coupling for small Higgs 
masses of around $120 \GeV$, but possibly not for higher ones~\cite{Djouadi:1999gv,
Gounaris:1979px,*Barger:1988jk,*Boudjema:1995cb,*Miller:1999ct,*Battaglia:2001nn,
*Castanier:2001sf,*Baur:2003gp}. The mass region below $140 \GeV$ is more difficult
for hadron colliders, because the dominant decay mode for the Higgs boson is a 
bottom quark--anti-quark pair which has large QCD backgrounds. 
Anyway, it is still accessible using rare decays~\cite{Baur:2003gp2}. 
The creation of the two Higgs bosons at an $e^+ e^-$ linear collider happens 
as double-Higgs strahlung in association with a $Z$ boson or as $WW$ double-Higgs
fusion in association with a $\bar{\nu}\nu$-pair as shown in
\fig{quartic:feynman_el} and \fig{quartic:feynman_dhf}, respectively.
For a Higgs mass of $120 \GeV$ the planned 
International Linear Collider (ILC) with a center-of-mass energy of $500 \GeV$ 
will be able to measure the trilinear Higgs coupling with an accuracy of $20 \%$ within 
one standard deviation after having accumulated a luminosity of $1~\text{ab}^{-1}$.
A proposed Compact Linear Collider (CLIC) with a center-of-mass energy of $3 \TeV$ 
finally could measure a $180 \GeV$ Higgs boson with a precision of $8 \%$ within 
one standard deviation using $5~\text{ab}^{-1}$ of integrated luminosity.

In general the proposed future generation of hadron (VLHC) and linear (CLIC) colliders
will be able to measure the trilinear Higgs coupling with an accuracy 
of $\Order{10 \%}$. A combination of both collider types thereby covers the whole
mass range where a Standard Model Higgs boson is expected to be found, as 
derived from electroweak precision analyses~\cite{Group:2005di}.

\section{Quartic Higgs coupling}

As the production of two Higgs bosons is needed for measuring the trilinear Higgs
coupling, three final-state Higgs bosons are necessary for a measurement of the
quartic Higgs coupling. 

Three-Higgs production at linear colliders has already been studied
in ref.~\cite{Accomando:2004sz}. 
It was found that even at CLIC the cross section is too
low, with only about five three-Higgs events per year being 
produced there at a center-of-mass energy of $10 \TeV$. Hence
it will be impossible to determine the quartic Higgs coupling at the next
generations of linear colliders.

\begin{figure}
\begin{center}
\subfigure[]{
\includegraphics[scale=0.4]{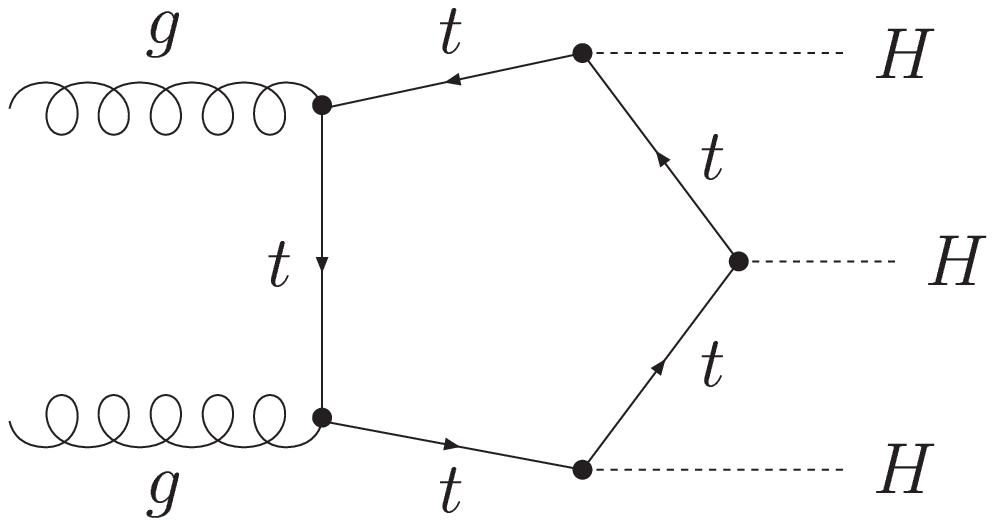}
}
\subfigure[]{
\includegraphics[scale=0.4]{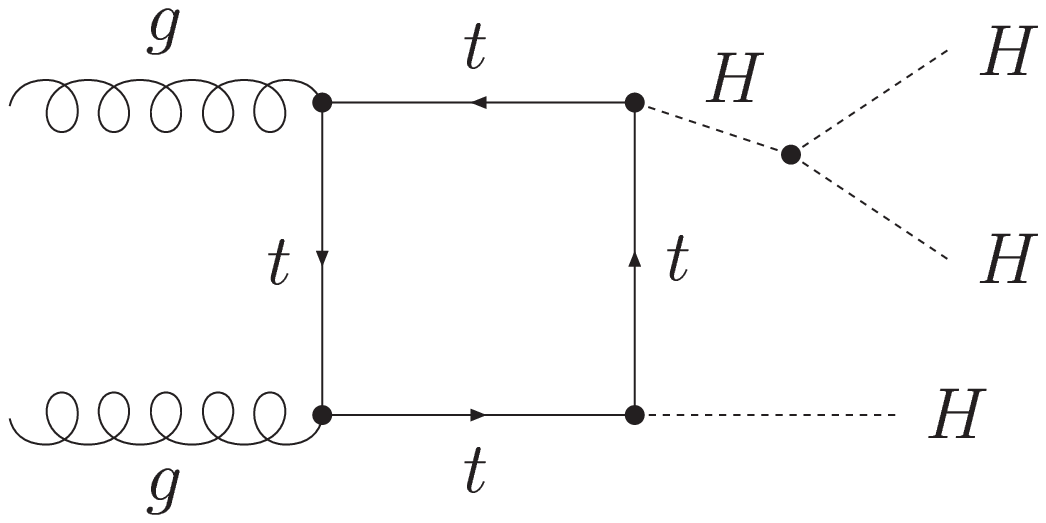}
}
\subfigure[]{
\includegraphics[scale=0.4]{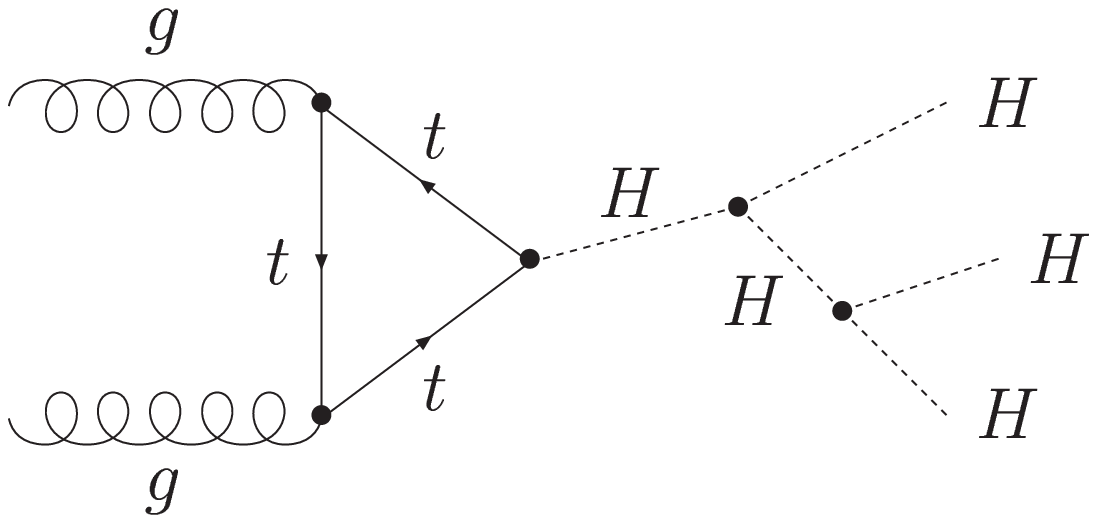}
}
\subfigure[]{
\includegraphics[scale=0.4]{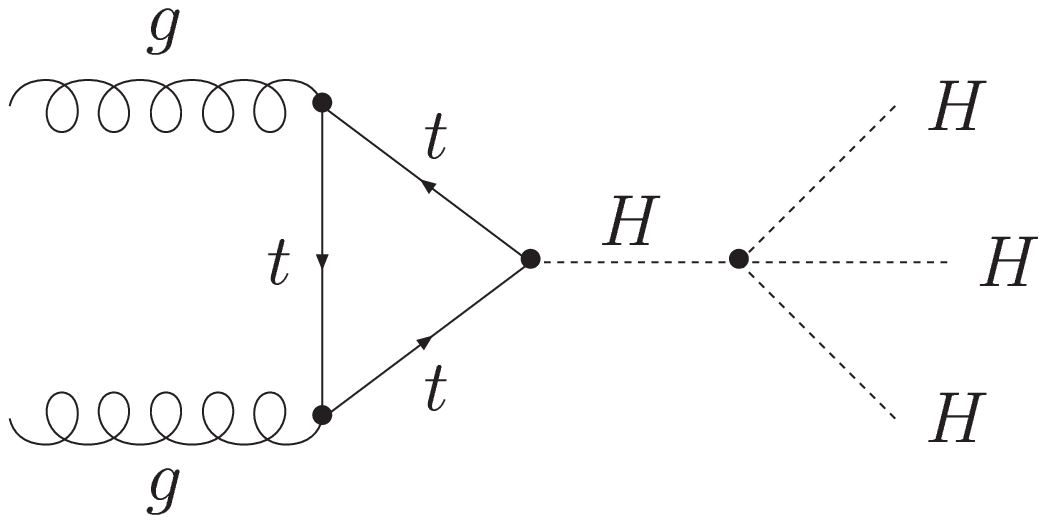}
}
\caption{Leading-order types of Feynman diagrams contributing to the 
  process $gg \rightarrow HHH$}
\label{quartic:feynman}
\end{center}
\end{figure}

At hadron colliders the three Higgs bosons are dominantly produced 
via gluon fusion and an intermediate top-quark loop, like in the 
two-Higgs case. Four distinct topologies 
appear as shown in \fig{quartic:feynman}: (a) continuum production of 
three Higgs bosons via a five-point top-quark loop, (b) production of two
Higgs bosons via a box-type loop and subsequent decay of one of the Higgs bosons
via the trilinear self-coupling into two Higgs bosons, and finally the production of 
one intermediate Higgs boson via a three-point loop. This can either (c) decay via 
a chain of two three-Higgs couplings or (d) through one quartic Higgs coupling.
Only the last diagram type contains the coupling we want to measure.

Looking at the diagrams it is clear that a precise knowledge of the trilinear
self-coupling is necessary to obtain results on the quartic self-coupling.
The process is also very sensitive to the top-quark Yukawa coupling
which must be known very well. In the numerical analysis we have also
included the diagrams where the top-quark loops are replaced by 
bottom-quark loops. The contribution of these diagrams is however 
less than one percent.

\begin{figure}
\begin{center}
\subfigure[LHC]{
\includegraphics[angle=270,scale=0.25]{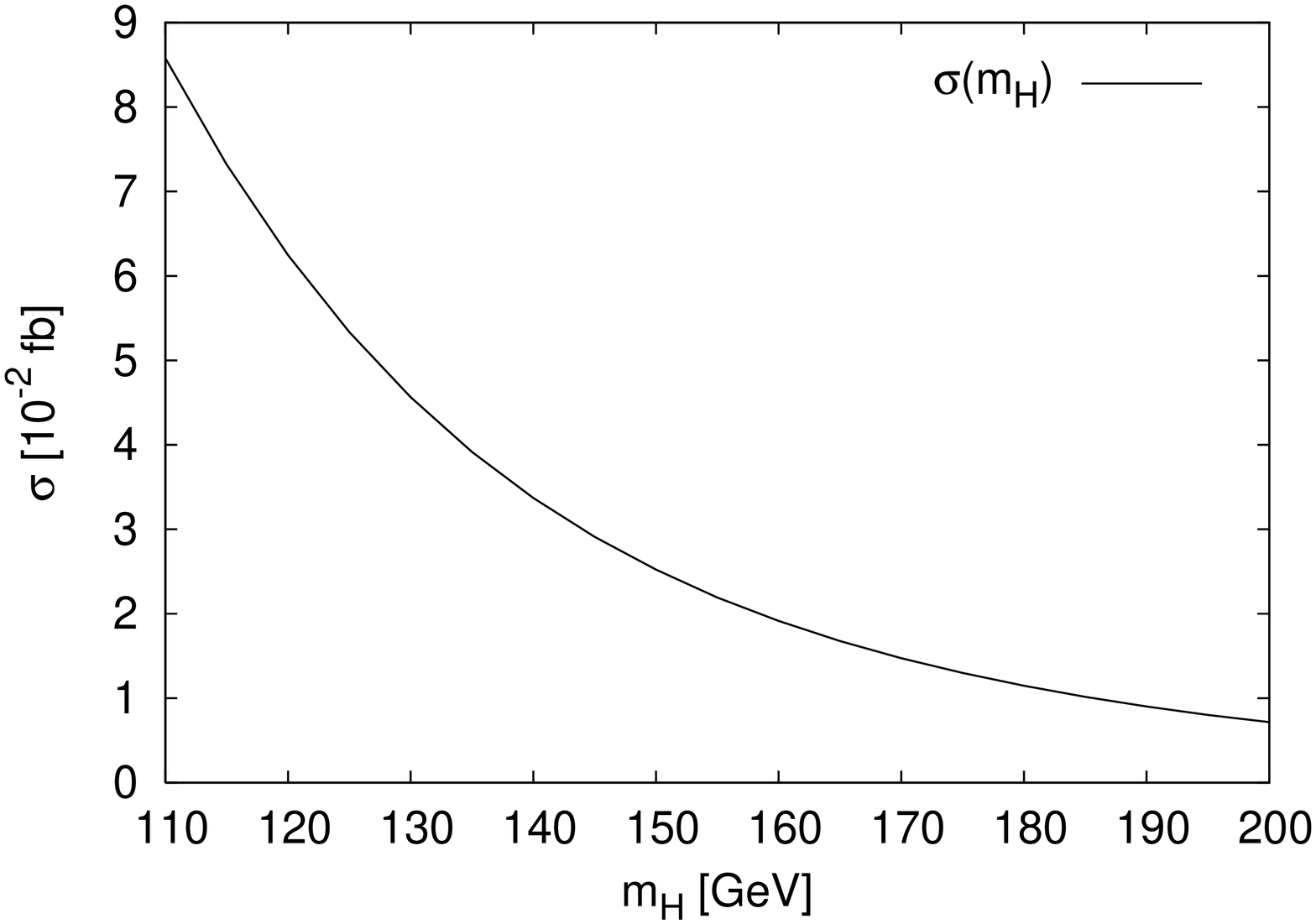}
}
\subfigure[VLHC]{
\includegraphics[angle=270,scale=0.25]{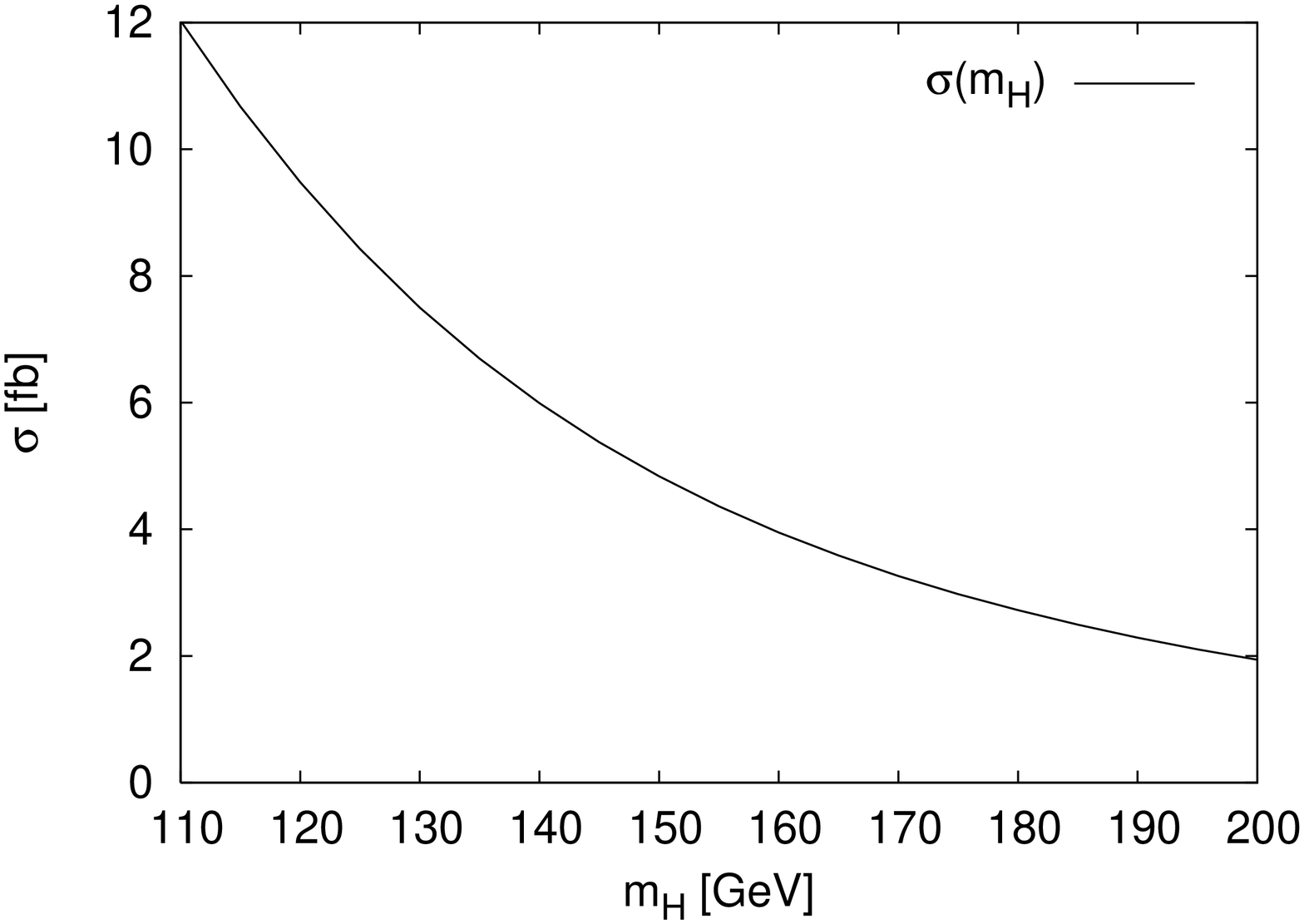}
}
\caption{Total hadronic cross section for the triple-Higgs production process 
 via gluon fusion with Standard Model couplings as a function of
 the Higgs boson mass.}
\label{quartic:higgsdep}
\end{center}
\end{figure}

The total cross section as function of the Higgs mass is shown in 
\fig{quartic:higgsdep} for the (a) LHC and a (b) $200 \TeV$ VLHC.
In the following the Higgs boson mass is set to $120 \GeV$ for all 
cross sections and plots.

\begin{figure}
\begin{center}
\subfigure[LHC]{
\includegraphics[angle=270,scale=0.43]{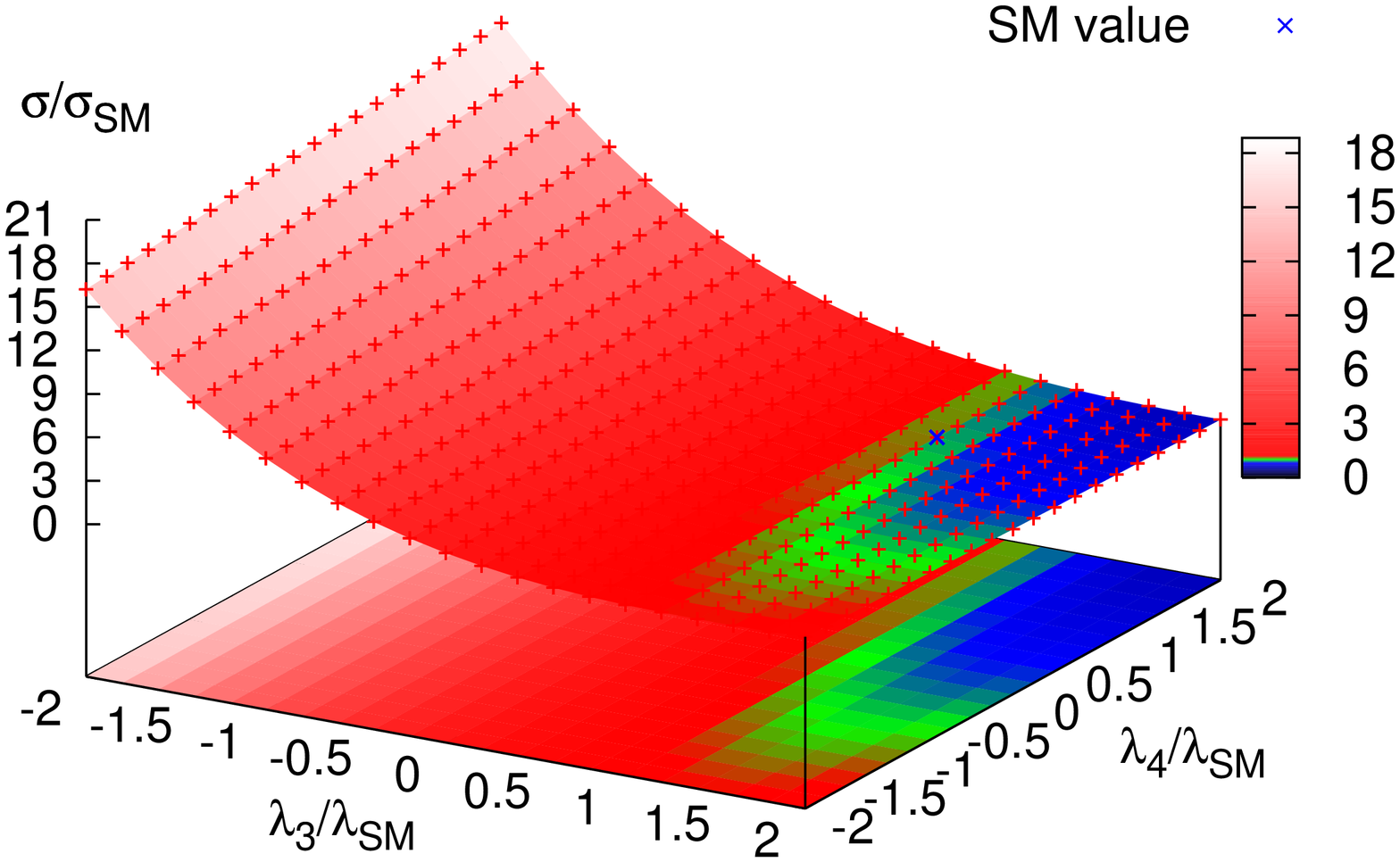}
}
\subfigure[VLHC]{
\includegraphics[angle=270,scale=0.43]{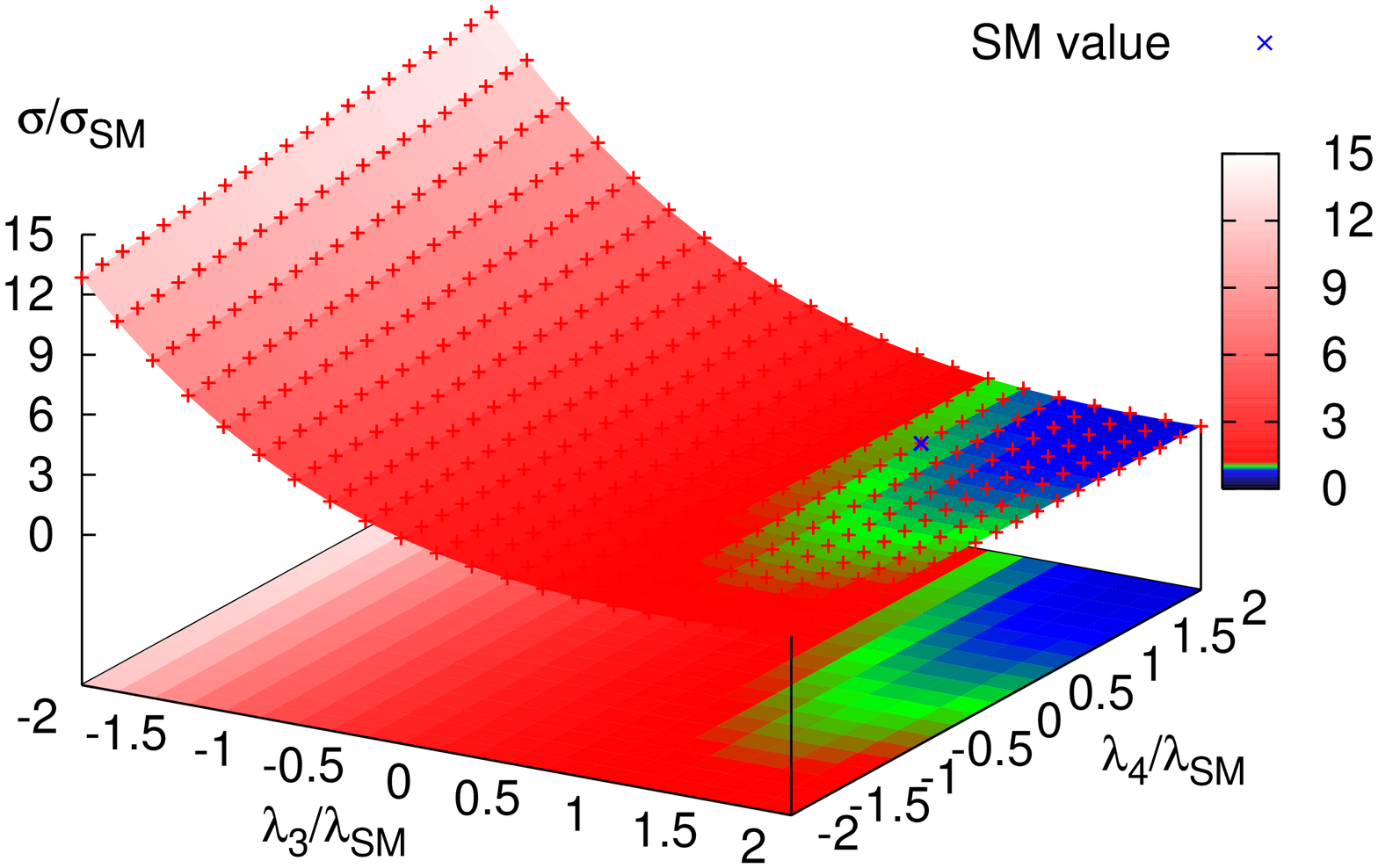}
}
\caption{Total hadronic cross section for triple-Higgs production via gluon fusion 
as a function of the trilinear and quartic Higgs coupling 
normalized to the Standard Model values. A color of green denotes the 
Standard Model value. A deviation of plus and minus 20 \% is signified by 
red and blue color, respectively. The maximum values obtained in the scanning
interval are colored white and black, respectively, using a linear color gradient
for intermediate values. The Standard Model point is additionally marked by 
a blue cross. A Higgs boson mass of $120 \GeV$ was used to obtain this plot.}
\label{quartic:higgs3d}
\end{center}
\end{figure}

\begin{figure}
\begin{center}
\subfigure[LHC]{
\includegraphics[angle=270,scale=0.43]{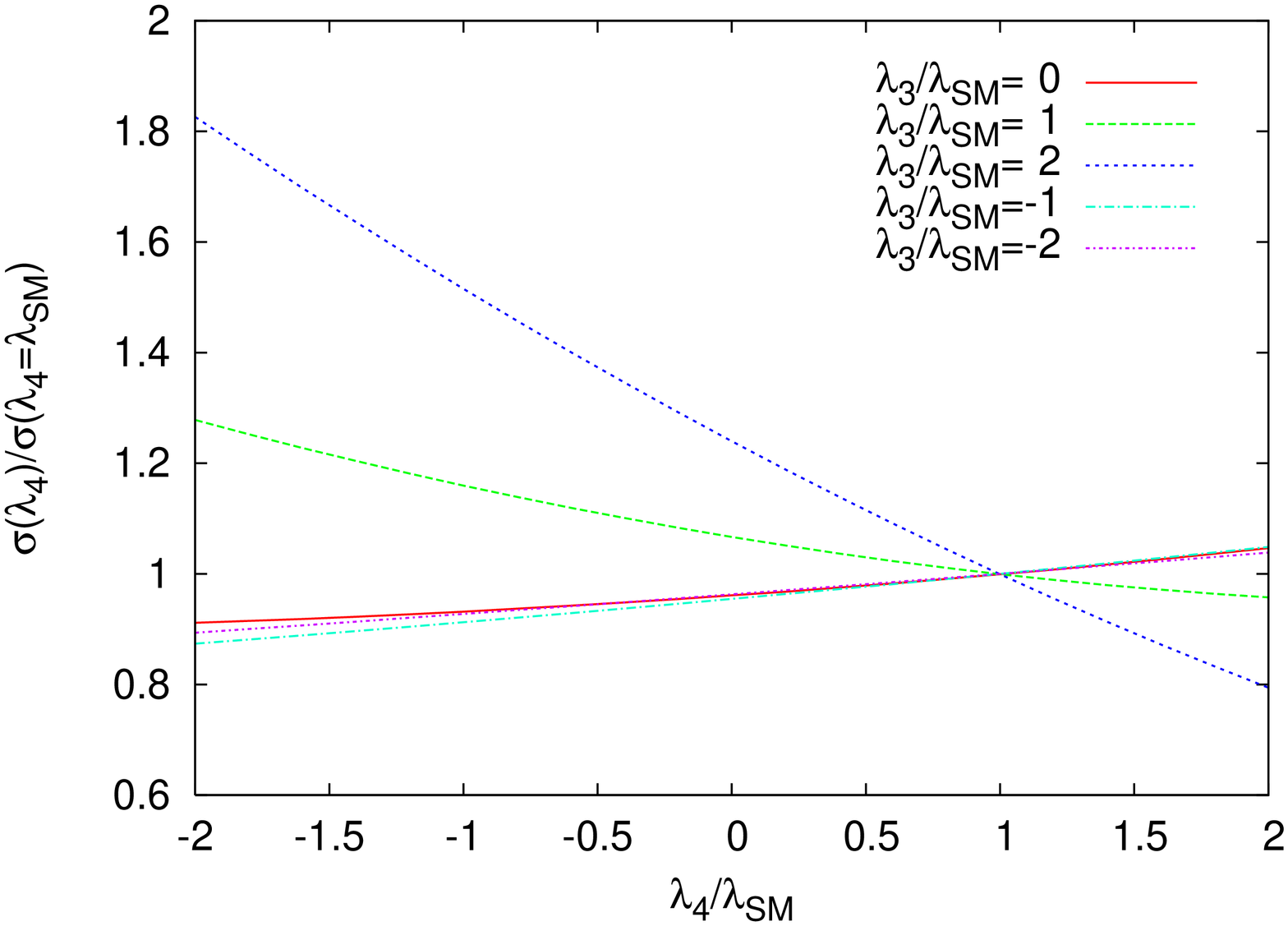}
}
\subfigure[VLHC]{
\includegraphics[angle=270,scale=0.43]{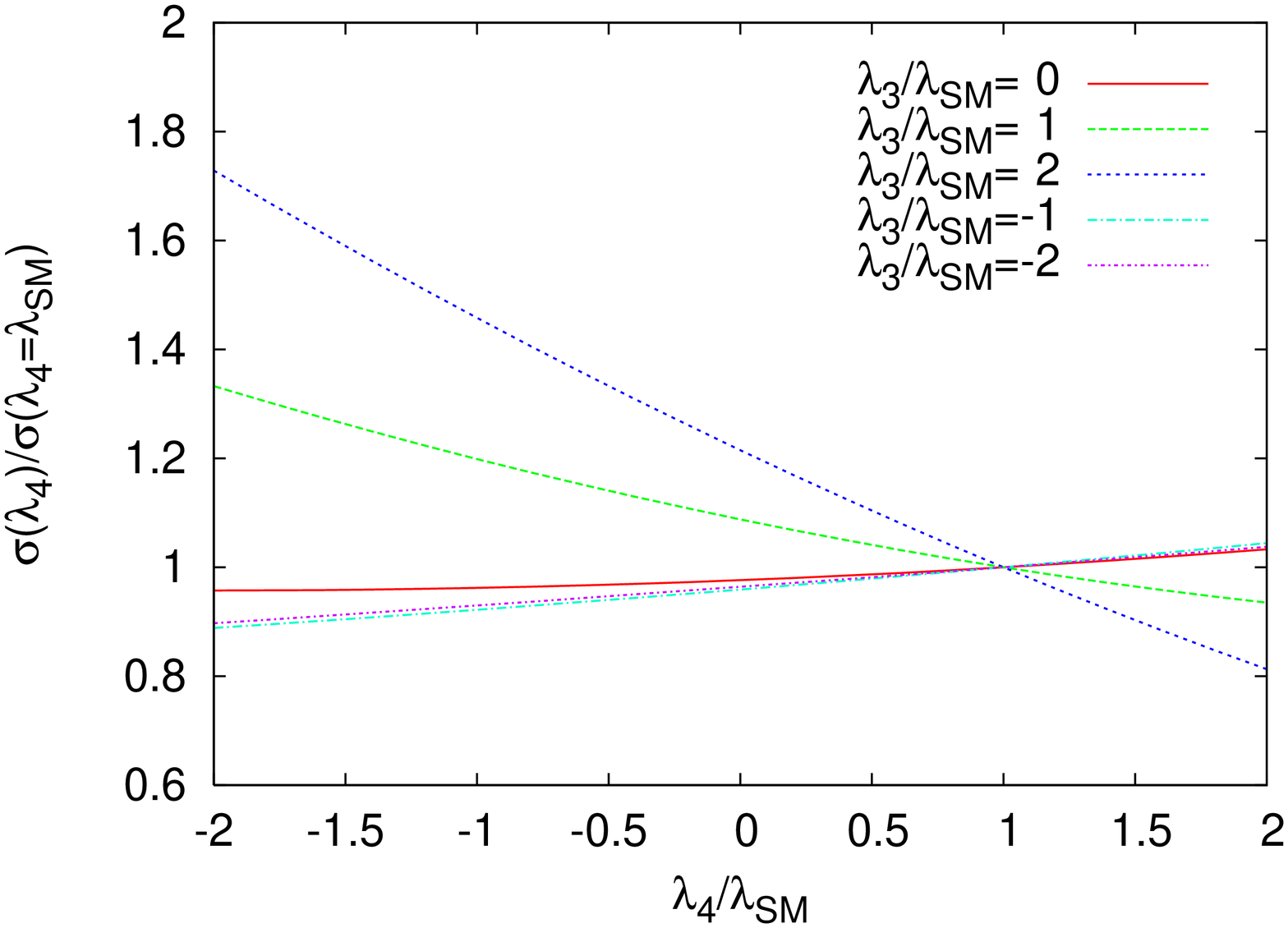}
}
\caption{Variation of the hadronic cross section for $HHH$-production 
as a function of $\lambda_4$ normalized to 
the value where $\lambda_4$ takes its Standard Model value. The Higgs mass was 
fixed to $120 \GeV$.}
\label{quartic:lambda4}
\end{center}
\end{figure}

In \fig{quartic:higgs3d} the dependence of the total cross section 
on the trilinear and quartic Higgs coupling is shown. 
The values of 
the trilinear and quartic self-couplings are varied between minus and 
plus two times the Standard Model value. 
One can clearly see the strong dependence on $\lambda_3$. The variation on 
$\lambda_4$ is much smaller, as one can see in more detail in \fig{quartic:lambda4}.
For positive values of $\lambda_3$ and $\lambda_4$ the variation of the cross section
stays below $20 \%$ at both the LHC and the VLHC. Including negative values of
$\lambda_4$ induces changes in the cross section of up to a factor 2.
For negative values of $\lambda_3$ the absolute variation as a function of $\lambda_4$
stays at the same order of magnitude. 
While the total cross section depends strongly on $\lambda_3$,
the relative variation
with $\lambda_4$ is heavily suppressed.

As the quartic Higgs coupling contributes only to the single diagram
\fig{quartic:feynman}(d) this behavior is expected. It is also reflected
in the partial contributions from the different diagram types. Taking into account 
all diagrams leads to a Standard Model cross section of 
$6.25 \cdot 10^{-2}~\text{fb}$ at the LHC. Using Standard Model couplings 
the five-(\fig{quartic:feynman}(a)), four-(\fig{quartic:feynman}(b)) and 
three-point(\fig{quartic:feynman}(c,d)) loop diagrams alone yield a cross section of 
$(17.07, 8.20, 0.46) \cdot 10^{-2}~\text{fb}$, respectively. The small 
size of the triangle-type
diagram results from a suppression factor by the intermediate Higgs propagator.
If one only takes this diagram type into account and sets either $\lambda_4$ 
or $\lambda_3$ to zero, a cross section of $0.17~\cdot 10^{-2}~\text{fb}$ from
the trilinear self-coupling and of $0.08 \cdot 10^{-2}~\text{fb}$ from the 
quartic self-coupling is obtained. 
For the VLHC the partial cross sections have similar ratios. So the diagram which
contains the quartic self-coupling is almost two orders of magnitude smaller 
than the total cross section. 

As one would expect from the evaluation of the trilinear Higgs coupling in two-Higgs
production~\cite{Baur:2002rb,*Baur:2002qd,*Dahlhoff:2005sz,*Gianotti:2002xx}
the interference between the pentagon and the box diagrams is indeed destructive
for positive values of $\lambda_3$. This 
results in the large increase of the cross section when $\lambda_3$ gets smaller and 
therefore the contribution from the box diagram diminishes as one can see 
in \fig{quartic:higgs3d}. For negative values of $\lambda_3$ the single three-Higgs
vertex in the box diagrams changes sign and therefore makes this interference
constructive, leading to the sharp rise of the cross section.
The trilinear and pentagon diagrams interfere constructively,
but because the triangle and box contributions have a more similar kinematic
configuration, the destructive interference between those two results in a slight
decrease of the cross section with growing $\lambda_4$. Only for $\lambda_3=0$,
where the box diagrams do not contribute any longer, the behavior of the cross 
section reverses and rises with increasing $\lambda_4$.
The relative signs of the different topologies can be understood analytically by using
the low-energy theorem for the leading form
factors~\cite{Ellis:1975ap,*Shifman:1979eb,*Spira:1995rr,*Kniehl:1995tn}.
An expansion is performed in the ratio $\tfrac{m_H}{m_t}$ with a partonic center 
of mass energy $\hat{s} \sim m_h^2$. The top-quark mass and the top-quark Yukawa 
coupling are both denoted by $m_t$.
These form factors are basically the squared matrix element without any couplings
or additional propagators which are not part of the loop integral. 
They are obtained starting from the top loop in the gluon self-energy. An additional
Higgs boson can be attached to the loop by using the following recursion relation:
\begin{equation}
F_{(n+1)H} = m_t^2 \frac{\partial}{\partial m_t}  \frac{F_{nH}}{m_t} \quad .
\end{equation}
This relation yields
\begin{equation}
F_{\text{pentagon}} = - F_{\text{box}} = F_{\text{triangle}} 
  = \frac23 + \Order{\frac{m_H^2}{m_t^2}} \quad .
\end{equation}
Therefore the structure of constructive and destructive interferences
in the diagram types is explained by the relative minus sign in front of 
the box-type term.

\begin{figure}
\begin{center}
\subfigure[LHC]{
\includegraphics[angle=270,scale=0.43]{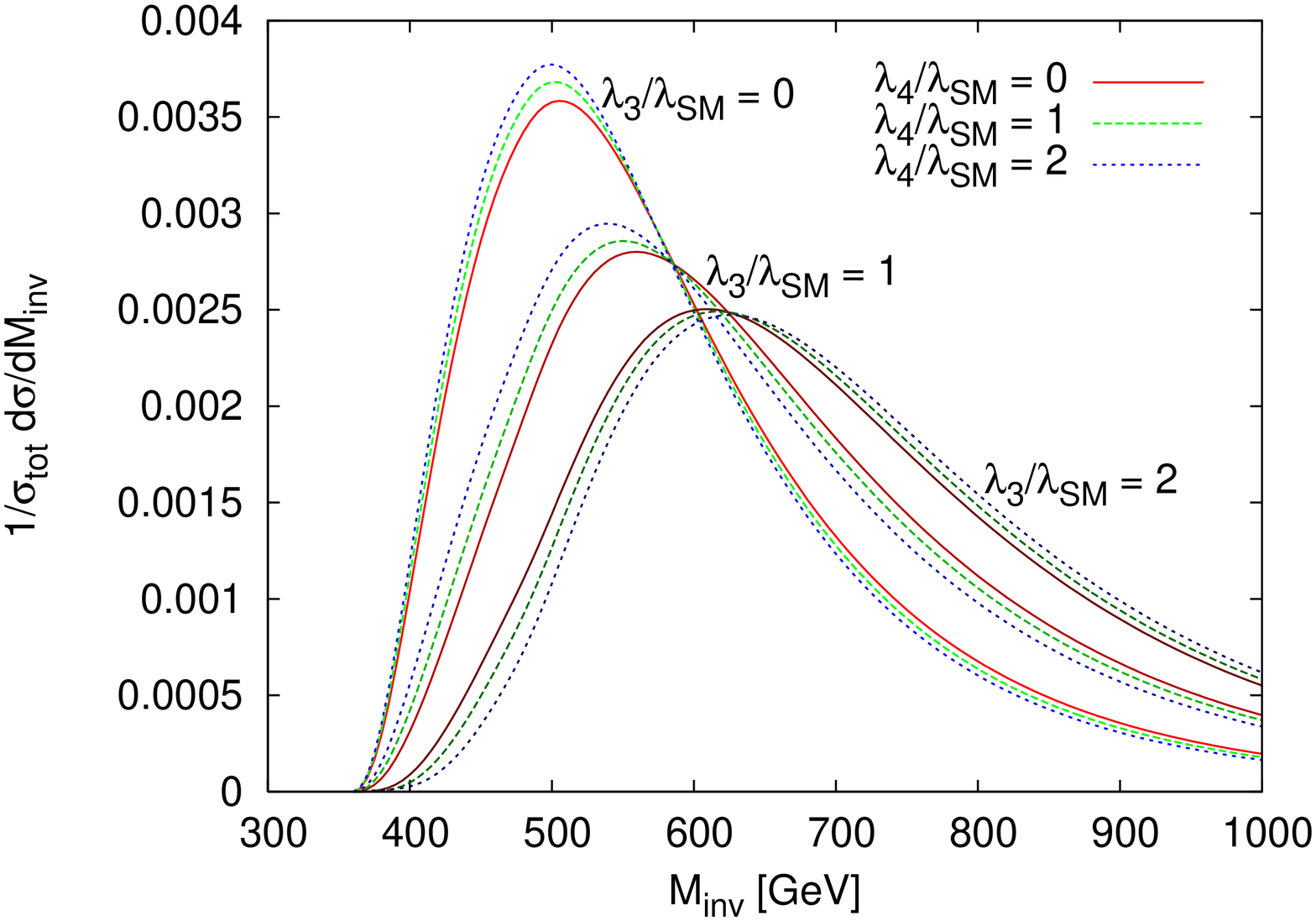}
}
\subfigure[VLHC]{
\includegraphics[angle=270,scale=0.43]{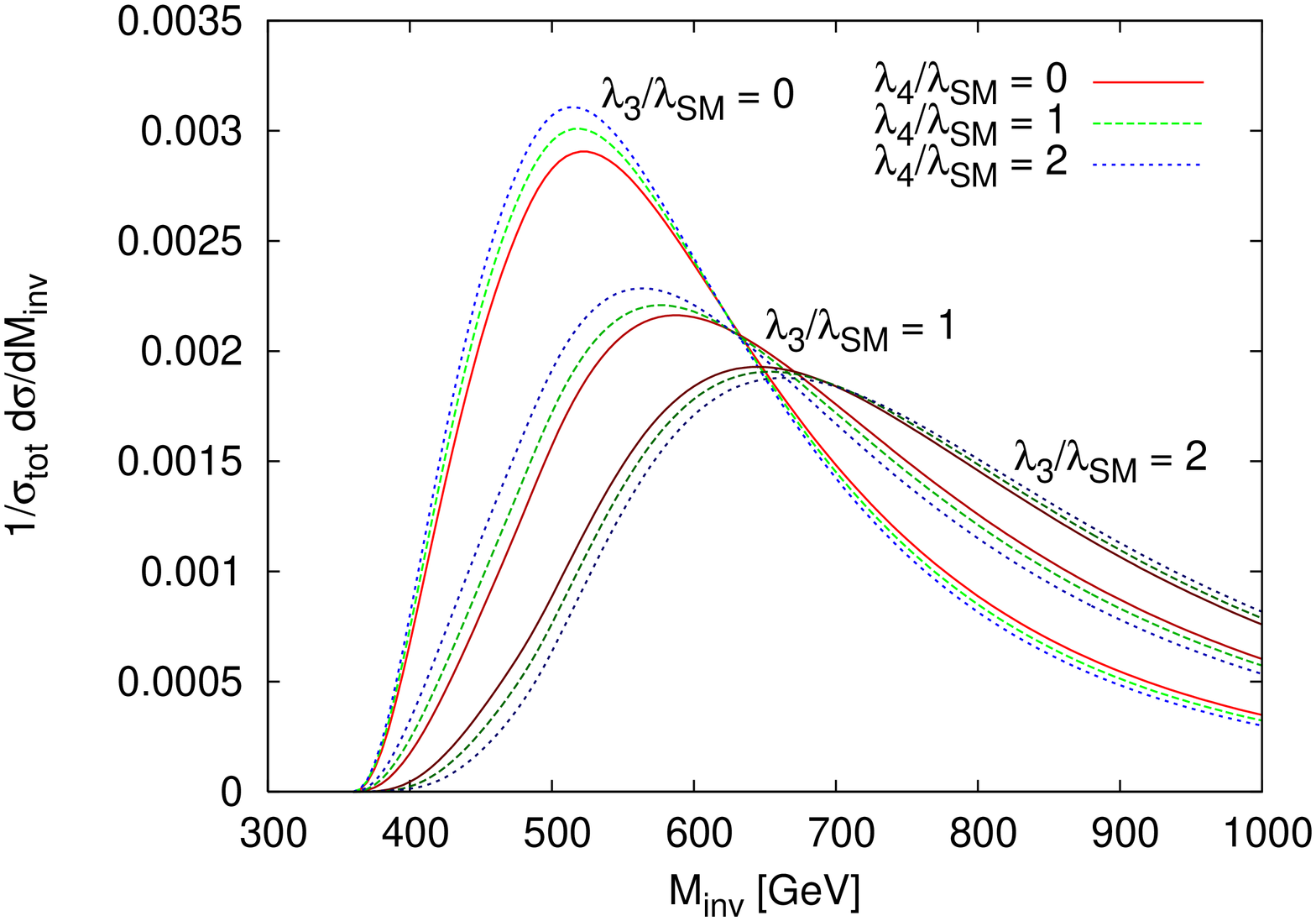}
}
\caption{Differential hadronic cross section for three-Higgs production 
with respect to the partonic center-of-mass energy,
normalized to the respective total cross section. The mass of 
the Higgs boson was set to $120 \GeV$. }
\label{quartic:diff}
\end{center}
\end{figure}

In the case of two-Higgs production the information on the value of 
$\lambda_3$ was not only encoded in the total cross section. Also the
differential hadronic cross section with respect to the invariant mass carries 
information on $\lambda_3$. The same is true for three-Higgs production.
In \fig{quartic:diff} the normalized cross section as a function of the 
partonic center-of-mass energy is shown.
The trilinear and quartic self-coupling both take the values $0$, $1$ and $2$ times
their respective Standard Model value. When varying $\lambda_3$ the position of the 
peak changes significantly and an extraction of this coupling is possible, as was already
found in the analysis of two-Higgs
production~\cite{Baur:2002rb,*Baur:2002qd,*Dahlhoff:2005sz,*Gianotti:2002xx}. 
When changing $\lambda_4$, and keeping $\lambda_3$ constant, the size of the shift is
about an order of magnitude smaller. Additionally for $\lambda_3=0$ the order of the 
$\lambda_4/\lambda_{\text{SM}}=0,1,2$ peaks is inverted which is due to
the different sign in the interference as explained above. 

The total hadronic cross section for triple-Higgs production via gluon fusion 
using Standard Model Higgs
self-couplings is $6.25 \cdot 10^{-2}~\text{fb}$ at the LHC for a Higgs boson 
with a mass of $120 \GeV$. This rate is too low to be measurable even for the 
high-luminosity mode of the LHC. At the VLHC with a center-of-mass energy of
$200 \GeV$ the cross section is $9.45~\text{fb}$ so three-Higgs production
might be observable at this future collider.
The rather strong dependence of the total cross section on $\lambda_3$,
especially for values smaller than the Standard Model value, allows to
extract the value of this coupling, thereby possibly confirming the 
two-Higgs production result. The variation in $\lambda_4$ is much smaller,
typically below $20\%$. Hence the extraction of this coupling is much harder.
If one takes into account the theoretical uncertainties from missing
higher-order corrections and the experimental error on the measurements
of $\lambda_3$ and the top-quark mass, the chances to be able to extract 
the quartic Higgs coupling are tiny.

Also in the differential cross section there is a clear effect on the peak position
when varying $\lambda_3$. This shift will be the mode to extract the trilinear coupling
in double-Higgs production. The size of the shift for a variation of $\lambda_4$ is
much smaller. If the errors on the measurements of $\lambda_3$ and the 
top-quark mass are again taken into account, the extraction of the 
quartic Higgs self-coupling looks challenging.

\chapter{Conclusions}
\label{concl} 

In this thesis production processes for Higgs bosons at hadron colliders 
were considered. 
In order to facilitate the computation of hadronic cross sections from a large
number of complicated parton processes, 
a computer code was developed. The calculation of 
hadronic cross sections, in particular for the production of 
supersymmetric Higgs bosons 
at the LHC in various processes was examined in the first part of this thesis. 
One-loop SUSY-QCD corrections, i.e.\ corrections with 
squarks and gluinos running in the loop, were calculated and the numerical 
results discussed.
In the second part triple-Higgs production in an effective theory was examined
and the question whether the quartic Higgs self-coupling
can be measured at hadron colliders, pursued.

The calculation of cross sections for processes which contain hundreds of 
single Feynman diagrams is not possible without the help of automated tools. 
For partonic cross sections there are already programs, like 
the packages \FeynArts{}, \FormCalc{} and \LoopTools{}. The latter one was 
extended, so that the five-point loop integrals are now included which makes a
calculation of $2 \rightarrow 3$ processes with these tools possible.
Additionally, the numerical stability of the loop integrals was improved.
To obtain hadronic cross sections, which are the observables which will be 
measured at the LHC, the partonic cross sections 
must be convoluted with the parton distribution
functions. Therefore a computer program, called \HadCalc{}, was written
which performs this task. Using this program it is possible to 
calculate both integrated and differential partonic and hadronic cross sections
of processes which were generated by \FormCalc{} beforehand. The cross 
sections can be differential with respect to the invariant mass of the final state,
or rapidity or transverse momentum of one of the outgoing particles.
Additionally, the possibility to apply cuts on the rapidity, the transverse
momentum, and the jet separation of the final-state particles was 
implemented. 

This program was then used to calculate SUSY-QCD corrections to various 
Higgs-boson-production processes at the LHC. The first considered process
was the production of a charged Higgs boson in association with a 
$W$ boson via bottom quark--anti-quark annihilation. It is known that the 
bottom-quark Yukawa coupling receives large one-loop corrections for 
large $t_\beta$. They are universal and can be parametrized via the 
variable $\Delta m_b$ and summed up to all orders in perturbation theory.
The numerical analysis showed that in the large-$t_\beta$ regime
this term indeed represents the dominant
contribution and the one-loop cross section is well approximated by the
$\Delta m_b$-corrected tree-level result. For small $t_\beta$, also other
terms play an important role. The leading subdiagram of this region was 
identified and its analytical behavior studied.
These contributions can reach a significant size, yielding
corrections of up to $50 \%$ for certain parameter combinations. 
It is a true one-loop result and cannot be taken into account by 
an effective tree-level coupling. 

Furthermore, the production of an $h^0$ via vector-boson fusion was 
calculated. This process has a clear experimental signature of two jets
in the forward region of the detector. In the theoretical analysis this 
phase-space region was selected by applying corresponding cuts.
The numerical size of the SUSY-QCD corrections to this process
is of $\Order{10^{-4}}$, and therefore significantly below the 
experimental uncertainty which LHC will be able to reach. 
This smallness of the corrections could be explained by
cancellations between different one-loop Feynman diagrams and 
the appearance of suppression factors. Hence, they do not induce a 
sufficiently large modification of the cross section, which would allow to
distinguish between the SM and the MSSM in the Higgs sector
in this way.
Additionally, $h^0$-production with two final-state jets, where one or 
two of the incoming partons are gluons, was considered. This
constitutes a background to the above-mentioned vector-boson-fusion
process. Thus for the calculation the same cuts were applied.
The total contribution of these processes is smaller than the total 
vector-boson-fusion cross section by more than four orders of 
magnitude, so they can be safely neglected 
in an analysis of $h^0$-production via vector-boson fusion.

Moreover, the associated production of an $h^0$ with a heavy,
i.e.\ bottom or top, quark--anti-quark pair was studied. These 
processes also form discovery channels for the Higgs boson.
Additionally, they can be used to measure the respective Yukawa 
couplings. In the case of bottom quarks the final-state jets are required to have 
large transverse momenta to avoid the appearance of large logarithms
and corresponding cuts on the phase space were applied.
For the top quarks no such cuts are necessary and therefore the calculation 
of the cross section was conducted over the full phase space.
Both times the SUSY-QCD corrections provide a significant contribution
to the total cross section, which must be taken into account. 
For final-state bottom-quarks the $\Delta m_b$-corrected tree-level cross
sections give a good estimate of the full one-loop result in large
regions of the MSSM parameter space. Only when the off-diagonal elements
of the sbottom mixing matrix become large, does this approximation break down.
Here other terms also give significant contributions, which lead to a sizable
change in the numerical result. As the sign of the two terms differs, the 
overall size of the one-loop corrections is reduced. Nevertheless, they provide 
a significant contribution to the total cross section, which can reach up 
to $40 \%$ for certain combinations of MSSM parameters. 
In the top-quark case the full one-loop calculation is never well approximated
by the $\Delta m_t$-corrected tree-level cross section. As the 
$\Delta m_t$ term contains a suppression of $\frac1{t_\beta}$ instead of
the $t_\beta$-enhancement of $\Delta m_b$, this behavior is expected.  
The total size of the corrections to $t\bar{t}h^0$-production is 
of the order of several percent.

In the last part of this dissertation the production of three Higgs bosons at 
hadron colliders was considered. This process can be used to extract the 
quartic Higgs self-coupling, thereby determining the Higgs potential. 
In this calculation not the MSSM was used as the underlying model, but an
effective theory
deduced from the Standard Model, where the three- and four-Higgs self-couplings
were left as free parameters. The numerical analysis showed that the cross section
is too small to be measured at the LHC. A future Very Large Hadron Collider with a 
projected center-of-mass energy of $200 \TeV$ would produce enough three-Higgs 
events. However, because of the interference structures of the different diagrams
which contribute to this final state, the extraction of the quartic Higgs coupling from 
the invariant-mass distribution will be seriously challenging. This is especially true 
if the theoretical uncertainties and the statistical errors on the measurement of the 
trilinear self-coupling and the top-quark mass are taken into account.

% The results obtained in this dissertation underline the necessity to perform full 
% one-loop calculations. Only when they are also included a reliable prediction
% of the cross section is possible.

\begin{appendix}
\chapter{Choice of Parameters}
\label{param}

\section{Standard Model Parameters}
\label{param:sm}

The parameters of the Standard Model have been measured by various experiments.
Their current best average is~\cite{Eidelman:2004wy}:
\begin{itemize}
\item Masses: 
  \begin{align*}
    \intertext{Quarks:}
    m_u &= 53.8 \textrm{ MeV} & m_c &= 1.5 \GeV & m_t &= 178 \GeV \\
    m_d &= 53.8 \textrm{ MeV} & m_s &= 150 \textrm{ MeV} & m_b &= 4.7 \GeV \\
    \intertext{Leptons:}
    m_e &= 510.999 \textrm{ keV} & m_\mu &= 105.658 \textrm{ MeV}
      & m_\tau &= 1.777 \GeV \\
    m_{\nu_e} &= 0 & m_{\nu_\mu} &= 0 & m_{\nu_\tau} &= 0 \\
    \intertext{Gauge bosons:}
    m_W &= 80.45 \GeV & m_Z &= 91.1875 \GeV & m_\gamma &= m_g = 0
  \end{align*}
\item Coupling constants
  \begin{align*}
    \alpha^{-1}\left(Q^2=0\right) &= 137.0359895 
      & \alpha_s^{\MSbar}(Q^2=m_Z^2) = 0.1172 \quad .
  \end{align*}
\end{itemize}
The masses of the first-generation quarks $m_u$ and $m_d$ are effective
parameters. They were chosen in a way that the vacuum polarization of the
photon, which was determined from experimental data via a dispersion relation
and is known more exactly, is correctly 
reproduced~\cite{Jegerlehner:1996ab,*Jegerlehner:2003ip}.

The calculations in this thesis were performed with the parameters
quoted above. Only the masses of the light quarks
$m_u$, $m_d$, $m_c$ and $m_s$ were set to zero exactly as their
contribution is negligible. Also the lepton masses are mentioned here only
for completeness, as they do not enter any diagram.

\section{SPA scenario of the MSSM}
\label{param:spa}

Even after adding experimental bounds and eliminating regions
of the parameter space which are disfavored by theoretical arguments 
the possible choices of the miscellaneous MSSM parameters are still plenty.
In order to unify the conventions used in the calculations and to allow for a 
comparison of the results of different working groups, an effort was made 
to apply the same conventions and use certain points in the MSSM parameter 
space as benchmark scenarios. For this the SPA (Supersymmetry Parameter
Analysis) conventions~\cite{spa} were established.

One of the favored reference points of the MSSM parameter space 
in the SPA conventions is called $\spa$. It is defined in
the minimal supergravity (mSUGRA) scenario. 
This scenario assumes that all parameters are real,
a unification of the gauge couplings happens at the GUT scale and the soft
su\-per\-sym\-me\-try-breaking terms are universal at the high-energy scale. 
Therefore the number of parameters is greatly
reduced to only four. They are a common scalar mass $M_0$, a 
common gaugino mass $M_{1/2}$, a common trilinear coupling $A_0$ and 
the ratio of the Higgs vacuum expectation values 
$t_\beta$. Additionally the sign of $\mu$ is not fixed
and can be chosen freely.
For $\spa$ these variables take the following values
\begin{gather}
M_0 = 70 \GeV  \quad\quad M_{1/2} = 250 \GeV  \quad\quad 
  A_0 = -300 \GeV \nonumber\\
t_\beta(\tilde{M}) = 10  \quad\quad \text{sign}(\mu) = +1 \quad .
\end{gather}
This parameter point was chosen in such a way that it is compatible with all
current experimental bounds.
The mass parameters are defined at the GUT scale and then evolved
via renormalization group equations (RGEs) to the SPA 
scale $\tilde{M} = 1 \TeV$ which is also the scale where $t_\beta$ is specified.

Using this procedure the MSSM parameters take the following values at the SPA
scale~\cite{spa,Fritzsche:2005da}
\begin{align}
t_\beta =& 10 & \mu =& 402.87 \GeV & m_A =& 431.02 \GeV \nonumber\\ 
M_1 =& 103.22 \GeV & M_2 =& 193.31 \GeV & M_3 =& 572.33 \GeV \nonumber\\
A_{u}^{1,2} =& -784.7 \GeV & A_{d}^{1,2} =& -1025.7 \GeV 
  & A_{e}^{1,2} =& -449.0 \GeV \nonumber \\
A_{u}^{3} =& -535.4 \GeV & A_{d}^{3} =& -938.5 \GeV &
  A_{e}^{3} =& -445.5 \GeV \nonumber\\
M_{\tilde Q}^{1,2} =& 526.9 \GeV & M_{\tilde U}^{1,2} =& 507.7 \GeV 
  & M_{\tilde D}^{1,2} =& 505.5 \GeV \nonumber \\
M_{\tilde Q}^{3} =& 471.3 \GeV & M_{\tilde U}^{3} =& 384.6 \GeV &
  M_{\tilde D}^{3} =& 501.3 \GeV \nonumber \\
M_{\tilde L}^{1,2} =& 181.3 \GeV & M_{\tilde R}^{1,2} =& 115.6 \GeV \nonumber \\
M_{\tilde L}^{3} =& 179.5 \GeV & M_{\tilde R}^{3} =& 109.8 \GeV \quad .
\end{align}
The RGE evolution leads to different values for the soft-breaking mass parameters 
and trilinear couplings of the first two and the third generation. In the above-mentioned
table the generation index is denoted by the superscript.

These parameters must be interpreted as parameters in the \DRbar{} renormalization
scheme. As most of the calculations in this thesis are done using OS renormalization
a further conversion step is necessary. Using these \DRbar{} parameters the masses
of all supersymmetric particles are computed at the one-loop level. Then a 
minimal set of these masses is chosen and the OS parameters are calculated
using the tree-level relations between parameters and masses.
%\footnote{The
%tree-level relations yield the OS parameters by construction because higher-order
%contributions to the masses are cancelled by appropriate counter terms.}
With this procedure the physical masses of the particles in both renormalization
schemes are equal. Thus a meaningful comparison of the results of calculations 
in both schemes is possible.

This conversion procedure yields the following MSSM parameters in the OS 
scheme~\cite{Fritzsche:2005da}:
\begin{align}
t_\beta =& 10 & \mu =& 399.26 \GeV & m_A =& 431.02 \GeV \nonumber\\ 
M_1 =& 100.11 \GeV & M_2 =& 197.55 \GeV & M_3 =& 612.85 \GeV \nonumber\\
A_{u}^{1,2} =& -784.7 \GeV & A_{d}^{1,2} =& -1025.7 \GeV 
  & A_{e}^{1,2} =& -449.0 \GeV \nonumber \\
A_{u}^{3} =& -535.4 \GeV & A_{d}^{3} =& -938.5 \GeV &
  A_{e}^{3} =& -445.5 \GeV \nonumber\\
M_{\tilde Q}^{1} =& 565.97 \GeV & M_{\tilde U}^{1} =& 546.78 \GeV 
  & M_{\tilde D}^{1} =& 544.95 \GeV \nonumber \\
M_{\tilde Q}^{2} =& 565.91 \GeV & M_{\tilde U}^{2} =& 546.84 \GeV 
  & M_{\tilde D}^{2} =& 544.97 \GeV \nonumber \\
M_{\tilde Q}^{3} =& 453.05 \GeV & M_{\tilde U}^{3} =& 460.52 \GeV &
  M_{\tilde D}^{3} =& 538.13 \GeV \nonumber \\
M_{\tilde L}^{1} =& 184.12 \GeV & M_{\tilde R}^{1} =& 118.02 \GeV \nonumber \\
M_{\tilde L}^{2} =& 184.11 \GeV & M_{\tilde R}^{2} =& 117.99 \GeV \nonumber \\
M_{\tilde L}^{3} =& 182.18 \GeV & M_{\tilde R}^{3} =& 111.29 \GeV \quad .
\end{align}
These are the parameters used for the calculations in this thesis 
when the $\spa$ parameter set is referred to.

\chapter{Basic Principles of Supersymmetry}
\label{appsusy}

\section{Poincar\'e group}
\label{appsusy:poincare}
Every point in four-dimensional space-time of Minkowski space is characterized
by a contravariant vector which is defined by
\begin{equation}
x^\mu = \left( x^0, x^1, x^2, x^3 \right) = \left( t, \vec{x}\right) 
\end{equation}
as generalized space coordinate. With the metric tensor
\begin{equation}
g_{\mu\nu} = g^{\mu\nu} = \diag\left( 1,-1,-1,-1 \right) 
\end{equation}
a covariant vector 
\begin{equation}
x_\mu = g_{\mu\nu} x^\nu = \left( t, -\vec{x}\right) 
\end{equation}
can also be defined.

Here and in the following Greek indices run from 0 to 3 and 
Latin ones from 1 to 3 except where denoted otherwise. 
Einstein's sum convention is implicitly assumed,
i.e.\ indices which appear once as covariant and once as contravariant index 
are summed over. Additionally natural units are used where $\hbar=c=1$.

Derivatives with respect to generalized space coordinates can be abbreviated 
as 
\begin{align}
\partial_\mu :=& \frac{\partial}{\partial x^\mu} 
= \left( \frac{\partial}{\partial t},\vec\nabla \right) & 
\partial^\mu :=& \frac{\partial}{\partial x_\mu}
= \left( \frac{\partial}{\partial t},-\vec\nabla \right)  \quad .
\end{align}
The momentum four-vector is defined as 
\begin{equation}
p^\mu = i \partial^\mu = \left( E,\vec{p} \right) \quad . 
\end{equation}

\section{Spinors}
\label{appsusy:spinors}

\subsection{Weyl spinors}
\label{appsusy:spinors:weyl}
Two-component anticommuting objects $\xi_\alpha$ which 
transform under a matrix $M$ of $SL(2,\Cmplx)$ as
\begin{align}
\xi_\alpha \rightarrow& {M_\alpha}^\beta \xi_\beta &
\bar{\xi}_{\dot\alpha} \rightarrow& 
  {{M^*}^{\dot\beta}}_{\dot\alpha} \bar{\xi}_{\dot\beta} \nonumber\\
\xi^\alpha \rightarrow& {\left( {M^{-1}}\right) _\beta}^\alpha \xi^\beta &
\bar{\xi}^{\dot\alpha} \rightarrow& 
  {{\left(M^{-1}\right)^*}^{\dot\alpha}}_{\dot\beta} \bar{\xi}^{\dot\beta} 
\end{align}
are called Weyl spinors. The spinor indices $\alpha$, $\dot\alpha$, $\beta$ and
$\dot\beta$ can take the values 1 and 2. Additionally the relations
\begin{align}
{\bar\xi}_{\dot\alpha} &\equiv \xi_\alpha^* &
\xi_\alpha &= \epsilon_{\alpha\beta} \xi^\beta
\end{align}
hold. On the right-hand side the two-dimensional totally antisymmetric tensor
\begin{equation}
\epsilon^{\alpha\beta} = -\epsilon_{\alpha\beta} = 
\begin{pmatrix} 0&1\\-1&0 \end{pmatrix} = 
\begin{cases}
1 & \text{for even permutations of $\left\lbrace 1,2 \right\rbrace$ } \\
-1 & \text{for odd permutations of $\left\lbrace 1,2 \right\rbrace$ } \\
0 & \text{else}
\end{cases}
\end{equation}
has been introduced. Along the same lines also three-dimensional
\begin{align}
\epsilon^{ijk} = \epsilon_{ijk} = 
\begin{cases}
1 & \text{for even permutations of $\left\lbrace 1,2,3 \right\rbrace$ } \\
-1 & \text{for odd permutations of $\left\lbrace 1,2,3 \right\rbrace$ } \\
0 & \text{else}
\end{cases}
\intertext{and four-dimensional versions }
\epsilon^{\mu\nu\rho\sigma} = -\epsilon_{\mu\nu\rho\sigma} = 
\begin{cases}
1 & \text{for even permutations of $\left\lbrace 0,1,2,3 \right\rbrace$ } \\
-1 & \text{for odd permutations of $\left\lbrace 0,1,2,3 \right\rbrace$ } \\
0 & \text{else}
\end{cases}
\end{align}
can be defined, which are needed later.

If one generalizes the definition of the Pauli matrices $\vec\sigma$ 
\begin{align}
\sigma^1 =& \begin{pmatrix} 0&1\\ 1&0 \end{pmatrix} &
\sigma^2 =& \begin{pmatrix} 0&-i\\ i&0 \end{pmatrix} &
\sigma^3 =& \begin{pmatrix} 1&0\\ 0&-1 \end{pmatrix}
\end{align}
to four dimensions via
\begin{align}
\sigma^\mu &= (1,\vec\sigma)  & {\bar\sigma}^\mu &= (1,-\vec\sigma) \quad , 
\end{align}
then the Dirac equation can be written in two-component notation as follows:
\begin{align}
\left( \bar\sigma_\mu p^\mu\right)^{\dot\alpha \beta} \xi_\beta &= 
  m {\bar\eta}^{\dot\alpha} & 
\left( \sigma_\mu p^\mu \right)_{\alpha \dot\beta} {\bar\eta}^{\dot\beta} &=
  m \xi_\alpha \quad .
\label{appsusy:dirac}
\end{align}

\subsection{Dirac and Majorana spinors}
\label{appsusy:spinors:dirac}
Out of two two-component Weyl spinors a four-component spinor 
\begin{equation}
\psi = \begin{pmatrix} \xi_\alpha \\ {\bar\eta}^{\dot\alpha}\end{pmatrix}
\label{appsusy:diracspinor}
\end{equation}
can be constructed. If $\eta=\xi$, $\psi$ is called a Majorana spinor, else
$\psi$ is called Dirac spinor.
The four-component equivalent to the Pauli matrices are the $\gamma$ matrices
which are defined via the Clifford algebra
\begin{equation}
\left\lbrace \gamma^\mu, \gamma^\nu \right\rbrace = 
\gamma^\mu \gamma^\nu + \gamma^\nu \gamma^\mu = 2 g^{\mu\nu} \quad .
\label{appsusy:clifford}
\end{equation}
Their hermitian conjugates are
\begin{equation}
{\gamma^\mu}^\dagger = \gamma^0 \gamma^\mu \gamma^0 \quad ,
\end{equation}
so $\gamma^0$ is hermitian and the $\gamma^i$ are antihermitian.
There are different representations of the $\gamma$ matrices 
which all fulfill \eq{appsusy:clifford}. The one corresponding to 
the form of the spinors in \eq{appsusy:diracspinor} is called
chiral representation and in that one the $\gamma$ matrices have the 
following form
\begin{equation}
\gamma_\mu = 
\begin{pmatrix}
0&{\sigma_\mu}_{\alpha\dot\beta}\\ {\bar\sigma}_\mu^{\dot\alpha\beta}&0
\end{pmatrix} .
\end{equation}
Additionally one defines $\gamma_5$ as
\begin{equation}
\gamma_5 = \gamma^5 = i \gamma^0 \gamma^1 \gamma^2 \gamma^3
\overset{\text{chiral}}{\underset{\text{representation}}{=}}
\begin{pmatrix} -\Id&0\\0&\Id \end{pmatrix} \quad ,
\end{equation}
for which the following relations hold:
\begin{align}
\left\lbrace \gamma^\mu, \gamma^5 \right\rbrace &= 0 &
\left( \gamma_5 \right)^2 &= 0 \quad .
\end{align}
Then the projection operators
\begin{align}
P_L &\equiv \frac12 \left( 1-\gamma_5 \right)  &
P_R &\equiv \frac12 \left( 1+\gamma_5 \right) 
\end{align}
yield the left- and right-handed part of a Dirac spinor, respectively
\begin{align}
\psi_L &= P_L \psi = \begin{pmatrix}\xi_\alpha \\ 0\end{pmatrix} &
\psi_R &= P_R \psi = \begin{pmatrix}0 \\ {\bar\eta}_{\dot\alpha} \end{pmatrix} \quad .
\end{align}
The $4\times4$ tensor matrices $\sigma^{\mu\nu}$ are constructed from 
the $\gamma$ matrices via
\begin{equation}
\sigma^{\mu\nu} = \frac{i}2 \left[ \gamma^\mu, \gamma^\nu \right] \quad .
\end{equation}
The spinor $\bar\psi$ which is adjoint to $\psi$ is defined as
\begin{equation}
\bar\psi = \psi^\dagger \gamma^0  = 
\begin{pmatrix} \eta^a, & {\bar\xi}_{\dot\alpha} \end{pmatrix} \quad .
\end{equation}
Starting from \eq{appsusy:dirac} the Dirac equation can now also be written 
in a four-component notation
\begin{equation}
\left( \gamma_\mu p^\mu - m \right) \psi \equiv 
\left( \fslash{p} - m \right) \psi = 0 \quad .
\end{equation}

In the following a few contraction identities and traces over $\gamma$ matrices 
are collected which are needed for the calculation of cross sections:
\begin{equation}
\begin{split}
\fslash{a}\fslash{b} &= a \cdot b - i \sigma_{\mu\nu} a^\mu b^\nu \\
\gamma^\mu \gamma_\mu &= 4 \\
\Tr\left({\Id}\right) &= 4 \\
\Tr\left(\gamma^\mu\right) &= 0 \\
\Tr\left(\gamma^5\right) &= 0 \\
\Tr\left(\gamma^\mu\gamma^5\right) &= 0 \\
\Tr\left(\gamma^\mu\gamma^\nu\right) &= 4 g^{\mu\nu} \\
\Tr\left(\gamma^\mu \gamma^\nu \gamma^5 \right) &= 0 \\
\Tr\left(\gamma^\mu\gamma^\nu\gamma^\rho\gamma^\sigma\right) &= 
4\left(g^{\mu\nu}g^{\rho\sigma}-g^{\mu\rho}g^{\nu\sigma}
      +g^{\mu\sigma}g^{\nu\rho}\right) \\
\Tr\left(\gamma^\mu\gamma^\nu\gamma^\rho\gamma^\sigma\gamma^5\right) &=
 -4i\epsilon^{\mu\nu\rho\sigma} \\
\Tr \bigl( \underbrace{\gamma^\mu \gamma^\nu \dots \gamma^\rho 
\gamma^\sigma}_{\text{odd number of $\gamma$'s}} \bigr) &= 0 \quad .
\end{split}
\end{equation}

\section{Grassmann variables}
\label{appsusy:grassmann}

For the definition of the supersymmetric anticommutation relations
in \chap{mssm:superfields} anticommuting numbers, so-called 
Grassmann numbers, were introduced.
The basic relation between two such numbers $\theta$ and $\xi$ is
\begin{equation}
\theta\xi = - \xi\theta \quad ,
\end{equation}
which is equivalent to
\begin{equation}
\left\lbrace \theta,\xi\right\rbrace = 0 \quad .
\end{equation}
The last relation shows the fermionic character of these numbers.
In particular, the square of any Grassmann number is zero.

Grassmann numbers form an Abelian group under the operation of addition.
The multiplication with ordinary complex numbers has the same properties as 
scalar multiplication of a vector, in particular the distributive law holds.

The integral over Grassmann numbers is defined as
\begin{equation}
\int \di\theta \left(  A + B \theta \right) = B \quad ,
\end{equation}
where $A$ and $B$ are complex numbers.
This leads to the expression for differentiation with respect to Grassmann numbers
\begin{equation}
\frac{\partial}{\partial\theta} \left(  A + B \theta \right) = B \quad .
\end{equation}
Complex Grassmann numbers can be built out of real and imaginary
parts in the same way as for ordinary complex numbers. It is convenient to define
the complex conjugation in such a way that the order of a product is reversed, 
as is done in hermitian conjugation of matrices:
\begin{equation}
\left( \theta \xi \right)^* = \xi^* \theta^* = - \theta^* \xi^* \quad .
\end{equation}
In the integral over complex Grassmann numbers $\theta$ and $\theta^*$ are 
treated as independent variables as the real and the imaginary part are 
independent of each other, so
\begin{equation}
\int \di\theta \di\theta^* \ \theta^* \theta = 1 \quad .
\end{equation}

\chapter{Phase-space parametrization}
\label{ps}

In this appendix the parametrization of the phase space
for $2\rightarrow2$ and $2\rightarrow3$ processes,
as it was used for the calculations of this thesis, is presented. 
It is the same parametrization which is also used 
in \FormCalc{}~\cite{Hahn:1998yk,Hahn:2005vh,FormCalcmanual}.
The parametrization is performed in the center-of-mass system of the 
two incoming particles, which define the beam axis and carry a center-of-mass
energy of $\sqrt{s}$.
For each final-state particle an integral over its three-momentum $\vec{k}$
occurs in the calculation of integrated cross sections. The energy $k^0$ of the 
particle is fixed by the on-shell condition 
$k^0 = \sqrt{|\vec{k}|^2 + m^2}$, where $m$ denotes the mass of the particle.
Four of these integrals are eliminated by global energy-momentum conservation.
In the following sections the parametrizations of the two- and three-particle
phase space are shown.

\section{Two-particle phase space}
\label{ps:2}
With two particles in the final state, 
labeled by the subscripts $3$ and $4$ in the following,
the phase-space integral can be written in terms of 
two angles. They are the azimuth angle $\phi$ and the polar angle $\theta$ with respect
to the beam axis. Because of rotational invariance around the beam axis 
the integration over $\phi$ is trivial
and amounts to a factor of $2\pi$. So the integral over the two-particle phase space
is given by
\begin{equation}
\int \di \Gamma_2 = \int_{-1}^{1} \di c_\theta \, \frac1{8\pi} 
  \frac{|\vec{k}_3|}{\sqrt{s}} \quad ,
\end{equation}
where
\begin{equation}
|\vec{k}_{3}|^2 = |\vec{k}_4|^2 = 
  \frac{s^2+m_3^4+m_4^2-2 m_3^2 s -2 m_4^2 s - 2 m_3^2 m_4^2}{4s}
\end{equation}
denotes the squared absolute value of the three-momentum of the final-state particles, 
$m_3$ and $m_4$ are their respective masses, and $\sqrt{s}$ specifies the 
center-of-mass energy of the incoming particles.

\section{Three-particle phase space}
\label{ps:3}

\begin{figure}
\begin{center}
\includegraphics[scale=0.8]{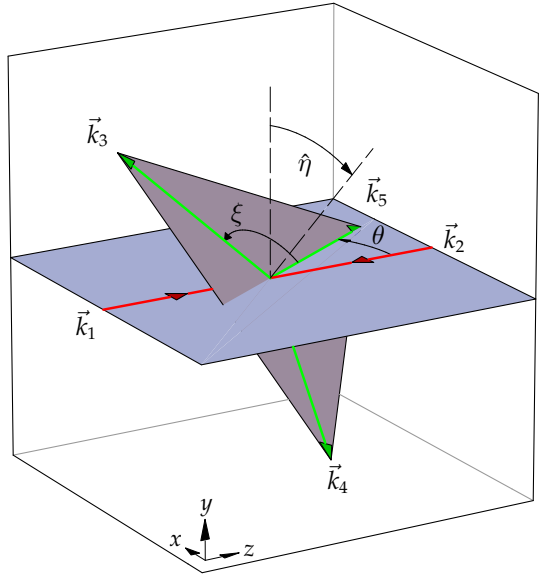}
\caption{Graphical representation of the variables used in the parametrization
  of the $2\rightarrow 3$ phase space. The figure is taken from
  ref.~\protect\cite{FormCalcmanual}.}
\label{ps:2to3}
\end{center}
\end{figure}

For the three-particle phase space, where the outgoing particles are labeled by the 
indices $3$, $4$ and $5$, five independent integration variables remain
after global energy-momentum conservation has been applied. They are the energies
$k_3^0$ and $k_5^0$, the azimuth angle $\phi$ and the polar angle $\theta$ of the
fifth particle with respect to the beam axis, and the angle $\hat\eta$ which rotates
particle $3$ out of the plane defined by particle $5$ and the beam axis. A graphical
representation of the angles is given in \fig{ps:2to3}. 

The four-momenta of the outgoing particles have the following explicit form
\begin{align}
k_3 &= (k_3^0,|\vec{k}_3| \vec{e}_3) &
k_4 &= (\sqrt{s}-k_3^0-k_5^0,-\vec{k}_3 - \vec{k}_5) \nonumber\\
k_5 &= (k_5^0,|\vec{k}_5| \vec{e}_5) \quad , 
\end{align}
with
\begin{align}
\vec{e}_3 &= 
  \begin{pmatrix} c_\theta c_{\hat\eta} s_\xi + s_\theta c_\xi \\ 
  s_{\hat\eta} s_\xi \\
  c_\theta c_\xi - s_\theta c_{\hat\eta} s_\xi \end{pmatrix} & 
\vec{e}_5 &= 
  \begin{pmatrix} s_\theta \\ 0 \\ c_\theta \end{pmatrix} \quad .
\end{align}
The angle $\theta$, which is also plotted in the figure, is defined over
\begin{equation}
c_\theta = 
  \frac{(\sqrt{s}-k_3^0-k_5^0)^2 - m_4^2 - |\vec{k}_3|^2 - |\vec{k}_5|^2 }%
    {2 |\vec{k}_3| |\vec{k}_5|} \quad .
\end{equation}
$m_i$ again denotes the mass of the respective particle $i$ 
and $\sqrt{s}$ is the center-of-mass energy of the initial-state particles.
Due to axial symmetry the trivial integration over $\phi$ can be carried out 
immediately and yields a factor of $2\pi$.

Then the parametrization of the three-particle phase space takes the following form
\begin{align}
\int \di \Gamma_3 &= 
  \int_{m_5}^{(k_5^0)^{\max}} \di k_5^0
  \int_{(k_3^0)^{\min}}^{(k_3^0)^{\max}} \di k_3^0
  \int_{-1}^{1} \di c_\theta
  \int_{0}^{2 pi} \di \hat\eta \,
    \frac1{64 \pi^3} \quad , 
\end{align}
where the integration limits are given by
\begin{equation}
(k_5^0)^{\max} = \frac{\sqrt{s}}2 - \frac{(m_3+m_4)^2 - m_5^2}{2 \sqrt{s}}
\end{equation}
and
\begin{equation}
(k_3^0)^{\max,\min} = \frac1{2\tau} \left[ \sigma (\tau + m_+ m_-) \pm 
  |\vec{k}_5| \sqrt{(\tau - m_+^2)(\tau - m_-^2)} \right] \quad , 
\end{equation}
using
\begin{align}
\sigma &= \sqrt{s} - k_5^0 & \tau &= \sigma^2 - |\vec{k}_5|^2 & 
  m_{\pm} &= m_3 \pm m_4 \quad .
\end{align}

\chapter{Loop Integrals}
\label{loopint}

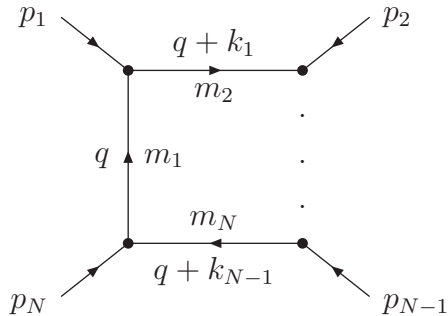
\begin{figure}
%%%
% taken verbatim from T. Hahn, LoopTools-2.1 manual, Sep 16 2005
%%%
\begin{center}
\unitlength=1bp%
\begin{picture}(130,125)(0,0)
\ArrowLine(5,10)(30,30)
\ArrowLine(5,115)(30,95)
\ArrowLine(120,115)(95,95)
\ArrowLine(120,10)(95,30)
\ArrowLine(95,30)(30,30)
\ArrowLine(30,30)(30,95)
\ArrowLine(30,95)(95,95)
\Vertex(30,30){2}
\Vertex(30,95){2}
\Vertex(95,30){2}
\Vertex(95,95){2}
\multiput(95,44)(0,17){3}{\makebox(0,0){$.$}}
\Text(0,115)[r]{$p_1$}
\Text(125,115)[l]{$p_2$}
\Text(125,7)[l]{$p_{N - 1}$}
\Text(0,7)[r]{$p_N$}
\Text(23,62)[r]{$q$}
\Text(62,100)[b]{$q + k_1$}
\Text(62,25)[t]{$q + k_{N - 1}$}
\Text(36,62)[l]{$m_1$}
\Text(62,90)[t]{$m_2$}
\Text(62,35)[b]{$m_N$}
\end{picture}
\end{center}
\caption{General one-loop diagram. The arrows denote the momentum flow.}
\label{loopint:general}
\end{figure}

When calculating quantum corrections to physical processes Feynman
diagrams appear which contain loops. The rules for evaluating Feynman
diagrams state that for every closed loop an integral over the loop
momentum appears in the expression for the amplitude.
In this thesis one-loop corrections are calculated and thus we are concerned only with
one-loop integrals.
The general one-loop integral which corresponds to the general N-point 
one-loop diagram depicted in \fig{loopint:general} 
is given by
\begin{equation}
\label{loopint:generaleq}
T_{\mu_1 \dots \mu_P}^N = \frac{\left( 2\pi\mu\right)^{4-D}}{i\pi^2}
  \int \di^D q \frac{q_{\mu_1} \dots q_{\mu_P}}%
    {\left[ q^2 - m_1^2 \right] \left[ \left( q+k_1 \right)^2 - m_2^2 \right] \dots 
      \left[ \left( q+k_{N-1} \right)^2 - m_N^2 \right] } .
\end{equation}
The notation and conventions used in this chapter correspond to that used 
in ref.~\cite{LoopToolsmanual}.
The loop momentum is denoted by $q$, the momenta $p_i$ are the momenta of 
the external propagators and the momenta $k_i$, which 
appear in the denominator, are related to the former via
\begin{align}
p_1 &= k_1,& p_2 &= k_2 - k_1, & \dots & & p_N &= k_N - k_{N-1} \nonumber\\
k_1 &= p_1,& k_2 &= p_1+p_2,& \dots & & k_N &= p_1 + \dots + p_N .
\end{align}
It is assumed implicitly that all propagators in the denominator have an 
infinitely small positive imaginary part $\epsilon^\prime$ which is set 
to zero only after the integration has been performed. Explicitly this 
is achieved by replacing the masses $m_i^2$ 
with $m_i^2 - i \epsilon^\prime$ everywhere.

The expression given in \eq{loopint:generaleq} is valid for both dimensional
regularization and dimensional reduction (see \chap{renorm:regul}). 
$D = 4 - 2\epsilon$ is the 
dimension of the integral, where $\epsilon$ is a small positive number which will be 
sent to zero at the end of the calculation.
The regularization parameter $\mu$ has the dimension of a mass and is introduced
to keep the dimension of the whole expression fixed when going from 
4 to $D$ dimensions.

Following ref.~\cite{'tHooft:1978xw} the loop integrals are denoted by capital letters
in ascending order, resulting in
\begin{center}
\begin{tabular}{ll}
$T^1=A$ & one-point loop integral, \\
$T^2=B$ & two-point loop integral, \\
$T^3=C$ & three-point loop integral, \\
$T^4=D$ & four-point loop integral, \\
$T^5=E$ & five-point loop integral, \dots .
\end{tabular}
\end{center}
Scalar integrals, which do not have
an index $\mu_i$, are denoted with an index $0$, e.g.\ $A_0$.

The scalar one-point integral $A_0$ can be calculated analytically and reads
\begin{equation}
A_0(m) = m^2\left( \Delta -\ln\frac{m^2}{\mu^2} +1\right) + \Order{\epsilon}
\end{equation}
where
\begin{equation}
\Delta = \frac1{\epsilon} - \gamma_E + \ln 4\pi 
\end{equation}
contains the divergent part of the loop integral with
the Euler-Mascheroni constant
\begin{equation}
\gamma_E \simeq 0.577215664901532\dots \quad .
\end{equation}
This one and expressions for the two-, three- and four-point loop integrals were
first given in ref.~\cite{'tHooft:1978xw} and further improved 
later~\cite{Beenakker:1988jr,Denner:1991qq}. Scalar loop integrals with five and 
more internal propagators can be reduced to four-point 
ones~\cite{Denner:1991kt,Denner:2002ii}.

The tensor integrals can be decomposed into linear combinations of Lorentz-covariant
tensors~\cite{Passarino:1979jh}. They consist of a basis which is formed of 
linearly independent momenta and the metric tensor $g^{\mu\nu}$, and 
coefficients which are Lorentz scalars. This decomposition is not unique. 
Here the momenta are chosen as the momenta $k_i$ appearing in the denominator. 
In this basis the coefficient functions are totally
symmetric in their indices. The decomposition for tensors up to rank four reads
\begin{align}
T^N_{\mu_1} &= \sum_{a=1}^{N-1} {k_a}_{\mu_1} T^N_a \\
T^N_{\mu_1 \mu_2} &= g_{\mu_1 \mu_2} T^N_{00}
 + \sum_{a,b=1}^{N-1} {k_a}_{\mu_1} {k_b}_{\mu_2} T^N_{ab} \\
T^N_{\mu_1 \mu_2 \mu_3} &= 
  \sum_{a=1}^{N-1} \left( g_{\mu_1 \mu_2} {k_a}_{\mu_3} + g_{\mu_2 \mu_3}
     {k_a}_{\mu_1} + g_{\mu_3 \mu_1} {k_a}_{\mu_2} \right) T^N_{00a} \nonumber\\
 &\phantom{=}+ \sum_{a,b,c=1}^{N-1} {k_a}_{\mu_1} {k_b}_{\mu_2}
    {k_c}_{\mu_3} T^N_{abc} \\
T^N_{\mu_1 \mu_2 \mu_3 \mu_4} &= 
  \left( g_{\mu_1 \mu_2} g_{\mu_3 \mu_4} + g_{\mu_1 \mu_3} g_{\mu_2 \mu_4}
  + g_{\mu_1 \mu_4} g_{\mu_2 \mu_3} \right)  T^N_{0000} \nonumber\\
 &\phantom{=}+ \sum_{a,b=1}^{N-1} \left( g_{\mu_1 \mu_2} {k_a}_{\mu_3} 
    {k_b}_{\mu_4} 
  + g_{\mu_1 \mu_3} {k_a}_{\mu_2} {k_b}_{\mu_4} 
  + g_{\mu_1 \mu_4} {k_a}_{\mu_2} {k_b}_{\mu_3} \right. \nonumber\\
  &\phantom{=}\qquad\quad\left.  + g_{\mu_2 \mu_3} {k_a}_{\mu_1} {k_b}_{\mu_4} 
  + g_{\mu_2 \mu_4} {k_a}_{\mu_1} {k_b}_{\mu_3} 
  + g_{\mu_3 \mu_4} {k_a}_{\mu_1} {k_b}_{\mu_2} \right) T^N_{00ab} \nonumber\\
 &\phantom{=}+ \sum_{a,b,c,d=1}^{N-1} {k_a}_{\mu_1} {k_b}_{\mu_2}
  {k_c}_{\mu_3} {k_d}_{\mu_4} T^N_{abcd} \quad .
\end{align}

Furthermore the coefficient functions of the tensorial loop integrals can be 
written as functions of the scalar integrals~\cite{Passarino:1979jh}.
This is known as Passarino-Veltman reduction scheme.
A complete set of equations for reducing loop integrals up to point rank $N=4$
and up to tensor dimension $N+1$ can be found for example in ref.~\cite{Denner:2002ii}.
These are all loop integrals which can appear in processes with up to four 
external legs.

In the reduction the inverse of the Gram matrix occurs. The Gram matrix $Z$
is a matrix which is built up from the momenta $k_i$ by $Z_{ij}=2 k_i k_j$.
This matrix can become singular. For up to four-point loop integrals this happens only 
at the borders of phase space. Care has to be taken when calculating phase 
space points close to the edges, as the computation can become numerically unstable.
A technique to improve stability will be presented in \app{numerics:gaussian}.

For loop integrals with five and more internal propagators the Gram matrix 
can become singular also at points inside the phase space. However the 
loop momenta are four-component Lorentz vectors, so a linear independent
combination of four of them spans the whole Minkowski space.
This allows one to eliminate the inverse Gram matrix. The reduction is done 
in such a way that $N$-point loop integrals are reduced to a combination of 
$(N-1)$-point loop integrals with the tensor rank increased by 
one~\cite{Denner:2002ii}. Recently a slightly different scheme has been found
which even reduces the tensor rank by one in the 
decomposition~\cite{Binoth:2005ff}. 

For phase space points close to those where the Gram matrix becomes singular
also expansions around vanishing determinant of the Gram matrix or 
methods which use numerical integration can be applied.
An overview of possible techniques with explicit formulae for loop functions
with point rank up to six was given in ref.~\cite{Denner:2005nn}.

For the numerical evaluation of the loop integrals the \program{LoopTools}
package~\cite{Hahn:1998yk,Hahn:2006MR}, which is based on 
\program{FF}~\cite{vanOldenborgh:1990yc}, was used.
In this package the stability of calculating the Passarino-Veltman
reduction was improved numerically with the method
of Gaussian elimination, which will be described in \app{numerics:gaussian}.
Additionally the five-point functions up to tensor rank four were
implemented based on ref.~\cite{Denner:2002ii} and the scalar 
four-point function amended according to ref.~\cite{Denner:1991qq} so it is
valid for all cases.
The numerical results of the code were compared to those of an independent
code from Dittmaier~\cite{Dittmaier:priv} and very good agreement could be found.
Moreover a Passarino-Veltman reduction of
the five-point tensor integrals was implemented.
Also here a comparison yielded excellent agreement except for points very 
close to the edges of phase space where the decomposition algorithm is 
known to become numerically unstable.

The explicit formulae for this decomposition are given below.
To shorten the notation some abbreviations are introduced:
\begin{align}
\bar\delta_{ij} =& 1 - \delta_{ij} = 
\begin{cases}
0 & \text{for $i=j$}\\
1 & \text{for $i\neq j$}
\end{cases}, &
i_j =& 
\begin{cases}
i & \text{for $i<j$} \\
i-1 & \text{for $i>j$}
\end{cases} \quad .
\end{align}
A number in brackets behind the loop integral denotes that the 
term in the denominator with a mass with this index is left out 
from the integrand, e.g. the normal scalar four-point integral
\begin{align}
D_0 &= \frac{\left( 2\pi\mu\right)^{4-D}}{i\pi^2} \cdot \nonumber\\
  & 
  \int \di^D q \frac{1}%
    {\left[ q^2 - m_1^2 \right] 
      \left[ \left( q+k_1 \right)^2 - m_2^2 \right] 
      \left[ \left( q+k_2 \right)^2 - m_3^2 \right]  
      \left[ \left( q+k_3 \right)^2 - m_4^2 \right] } \quad , \\
\intertext{where the second propagator is left out, becomes}
D_0\left(2\right) &= \frac{\left( 2\pi\mu\right)^{4-D}}{i\pi^2}
  \int \di^D q \frac{1}%
    {\left[ q^2 - m_1^2 \right] \left[ \left( q+k_2 \right)^2 - m_3^2 \right]  
      \left[ \left( q+k_3 \right)^2 - m_4^2 \right] } \quad .
\end{align}
The resulting integral is a loop integral with the point-rank
reduced by one, as one can see in the example above, 
which is a three-point loop integral. For integrals where the first
propagator is eliminated the integration momentum must be 
shifted by $q \rightarrow q-k_2$, so that the standard 
form~\eq{loopint:generaleq} is again obtained.

With this the tensorial coefficients of the five-point loop integrals are:
\begin{align}
E_i =& \sum_{n=1}^4 \left(Z^{(4)}\right)^{-1}_{in} S^1_{n} 
  \label{loopint:e1}\\
E_{ij} =& \sum_{n=1}^4 \left(Z^{(4)}\right)^{-1}_{in} S^2_{nj} 
  \label{loopint:e2}\\
E_{ijk} =& \sum_{n=1}^4 \left(Z^{(4)}\right)^{-1}_{in} S^3_{njk} 
  \label{loopint:e3}\\
E_{ijkl} =& \sum_{n=1}^4 \left(Z^{(4)}\right)^{-1}_{in} S^4_{njkl} 
  \label{loopint:e4}\quad , 
\end{align}
where $i,j,k,l=1,\dots,4$ and with
\begin{align}
f_n =& k_n^2 - m_{n+1}^2 + m_1^2 \\
S^1_{n} =& D_0\left( n+1\right) - D_0\left( 1\right) - f_n E_0 \\
S^2_{nj} =& D_{j_n}\left( n+1\right) \bar\delta_{nj}- D_j\left( 1\right) - f_n E_j \\
S^3_{njk} =& D_{{j_n}{k_n}}\left( n+1\right) \bar\delta_{nj}\bar\delta_{nk}
  - D_{jk}\left( 1\right) - f_n E_{jk} \nonumber\\
  &+ 2 \left(Z^{(4)}\right)^{-1}_{jk} 
    \left( D_{00}\left( n+1\right) - D_{00}\left( 1\right) \right) \\
S^4_{njkl} =& D_{{j_n}{k_n}{l_n}}\left( n+1\right) 
    \bar\delta_{nj}\bar\delta_{nk}\bar\delta_{nl}
  - D_{jkl}\left( 1\right) - f_n E_{jkl} \nonumber\\
  &+ 2 \left(Z^{(4)}\right)^{-1}_{jk} 
    \left( D_{00{l_n}}\left( n+1\right)\bar\delta_{nl} - D_{00l}\left( 1\right) \right)
    \nonumber\\
  &+ 2 \left(Z^{(4)}\right)^{-1}_{kl} 
    \left( D_{00{j_n}}\left( n+1\right)\bar\delta_{nj} - D_{00j}\left( 1\right) \right)
    \nonumber\\
  &+ 2 \left(Z^{(4)}\right)^{-1}_{lj} 
    \left( D_{00{k_n}}\left( n+1\right)\bar\delta_{nk} - D_{00k}\left( 1\right) \right)
\end{align}
and the Gram matrix
\begin{equation}
Z^{(4)} = 2
  \begin{pmatrix}
    k_1 k_1 & k_1 k_2 & k_1 k_3 & k_1 k_4 \\
    k_1 k_2 & k_2 k_2 & k_2 k_3 & k_2 k_4 \\
    k_1 k_3 & k_2 k_3 & k_3 k_3 & k_3 k_4 \\
    k_1 k_4 & k_2 k_4 & k_3 k_4 & k_4 k_4
  \end{pmatrix} .
\end{equation}

All coefficients which are multiplied by the metric tensor, i.e.\ have a 
$00$ in the index, vanish identically
\begin{equation}
E_{00} = E_{00i}=E_{0000}=E_{00ij}=0 .
\label{loopint:pvzero}
\end{equation}

The decomposition into coefficients multiplied by the metric tensor and 
such multiplied by the momenta is not unique as the four linearly
independent momenta $k_i$ span the whole Minkowski space and 
are related to the metric tensor by
\begin{equation}
g^{\mu\nu} = \sum_{i,j=1}^4 2 \left(Z^{(4)}\right)^{-1}_{ij} k_i^\mu k_j^\nu .
\label{loopint:fiveptmetrictensor}
\end{equation}

The formulae given in ref.~\cite{Denner:2002ii} use a different decomposition in order
to avoid inverse Gram matrices. This leads to a non-vanishing $E_{00}$ which 
is compensated by different $E_{ij}$ and the same happens for higher-tensor ranks.

For a numerical comparison the two expressions must be transformed into
each other by exploiting the relation \eq{loopint:fiveptmetrictensor}:
\begin{align}
E_{ij}^{PV} =& E_{ij}^{D} + 2 \left(Z^{(4)}\right)^{-1}_{ij} E_{00}^{D} \\
E_{ijk}^{PV} =& E_{ijk}^{D} 
  + 2 \left(Z^{(4)}\right)^{-1}_{ij} E_{00k}^{D} 
  + 2 \left(Z^{(4)}\right)^{-1}_{jk} E_{00i}^{D} 
  + 2 \left(Z^{(4)}\right)^{-1}_{ki} E_{00j}^{D}\\
E_{ijkl}^{PV} =& E_{ijkl}^{D} 
  + 2 \left(Z^{(4)}\right)^{-1}_{ij} E_{00kl}^{D}
  + 2 \left(Z^{(4)}\right)^{-1}_{ik} E_{00jl}^{D}
  + 2 \left(Z^{(4)}\right)^{-1}_{il} E_{00jk}^{D}\nonumber\\
 &+ 2 \left(Z^{(4)}\right)^{-1}_{jk} E_{00il}^{D}
  + 2 \left(Z^{(4)}\right)^{-1}_{jl} E_{00ik}^{D}
  + 2 \left(Z^{(4)}\right)^{-1}_{kl} E_{00ij}^{D}\nonumber\\
 &+ 4 \left( \left(Z^{(4)}\right)^{-1}_{ij} \left(Z^{(4)}\right)^{-1}_{kl}
  + \left(Z^{(4)}\right)^{-1}_{ik} \left(Z^{(4)}\right)^{-1}_{jl}
  + \left(Z^{(4)}\right)^{-1}_{il} \left(Z^{(4)}\right)^{-1}_{jk}
  \right) E_{0000}^{D} \quad .
\end{align}
In the equations above the expressions from~\cite{Denner:2002ii} are denoted by a 
superscript $D$ and the ones from the Passarino-Veltman reduction by a 
superscript $PV$. 

\chapter{Numerical Methods}
\label{numerics}

The numerical calculation of cross sections is only possible
with the aid of computer programs. Computers can do
floating point operations only with a finite precision so
rounding errors occur inevitably in many steps of the program.
Expressions which are still valid analytically might give a 
numerical result which is utter nonsense. Therefore not only the 
analytical correctness must be checked when implementing
algorithms, but also that the code is numerically stable.

\section{Gaussian Elimination}
\label{numerics:gaussian}

When calculating one-loop integrals, Gram matrices occur which contain
scalar products of the momenta, as was shown in \chap{loopint} .
For loop integrals with up to four external legs the inverse of the Gram matrix
has to be calculated. This matrix can become singular at the edges
of phase space, i.e.\ for forward scattering or at the production threshold.
Already close to the edge naive matrix inversion can become unstable
as will be shown below.

%%% Problem: A.x=b
As is often the case in problems where the inverse of a matrix appears
in the analytical expression the inverse of the Gram matrix itself 
actually is not needed in the Passarino-Veltman reduction scheme.
This can be seen easily for example from the expression for the five-point loop
integrals given in \chap{loopint}. Left-multiplying 
eqs.~(\ref{loopint:e1})-(\ref{loopint:e4}) with the Gram matrix $Z^{(4)}$ 
reduces the problem  to the problem of solving a system of linear 
equations of the form
\begin{align}
A x &= b,
\label{lineqn}
\end{align}
where $A$ is an $n\times n$ matrix and $x$ and $b$ are vectors of dimension $n$. 
$A$ and $b$ are input parameters and $x$ is the solution vector of the system
of equations.
We are going to consider the most general case here, that all components
of the matrix and the vectors can be complex numbers and $b$ is a 
generic vector. The physical case of real matrices $A$ then follows 
directly from this as a special case.

%%% Formeller Weg x=A^{-1}.b instabil
Let us first consider the analytic way of multiplying the input vector
with the inverse matrix. Then \eq{lineqn} can be written as a function $f$
\begin{align}
f:\quad x &= A^{-1} b \quad .
\label{lineqninv}
\end{align}
Calculating the solution vector $x$ in this manner however leads
to numerical instabilities. Let us first decompose $f$ into the two
partial steps
\begin{alignat}{2}
g:\quad&& A &\mapsto A^{-1} \\
h:\quad&& A^{-1} &\mapsto x
\end{alignat}
and assume the ideal case that both partial steps can be calculated in 
a numerically stable way.
Then a stability analysis~\cite{Bornemann:2004} yields,
that the error on $x$ is proportional to $\kappa(A)$, the condition of $A$.
The condition is defined as
\begin{align}
\kappa(A) &= ||A|| \cdot ||A^{-1}|| 
 = \left( \min\left\{ \frac{||\Delta A||}{||A||}: 
            A+\Delta A \text{ singular} \right\} \right)^{-1},
\end{align}
where $||\cdot||$ denotes a matrix norm.
In a geometrical interpretation this is the distance of $A$ to a 
singular matrix for which \eq{lineqn} has no longer a
unique solution.

The numerical evaluation of mathematical expressions was done
in \texttt{double precision} in this thesis, i.e.\ floating point arithmetics
with double precision as defined in ref.~\cite{IEEE754}. 
These numbers offer about sixteen valid digits, so that for a 
condition $\kappa(A)\ge 10^{16}$ 
the matrix cannot be distinguished any longer numerically from a 
singular matrix. One also says that the matrix is numerically singular.
When calculating one-loop integrals this case occurs at the edges
of phase space.

Even earlier the error $\Delta x$ of the solution vector $x$ increases,
\begin{align}
\frac{||\Delta x||}{||x||} \approx \kappa(A) \cdot 10^{-16},
\end{align}
and the result becomes inaccurate. Such a behavior could 
indeed be observed while doing the numerical evaluation 
of the vector boson fusion processes~(\chap{vbf}).

%%% LR-Zerlegung
To avoid this problem, one decomposes the matrix $A$ into a
unipotent, i.e. whose diagonal contains 1, lower triangular matrix $L$
and a non-singular upper triangular matrix $R$
\begin{align}
A &= L R = 
\begin{pmatrix}
1&0&0&\dots&0 \\
*&1&0&\dots&0 \\
*&*&1&\ddots&0 \\
\vdots&\vdots&\ddots&\ddots&0\\
*&*&\dots&*&1
\end{pmatrix}
\begin{pmatrix}
*&*&*&\dots&* \\
0&*&*&\dots&* \\
0&0&*&\dots&* \\
\vdots&\vdots&\ddots&\ddots&\vdots\\
0&0&\dots&0&*
\end{pmatrix}.
\end{align}
$*$ here is a place holder for an arbitrary complex number.
The algorithm partitions the matrices such that
\begin{align}
\begin{pmatrix}
\alpha&u^T\\v&A_*
\end{pmatrix}
&=
\begin{pmatrix}
1&0\\w&L_*
\end{pmatrix}
\begin{pmatrix}
\alpha&u^T\\0&R_*
\end{pmatrix}
\end{align}
where
\begin{align}
\alpha &\in \Cmplx;& u,v,w &\in \Cmplx^{n-1};& A_*, L_*, R_* &\in
\Cmplx^{(n-1)\times(n-1)}
\end{align}
so
\begin{align}
w &= \frac{v}{\alpha} \\
L_* R_* &= A_* - w u^T .
\end{align}
$L_*$ and $R_*$ are again a unipotent lower and non-singular upper 
triangular matrix, respectively. In the next step of the iteration 
the matrix $L_* R_*$ has to be partitioned in this way. After applying this 
procedure recursively one obtains the desired decomposition. 

%%% geht nicht immer => (Spalten-)Pivotisierung
If $\alpha$ happens to be zero, the algorithm breaks down.
Therefore, not $A$ is decomposed, but $P A$, where $P$ 
is an $n\times n$ permutation matrix and chosen in such a way,
that in every iteration step the first row and the row whose
first column contains the largest element by absolute value
are swapped. This method is called partial pivoting.
It can be shown that with partial pivoting every non-singular matrix $A$
can be decomposed into $L$ and $R$ and the decomposition
is unique.
%%% Skalenfaktor ???
%%% Vorwärts-/Rückwärtssubstitution
The system of linear equations now has the form
\begin{align}
L R x &= P^{-1} b .
\end{align}
To solve it one first solves the system
\begin{align}
L y &= P^{-1} b
\end{align}
with an auxiliary vector $y$. Because of the triangular structure
of $L$ this can be done very easily by a recursive forward substitution
\begin{align}
y_i &= (P^{-1}b)_i - \sum_{j=1}^{i-1} y_j   & i=1\rightarrow n \quad .
\end{align}
The arrow denotes the order in which the $y_i$ must be calculated.
Finally,
\begin{align}
R x &= y
\end{align}
is solved by backward substitution and one obtains
the solution vector $x$
\begin{align}
x_i &= y_i - \sum_{j=i+1}^{n} x_j   & i=n\rightarrow 1 \quad .
\end{align}
The algorithm presented here is known in the literature by the
name of Gaussian elimination with partial 
pivoting~\cite{Bornemann:2004,numerik:dh}.

%%% im wesentlichen stabil (abhängig von Konditionszahl)
An error analysis yields that the error on $x$ is determined by
\begin{align}
\kappa &= \frac{||L||\cdot||R||}{||P A||} \\
       &\le ||L|| \cdot ||L^{-1}|| = \kappa(L)
\end{align}
If one chooses the norm as the maximum norm, the inequality
$|L_{ij}| \le 1$  holds because $L$ is unipotent and 
partial pivoting was used. Therefore
\begin{align}
||L||_\infty &\le n \\
||L^{-1}||_\infty &\le 2^{n-1} \quad\cite{wilkinson:321076},
\intertext{so}
\kappa &\le n \cdot 2^{n-1} \quad .
\end{align}
When calculating loop integrals only matrices of dimension $n\le4$ occur, 
so the calculation of $x$ in this manner is absolutely stable.

\chapter{Manual of the \texorpdfstring{\HadCalc{}}{HadCalc} Program}
\label{hadcalc}

For the calculation of hadronic cross sections a computer code, called \HadCalc{}, 
was written (see \chap{hadWQ:hadcalc}). In this appendix
the manual of the program is presented.

\section{Prerequisites and Compilation}
\subsection{Prerequisites}

The following programs are required for compiling and running HadCalc
and must be installed:
\begin{itemize}
\item a Fortran compiler compliant with the Fortran77 standard,
\item a C compiler conforming to ANSI-C,
\item GNU make,
\item FormCalc 4 \cite{Hahn:1998yk},
\item one of the two following packages that include sets of
      parton distribution functions from various groups
  \begin{itemize}
  \item PDFLIB (CERN Computer Program Library entry W5051) \cite{PDFlibmanual}, or
  \item LHAPDF \cite{LHAPDFmanual}.
  \end{itemize} 
\end{itemize}
Additionally, support for the following two programs is integrated into
\HadCalc
\begin{itemize}
 \item FeynHiggs 2.1beta or newer \cite{Heinemeyer:1998yj},
 \item Condor workload management system for compute-intensive jobs.
\end{itemize}

\subsection{Configuration and Compilation}
First the partonic subprocess must be generated and prepared 
by following the instructions in the
FormCalc4 manual. Especially the definitions in process.h have to be
updated correctly as \HadCalc{} relies on those. It is not necessary to fill
in correct MSSM parameters or tune integration parameters, however.

Then the distribution file \texttt{HadCalc-0.5.tar.gz} should be unpacked.
As next step change into its
subdirectory and run \command{configure} from there. The following configure
options are mandatory:\\*
\begin{tabular}{lp{0.5\textwidth}}
-{}-with-partonprocess=DIR & This is the location of the \FormCalc{}-generated
partonic output. \\
-{}-with-processtype=$mn$ & By this option the processtype is fixed, 
specified by the number of
incoming particles $m$ and the number of outgoing particles $n$. Note
that $m$ and $n$ form a single number, i.e. for a $2\rightarrow2$
process one would write -{}-with-processtype=22. Currently, $2\rightarrow1$,
$2\rightarrow2$ and $2\rightarrow3$ is implemented and can be entered here.\\
-{}-with-parton1=$i$ & The type of the first parton is specified by $i$, 
given as the PDG flavor code~\cite{Eidelman:2004wy} 
(see table~\ref{PDGflavourcode}). \\
-{}-with-parton2=$i$ & Similarly, this is the PDG flavor code for the
second parton.
\end{tabular}
\begin{table}
\begin{tabular}{ll}
PDG flavor code & Particle \\\hline
0 & gluon $g$ \\
1 & down quark $d$ \\
2 & up quark $u$ \\
3 & strange quark $s$ \\
4 & charm quark $c$ \\
5 & bottom quark $b$ \\
6 & top quark $t$ \\
-1 & down anti-quark $\bar{d}$ \\
-2 & up anti-quark $\bar{u}$ \\
-3 & strange anti-quark $\bar{s}$ \\
-4 & charm anti-quark $\bar{c}$ \\
-5 & bottom anti-quark $\bar{b}$ \\
-6 & top anti-quark $\bar{t}$ 
\end{tabular}
\caption{PDG flavor codes}
\label{PDGflavourcode}
\end{table}

\pagebreak %%% XXX LAYOUT XXX %%%
\noindent Additionally the following options are recognized by configure and 
enable optional features:\\*
\begin{tabular}{lp{0.5\textwidth}}
-{}-enable-antiproton1 & Hadron 1 is an anti-proton instead of a proton. \\
-{}-enable-antiproton2 & Hadron 2 is an anti-proton instead of a proton. \\
-{}-with-condor[=DIR] & Link the final code with the Condor workload management
system libraries. If the binary is not in the standard search path of your
shell, its location can be specified with the optional DIR argument. \\
-{}-with-feynhiggs[=DIR] & Link the final code with the FeynHiggs library.
This is mandatory if the \FormCalc{} option to compute the MSSM-Higgs sector
via FeynHiggs is chosen. The optional DIR
specifies the location of the FeynHiggs library libFH.a, if it is not in
the standard search path of the compiler. \\
-{}-with-looptools=DIR & If LoopTools is not in the standard search path
of the compiler, its location can be specified here.\\
-{}-with-lhapdf[=DIR] & Use LHAPDF for the parton distribution functions.
If the LHAPDF library is not in the standard search path, its 
location can be given by the optional DIR argument. The PDF data is assumed to
be found at the same place.\\
-{}-with-pdflib[=DIR] & Use PDFlib for the parton distribution functions.
If the PDFlib library is not in the standard search path
and the CERNlib environment variables \$CERN and
\$CERN\_LEVEL are not set, the DIR argument designates where it can
be found.
\end{tabular}

\noindent Only one of the last two options can be given on the command line.  If neither
-{}-with-lhapdf nor -{}-with-pdflib was given, configure first tries to find
LHAPDF and, if this fails, probes the existence of PDFlib.

After having run \command{configure}, a call to \command{make} compiles the
program.  When it successfully finishes, a binary called \command{HadCalc} has been
created in the current path.

\section{Running the program}
\label{hadcalc:running}
The program is simply started by running \command{./HadCalc}. 
It will then present
a menu which allows one to tune various settings and start the calculation
of cross sections. The following subsections describe the possible
settings in detail.
An item is chosen by typing the number shown in brackets before the
item and pressing ``Enter''. In every menu ``(0)'' exits the submenu or, for
the top level menu, quits the program.
Invalid input is ignored and an error message is written
on the screen.

\subsection{Physics parameters}
This submenu sets the parameters of the MSSM and related things and is
divided in three further submenus.

\subsubsection{MSSM parameters}
\label{MSSMparameters}

Here all values which correspond to parameters of the MSSM are set.

First let us look at menu item 16. This decides whether the program
should use a common mass $M_{SUSY}$ in the sfermion sector, or if
individual values for the left-handed squarks and sleptons and the
right-handed sups, sdowns and selectrons are allowed. 
Depending on this flag either the MS* variables cannot be set (because
they are fixed at $M_{SUSY}$) or $M_{SUSY}$ itself cannot be set
(because it is irrevelant and not used in the computation). When
choosing a common SUSY mass scale, the settings in the MS* variables
are retained and restored when deselecting this option. 

All other parameter settings can be in two states. They can either have a
fixed setting, then this value is used for all calculations. Or their
value can be running. In this case a lower and upper bound and the number of
 intermediate intervals must be chosen. Then the computation of the cross
section is done (``intermediate intervals'' + 1) times, with the value of the
parameter increasing from lower bound to upper bound\footnote{Despite its name,
the lower bound can be mathematically larger than the upper bound; then the
value of the parameter is decreasing.}. The distance between two values is
equal for the setting ``linear'' and exponentially increasing for the setting
``logarithmic'', i.e. the values are closer at the lower bound and they 
have equal distance again when plotting them on a logarithmic scale.
A behavior vice versa with values closer at the upper bound can be easily
achieved by exchanging lower and upper bound. If more than one
parameter is chosen to be running, the iteration loops are nested, with the
first parameter varying fastest.

\subsubsection{Kinematic parameters}

In this menu all parameters are set which are related to kinematic
variables of the process.

The underlying parameter of items 3 and 4 depends on the type of process. 
For processes with two particles in the final state, this is the angle
$\theta$ between the two outgoing particles, for those with three
final particles, it denotes the energy $k_5^0$ of the fifth particle, 
which is the third final-state particle. 
The menu items 8, 12, 14 and 15, which refer to the fifth particle, are
ignored for $2\rightarrow2$ processes and cannot be changed.

The settings of the parameters are possible in the same way as already
described in the previous item.

\subsubsection{Scale parameters}

This menu sets the renormalization and factorization scale of
the process in the same way as described above. 
A negative number for the renormalization scale has a special
meaning. Then the sum of the masses of the final-state particles
is taken, multiplied with the absolute value of the setting, 
and this number is taken as the renormalization scale.
Additionally it can be
chosen that both renormalization and factorization scale are 
always set to the same value.

\subsubsection{Show ModelDigest (FormCalc)}

Finally this choice invokes \FormCalc's ModelDigest function, which takes the
parameters as an input and calculates the physical masses of the
particles. Thereby it applies lower bounds on the masses established by
experiment and refuses the calculation if these bounds are violated. The
calculated cross section will also be zero in that case. The
\FormCalc{} manual contains a more detailed explanation of this function.
There it is also described how the check for the violation of experimental
bounds can be switched off by flipping a switch in FormCalc's process.h.

\subsection{PDF parameters}
The set used for the calculation of the parton distribution functions is chosen 
by this submenu. The layout and choices presented depends on whether
LHAPDF or PDFlib is used. For PDFlib three numbers must be entered. 
The first denotes the type of parton distribution functions and is  
$1$ for proton PDFs. The second number specifies the respective
group which has performed the fit to the experimental data and the third
number chooses a specific PDF set. When using LHAPDF a string must be 
entered that directly specifies the filename of the PDF set in the LHAPDF
subdirectory.

\subsection{Integration parameters}
This submenu chooses the integration routine and sets its parameters.
Currently there are six integration routines available:\\[\baselineskip]
\begin{tabular}{ll}
  GAUSSKR & One-dimensional Gauss-Kronrod algorithm \\
  GAUSSAD & One-dimensional adaptive Gauss algorithm \\
  DCUHRE & Multi-dimensional adaptive Gauss algorithm \\
  VEGAS & Monte Carlo integration algorithm \\
  SUAVE & Subregion adaptive Monte Carlo integration algorithm \\
  DIVONNE & Monte Carlo integration via stratified sampling \\
\end{tabular}\\[\baselineskip]
The last four algorithms are part of the CUBA 
library~\cite{Hahn:2004fe}. In the following only a short description of
the possible parameter settings is given. The technical details of these algorithms 
and the precise impact of the variables are described in the CUBA manual 
and shall not be repeated here.

The GAUSSKR and GAUSSAD algorithms can only handle one-dimensional
integrals.  If multi-dimensional integrals are attempted to be
computed, VEGAS is used as a fallback. In contrast the DCUHRE and DIVONNE
algorithms cannot handle one-dimensional integrals. There the GAUSSKR
algorithm is used instead. In both cases a warning is printed on the
screen.

All integration routines share these two variables:
\begin{itemize}
\item relative error: the desired relative error
\item absolute error: the desired absolute error
\end{itemize}
Additionally, the following variables are available for one or more of
the routines. Which ones these are is denoted in square brackets
after the entry.
\begin{itemize}
\item maximum \# of points: the maximum number of function evaluations
                            used [GAUSSAD, DCUHRE, VEGAS, SUAVE, DIVONNE]
\item \# of points for starting: the initial number of points per
                                 iteration [VEGAS]
\item increase in \# of points: the number of points the previous number
                                is incremented for the next iteration
                                [VEGAS]
\item \# of points for subdivision: the number of points used to 
                                    sample a subdivision [SUAVE]
\item flatness \# for splits: the type of norm used to compute the 
                              fluctuation of a sample [SUAVE]
\item \# of passes: the number of passes after which the partitioning
                    phase terminates [DIVONNE]
\item key 1: determination of sampling in the partitioning phase
             [DIVONNE]
\item key 2: determination of sampling in the final integration phase
             [DIVONNE]
\item key 3: sets the strategy for the refinement phase [DIVONNE]
\item maximum $\chi^2$ for subregion: the maximum $\chi^2$ value a single
		    subregion is allowed to have in the
                                    final integration phase [DIVONNE]
\item minimum deviation for split: a bound which determines whether it
                                   is worthwhile to further examine a
		   region that failed the $\chi^2$ test [DIVONNE]
\end{itemize}

\subsection{Amplitude switches}

This submenu sets the type of diagrams used for the computation and how the
cuts should be applied. The value of the cuts is set in the parameter
section and was already described there.

The first entry decides whether the tree-level and the one-loop result
shall be computed in one go or only one of them. 
Possible choices are ``Tree only'', ``Tree+Loop'' and ``Loop only''.
Which way is better depends on the concrete circumstances and features
of the problem. Computing both at the same time might save computation
time, but the integration routine has to optimize its choices for both
at the same time, which might lead to sub-optimal performance. On the
other hand it is not too likely that there are problematic regions in
the tree-level cross section which are no longer present in the one-loop
computation, so normally this procedure gives satisfactory results.
If only one cross section is computed, the value of the other one is set
to zero.

%%% XXX: Commented out in the source code - gives compilation error
%%% if diagqcd isn't defined in the partonic process
%The second entry illustrates how user-defined types can be introduced
%into the program. In this case all diagrams in the partonic cross
%section have a factor of diagqcd or diagnoqcd, depending on whether the
%corresponding diagram belongs to the QCD corrections or to the
%electroweak ones. With this switch it is possible to choose for example
%only QCD diagrams.

The remaining entries decide if and how the cuts on rapidity, transverse
momentum and jet separation should be applied. It is either possible to have
the particle, or a pair of particles in case of the jet separation, fulfill a
cut, violate it or ignore the cut altogether. Since \HadCalc{} relies on
\FormCalc{} for the partonic process and implementation details, for the cuts
for particle three in the $2\rightarrow 2$ case and particle five in the
$2\rightarrow 3$ case it cannot be chosen that the rapidity and transverse
momentum cut is violated, but they always have to be fulfilled. They can,
however, be switched off by setting the relevant entry in the parameters
section to zero.

\subsection{Input/Output options}

This submenu allows one to read in a set of parameters from a file and
specify where and how to write the calculation output.

To read in a set of parameters a parameter specification must have been written
into a file and this filename then has to be entered here.
All possible variables which can be set in such a file are given in
section~\ref{hadcalc:inputtoken}.
There are three basic types of variables. Those which specify a 
parameter can either take four comma-separated values that are
the lower and upper bound, the behavior with respect to increments, i.e.\ linear
or logarithmic, and the number of intermediate intervals, or a single number
denoting its fixed value. The ones of type \var{boolean} turn on a certain 
switch and take no arguments. All remaining ones take a single argument
and the variable is assigned to this parameter.

In the following also a formal definition of the parsing rules is given:
\begin{itemize}
\item The file is read line by line.
\item White space at the beginning of a line is ignored.
\item Empty lines are ignored.
\item Lines starting with the character ``\#'' (after optional white space)
      are comment lines and ignored.
\item The first token which is separated by white space from the rest of
      the line is extracted. This token has to be a token from the list 
      of valid tokens in section~\ref{hadcalc:inputtoken}.
\item If the token type is boolean, its associated parameter is set.
\item If the token type is integer, an attempt to read an integer value is made
          and if it succeeds, this is assigned to the associated parameter.
\item If the token type is double, an attempt to read a double precision 
          floating-point number is made
          and if it succeeds, this is assigned to the associated parameter.
\item If the token type is string, the second token is assigned to the 
          associated parameter.
\item If the token type is parameter, the following actions happen:
\begin{itemize}
\item An attempt to read four comma-separated double precision
      floating-point numbers is made. 
\item If this attempt succeeds, the four numbers are assigned to 
      lower bound, upper bound, log and number of intermediate
      intervals of the
      parameter. log means linear increase if this variable is zero and
      exponential one otherwise.
\item If this does not succeed, an attempt to read a single double
      precision floating point number is made.
\item If this succeeds, this number is the constant value of the
      parameter.
\item If this also does not succeed, the line is flagged as not
      parsable.
\end{itemize}
\item For lines not parsable by the rules above a warning message is
      printed and their content is ignored.
\end{itemize}

Furthermore some integration routines offer the possibility to write
intermediate results or progress report to the screen. This is turned on
with \var{Verbose integration output}. For hadronic cross sections this
also enables writing PDFlib statistics on the screen at the end of the
calculation.

Finally one can choose whether the calculation results will be
written to the screen or into a file. In the latter case the variable
\var{outputstring} describes which elements should be written to the
output file. The form of this variable together with the valid tokens is
described in section~\ref{hadcalc:outputstring}. The output-file format starts
with a ``\#''-quoted header with a file identification and the content of
\var{outputstring}. Then, each on a line by itself, for every scanned
parameter point the values defined in \var{outputstring} are written,
separated from each other by a space.

\subsection{Amplitude calculation}

This submenu finally allows one to choose the cross section one wants to
compute and does the calculation. During the following integration 
the process may be interrupted with ``Ctrl-C'', after which it aborts the current
calculation and jumps back into the main menu. Due to restrictions imposed by
Condor this feature is not available if \HadCalc{} was configured with the
option -{}-with-condor. Here pressing ``Ctrl-C'' aborts the whole \HadCalc{}
program.

\section{Allowed tokens in input files}
\label{hadcalc:inputtoken}

The following list shows all token names that may appear in an input
file together with its associated type. The tokens are not case-sensitive.
Thereby \var{parameter} means
that the variable can either be followed by four comma-separated values that denote
the lower and upper bound, whether the increase is linear or logarithmic, and
the number of intermediate intervals, or a single number that is the fixed
value of this parameter. \var{boolean} means that a specific behavior is 
switched on. There is a corresponding separate token that switches the same
behavior off again. \var{double} and  \var{integer} tokens take a 
single double-precision or integer value, respectively, as input. \var{string}
assigns the remainder of the line to the parameter. Finally \var{preselected}
takes special values as an argument. The possible choices for each of these
ones were discussed during the description of the menus given in 
section~\ref{hadcalc:running}. Any settings referring to particle $5$ are relevant
only for $2\rightarrow 3$ processes and will be silently ignored otherwise.
\pagebreak
\begin{longtable}{l@{~:~}lp{35ex}}
token & type & description \\\hline\endhead
\param{MA0} & \var{parameter}
 & mass of the CP-odd Higgs boson \\
\param{TB} & \var{parameter} & ratio of the Higgs vacuum expectation values \\
\param{MUE} & \var{parameter} & $\mu$ parameter in the Higgs sector \\
\param{MSusy} & \var{parameter} & common SUSY mass scale \\
\param{MSQ} & \var{parameter} & mass parameter of the left-handed squarks \\
\param{MSU} & \var{parameter} & mass parameter of the right-handed 
 sup-like squarks \\
\param{MSD} & \var{parameter} & mass parameter of the right-handed
 sdown-like squarks \\
\param{MSL} & \var{parameter} & mass parameter of the left-handed
 sleptons \\
\param{MSE} & \var{parameter} & mass parameter of the right-handed
 selectron-like sleptons \\
\param{A\_t} & \var{parameter} & trilinear coupling of the sup-like squarks \\
\param{A\_b} & \var{parameter} & trilinear coupling of the sdown-like squarks\\
\param{A\_tau} & \var{parameter} & trilinear coupling of the selectron-like sleptons\\
\param{M\_1} & \var{parameter} & U(1)$_Y$ gaugino mass \\
\param{M\_2} & \var{parameter} & SU(2)$_L$ gaugino mass \\
\param{MGl} & \var{parameter} & SU(3)$_C$ gaugino mass \\
\param{SQRTS} & \var{parameter} 
 & square root of the hadronic center of mass energy \\
\param{SQRTSHAT} & \var{parameter}
 & square root of the partonic center of mass energy \\
\param{THETA} \footnotemark
 & \var{parameter} & angle between the two outgoing particles (in degrees) 
  \footnotemark\addtocounter{footnote}{-2}\\
\param{THETACUT} \footnote{only for $2\rightarrow2$ processes}
 & \var{parameter} & cut on the angle between the two outgoing particles 
                     (in degrees)\\
\param{K50} \footnotemark\addtocounter{footnote}{-1}
 & \var{parameter} & energy of the third outgoing particle \\
\param{K50CUT} 
 & \var{parameter} & cut on the energy of the third outgoing particle \\
\param{PTRANS} \footnotemark\addtocounter{footnote}{-1}
 & \var{parameter} & transverse momentum \\
\param{PTRANS3CUT} 
 & \var{parameter} & cut on the transverse momentum of particle 3 \\
\param{PTRANS4CUT} 
 & \var{parameter} & cut on the transverse momentum of particle 4 \\
\param{PTRANS5CUT} 
 & \var{parameter} & cut on the transverse momentum of particle 5 \\
\param{RAPID} \footnote{only for differential cross sections}
 & \var{parameter} & rapidity \\
\param{RAPID3CUT} 
 & \var{parameter} & cut on the rapidity of particle 3 \\
\param{RAPID4CUT} 
 & \var{parameter} & cut on the rapidity of particle 4 \\
\param{RAPID5CUT} 
 & \var{parameter} & cut on the rapidity of particle 5 \\
\param{DELTAR34CUT} 
 & \var{parameter} & cut on the distance between particles 3 and 4 \\
\param{DELTAR35CUT} 
 & \var{parameter} & cut on the distance between particles 3 and 5 \\
\param{DELTAR45CUT}
 & \var{parameter} & cut on the distance between particles 4 and 5 \\
\param{RENSCALE} & \var{parameter} & renormalization scale \\
\param{FACTSCALE} & \var{parameter} & factorization scale \\
\param{CommonSUSYMassScale} & \var{boolean} 
 & choose a common SUSY mass scale \\
\param{NoCommonSUSYMassScale} & \var{boolean} 
 &  do not choose a common SUSY mass scale \\
\param{CommonRenFactScale} & \var{boolean} 
 & choose a common remormalization and factorization scale \\
\param{NoCommonRenFactScale} & \var{boolean} 
 &  do not choose a common remormalization and factorization scale \\
\param{AMPLITUDE} & \var{preselected} 
 &  choose which amplitude(s) to calculate \\
\param{Ptrans3>cut} & \var{boolean} 
 & require the transverse momentum of particle 3 to be larger than the cut \\
\param{Ptrans3<cut} & \var{boolean} 
 & require the transverse momentum of particle 3 to be smaller than the cut \\
\param{Ptrans3nocut} & \var{boolean} 
 & disable cut on the transverse momentum of particle 3 \\
\param{Rapid3>cut} & \var{boolean} 
 & require the rapidity of particle 3 to be larger than the cut \\
\param{Rapid3<cut} & \var{boolean} 
 & require the rapidity of particle 3 to be smaller than the cut \\
\param{Rapid3nocut} & \var{boolean} 
 & disable cut on the rapidity of particle 3 \\
\param{Ptrans4>cut} & \var{boolean} 
 & require the transverse momentum of particle 4 to be larger than the cut \\
\param{Ptrans4<cut} & \var{boolean} 
 & require the transverse momentum of particle 4 to be smaller than the cut \\
\param{Ptrans4nocut} & \var{boolean} 
 & disable cut on the transverse momentum of particle 4 \\
\param{Rapid4>cut} & \var{boolean} 
 & require the rapidity of particle 4 to be larger than the cut \\
\param{Rapid4<cut} & \var{boolean} 
 & require the rapidity of particle 4 to be smaller than the cut \\
\param{Rapid4nocut} & \var{boolean} 
 & disable cut on the rapidity of particle 4 \\
\param{DeltaR34>cut} & \var{boolean} 
 & require the jet separation between particles 3 and 4 to be larger than the cut \\
\param{DeltaR34<cut} & \var{boolean} 
 & require the jet separation between particles 3 and 4 to be smaller than the cut \\
\param{DeltaR34nocut} & \var{boolean} 
 & disable the cut on the jet separation between particles 3 and 4 \\
\param{DeltaR35>cut} & \var{boolean} 
 & require the jet separation between particles 3 and 5 to be larger than the cut \\
\param{DeltaR35<cut} & \var{boolean} 
 & require the jet separation between particles 3 and 5 to be smaller than the cut \\
\param{DeltaR35nocut} & \var{boolean} 
 & disable the cut on the jet separation between particles 3 and 5 \\
\param{DeltaR45>cut} & \var{boolean} 
 & require the jet separation between particles 4 and 5 to be larger than the cut \\
\param{DeltaR45<cut} & \var{boolean} 
 & require the jet separation between particles 4 and 5 to be smaller than the cut \\
\param{DeltaR45nocut} & \var{boolean} 
 & disable the cut on the jet separation between particles 4 and 5 \\
\param{INTMETHOD} & \var{preselected} 
 & choose the integration routine \\
\param{EPSABS} & \var{double} 
 & absolute integration error \\
\param{EPSREL} & \var{double} 
 & relative integration error \\
\param{MAXPTS} & \var{integer} 
 & maximum number of points \\
\param{VSTARTPTS} & \var{integer} 
 & number of points for starting \\
\param{VINCREASE} & \var{integer} 
 & increase in number of points \\
\param{SNNEW} & \var{integer} 
 & number of points for subdivision \\
\param{SFLATNESS} & \var{integer} 
 & flatness number for splits \\
\param{MAXDPASS} & \var{integer} 
 & number of passes in partitioning phase \\
\param{DKEY1} & \var{integer} 
 & Divonne key 1 \\
\param{DKEY2} & \var{integer} 
 & Divonne key 2 \\
\param{DKEY3} & \var{integer} 
 & Divonne key 3 \\
\param{DBORDER} & \var{double} 
 & border of the integration region \\
\param{MAXDCHISQ} & \var{double} 
 & maximum $\chi^2$ for subregion \\
\param{MINDDEV} & \var{double} 
 & minimum deviation for split \\
\param{VERBOSITY} & \var{integer} 
 & verbosity of integration output \\
\param{PDFTYPE} & \var{double} 
 & type of the PDF [PDFlib] \\
\param{PDFGROUP} & \var{double} 
 & group of the PDF [PDFlib] \\
\param{PDFSET} & \var{double} 
 & set of the PDF [PDFlib] \\
\param{PDFPATH} & \var{string} 
 & path where the PDF files are [LHAPDF] \\
\param{PDFNAME} & \var{string} 
 & name of the PDF [LHAPDF] \\
\param{ScreenOutput} & \var{boolean} 
 & print output on the screen \\
\param{OUTPUTFILE} & \var{string} 
 & print output into file \\
\param{OUTPUTSTRING} & \var{string} 
 & parameters to print in output (see section \ref{hadcalc:outputstring}) \\
\end{longtable}

\section{Allowed variable names for \var{outputstring}}
\label{hadcalc:outputstring}

The following list shows all variable names that may appear in
\var{outputstring}. The individual entries are separated from each other
by spaces. Variables with the dimension of a mass are output in GeV.
Note that these names are case-sensitive.

\begin{longtable}{l@{~:~}lp{50ex}}
Name & \multicolumn{2}{l}{Parameter description} \\\hline\endhead
\param{MA0} & $m_{A_0}$ & mass of the CP-odd Higgs boson \\
\param{TB} & $\tan\beta$ & ratio of the Higgs vacuum expectation values \\
\param{MUE} & $\mu$ & $\mu$ parameter in the Higgs sector \\
\param{MSusy} & $m_{SUSY}$ & common SUSY mass scale \\
\param{MSQ} & $m_{\tilde q}$ & mass parameter of the left-handed squarks \\
\param{MSU} & $m_{\tilde u}$ & mass parameter of the right-handed 
 sup-like squarks \\
\param{MSD} & $m_{\tilde d}$ & mass parameter of the right-handed
 sdown-like squarks \\
\param{MSL} & $m_{\tilde l}$ & mass parameter of the left-handed
 sleptons \\
\param{MSE} & $m_{\tilde e}$ & mass parameter of the right-handed
 selectron-like sleptons \\
\param{A\_t} & $A_t$ & trilinear coupling of the sup-like squarks\\
\param{A\_b} & $A_b$ & trilinear coupling of the sdown-like squarks\\
\param{A\_tau} & $A_\tau$ & trilinear coupling of the selectron-like sleptons\\
\param{M\_1} & $M_1$ & U(1)$_Y$ gaugino mass \\
\param{M\_2} & $M_2$ & SU(2)$_L$ gaugino mass \\
\param{MGl} & $m_{\tilde g}$ & gluino mass \\
\param{SQRTS} 
 \footnote{only relevant for the computation of hadronic cross sections}
 & $\sqrt{S}$ & square root of the hadronic center-of-mass energy \\
\param{SQRTSHAT} 
 \footnote{only relevant for the computation of partonic cross sections and 
  differential hadronic cross section with respect to invariant mass}
 & $\sqrt{\hat{s}}$ & square root of the partonic center-of-mass energy \\
\param{THETA} \footnotemark\addtocounter{footnote}{-1} 
 & $\theta$ & angle between the two outgoing particles (in degrees)\\
\param{THETACUT} \footnote{only for $2\rightarrow2$ processes} 
 & $\theta_{cut}$ & cut on the angle between the two outgoing particles 
                   (in degrees)\\
\param{K50} \footnotemark\addtocounter{footnote}{-1} 
 & $k_5^0$ & energy of the third outgoing particle \\
\param{K50CUT} \footnote{only for $2\rightarrow3$ processes} 
 & ${k_5^0}_{cut}$ & cut on the energy of the third outgoing particle \\
\param{PTRANS} 
 & $p_{trans}$ & transverse momentum \\
\param{PTRANS3CUT} 
 & ${p^3_{trans}}_{cut}$ & cut on the transverse momentum of particle 3 \\
\param{PTRANS4CUT} 
 & ${p^4_{trans}}_{cut}$ & cut on the transverse momentum of particle 4 \\
\param{PTRANS5CUT} 
 & ${p^5_{trans}}_{cut}$ & cut on the transverse momentum of particle 5 \\
\param{RAPID} 
 & $\eta$ & rapidity \\
\param{RAPID3CUT} 
 & $\eta^3_{cut}$ & cut on the rapidity of particle 3 \\
\param{RAPID4CUT} 
 & $\eta^4_{cut}$ & cut on the rapidity of particle 4 \\
\param{RAPID5CUT} 
 & $\eta^5_{cut}$ & cut on the rapidity of particle 5 \\
\param{DELTAR34CUT} 
 & ${\Delta R^{34}}_{cut}$ & cut on the distance between particles 3 and 4 \\
\param{DELTAR35CUT} 
 & ${\Delta R^{35}}_{cut}$ & cut on the distance between particles 3 and 5 \\
\param{DELTAR45CUT} 
 & ${\Delta R^{45}}_{cut}$ & cut on the distance between particles 4 and 5 \\
\param{RENSCALE} & $\mu_R$ & renormalization scale \\
\param{FACTSCALE} & $\mu_F$ & factorization scale \\
\param{Mh0} & $m_{h^0}$ & mass of the lighter CP-even Higgs boson \\
\param{MH0} & $m_{H^0}$ & mass of the heavier CP-even Higgs boson \\
\param{MHpm} & $m_{H^\pm}$ & mass of the charged Higgs boson \\
\param{MCha(1)} & $m_{\chi_1}$ & mass of the lighter chargino \\
\param{MCha(2)} & $m_{\chi_2}$ & mass of the heavier chargino \\
\param{MNeu(1)} & $m_{\chi^0_1}$ & mass of the lightest neutralino \\
\param{MNeu(2)} & $m_{\chi^0_2}$ & mass of the second-lightest neutralino \\
\param{MNeu(3)} & $m_{\chi^0_3}$ & mass of the second-heaviest neutralino \\
\param{MNeu(4)} & $m_{\chi^0_4}$ & mass of the heaviest neutralino \\
\param{MGl} & $m_{\tilde g}$ & mass of the gluino \\
\param{MSn(1)} & $m_{\tilde{\nu}_e}$ & mass of the electron sneutrino \\
\param{MSn(2)} & $m_{\tilde{\nu}_\mu}$ & mass of the muon sneutrino \\
\param{MSn(3)} & $m_{\tilde{\nu}_\tau}$ & mass of the tau sneutrino \\
\param{MSl(1)} & $m_{\tilde{e}_1}$ & mass of the lighter selectron \\
\param{MSl(2)} & $m_{\tilde{\mu}_1}$ & mass of the lighter smuon \\
\param{MSl(3)} & $m_{\tilde{\tau}_1}$ & mass of the lighter stau \\
\param{MSL(1)} & $m_{\tilde{e}_2}$ & mass of the heavier selectron \\
\param{MSL(2)} & $m_{\tilde{\mu}_2}$ & mass of the heavier smuon \\
\param{MSL(3)} & $m_{\tilde{\tau}_2}$ & mass of the heavier stau \\
\param{MSu(1)} & $m_{\tilde{u}_1}$ & mass of the lighter sup \\
\param{MSu(2)} & $m_{\tilde{c}_1}$ & mass of the lighter scharm \\
\param{MSu(3)} & $m_{\tilde{t}_1}$ & mass of the lighter stop \\
\param{MSU(1)} & $m_{\tilde{u}_2}$ & mass of the heavier sup \\
\param{MSU(2)} & $m_{\tilde{c}_2}$ & mass of the heavier scharm \\
\param{MSU(3)} & $m_{\tilde{t}_2}$ & mass of the heavier stop \\
\param{MSd(1)} & $m_{\tilde{d}_1}$ & mass of the lighter sdown \\
\param{MSd(2)} & $m_{\tilde{s}_1}$ & mass of the lighter sstrange \\
\param{MSd(3)} & $m_{\tilde{b}_1}$ & mass of the lighter sbottom \\
\param{MSD(1)} & $m_{\tilde{d}_2}$ & mass of the heavier sdown \\
\param{MSD(2)} & $m_{\tilde{s}_2}$ & mass of the heavier sstrange \\
\param{MSD(3)} & $m_{\tilde{b}_2}$ & mass of the heavier sbottom \\
\param{TREE} & $\sigma^0$ & tree-level cross section \\
\param{LOOP} & $\sigma^1$ & one-loop cross section \\
\param{TREEERR} & $\sigma(\sigma_0)$ 
 & integration error of the tree-level cross section \\
\param{LOOPERR} & $\sigma(\sigma_1)$  
 & integration error of the one-loop cross section \\
\param{TREEPROB} & $\chi^2(\sigma(\sigma_0))$ 
 & probability of the integration error of the tree-level cross section \\
\param{LOOPPROB} & $\chi^2(\sigma(\sigma_1))$  
 & probability of the integration error of the one-loop cross section \\
\param{NREGIONS}\footnotemark\addtocounter{footnote}{-1} & 
 & number of regions used for integration \\
\param{NEVAL}\footnotemark\addtocounter{footnote}{-1} & 
 & number of function evaluations used for integration \\
\param{FAIL}\footnote{only relevant for some integration routines} &   
 & a non-zero value indicates that the desired accuracy could not be reached
\end{longtable}

\phantomsection
\bibliographystyle{mrauchplain_mcite}
\addcontentsline{toc}{chapter}{Bibliography}
\bibliography{papers}

\end{appendix}

\cleardoublepage
\chapter*{}
\vspace*{-4.8cm}
{\Huge\bfseries\noindent Acknowledgments}
\addcontentsline{toc}{chapter}{Acknowledgments}
\phantomsection
\label{ack}
\thispagestyle{empty}
\enlargethispage*{10cm}
\vspace{2\baselineskip}

\noindent First of all, I would like to cordially thank my PhD advisor,
Prof.~Wolfgang~Hollik, for 
his continuous support and the possibility to do my PhD thesis in his working group.
Despite his many obligations as one of the directors of the 
Max-Planck-Institut f\"ur Physik, he has always found the time to answer 
my questions, discuss new results, and give many further valuable 
suggestions and advice.

I would also like to thank Prof.~Manuel Drees for initially taking over the 
formal supervision of my PhD.

I am also indebted to Tilman Plehn for his support and for the many 
fruitful discussions about Higgs physics and collider phenomenology. 
The work on our joint paper has always been much fun.
A big thanks as well to Heidi Rzehak 
for answering all my questions about renormalization in 
different schemes, for many valuable discussions and
for suggesting to me to join such a great working group.
Also Pietro Slavich
% , with whom I had the pleasure to share an office for one and a half year, 
has taught me many details about quantum field theory 
and in particular about the universal corrections to the bottom-quark Yukawa coupling.
Thomas Hahn has always had an answer for even the trickiest challenges in
programming, regardless whether the problems occurred in 
Mathematica, Fortran, UNIX-shell or \LaTeX{}. 

Stefan Dittmaier has been of great help during the implementation of the five-point
loop integrals and has also kindly provided his code for a numerical
comparison.
The development of \HadCalc{} was greatly influenced by Oliver Brein, 
who has told me all details about calculating hadronic cross sections and whose 
private computer code for the calculation of hadronic cross sections has triggered 
my work on \HadCalc{}.
I would also like to thank Martin F\"urst and Christian Schappacher for their feedback
on the program, the bug reports and the helpful suggestions to make the program
more useful and user-friendly.
Also Cailin Farrell, Axel Bredenstein and Thomas Fritzsche have invested much time
in answering all my questions about their respective work, from effective field
theories over Monte Carlo generators to the renormalization of the MSSM.
The great atmosphere at the institute and all other colleagues, which I have 
not mentioned so far and which would go beyond the scope of these
acknowledgments, have contributed significantly to the success of this thesis as well.

Performing numerical calculations would not have been possible without 
the excellent maintenance of the computer system by 
Thomas Hahn, Peter Breitenlohner and all other system administrators
of the MPI and the Rechenzentrum Garching der MPG, where parts of the calculations 
were carried out.

Many thanks also to Heidi Rzehak, Ulrich Meier, Michael Pl\"umacher, 
Christoph Rei\ss{}er and Axel Bredenstein for proof-reading several chapters 
of my thesis each and contributing many useful corrections and suggestions, 
and especially to Arne Weber, who has not only carefully read large parts of 
this dissertation and given helpful remarks, but also greatly 
improved its linguistic aspect.

Finally, I would like to thank my parents for their support.
\textcolor{white}{And last, but definitely not least, a cordial thanks
also to Kathrin Hochmuth for her support, her great help while explaining
neutrino- and astrophysics to me, and many joint tea breaks.}

\end{document}